\documentclass[12pt,a4paper]{article}

\usepackage{soul}        

\usepackage{amsmath,amssymb,epsf,epsfig}
\usepackage{array}
\usepackage{fancybox}
\allowdisplaybreaks[1]
\usepackage{slashed}
\usepackage{tocbibind}
\usepackage{lipsum}
\usepackage[usenames]{color}
\usepackage{amsfonts,bm}
\usepackage{graphicx}
\usepackage{enumerate}
\usepackage{geometry}
\usepackage{tikz}
\usetikzlibrary{arrows}
\usepackage{parskip}     
\usepackage[
      colorlinks=true,
      linkcolor=black,
      urlcolor=blue,
      filecolor=blue,
      citecolor=black,
      pdfstartview=FitH,
      pdftitle={},
      pdfauthor={},
      pdfsubject={},
      pdfkeywords={},
      pdfpagemode=UseNone,
      bookmarksopen=true,
      ]{hyperref}
\usepackage[all]{hypcap}     

\numberwithin{equation}{section}

\newcommand{\be}{\begin{equation}}
\newcommand{\ee}{\end{equation}}
\newcommand{\bea}{\begin{eqnarray}\displaystyle}
\newcommand{\eea}{\end{eqnarray}}

\def\beq{\begin{equation}}
\def\eeq{\end{equation}}
\def\beqa{\begin{eqnarray}}
\def\eeqa{\end{eqnarray}}
\def\bet{\begin{tabular}}
\def\eet{\end{tabular}}
\def\bs{\begin{split}}
\def\es{\end{split}}

\def\k{\kappa}

\def\one{{\hbox{\kern+.5mm 1\kern-.8mm l}}}
\def\zero{{\hbox{0\kern-1.5mm 0}}}

\definecolor{orange}{rgb}{1,0.5,0}

\newcommand{\bra}[1]{{\langle {#1} |\,}}
\newcommand{\ket}[1]{{\,| {#1} \rangle}}
\newcommand{\braket}[2]{\ensuremath{\langle #1 | #2 \rangle}}

\newcommand{\bean}{\begin{eqnarray*}}
\newcommand{\eean}{\end{eqnarray*}}

\begin{document}

\begin{titlepage}

\thispagestyle{empty}
 \newgeometry{left=5.57cm, right=1.15cm, top=2.8cm, bottom=3cm}
    \begin{tikzpicture}[remember picture, overlay]
        \node[inner sep=0] at (current page.center)
        {\includegraphics[width=\paperwidth,height=\paperheight]{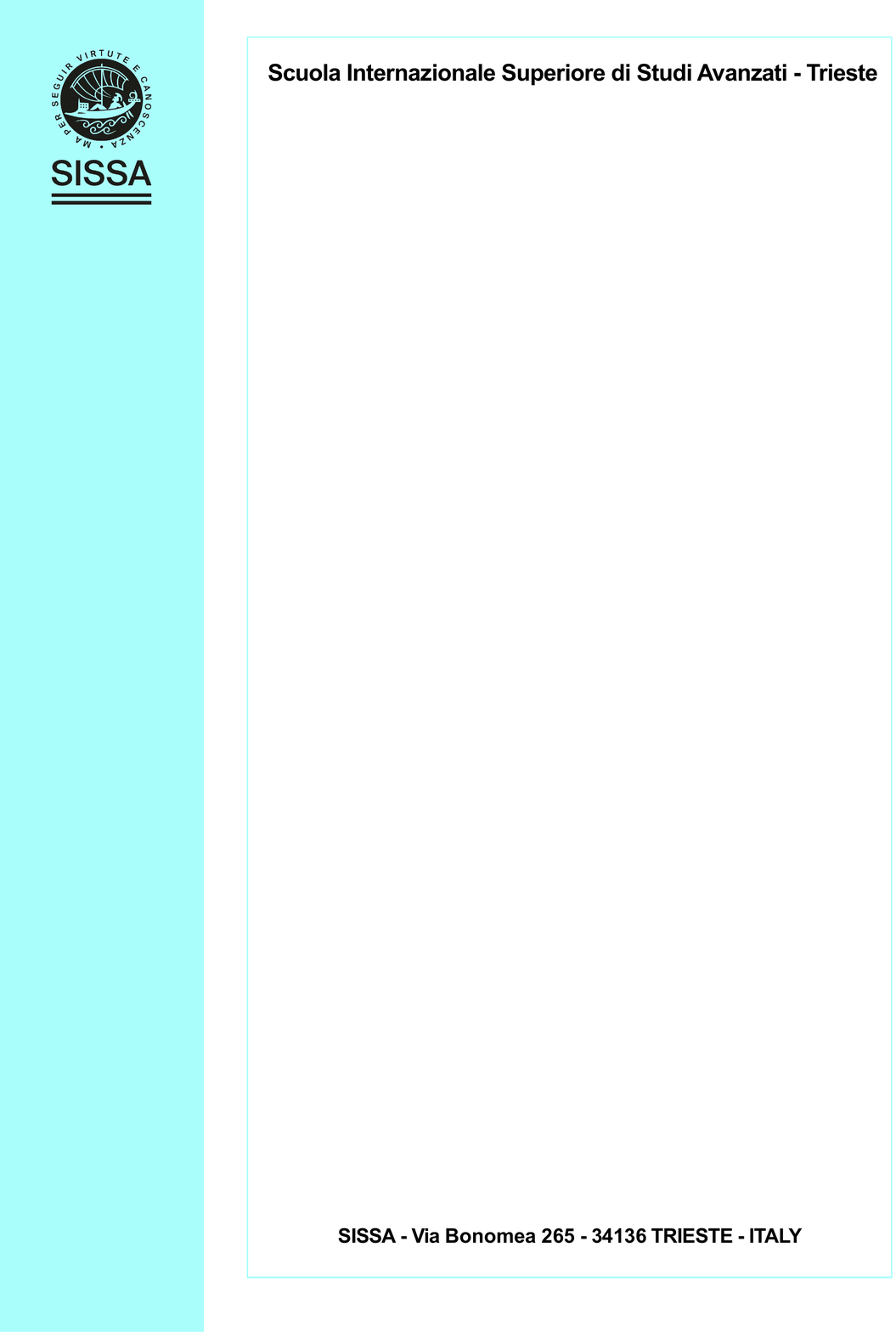}};
    \end{tikzpicture}
    \begin{center}
        {
            \vspace*{1.5cm}
            \LARGE
            \scalebox{0.9}{\textsc{Doctoral Thesis}}\\
        }
        \vspace*{1.9cm}
        {\huge
        
\rule{0.9\textwidth}{1pt}

\vspace*{3mm}

        \scalebox{1.2}{FQHE and $tt^{*}$ Geometry }     
      
        }
        \vspace*{4mm}

        
\rule{0.9\textwidth}{1pt}

        \vspace*{2.7cm}
        {
          \Large
          \begin{tabular}{ll}
              \textsc{Candidate}~: & Riccardo Bergamin\\
              \textsc{Supervisor}~: & Sergio Cecotti 
               
        \end{tabular}
          \\
        }
  
   \vspace*{2.7cm}
        {
          \Large
          \begin{tabular}{ll}
              \textsc{Opponents}~: & Michele Del Zotto \\
               & Cumrun Vafa
               
        \end{tabular}
          \\
        }      
        
        \vfill

        {
         \large 
         \textsc{Academic Year 2018 -- 2019}
        }

    \end{center}
    \restoregeometry
   \end{titlepage}

\begin{center}
{\large{
Abstract} }
\end{center}
\thispagestyle{empty}

Cumrun Vafa proposed in \cite{rif1} a new unifying model for the principal series of FQHE which predicts non-Abelian statistics of the quasi-holes. The many-body Hamiltonian supporting these topological phases of matter is invariant under four supersymmetries. In the thesis we study the geometrical properties of this Landau-Ginzburg theory. The emerging picture is in agreement with the predictions in \cite{rif1}. The $4$-SQM Vafa Hamiltonian is shown to capture the topological order of FQHE and the $tt^{*}$ monodromy representation of the braid group factors through a Temperley-Lieb/Hecke algebra with $q = \pm \exp(\pi i/\nu)$. In particular, the quasi-holes have the same non-Abelian braiding properties of the degenerate field $\phi_{1,2}$ in Virasoro minimal models. Part of the thesis is dedicated to minor results about the geometrical properties of the Vafa model for the case of a single electron. In particular, we study a special class of models which reveal a beautiful connection between the physics of quantum Hall effect and the geometry of modular curves. Despite it is not relevant for phenomenological purposes, this class of theories has remarkable properties which enlarge further the rich mathematical structure of FQHE. 

\vspace{5cm}

The thesis is based on the following papers: \\ \\ 
R. Bergamin and S. Cecotti, `` FQHE and $tt^{*}$ geometry '', 	arXiv:1910.05022 [hep-th], 2019. \\ \\
R. Bergamin, ``$tt^{*}$ Geometry of Modular Curves'', arXiv:1803.00489 [hep-th], \\ published in JHEP 1908 (2019) 007.

\newpage

\section*{Acknowledgements}

Voglio ringraziare Mamma, Pap\'a e Stefano per tutto quel che hanno fatto per me in questi anni di studi. Un supporto non solo morale, ma anche alimentare. Una menzione speciale va infatti alle torte salate e dolci che mi ritrovavo in valigia ogni domenica sera di ritorno a Trieste. Di questo anche i miei coinquilini ringraziano sentitamente. \newline
Ringrazio di cuore tutti i coinquilini con cui ho abitato la Geppa in questi quattro anni intensissimi: Sofy, Bruno, Ane, Ale, Mizu, StronzEle, Milly, Segret, Leo, Fra, Carly e Skodella. Li ringrazio per il loro grande affetto e supporto morale, nonch\'e per tutti i bei momenti trascorsi assieme come una vera famiglia. \newline 
Ringrazio di cuore anche Giovanni, Sara, Luca, Matteo, Giulia, Alessandro, Andrea e Jenny per la loro amicizia incondizionata. 
Perch\'e anche se trascorro molto tempo lontano da casa, al mio ritorno trovo sempre delle persone su cui contare.\newline
Ringrazio anche Fabio e Cristina per la loro grande ospitalit\'a. Quella di Fabio \'e la naturale evoluzione dell' osmiza: se magna, se beve e se ride, ma al posto del vinello scrauso trovi Whisky americano di grande qualit\'a. \newline
Ci tengo molto a ringraziare i miei colleghi ed amici in SISSA per il tempo trascorso assieme e le interessanti discussioni scentifiche. In particolare ringrazio Giulio Ruzza e Matteo Caorsi, che sono sicuramente i ragazzi che ho importunato di pi\'u con le mie domande. \newline 
Voglio ringraziare il prof. Sergio Cecotti per avermi fatto conoscere l'argomento del fractional quantum Hall effect ed in generale delle fasi topologiche della materia, che trovo uno dei pi\'u interessanti argomenti della fisica teorica di oggi. Lo ringrazio in particolare di avermi reso partecipe di nuove ed interessanti scoperte al riguardo. Ringrazio anche tutti i professori che mi hanno dato tempo e disponibilit\'a, oltre al personale della caffetteria e la mensa, sempre allegro, gentile e disponibile. \newline
Ringrazio infine i ragazzi della palestra Audace per avermi fatto conoscere il crossfit e per la loro grande simpatia, disponibilit\'a ed entusiasmo durante gli intensi allenamenti. Come recita il famoso detto: ``Mens sana in corpore sano''.

\newpage

\tableofcontents

\section{Introduction and Overview}

The theory of topological phases of matter is one of the most interesting subject in theoretical physics. The remarkable discovery of these quantum states showed the limit of the Landau-Ginzburg theory of phase transitions. The topological universality classes cannot be classified with the concepts of symmetry and symmetry breaking: we can have either states in the same topological class but with different symmetries, or states in inequivalent topological classes but with the same symmetries. These quantum phases of matter encode a new kind of order, the topological order, which is not associated to any local order parameter, but rather to global non-local observables \cite{wen}. From these considerations one can understand the importance of the discovery of quantum Hall effect. A system of interacting electrons moving on a $2d$ surface in a strong magnetic field at low temperature exhibits very surprising properties like the quantization of the Hall conductivity

\begin{equation*}
\sigma_{xy}=\frac{e^{2}}{2\pi \hslash}\nu.
\end{equation*}

Initially $\nu$ was found to be integer. The quantization of a physical quantity is not new in quantum mechanics, but in this context it acquires a new meaning. The robustness of the conductivity under deformations of the magnetic field within a certain range reveals the topological nature of the Hall states.
Subsequently, it was found that $\nu$ can also assume very specific rational values. The most prominent fractions experimentally are $\nu = 1/3$ and $\nu = 2/5$, but many others have been observed. This quantum number has the interpretation of filling fraction of the lowest Landau levels. It is for a rational $\nu$ that the most interesting phenomena of the quantum Hall effect happen. It turns out that the charged excitations of these systems, the quasi-holes, carry a fraction of the charge of the electron, and, more remarkably, they behave like particles of anyonic statistics. From a theoretical point of view what provides the connection between quantum mechanics and geometry is the Berry's connection. The definition of this mathematical object arises from the Schroedinger equation when we study the adiabatic evolution of quantum systems. The Berry's holonomy induced by taking a quasi-hole around another one contains the informations about the statistics of these quantum objects. In chapter \ref{chapter1} we briefly review the theory of quantum Hall states and the relation between topological order and anyonic particles.\\ Despite the discovery of FQHE \cite{discovery} dates back more than thirty years ago, saying if the quasi-holes are abelian or non-abelian particles still represents a challenging problem for both theorists and experimentalists. From the theoretical point of view, many models have been developed to explain the observed filling fractions. Among these, the Laughlin's proposal \cite{ab1} and the idea of hierarchy states of Haldane and Halperin \cite{ab2,ab3}, as well as Jain's composite fermion theory \cite{ab4}, predict abelian anyonic statistics for the principal series of FQHE. Also the possibility of non-abelian statistics has been explored by several models for other filling fractions \cite{nab1,nab2,nab3,nab4}. More recently, C.Vafa proposed in \cite{rif1} a unifying model of FQHE which leads to new predictions for the statistics of the quasi-holes. He claims that the effective theory of FQH systems with 

\begin{equation}
\nu=\frac{n}{2n\pm 1}
\end{equation}

can be realized in the framework of the AGT correspondence \cite{rif3}. The compactification of the six dimensional $\mathcal{N}=(2,0)$ theory of $A_{1}$ type on a punctured Riemann surface leads to a correspondence between the Nekrason partition functions \cite{rif24} of certain $SU(2)$ gauge theories and the correlators of the Liouville CFT. In this set up the punctured Riemann surface is identified with the target manifold of FQHE, where the punctures correspond to insertions of quasi-holes, and the Liouville conformal blocks are the wave functions of FQHE. In Appendix \ref{classS} we provide a short review about the Gaiotto theory and the relation between topological string amplitudes and Liouville chiral blocks which inspires the Vafa proposal. This construction motivates in addition a microscopic description of FQHE states in terms of a $\mathcal{N}=4$ supersymmetric Hamiltonian. Choosing the plane as Riemann surface, we have a Landau-Ginzburg model with superpotential

\begin{equation}\label{susymod}
\mathcal{W}(z)= \sum_{i=1}^{N}\left( \sum_{a=1}^{n}\log (z_{i}-x_{a})-\sum_{k=1}^{M}\log(z_{i}-\zeta_{k}) \right)  + \frac{1}{\nu}\sum_{i<j}\log (z_{i}-z_{j}),
\end{equation}

where $z_{i},i=1,...,N $ are the electron coordinates and $x_{a},\zeta_{\alpha}$ are respectively the positions of quasi-holes and magnetic fluxes. The term $\log(z-x_{a})$ is the two dimensional Coulombic potential which describes the interaction between an electron and a point-like source at $x_{a}$ with charge $1$, while the term $\sum_{i<j}\log (z_{i}-z_{j})$ keeps track of the Coulomb repulsion between electrons. In order to reproduce the large macroscopically uniform magnetic field we have to consider a uniform distribution of the flux sources $\zeta_{k}$ in $\mathbb{C}$.\\ In this thesis we want to study the geometrical properties of the model \ref{susymod} and verify the predictions of \cite{rif1}. In chapter \ref{wedisc} we analyze the Vafa Hamiltonian in relation to microscopic physics of FQHE. We show that the degeneracy of the lowest Landau level of an electron moving on a generic Riemann surface in a uniform magnetic field $B$ can be mapped to the degeneracy of a supersymmetric system with four supercharges. Considering the complex plane as an example, in the holomorphic gauge a generic state in the lowest Landau level can be written as 

\begin{equation}
\Psi(z)= f(z)e^{-B \vert z \vert^{2}},
\end{equation}

where $f(z)$ is an holomorphic function. In the correspondence the holomorphic part of the wave function defines canonically an element of the chiral ring $\mathcal{R}=\mathbb{C}[z]/\partial W$ of a $\mathcal{N}=4$ LG model with superpotential $W$. This space has the structure of a Frobenious algebra and its elements label the vacua of the supersymmetric system \cite{rif14}. Moreover, we argue that any Hamiltonian describing the motion in a plane of many electrons coupled to a strong magnetic field is described (at the level of topological order) by Vafa's $\mathcal{N}=4$ Hamiltonian independently of the details of the interactions between the electrons. This result implies that the LG model with superpotential \ref{susymod} represents the correct universality class of the fundamental many-electron theory.\\
Varying the parameters in $W$ we get a Berry's connection $D$ on the bundle of vacua which satisfies a set of equations called $tt^{*}$ geometry \cite{rif10}. One can define in terms of $D$ the $tt^{*}$ Lax connection 

\begin{equation}
\nabla_{\zeta}= D+\frac{1}{\zeta}C, \hspace{1cm} \overline{\nabla}_{\zeta}= \overline{D}+ \zeta\overline{C},
\end{equation}

where $C,\overline{C}$ denotes the action of chiral and antichiral operators on the vacua and $\zeta \in \mathbb{C}^{\times}$ is an arbitrary parameter. The $tt^{*}$ equations can be rephrased as flatness conditions for $\nabla_{\zeta},\overline{\nabla}_{\zeta}$. This connection admits flat sections $\Psi_{i}$ which satisfy \cite{rif23,rif11}

\begin{equation}
\nabla_{\zeta}\Psi_{i}= \overline{\nabla}_{\zeta}\Psi_{i}=0.
\end{equation}

The advantage of having a $4$-susy Hamiltonian in the same universality class of FQHE is that we can use the tools of $tt^{*}$ geometry to study the statistics of the quasi-holes. According to the Vafa's program, the topological order of FQHE is captured by the parallel transport of the flat connection $\nabla_{\zeta}$. We can project the coupling constant space of the model on the configuration space of $N$ identical particles on the plane

\begin{equation}
Y_{n}=C_{n}/S_{n},
\end{equation}

where 

\begin{equation}
C_{n}=\left\lbrace (x_{1},...,x_{n}) \in \mathbb{C}^{n} \vert x_{i} \neq x_{j} \ \mathrm{for} \ i \neq j \right\rbrace
\end{equation}

is the space of $n$ ordered distinct points and $S_{n}$ is the permutation group of $n$ objects. Hence, the $tt^{*}$ Lax connection restricted to the vacuum bundle 

\begin{equation}
\mathcal{V}\rightarrow Y_{n}
\end{equation}

provides a monodromy representation of the braid group of the quasi-holes $B_{n}=\pi_{1}(Y_{n})$. In the chapters \ref{reviewtt},\ref{chap5},\ref{chap6} we discuss all the necessary tools to compute the $tt^{*}$ monodromy representation of the Vafa model. The chapter \ref{reviewtt} contains a review of supersymmetric quantum mechanics and basics of $tt^{*}$ geometry that the reader can find in literature. The chapters \ref{chap5},\ref{chap6} cover more advanced topics in $tt^{*}$ geometry which are relevant for our problem. In chapter \ref{chap5} we discuss the concept of statistics in $tt^{*}$ geometry. In particular, since the electrons are fermionic particles, we study in full generality the fermionic sector of a LG model of $N$ identical particles. We also discuss the formulation of $tt^{*}$ geometry in models with a non simply connected target manifold and multivalued superpotential. The Vafa superpotential \ref{susymod} belongs to this family of theories. In order to treat properly these models one has to introduce the concept of covering spaces and extend the vacuum space with the so called ``$\theta$-sectors''. The common denominator of the chapter is the $tt^{*}$ functoriality, which turns out to be a very powerful tool to generate isomorphisms between quantum theories. In chapter \ref{chap6} we study a special class of $tt^{*}$ geometries which we call \\`` very complete''. These models are quite peculiar. In the language of $2d$ $\mathcal{N}=(2,2)$ theories, being very complete means that all the operators in the chiral ring of the UV conformal fixed point are IR relevant or marginally non-dangereous and there are no wall-crossing phenomena. We show that in this case the UV limit of the Berry connection is an $SL(2,\mathbb{C})$ Knizhnik-Zamolodchikov connection \cite{rif65,rif67} on $C_{n}$ with global coordinates given by the critical values of the superpotential $w_{i}$. The UV limit consists in rescaling the critical coordinates $w_{i}\rightarrow \beta w_{i}$ and taking $\beta\rightarrow 0$. In this regime the Lax connection and the Berry connection coincide and, if the model is symmetric under permutations of the $w_{i}$, the UV Berry connection provides a unitary representation of the full braid group $B_{n}$.\\ In chapter \ref{chap7} we study the $tt^{*}$ geometry of the Vafa Hamiltonian. We show that the model is very complete and symmetric. The corresponding UV Berry connection takes the form of a Kohno connection

\begin{equation}
\mathcal{D}=d+ \lambda \sum_{i<j}s_{\ell}^{i}s_{\ell}^{j} \frac{d(w_{i}-w_{j})}{w_{i}-w_{j}},
\end{equation}

acting on the space $V^{n+M}=\bigotimes_{i} V_{i}$ with $V_{i}\simeq \mathbb{C}^{2},i=1,...,n+M$ and $s_{\ell}^{i}, \ell=1,2,3$ is the $su(2)$ generator acting on the $V_{i}\simeq \mathbb{C}^{2}$ factor, namely 

\begin{equation}
s_{\ell}^{i}= 1 \otimes....\otimes 1 \otimes \frac{1}{2}\sigma_{\ell} \otimes 1 \otimes ....\otimes 1.
\end{equation}

It is shown in \cite{rif65} that the monodromy representation of the flat connection above is a Hecke algebra representation of the braid group $B_{n+M}$ which factorizes through the Temperley-Lieb algebra $A_{n+M}(q)$ with 

\begin{equation}
q= \exp (\pi i \lambda).
\end{equation}

One can restrict to the monodromy representation of $B_{n}$ with the projection $p:Y_{n+M}\rightarrow Y_{n}$. 
It turns out that the parameter $\lambda$ is related to the filling fraction $\nu$ by 

\begin{equation}
q^{2}=e^{2\pi i /\nu},
\end{equation}

which gives the two possibilities $q=\pm e^{i\pi/\nu}$. It is believed that $3d$ topological Chern-Simons theories are effective description of FQHE states. Since the braiding of conformal blocks in $2d$ WZW models is the same of the Wilson lines in non-Abelian Chern-Simons theories, it is natural to require the UV Berry connection to be a Knizhnik-Zamolodchikov connection for $SU(2)$ current algebra with level $k$ quantized in integral units, namely 

\begin{equation}
\lambda=\pm \frac{2}{k+2}, \hspace{0.5cm} k \in \mathbb{Z}.
\end{equation}

This condition leads to the determine the values of the filling fraction which are consistent with $tt^{*}$ geometry. In the case of $q=e^{i\pi/\nu}$ we get

\begin{equation}
\nu= \frac{b}{2 b \pm 1}, \  b \in \mathbb{N}, \hspace{1cm} \nu= \frac{b}{2(b \pm 1)}, \ b \ \mathrm{odd},
\end{equation}

where the first one corresponds to the principal series of FQHE. From point of view of \cite{rif1}, the element $\sigma_{i}^{2}$ of the pure braid group for the principal series has two distinct eigenvalues, in correspondence with the two different fusion channels of the $\phi_{1,2}$ operator in the minimal $(2n, 2n\pm1)$ Virasoro model. The ratio of the two eigenvalues is

\begin{equation}
q^{2}=\frac{\exp [ 2\pi i (h_{1,3}-2h_{1,2} ) ] }{\exp [ 2\pi i (h_{1,1}-2h_{1,2})] }=\exp(2\pi i/\nu).
\end{equation}

As predicted in \cite{rif1}, the quasi-holes have the same non-Abelian braiding properties of the degenerate fields $\phi_{1,2}$ in Virasoro minimal models. On the other hand, the less natural solution $q=-e^{i\pi/\nu}$ gives other two series of filling fractions.
These are respectively

\begin{equation}
\nu= \frac{m}{m+2}, \hspace{0.5cm} m=k+2 \in \mathbb{N}\geq 2, \hspace{1cm} \nu= \frac{m}{3m-2}, \hspace{0.5cm} 
m=k+2\geq 2,
\end{equation}

where the first series contains the values of $\nu$ corresponding to the Moore-Read \cite{nab2} and Read-Rezayi models \cite{nab3}.\\ The reader can find in appendix other minor results about the geometrical properties of the Vafa model for the case of a single electron. In particular, in \ref{papermodjhep} we study a special class of models which reveal a beautiful connection between the physics of quantum Hall effect and the geometry of modular curves. The analysis is based on \cite{rif54}. Despite it is not relevant for phenomenological purposes, this class of theories has remarkable properties which enlarge further the rich mathematical structure of FQHE. Among the main results, the theorems about the cusps counting and classification are recovered in a physical language. From our investigation of this family of models, the algebraic properties of the modular curves emerge in an elegant manner.

\section{Topological Order, FQHE and Anyons}\label{chapter1}

\subsection{ A New Kind of Order}

The fractional quantum Hall effect opened a new chapter in condensed matter physics. For a long time, before the discovery of the quantum Hall states, Landau's symmetry breaking theory defined the fundamental paradigm of many-body physics. We know that at sufficiently high temperature matter is in form of gas. In this regime the particles are weakly interacting and the motion of a single constituent is not influenced by the other ones. When temperature decreases the particles become more and more correlated and start to develop a regular pattern. In this regime we observe an emergent collective behaviour revealing an internal structure, which we also call order. The concept of order allows to classify different states of matter and is intimately related to the concept of phase transition. One says that two many-body states has the same order if we can change one states into the other with a smooth deformation of the Hamiltonian. If this is not possible we encounter a phase transition. The definition of order organizes the states of matter in equivalence classes which are said universality classes in the Wilsonian language. According to Landau's theory there is a deep relation between order and symmetry. Different orders are associated with different symmetries, which are described by local parameters, and a phase transition involves the breaking of some symmetry. This picture is able to explain a large class of experimentally observed states of matter and predict the existence of gaplessness excitations in the materials due to symmetry breaking. The existence of the quantum Hall states reveals the limits of Landau's paradigm and shows the existence of a new type of order, usually called topological order \cite{wen}. These states of matter are created by confining electrons on a $2d$ interface between two different semiconductors under strong magnetic fields and low temperature. In this regime the electrons are strongly correlated and behave collectively like a quantum liquid. The topological phases are not classified by symmetries and cannot be described by local order parameters. Indeed, all the quantum Hall states have the same symmetries and a phase transition between them does not involve any symmetry breaking. One can define a topological universality class as a family of Hamiltonians with gapped spectrum which can be smoothly deformed to each other without the emergence of gappless excitations. The topological order can be characterized by global, non-local observables, which are robust under any local perturbations that can break symmetries. These are essentially the degeneracy of the ground state, which depends only on the topology of the target space, and the non-local behaviour of the quasi-hole and quasi-particle excitations.   

\subsection{The Quantum Hall Effect: General Setting and Phenomenology}

We recall some basic facts about the phenomenology of quantum Hall effect. More details can be found in \cite{wen,tong} and references there. The Hall effect was originally discovered by Edwin Hall in 1879. The physical set-up is very simple to construct: one has to take a bunch of electrons, restrict them to move on a two dimensional plane and turn on an electric field $\vec{E}$ and magnetic field $\vec{B}$ respectively parallel and orthogonal to the plane. The classical Hall effect is simply the consequence of the motion of charged particles in a magnetic field. In a static regime the electric force acting on the charge carriers is balanced by the Lorentz force of the magnetic field. In natural units we have the condition

\begin{equation}
q_{e}\vec{E}= q_{e} \vec{v} \times \vec{B},
\end{equation}

where $q_{e}=-e$ is the charge of the electron and $\vec{v}$ is the velocity of the particles. Denoting with $n$ the density of the electrons, a current $\vec{j}=n\vec{v}$ is made to flow in the normal direction to the electric field. Moreover, the norms $E,j$ are related by 

\begin{equation}\label{ohm}
E=j \frac{B}{n q_{e} c}=j \frac{h}{q_{e}^{2}} \frac{1}{\nu}
\end{equation}

where $h,c$ are respectively the Planck constant and the speed of light, while

\begin{equation}
\nu= \frac{nhc}{q_{e}B}= \frac{n}{B/\Phi_{0}}=\frac{\mathrm{number \ of \ particles}}{\mathrm{number \ of \ flux \ quanta }}
\end{equation}

is the filling fraction. In the above equality we introduced $\Phi_{0}=\frac{hc}{q_{e}}$ as the quantum unit of magnetic flux. Recalling the Ohm's law 

\begin{equation}
\vec{E}= \rho \vec{j}
\end{equation}

where 

\begin{equation}
\rho= \begin{pmatrix}
\rho_{xx} & \rho_{xy} \\ 
-\rho_{xy} & \rho_{yy}
\end{pmatrix}
\end{equation}

is the resistivity tensor, from \ref{ohm} we obtain the relations

\begin{equation}
\rho_{xx}=0, \hspace{1cm} \rho_{xy}= \frac{h}{q_{e}^{2}}\frac{1}{\nu}.
\end{equation}

\begin{figure}[htbp] 
\centering 
\includegraphics[scale=0.5]{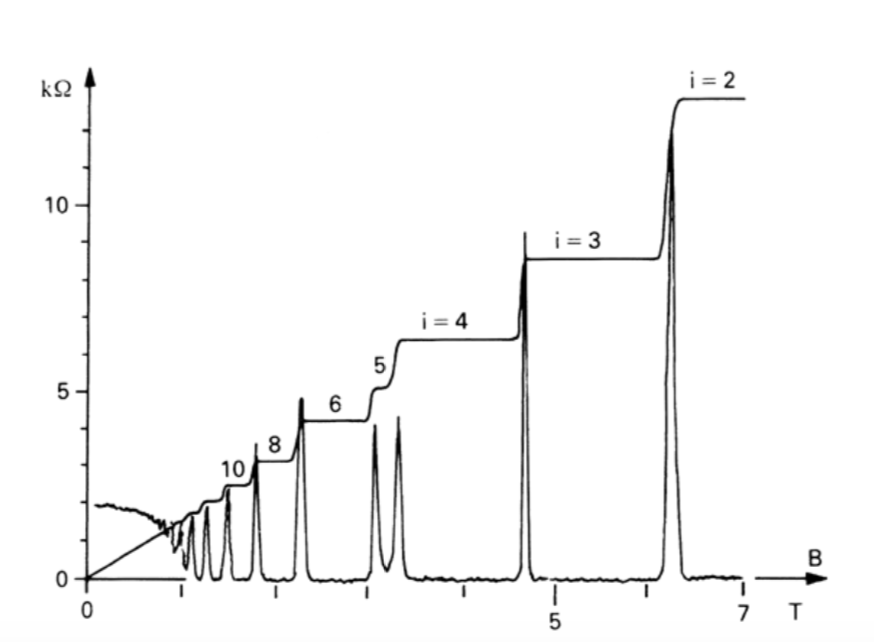}
\caption{Plateau structure} 
\label{plateau} 
\end{figure}

According to the classical theory, the Hall resistance $\rho_{xy}$ is proportional to the magnetic field at fixed electron density. Indeed, experimentally one finds that $\rho_{xy}\propto B$ at weak fields. However, in the early 1980s it was found that for strong magnetic fields ($\sim 10$ T) and very low temperatures ($\sim 1$ K) the Hall resistance develops a plateau structure as in Figure \ref{plateau}. In particular, the filling fraction $\nu$ is quantized in this regime and labels the different plateaus. These new states of matter behave like an incompressible fluid of uniform density. Initially $\nu$ was found to be an integer. This phenomenon is called integer quantum Hall effect (IQHE) and was discovered by von Klitzing in 1980 \cite{vonk}. In the experiment the electrons are confined on an interface between two different semiconductors which play the role of $2d$ plane. The quantization of a physical quantity is rather common at the microscopic level, but the case of the filling fraction is different from the usual quantization phenomena in quantum mechanics. The profile of the resistivity shows that the quantum Hall states are robust under deformations of the magnetic field within a certain range. Moreover, the plateau spectrum turns out to be indipendent from the local, microscopic details of the material. These features reveal the topological nature of these phases of matter. 

Subsequently, it was found that $\nu$ can also take very specific rational values. This phenomenon is called fractional quantum Hall effect (FQHE) and was discovered by Tsui and Stormer in 1982 \cite{discovery}. The most prominent fractions experimentally are $\nu = 1/3$ and $ \nu=2/5 $, but many other fractions have been seen. The majority of them has odd denominator and can be recasted in the so called principal series $\nu=\frac{n}{2n\pm 1}$. It is in these states that the most remarkable things happen. The charged excitations of the Hall fluid, the quasi-holes, carry a fraction of the charge of the electron, as if the electron split itself into several pieces. It is not just the charge of the electron that fractionalises: this happens to the statistics of the electron as well. It is known that quantum particles in two spatial dimensions can have statistics which does not correspond to the classical bosonic or fermionic one. After the braiding of two identical particles the wave function of the many-body system picks up a phase $e^{i\theta}$, where the parameter $\theta$ determines the statistics. The statistics of the particles in two dimensions are

\begin{equation}
\begin{cases} 
\theta=0 \hspace{1.8 cm} \mathrm{boson} \\ 
0< \theta < \pi \hspace{1cm} \mathrm{anyon}\\ 
\theta= \pi \hspace{1.8cm}   \mathrm{fermion}\end{cases}
\end{equation}

In more complicated examples even this description breaks down: the resulting objects are called non-Abelian anyons and provide physical realization of non-local entanglement in quantum mechanical systems. In this case the Hilbert space of the system is degenerate and a braiding operation induces a unitary transformation on a generic state. While the fractional charge of quasi-holes has been measured experimentally, a direct detection of their statistics is more challenging. It is confirmed that the quasi-holes of the fractional Hall states do not posses classical statistics, but if they are Abelian or non-Abelian anyons is still an open question.

\subsection{Landau Levels}\label{eleclandau}

The quantum Hall effect is based at the microscopic level on the dynamics of charged particles in a magnetic field. The Hamiltonian of a quantum particle in a constant, uniform magnetic field has discrete spectrum. The structure of the energy levels, known as Landau levels, allows to describe the plateaus at integer values of the filling fraction \cite{wen}. In particular we are interested in the lowest Landau level, namely the ground state. We ignore for the moment the Coulomb interaction and assume the electrons to be spin polarized. An electron of mass $m$ in a uniform magnetic field is described by the Hamiltonian

\begin{equation}
H=-\frac{1}{2m} (\partial_{i}-iq_{e}A_{i})^{2}, \hspace{1cm} \hslash=c=1,
\end{equation}

where we choose the symmetric gauge

\begin{equation}
(A_{x},A_{y})= \frac{B}{2} (-y,x), \hspace{1cm} B= \partial_{x}A_{y}-\partial_{y}A_{x}.
\end{equation}

To find the energy levels is convenient to introduce the complex coordinate $z=x+iy$ and the holomorphic and anti-holomorphic derivatives

\begin{equation}
\partial_{z}= \frac{1}{2}(\partial_{x}-i\partial_{y}), \hspace{1cm} \partial_{\bar{z}}= \frac{1}{2}(\partial_{x}+i\partial_{y}).
\end{equation}

One can rewrite the Hamiltonian in complex coordinates as 

\begin{equation}
H=-\frac{1}{m}\left( D_{z}D_{\bar{z}}+D_{\bar{z}}D_{z}\right) 
\end{equation}

where $D_{z},D_{\bar{z}}$ are the covariant derivatives 

\begin{equation}
\begin{split}
& D_{z}= \partial_{z}-iq_{e}A_{z}, \hspace{1cm} A_{z}= \frac{1}{2}(A_{x}-iA_{y})=\frac{B}{4i}\bar{z}, \\ 
& D_{\bar{z}}= \partial_{\bar{z}}-iq_{e}A_{\bar{z}}, \hspace{1cm} A_{\bar{z}}= \frac{1}{2}(A_{x}+iA_{y})=-\frac{B}{4i}z.
\end{split}
\end{equation}

One can exploit the commutation relation $\left[D_{z},D_{\bar{z}} \right]=\frac{1}{2}m\omega_{B}$, where $\omega{B}=q_{e}B/m$ is the cyclotron frequency, to simplify the expression of $H$ as 

\begin{equation}
H=-\frac{2}{m} D_{z}D_{\bar{z}} + \frac{1}{2}\omega_{B}.
\end{equation}

The constant $E_{0}=\frac{1}{2}\omega_{B}$ is the energy of the first Landau level. A wave function $\Psi(z,\bar{z})$ in this subspace satisfies the condition 

\begin{equation}
D_{\bar{z}}\Psi(z,\bar{z})=0.
\end{equation}

Assuming $q_{e}B>0$ we introduce the magnetic length $l_{B}$ defined by

\begin{equation}
l_{B}^{2}= \frac{1}{q_{e}B} 
\end{equation}

which represents the characteristic length scale governing the magnetic phenomena in the quantum regime.
The Hamiltonian above is conjugated by the relation 

\begin{equation}
H=e^{-\vert z \vert^{2}/4l_{B}^{2}} \tilde{H} e^{+\vert z \vert^{2}/4l_{B}^{2}} 
\end{equation}

to the operator 

\begin{equation}
\tilde{H}=-\frac{2}{m} \tilde{D}_{\bar{z}}\tilde{D}_{z} + E_{0}
\end{equation}

where 

\begin{equation}
\begin{split}
& \tilde{D}_{z}= e^{+\vert z \vert^{2}/4l_{B}^{2}} D_{z} e^{-\vert z \vert^{2}/4l_{B}^{2}} = \partial_{z}-\frac{1}{2l_{B}^{2}}\bar{z}, \\ & \tilde{D}_{\bar{z}}= e^{+\vert z \vert^{2}/4l_{B}^{2}} D_{\bar{z}} e^{-\vert z \vert^{2}/4l_{B}^{2}} = \partial_{\bar{z}}.
\end{split}
\end{equation}

We note that the non unitary transformation $e^{+\vert z \vert^{2}/4l_{B}^{2}}$ makes the anti-holomorphic part of the gauge connection vanishing. Rewriting a generic state in the Hilber space as 

\begin{equation}
\Psi(z,\bar{z})= \tilde{\Psi}(z,\bar{z}) e^{-\vert z \vert^{2}/4l_{B}^{2}},
\end{equation}

where $\tilde{\Psi}(z,\bar{z})$ is a smooth function, the condition to be in the lowest Landau level becomes 

\begin{equation}
\tilde{D}_{\bar{z}}\tilde{\Psi}(z,\bar{z})= \partial_{\bar{z}}\tilde{\Psi}(z,\bar{z})=0,
\end{equation}

namely $\tilde{\Psi}(z,\bar{z})$ must be an holomorphic function on the complex plane. A natural basis of holomorphic functions is given by the monomials $z^{k}$ with $k\geqslant 0$. Hence, a basis of vacua is 

\begin{equation}
\Psi_{0;k}= z^{k}e^{-\vert z \vert^{2}/4l_{B}^{2}}.
\end{equation}

These wave functions have a circular shape and $k$-th state is peaked at the radious $r_{k}=\sqrt{2k}l_{B}$.
Noting that the ring of the $k$-th vacuum encloses $m$ flux quanta $\Phi_{0}=2\pi l_{B}^{2}$, we learn that there is one state for every flux quantum. Hence, the number of states in the lowest Landau level is equal to the number of flux quanta $\Phi/\Phi_{0}$, where $\Phi$ is the magnetic flux of $B$. One can obtain the wave functions corresponding to the $n$-th excited Landau levels as follows

\begin{equation}
\Psi_{n;k}= D_{z}^{n} \Psi_{0;k},
\end{equation}

where the energy of the $n$-th state is $E_{n}=(\frac{1}{2}+n)\omega_{B}$. It is clear that all the Landau levels have the same degeneracy.\\ In the case of $N$ free electrons in a uniform magnetic field one has to fill the Landau levels with Fermi statistics. It is not difficult to guess that for a quantum Hall state at an integer filling fraction $\nu=n \in \mathbb{N}$ the first $n$ Landau levels are completely filled. The finite gap $\Delta=\omega_{B}$ for the excitations explains why at low temperature we observe these plateau, while the Pauli exclusion principle explains the incompressibility of the Hall fluid. The wave function for the $\nu=1$ state is 

\begin{equation}
\Psi(z_{1},...,z_{N})= \begin{vmatrix}
1 & 1 & ... \\  z_{1} & z_{2} &  ... \\ z_{1}^{2} & z_{2}^{2} & ...\\ . & . & . 
\end{vmatrix} e^{-\sum_{i}\vert z_{i} \vert^{2}/4l_{B}^{2}}= \prod_{i<j}(z_{i}-z_{j})e^{-\sum_{i}\vert z_{i} \vert^{2}/4l_{B}^{2}}
\end{equation}

where $N$ must be equal to the number of flux quanta. The above wave function describes a circular droplet with radious $R=2\pi N l_{B}^{2}$ and uniform density. Indeed, each electron occupies an area $\pi r_{k+1}^{2}-\pi r_{k}^{2}=2\pi l_{b}^{2}$ and so the density has the constant value $\rho=1/2\pi l_{B}^{2}$. For magnetic fields $B \sim 1 \ T$ we have $\ell_{B}\sim 10^{-8} m$. In quantum Hall effect experiments the length scale of the sample is of order $R\sim cm $ and so the typical number of electrons is $N\sim 10^{12}$.

\subsection{The Laughlin Wave Function}\label{laughlin}

In order to describe the fractional states one can try to apply the logic that we used in the integer case. Let us consider the states with $ 0<\nu<1$. For such values of the filling fraction we naturally conclude that the lowest Landau level is partially filled by the electrons, whose number must be $N=\nu \mathcal{N}$, where $\mathcal{N}=\Phi/\Phi_{0}$ is the capacity of the lowest Landau level. It is immediate to see that the free electrons description cannot work in the fractional case. Indeed, if we fill a fraction of the lowest Landau level with Fermi statistics, in the thermodynamical limit we get a hugely degenerate ground state. The number of states is $\begin{pmatrix} \mathcal{N} \\ \nu \mathcal{N} \end{pmatrix}$, which is approximately $ \left( \frac{1}{\nu}\right) ^{\nu \mathcal{N}} \left( \frac{1}{1-\nu}\right) ^{(1-\nu) \mathcal{N}}\sim 10^{10^{12}} $ for $\mathcal{N}\sim 10^{12}$. Hence, it is clear that the fractional quantum Hall states are strongly correlated system in which the Coulomb interaction plays a fundamental role. This partially removes the degeneracy of the lowest Landau level in such a way that we can observe plateaus at fractional valules of $\nu$. Solving the Schoredinger equation and find the exact wave function for these systems is hopeless. The first approach to describe these plateaus is due to Laughlin \cite{ab1}. He proposed a wave function for the $\nu=\frac{1}{m}$ states, with $m$ an odd integer. His guess is based on some physical considerations. First of all, any wave function in the lowest Landau level must have the form 

\begin{equation}
\Psi(z_{1},...,z_{N})= f(z_{1},...,z_{N})e^{-\sum_{i}\vert z_{i} \vert^{2}/4l_{B}^{2}},
\end{equation}

where $f(z_{1},...,z_{N})$ must be holomorphic and anti-symmetric under the exchange of any two particle positions because of the Fermi statistics. Moreover, this wave function should describe a fluid of uniform density on a disk of radious $R=\sqrt{2mN}l_{B}$, compatibly with the fact that the number of states in the full Landau level is $\mathcal{N}=mN$. The Laughlin proposal is 

\begin{equation}\label{Laugh}   
\Psi(z_{1},...,z_{N})= \prod_{i<j}(z_{i}-z_{j})^{m}e^{-\sum_{i}\vert z_{i} \vert^{2}/4l_{B}^{2}}.
\end{equation}

We see that the highest power for each electron coordinate appearing in $f(z_{1},...,z_{N})$ is $m(N-1)$. This means that the maximum radious for each particle is $R\sim \sqrt{2mN}l_{B}$, where we can replace $N-1$ with $N$ in the thermodynamical limit. Correspondingly, the area of the droplet is $A \sim 2\pi m N l_{B}^{2}$ and the density of the Hall fluid is $\rho=\frac{1}{2\pi m l_{B}^{2}}$. The Laughlin wave function captures the relevant physics of the fractional quantum Hall states at the filling fractions $\nu=1/m$, but is not the exact ground state of the many-body Hamiltonian. However, it is possible to construct an Hamiltonian whose ground state is precisely given by the Laughlin state \cite{tong}. According to the Laughlin proposal, this Hamiltonian should be in the same universality class of the FQHE Hamiltonian for $N=\nu\mathcal{N}$ electrons and support the same topological phase as ground state. This motivates why one can use the above wave function to describe the $\nu=1/m$ states rather than the exact one.\\ Following the Laughlin proposal, other models have been developed to explain and describe the other filling fractions, such as the construction of Hierarchy states proposed by Haldane and Halperin \cite{ab2,ab3}. We will explain this idea in section \ref{effective}.

\subsection{Berry's Connection}

Topological phases encode the non trivial low energy dynamics of certain gapped systems below the energy scale of the minimal excitation. By definition they cannot have local propagating degrees of freedom and what distinguish them from the trivial phase are their geometrical properties. The mathematical object which puts in relation quantum mechanics with geomery and topology is the Berry's connection \cite{berry}. Since it plays a central role in the study of FQHE we briefly recall its definition (see \cite{tong,wilczek} for a more detailed review). Let us consider an Hamiltonian 

\begin{equation}
H(x_{i};\lambda_{j})
\end{equation}

where $x_{i}$ are the degrees of freedom of the system, as for instance the positions and spins of the particles, and $\lambda_{j}$ is a set of parameters determined by some external apparatus that an observer can vary. These parameters can be thought as coordinates on a space of couplings $\mathcal{M}$ which parametrizes a family of theories. We assume that, apart from some isolated critical points, for every choice of $\lambda_{j}$ the spectrum of the Hamiltonian is gapped and the ground state has degeneracy $N\geqslant 1$. One defines the vacuum bundle $\mathcal{V} \rightarrow \mathcal{M}$ where the fiber $\mathcal{V}_{\lambda}$ at a point $\lambda_{j}$ is the vacuum space of $H(x_{i};\lambda_{j})$. Let us a set a local frame

\begin{equation}
\ket{a;\lambda}, \hspace{1cm} a=1,...,N.
\end{equation}

The definition of the Berry's connection arises when we study the adiabatic evolution of the system. The adiabatic theorem states that, if we prepare the quantum system in a certain vacuum $\ket{a;\lambda}$ for certain $\lambda_{j}$ and vary slowly the parameters compared to the energy gap of the excitations, at every point $\lambda_{j}^{\prime}$ of the path in the space of couplings the perturbed state $\ket{\psi_{a}(\lambda^{\prime})}$ is in the ground state $\mathcal{V}_{\lambda^{\prime}}$. More precisely, the adiabatic ansatz for the evolved state is

\begin{equation}\label{ansatz}
\ket{\psi_{a}(\lambda^{\prime})}= U_{ab}(\lambda^{\prime})\ket{b;\lambda^{\prime}}
\end{equation}

where $U\in U(N)$ since time evolution in quantum mechanics is unitary. We see from this formula that a vacuum state is parallely transported along a certain in path in the space of couplings. Plugging the adiabatic ansatz in the Schoredinger equation provides the definition of a unitary connection on the vacuum bundle which generates the time evolution. This is the Berry's connection 

\begin{equation}
(A_{i})_{ba}=-i \bra{a;\lambda} \partial_{\lambda_{i}}\ket{b;\lambda}.
\end{equation}

The unitary matrix introduced in \ref{ansatz} contains also the dynamical phase $e^{-i\int d\lambda E_{o}(\lambda)}$ which we always have in the time evolution of a quantum state. One can ignore this contribution or simply set $E_{o}(\lambda)=0$ by adding a constant to the Hamiltonian. It is interesting in general to evaluate the transformation of a state after a closed loop in the space space of couplings. The associated unitary matrix is the Berry's holonomy 

\begin{equation}\label{holonomy}
U=\mathcal{P} \exp \left(    -i \oint A_{i}d\lambda^{i}    \right),
\end{equation}

where $\mathcal{P}$ stands for path ordering. In the case of degenerate ground state the Berry's connection is non-Abelian and the associated holonomy group is a subgroup of $U(N)$ with $N>1$. If the system has a unique vacuum state $\ket{\lambda}$, the Berry's connection is Abelian and the holonomy \ref{holonomy} reduces to the Berry's phase 

\begin{equation}
e^{i\gamma}= \exp \left(  -i\oint A_{i}(\lambda)d\lambda^{i}  \right),
\end{equation}

with $A_{i}(\lambda)= -i \bra{\lambda} \partial_{\lambda_{i}}\ket{\lambda}$.

\subsection{Anyonic Particles}

Topological phases are beleaved to support anyonic particles. These have not to be thought as fundamental particles existing in nature. Indeed, they are artificially realized as charge excitations of quantum liquids. The fusion and braiding rules of anyons are among the geometrical observables which characterize the topological order. Moreover, as we will discuss in the next section, the existence of anyonic particles is deeply related to the topological degeneracy of the ground state. Given their importance in the phenomenology of topological phases, we spend some words to recall some basic properties about a theory of anyons. We know that in $d=3$ spatial dimensions we can have only fermionic and bosonic statistics. This is related to the topology of the configuration space of $N$ identical particles in three dimensions. Denoting with $r_{i},i=1,...,N$ the positions of the particles in $\mathbb{R}^{3}$, the space is $\left( \mathbb{R}^{3N}\setminus\left\lbrace r_{i}=r_{j} \right\rbrace \right)/ \mathcal{S}_{N} $, where $\mathcal{S}_{N}$ is the permutation group of $N$ objects. The fundamental group of this space is $\mathbb{Z}_{2}$. This means that all the possible ways to exchange two particles are topologically equivalent, while a double exchange is topologically trivial. In quantum mechanics the exchange of particles acts as unitary operator on the wave function $\psi(r_{1},r_{2},...,r_{N})$. Hence, the statistics are in correspondence with unitary irreducible representations of $\mathbb{Z}_{2}$. These are one dimensional and the unique possibilities are given by

\begin{equation}
\psi(r_{1},r_{2},...,r_{N})= \pm \psi(r_{2},r_{1},...,r_{N})
\end{equation}

where $\pm 1$ corresponds respectively to the bosonic and fermionic statistics.\\ In two spatial dimensions the classification of statistics is much richer. Let us consider $N$ particles on the plane sitting along a line and order them. The image in space-time of the particle worldlines under an exchange of their order is called a braid. One distinguishes braids by their topological class, which means that two braids are considered the same if we can smoothly change one into the other without crossing of the worldlines. Such braidings form an infinite group called braid group $B_{N}$. This coincide with the fundamental group of the confirguration space of $N$ identical particles on the plane. This is $\left( \mathbb{C}^{N}\setminus\left\lbrace z_{i}=z_{j}\right\rbrace \right)  / \mathcal{S}_{N}$, where $z_{i}$ are the coordinates of the particles in $\mathbb{C}$. The braid group of $N$ objects is generated by $N-1$ operators $R_{1},...,R_{N-1}$, where $R_{i}$ exchanges the $i$-th and $(i+1)$-th particles in an anti-clockwise direction. These satisfy the defining relations 

\begin{equation}\label{yb1}
R_{i}R_{j}=R_{j}R_{i} \hspace{1cm} \vert i-j \vert >2
\end{equation}

and the Yang-Baxter equations

\begin{equation}\label{yb2}
R_{i}R_{i+1}R_{i}= R_{i+1}R_{i}R_{i+1},       \hspace{1cm}    i=1,...,N-1.
\end{equation}

Different statistics in two dimensions are associated with unitary irreducible representations of the braid group. Particles that form a one dimensional representation of $B_{N}$ are called Abelian anyons. In this case the braid matrices act on the wave function as phases 

\begin{equation}
R_{i}=e^{i\pi \alpha_{i}}.
\end{equation}

The Yang-Baxter equations for these operators require that $e^{i\pi \alpha_{i}}=e^{i\pi \alpha_{i+1}}$, implying that all identical particles have the same phase. If $\alpha=0,1$ we recover the classical statistics, while for $0<\alpha<1$ we have fractional statistics.\\ The most general case is given by non-Abelian anyons. The Hilbert space of these objects is a non-Abelian representation of the braid group and can be characterized in an abstract way with the idea of fusion. One introduce a set of variables $a,b,c....$ which denote the different type of anyons in our theory. These objects satisfy a multiplicative algebra, usually called fusion algebra, which reads 

\begin{equation}
a\star b= \sum_{c}N_{ab}^{c} c
\end{equation}

where $N_{ab}^{c}$ are non-negative integers telling how many different ways there are to get the anyon $c$ when we bring together two anyons $a$ and $b$. The fusion algebra is commutative and associative, implying that the order in which we fuse anyons is irrelevant. Each anyonic species identifies an irreducible representations of the braid group. Hence, the Hilbert of two anyons $a,b$ decomposes in irreducible representations of dimension $N_{ab}^{c}$ corresponding to the fusion channels $c$ appearing in $a\star b$. The anyons are said non-Abelian if $N_{ab}^{c}\geq 2$ for some $c$. The vacuum of the theory corresponds to the identity element of the algebra $1$ and satisfies 

\begin{equation}
a\star 1=a
\end{equation}

for any $a$. This relation implies that the Hilbert space of a single anyon is always one dimensional. Hence, we see that a single anyonic particle does not have any internal degree of freedom. The information contained in the Hilbert space $H_{ab}$ of two non-Abelian anyons $a,b$ is a property of the pair and cannot be associated to a single particle. We learn that anyonic particles are mutually non local objects and the corresponding Hilbert space is said topological since the stored information is related to global properties of the system, namely how the anyons braid.

\subsection{Charged Excitations of The Laughlin State}

The Berry's connection allows to study the topological properties of the quantum Hall states and in particular the physics of the charged exctitations. One can introduce $M$ quasi-holes in the Hall fluid at positions $\zeta_{a}, a=1,...,M$. These are created experimentally by inserting thin solenoids in the fluid and have a three dimensional nature of magnetic fluxes. However, from the $2d$ perspective they behave like point like particles. A Laughlin state $\nu=1/m$ with $M$ quasi-holes is described by the wave function

\begin{equation}
\Psi(z_{1},...,z_{N};\zeta_{1},...,\zeta_{M})=\prod_{i,a}(z_{i}-\zeta_{a}) \prod_{i<j}(z_{i}-z_{j})^{m}e^{-\sum_{i}\vert z_{i} \vert^{2}/4l_{B}^{2}}.
\end{equation}

A remarkable property of the quasi-holes is that they carry a fraction of the electron charge. If the elecron has charge $q_{e}=-e$, these particles have charge $e^{\ast}=e/m$. An euristic justification of this fact is that putting $m$ quasi-holes in the same position is equivalent to fix the position of one electron, making it non-dynamical and creating a hole in the fluid. Hence, if we inject a hole into the quantum Hall fluid, this splits into $m$ independent quasi-particles.\\ A more rigorous way to show the fractionalization of the electron charge requires to compute the Berry's connection associated with the Laughlin state. One can interpret the positions of the quasi-holes $\zeta_{j}$ as coordinates in the space of couplings and denote with $\ket{\zeta_{1},...,\zeta_{M}}$ a frame on the vacuum bundle which satisfies $\braket{z_{1},...,z_{N}}{\zeta_{1},...,\zeta_{M}}= \Psi(z_{1},...,z_{N};\zeta_{1},...,\zeta_{M})$. The calculation of the Berry's connection is done in detail in \cite{tong} in the normalized frame 

\begin{equation}
\ket{\psi}=\frac{1}{\sqrt{Z}}\ket{\zeta_{1},...,\zeta_{M}}
\end{equation}

where $Z=\braket{\zeta_{1},...,\zeta_{M}}{\zeta_{1},...,\zeta_{M}}$. It is assumed in the computation that the quasi-holes are not brought too close to each other. The Abelian connection over the configuration space of $M$ quasi-holes is 

\begin{equation}
A_{\zeta_{i}}=-\frac{i}{2m}\sum_{j\neq i} \frac{1}{\zeta_{i}-\zeta_{j}} + \frac{i\bar{\zeta_{i}}}{4 m l_{B}^{2}}, \hspace{1cm} A_{\bar{\zeta_{i}}}= \overline{A_{\zeta_{i}}}.
\end{equation}

From the Berry connection one can exctract several informations about the physics of these particles. If we move a quasi-hole $\zeta$ along a closed path $C$ which do not enclose any other particle the corresponding Berry phase is 

\begin{equation}
e^{i\gamma}= \exp \left( -i\oint_{C} A_{\zeta}d\zeta + A_{\bar{\zeta}}d\bar{\zeta} \right) = \exp \left(  i\frac{e\Phi }{m} \right) ,
\end{equation}

where $\Phi$ is the magnetic flux enclosed by $C$. In this case the phase $\gamma$ has the interpretation of Aharonov-Bohm phase $\gamma=e^{\ast}\Phi$ picked up by a charged particle in a magnetic field. Hence, we see that a quasi-hole carries fractional charge $e^{\ast}=e/m$.\\ Another remarkable property of the quasi-holes of the Laughlin state is that they behave like Abelian anyons. The statistics is encoded in the flat part of the connection. The corresponding Berry phase provides a representation of the braid group of $M$ quasi-holes. Ignoring the Aharonov-Bohm contribution, if we take a quasi-hole in $\zeta_{1}$ along a closed loop $C$ encircling another quasi-hole in $\zeta_{2}$ we get the following phase

\begin{equation}
e^{i\pi \alpha}=\exp \left( - \frac{1}{2m}\oint_{C}     \frac{d\zeta_{1}}{\zeta_{1}-\zeta_{2}}    + \mathrm{h.c} \right) =e^{2\pi i/m}.
\end{equation}

Mooving a particle around another one is equivalent to exchange them two times. Hence, the phase picked up by the wave function after a single braid is 

\begin{equation}
\alpha=\frac{1}{m}.
\end{equation}

In the case of $m=1$, namely the integer plateau at $\nu=1$, the quasi-holes behave like fermionic particles, while in a generic Laughlin state with $m>1$ the quasi-holes have Abelian fractional statistics.\\ An important property of fractional quantum Hall states is also the topological degeneracy of the ground state, namely the number of ground states depend on the topology of the manifold. This is one of the main future of topological phases and is deeply related to the existence of anyonic particles. This feature becomes manifest if we put the system on a compact manifold of non trivial topology. Let us put the system on a torus and consider the following process. We create from the vacuum a quasi-particle – quasi-hole pair, where the quasi particle is a charged excitation of the Hall fluid with opposite charge $e^{\ast}=-e/m$. We then separate the particles and take them around one of the two different cycles of the torus before making them annihilate. We denote with $T_{1},T_{2}$ the operators which implement the process for the first and second cycle of the torus. Composing these operations one can see that $T_{1}T_{2}T_{1}^{-1}T_{2}^{-1}$ is equivalent to take one anyon around another. Hence, the torus generators satisfy the algebra

\begin{equation}
T_{1}T_{2}=e^{2\pi i/M}T_{2}T_{1}.
\end{equation}

The Hilbert space must be an irreducible representation of this algebra and has dimension $m$. The generalisation of this argument to an arbitrary genus-g Riemann surface shows that the ground state must have degeneracy $m^{\mathrm{g}}$. This number depends only on the topology and not on the local details of the manifold. One can also construct the analog for the Laughlin states on the torus using the Jacobi theta functions, checking that there are exactly $m$ ground states. We note also that the integer plateau given by $m=1$ corresponds to the trivial topological phase.

\subsection{Effective Theory of Laughlin States}

Different quantum Hall states support inequivalent theories of anyonic particles, with different fractional charge and fractional statistics. This means that, despite these states have the same symmetry, they belong to different quantum phases and have different topological order. An efficient and systematic way to study topological order is to construct an effective low energy theory for the FQH states. This captures the topological properties of FQH liquids and provide a universal characterization of topological order. To construct this theory one follows the hydrodynamical approach \cite{wen}. Below the scale of the minimal excitation, electrons in a strong magnetic field and low temperature form a quantum liquid which we describe with a vector field $J^{\mu}, \mu=0,1,2$, where $J^{0}$ is the electron charge density and $J^{i},i=1,2$ are the components of the electron current density. A Laughlin state $\nu=1/m$ has charge density $J^{0}=-e/2\pi l_{B}^{2}= e^{2} \nu B/2\pi$ and the charge current induced by an external electrostatic field $E_{i}$ is $J^{i}=\sigma_{xy} \epsilon^{ij} E_{j}$, where $\sigma_{xy}= \frac{\nu e^{2}}{2\pi}$ is the Hall conductance. Hence, the vector field $J_{\mu}$ has the following response to a variation of the electromagnetic field 

\begin{equation}
\delta J^{\mu}=\sigma_{xy}\epsilon^{\mu\nu\lambda}\partial_{\nu}\delta A_{\lambda}.
\end{equation}

We want to write an action giving the above equation as equation of motion. The electron density can be parametrized in term of a $U(1)$ gauge field $a_{\mu}$ as 

\begin{equation}\label{current}
J^{\mu}=\frac{e^{2}}{2\pi}\partial_{\nu} a_{\lambda} \epsilon^{\mu \nu \lambda}.
\end{equation}

This current satisfies automatically the conservation law $\partial_{\mu}J^{\mu}=0$. The gauge field $a_{\mu}$ is the emergent topological degree of freedom of the FQH liquid. The Lagrangian describing the dynamics of this field and reproducing the above equation of motion is 

\begin{equation}
\mathcal{L}(a_{\mu}, A_{\nu})=-\frac{e^{2} m}{4\pi}a_{\mu}\partial_{\nu}a_{\lambda} \epsilon^{\mu\nu\lambda} + \frac{e^{2}}{2\pi}A_{\mu}\partial_{\nu}a_{\lambda}\epsilon^{\mu \nu \lambda}.
\end{equation}

The first piece of the Lagrangian is the well known Abelian Chern-Simons term. As one could expect, the effective theory describing the geometrical properties of FQH states is a topological field theory in $2+1$ dimensions. The identification of the inverse filling fraction $\frac{1}{\nu}=m$ with the Chern-Simons coupling, which is quantized by the requirement of gauge invariance at quantum level, leads to a correspondence between Chern-Simons levels and FQH states.\\ To provide a complete description of the FQH liquids one has to include in the Lagrangian the quasi-hole and quasi-particle excitations. This requires to introduce another current $j_{\mu}$ which couples to $a_{\mu}$. The Lagrangian gains a new term 

\begin{equation}
\Delta \mathcal{L}= a_{\mu}j^{\mu}.
\end{equation}

The gauge invariance of the action is preserved if the current is conserved: $\partial_{\mu}j^{\mu}=0$. Turning off the electromagnetic field $A_{\mu}$, the equation of motion is 

\begin{equation}
\frac{e^{2}}{2\pi}f_{\mu \nu}= \frac{1}{m} \epsilon_{\mu\nu\rho} j^{\rho}.
\end{equation}

Let us place a static quasi-hole in the origin. The current is given by $j^{0}=e\delta^{2}(x), j^{1}=j^{2}=0$. The equation of motion becomes 

\begin{equation}
\frac{1}{2\pi} f_{12}=\frac{1}{em} \delta^{2}(x).
\end{equation}

We see that the effect of the Chern-Simons term is to attach a magnetic flux 

\begin{equation}
\Phi=\frac{2\pi}{em} 
\end{equation}

to a particle of charge $q=e$. Consequently, If we take a quasi-hole around another one we get an Aharonov-Bohm phase $e^{iq \Phi}=e^{2\pi i/m}$, which we interpret as the statistical phase generated by a double exchange of the particles. This result agrees with the fractional statistics that we find for the the Laughlin state $\nu=1/m$. Using the definition of electromagnetic current density in \ref{current} we also get the equality

\begin{equation}
J^{0}=\frac{e^{2}}{2\pi}f_{12}= \frac{e}{m}\delta^{2}(x),
\end{equation}

from which we correctly the recover the fractionalization of the quasi-hole charge $e^{\ast}=\frac{e}{m}$.\\ Recent reviews about the Chern-Simons approach to FQHE can be found in \cite{tong,witten}.

\subsection{Other Filling Fractions}\label{effective}

So far we discussed only FQH liquids at the fractions $\nu=\frac{1}{m}$. As we anticipated in section \ref{laughlin} one can describe the other plateaus with the idea of hierarchical states proposed by Haldane and Halperin. One starts with a Laughlin state at $\nu=1/m$ and vary the magnetic field to make the system change the plateau. Given their nature of magnetic fluxes, we can increase or decrease the magnetic field by injecting respecticely quasi-particles or quasi-holes in a $\nu=1/m$ state such that they form themselves a Laughlin condensate. The effective Lagrangian of the model is modified as follows \cite{tong,wen}. Let us set $e=1$ for simplicity. One writes the quasi-holes current as

\begin{equation}\label{type}
j_{\mu}= \frac{1}{2\pi}\epsilon^{\mu\nu\rho}\partial_{\nu}\tilde{a}_{\rho},
\end{equation}

where $\tilde{a}_{\rho} $ is another emergent Chern-Simons field describing the quasi-hole condensate. The current couples to $a_{\mu}$ and the new Lagrangian reads 

\begin{equation}\label{previousmodel}
\mathcal{L}(a_{\mu},\tilde{a}_{\nu})= \frac{1}{2\pi}A_{\mu}\partial_{\nu}a_{\lambda}\epsilon^{\mu \nu \lambda}-\frac{m}{4\pi}a_{\mu}\partial_{\nu}a_{\lambda} \epsilon^{\mu\nu\lambda} + \frac{1}{2\pi}a_{\mu}\partial_{\nu}\tilde{a}_{\lambda}\epsilon^{\mu \nu \lambda} - \frac{\tilde{m}}{4\pi}\tilde{a}_{\mu}\partial_{\nu}\tilde{a}_{\lambda},
\end{equation}

where $\tilde{m}$ is an integer. To compute the Hall conductivity one has to solve the equations of motion for $\tilde{a}_{\mu}$ and then for $a_{\mu}$. One finds that this theory describes a Hall state with filling fraction 

\begin{equation}
\nu= \frac{1}{m-\frac{1}{\tilde{m}}}.
\end{equation}

We can compute the charge and statistics of quasi-holes in this new state. There are two type of excitations for this fluid, the ones which couple to $a_{\mu}$ and the ones which couple to $\tilde{a}_{\mu}$. For the first type we find 

\begin{equation}
m f_{12}-\tilde{f}_{12}= 2\pi \delta^{2}(x), \hspace{1cm}  \tilde{m}\tilde{f}_{12}-f_{12} =0,
\end{equation}

from which we get 

\begin{equation}
f_{12}= \frac{2\pi}{m-\frac{1}{\tilde{m}}} \delta^{2}(x).
\end{equation}

For the second type we have

\begin{equation}
m f_{12}-\tilde{f}_{12}=0 \hspace{1cm} \tilde{m}\tilde{f}_{12}-f_{12}= 2\pi \delta^{2}(x)
\end{equation}

which gives 

\begin{equation}
f_{12}= \frac{2\pi}{ m\tilde{m}-1}.
\end{equation}

If we set for instance $m = 3  $ and $\tilde{m} = 2$ we find the state $\nu=2/5$, which is among the most prominent among the observed plateau. The resulting charges of the quasi-holes are $e^{\ast} = 2/5$ and $e^{\ast} = 1/5$, which have been measured experimentally. One can now repeat this construction: the quasi-particles of the new state form a condensate described another gauge field, for which we introduce a new Chern-Simons term and current of type \ref{type} which couples to $\tilde{a}_{\mu}$. Iterating this procedure we obtain Abelian quantum Hall states with filling fraction 

\begin{equation}
\nu=\frac{1}{m-\frac{1}{\tilde{m_{1}} - \frac{1}{\tilde{m_{2}}-....}}}.
\end{equation}

Using the idea of hierarchy one can write down the most general Abelian quantum Hall state \cite{wen,tong}. We consider $N$ emergent gauge fields $a_{\mu}^{i}$, with $i=1,...,N$. The effective theory for these fields is 

\begin{equation}
\mathcal{L}(a_{\mu}^{i})= \frac{1}{4\pi}K_{ij} \epsilon^{\mu\nu\rho}a_{\mu}^{i}\partial_{\nu}a_{\rho}^{j}+\frac{1}{2\pi}t_{i}\epsilon^{\mu\nu\rho} A_{\mu}\partial_{\nu}a_{\nu}^{i}
\end{equation}

where the $K$ matrix specifies the Chern-Simons coupling and the $t$ vector tells which linear combination of currents play the role of electron current. So far we only considered the so called single-layer FQH states, where the electrons are confined a single $2d$ interface. This setting is described by the choice $t=(1,0,0,...,0)$, namely we have a single charge carrier. However, one can also make samples with multiple layers of interfaces. The electrons confined on different layers under a strong magnetic filed form a multi-layer FQH state. The topological properties of the model are encoded in $K$ and $t$. The Hall conductance can be computed by integrating out the gauge fields and is given by 

\begin{equation}
\left( K^{-1}\right)^{ij} t_{i}t_{j},
\end{equation}

while the charge of the quasi-holes which couple to $a_{\mu}^{i}$ is 

\begin{equation}
(e^{\star})^{i}= \left( K^{-1}\right)^{ij}t_{j}.
\end{equation}

The statistics between quasi-holes which couple to $a^{i}$ and $a^{j}$ is 

\begin{equation}
\alpha^{ij}= \left( K^{-1}\right)^{ij}.
\end{equation}

Moreover, one can show that the degeneracy of the ground state on a surface of genus-g is $\vert \mathrm{det} K \vert^{g}$. When can recover the single-layer model given in \ref{previousmodel} with the choice 

\begin{equation}
K=\begin{pmatrix} 
m & -1 \\ -1 & \tilde{m}
\end{pmatrix} \hspace{1cm }
t=(1,0).
\end{equation}

It is possible to write down a wave function for the hierarchy states, also in the multi-layer case, which extends the Laughlin proposal to more general filling fractions. The interested reader can refer to \cite{wen} for the details of the construction.\\ An alternative way to describe the hierachy states is given by the composite fermion theory proposed by Jain \cite{ab4}. The basic proposal is that FQH states for electrons can be thought as integer Hall states for new weakly interacting degrees of freedom. These are the composite fermions, namely electrons bounded to magnetic flux vortices. One can find a discussion about this alternative model in \cite{tong}. Despite the approach is very different from the hierarchy theory, the composite fermion approach reproduces the same filling fractions. Among these we get in particular the principal series 

\begin{equation}
\nu=\frac{n}{2n \pm 1},
\end{equation}

also known as Jain's series. As the hierarchy model, the composite fermion theory predicts Abelian fractional statistics for the quasi-holes.\\ The possibility of non-Abelian statistics has been explored for certain FQH states at higher Landau levels \cite{nab1,nab2,nab3,nab4}. Among these the most prominent is observed at $\nu=5/2$, which consists of fully filled lowest Landau level for both spin up and spin down electrons, followed by a spin polarized Landau level at half filling. This is also known as Moore-Read state and is believed to support non-Abelian anyons of Ising type \cite{tong}. One can construct effective low energy theories also for non-Abelian quantum Hall states. These naturally involve emergent non-Abelian Chern-Simons fields associated with certain non-Abelian gauge groups.

\subsection{Theory of Edge States}\label{theoryofedgestates}

It has been noticed in \cite{nab2} a remarkable connection between wave functions of FQH states and correlators of certain $2d$ conformal field theories. Let us consider a $\nu=1/m$ Laughlin state with $N$ electrons at positions $z_{i},i=1,...,N$ and $M$ quasi-holes at positions $\zeta_{a},a=1,..,M$. We consider the $c=1$ theory describing a free scalar field $\phi(z)$ and associate to electrons and quasi-holes respectively the chiral vertex operators $V(z_{i})=\exp\left( i\phi(z_{i})/\sqrt{\nu}\right) $ and $ W(\zeta_{a})=\exp \left(i\sqrt{\nu}\phi(\zeta_{a}) \right) $. The holomorphic part of the Laughlin wave function is captured up to a prefactor depending on the $\zeta_{a})$ by the correlator 

\begin{equation}
\langle \prod_{i,a} V(z_{i})W(\zeta_{a})\rangle= \prod_{i,a}(z_{i}-\zeta_{a}) \prod_{i<j}(z_{i}-z_{j})^{\frac{1}{\nu}}.
\end{equation}

The contribution of the magnetic field can be included in this set up as follows. To be compatible with the holomorphicity of the correlator, instead of a uniform $B$ field, one considers a lattice of fundamental units of magnetic flux $\Phi(a)=a^{2}/2\pi l_{B}^{2}ì$ at positions $\zeta_{m,n}(a)=am+ian$, $\ m,n \in \mathbb{Z}$, where $a$ is a real parameter. One has to insert in the correlators the vertex operators

\begin{equation}\label{vertex}
\prod_{\Lambda(a)\cap D(R)} \exp\left(-i  \Phi(a)\sqrt{\nu}\phi(\zeta_{m,n}(a)) \right), 
\end{equation}

where the lattice $\Lambda(a)= a\mathbb{Z}+ia\mathbb{Z}$ is contained in a disk $D(R)$ of radius $R=\sqrt{2mN}l_{B}$. It is manifest in this language that a point-like charge and a magnetic flux unit have the same nature in the two dimensional physics. Hence, the FQH wave function gains the new term

\begin{equation}\label{term}
\prod_{i,\Lambda(a)\cap D(R)} \frac{1}{(z_{i}-\zeta_{m,n}(a))}= \exp \left( -\Phi(a)\sum_{i,\Lambda(a)\cap D(R)} \log (z_{i}-\zeta_{m,n}(a))  \right). 
\end{equation}

We know that a correlator of chiral primaries $\langle \prod_{i}e^{i\alpha_{i}\phi(z_{i})} \rangle$ is non-vanishing only if the neutrality condition $\sum_{i}\alpha_{i}=0$ is satisfied. With the inclusion of the magnetic lattice in the correlator one finds the relation

\begin{equation}
N= \nu (\mathcal{N}-M),
\end{equation}

where $\mathcal{N}=\sum_{\Lambda(a)\cap D(R)}\Phi_{a}= R^{2}/2l_{B}^{2}$ is the number of flux units contained in $D(R)$, namely the capacity of the lowest Landau level. In this framework the magnetic field plays the role of neutralizing background charge for the correlator. We see that the above equality reproduces the correct relation between the number of electrons and the magnetic flux, including also the contribution of the quasi-holes.\\
In order to recover the uniform constant $B$ field one should take the continuous limit of the lattice and then omit terms suppressed by $\vert z_{i} \vert /R\ll 1$. The immaginary part of the sum $-\Phi(a)\sum_{\Lambda(a)\cap D(R)} \log (z-\zeta_{m,n}(a))$ oscillates very rapidly for $a\ll 1$ and the average of the oscillations is vanishing in the limit $a\rightarrow 0 $. Instead, the real part coincides with the $2$-dimensional electrostatic potential generated by a discrete set of charges, which tends to a continuous distribution for $a\ll 1$. So, we obtain the limit 

\begin{equation}
-\Phi(a)\sum_{\Lambda(a)\cap D(R)} \log (z-\zeta_{m,n}(a)) \xrightarrow{ a\rightarrow 0, \ R\rightarrow \infty} -\sum_{i} \vert z \vert ^{2}/4l_{B}^{2},
\end{equation}

which reproduces the non-holomorphic part of the FQHE wave functions.\\ 
The relation between FQHE and $2d$ CFT is not accidental. The $c=1$ chiral theory describes the edge excitations of the Abelian Chern-Simons theory \cite{wen,tong}. Contrary to the bulk topological theory, which is gapped by definition, the edge theory is a conformal chiral liquid with gapless excitations. This bulk to boundary correspondence is one of the main property of topological phases and apply also to non-Abelian quantum Hall states. The physics of the edge modes is a reflection of the topological order in the bulk and provides the most powerful way tool to classify topological universality classes. It turns out that the edge excitations have the same spectrum of the quasi-particle and quasi-hole excitations of the fluid. The role of the different kinds of anyons is now played by the different representations of the conformal algebra that appear in a given conformal field theory. Each of these representations is labelled by a primary operator $O_{i}$. In classifying topological order we consider rational conformal field theories, since they have a finite number of primary operators. The CFT algebra encodes also the concept of fusion. Given two operators $O_{i},O_{j}$, the operator product expansion (OPE) between them can contain other representations labelled by some $O_{k}$. The OPEs are naturally interpreted as the fusion rules satisfied by the anyonic particles. One can write the OPE as 

\begin{equation}
O_{i}\star O_{j}=\sum_{k}N_{k}^{ij}O_{k}
\end{equation}

where $N_{ij}^{k}$ are integers.\\ The CFT description of FQHE provides also the definition of braiding matrices. In general a CFT contains both chiral and anti-chiral modes. A correlation function of primary operators can be written as  

\begin{equation}
 \langle \prod_{i} O(z_{i},\bar{z}_{i}) \rangle= \sum_{p} \vert \mathcal{F}_{p}(z_{i}) \vert^{2},
\end{equation}

where $\mathcal{F}_{p}(z_{i})$ are multi-branched holomorphic functions of the $z_{i}$ depending on the set of operators inserted in the correlator. These are also known as conformal blocks and provide a basis of quantum Hall wave functions. As we exchange the positions $z_{i}$ of the quasi-particles the conformal blocks will be analytically continued onto different branches. The result can be written as some linear combination of the original functions, from which one derives the braiding properties of the anyons.

\subsection{A New Proposal}

According to the previous discussion, to study the topological properties of FQHE is enough to identify the correct CFT. This encodes all the necessary physical informations to characterize the universality classes of FQH states, such as fusion and braiding rules of the quasi-particles excitations. Moreover, it provides a systematic way to construct the quantum Hall wave functions without any guesswork. Following this approach, it has been recently proposed by Vafa a new unifying model for the principal series of FQHE systems with $\nu=\frac{n}{2n\pm 1}$ \cite{rif1}. The identification of the CFT describing the Hall states comes from consistency conditions between the above description and the CFT paradigms. The starting point is again the Laughlin state $\nu=1/m$, which according to the previous models is described by the $c=1$ CFT. The wave function interpretation of the correlators implies that we have to integrate them over the electron positions $z_{i}$. However, in the context of a CFT this is allowed only if the vertex operator $\exp(i\phi(z_{i})/\sqrt{\nu})$ has dimension $1$. This can be achieved by modifying the $c=1$ theory with the addition of a background term $Q^{\prime}\mathcal{R}\phi$ to the scalar action, where $\mathcal{R}$ is the Ricci curvature and $Q^{\prime}$ is the background charge. This is given by 

\begin{equation}
\frac{Q^{\prime}}{\sqrt{2}}= Q= \frac{1}{b} + b= i\left( \sqrt{2\nu}-\frac{1}{\sqrt{2\nu}} \right),
\end{equation}

where 

\begin{equation}
b=\frac{-i}{\sqrt{2\nu}}.
\end{equation}

Hence, it turns out that FQHE systems are described by a Liouville theory with action \cite{rif4,rif5}

\begin{equation}
S=\int d^{2}z \left[  \frac{1}{8\pi} \partial\phi\bar{\partial} \phi + iQ^{\prime}R\phi    \right]. 
\end{equation}

The central charge of the theory reads

\begin{equation}
c=1+6Q^{2}=1-3\frac{(2\nu-1)^{2}}{\nu}.
\end{equation}

We relax the restriction $\nu=\frac{1}{m}$ and consider a generic rational filling fraction $\nu=\frac{n}{m}$. We obtain 

\begin{equation}
c=1-6\frac{(2n-m)^{2}}{2nm}
\end{equation}

which is the central charge of the $2d$ CFT minimal model $(2n,m)$ \cite{bel}. Here $m$ needs to be odd in order to have $2n$ and $m$ relatively prime. Moreover, since the edge modes in the FQHE are supposed to have correlations which fall off with the distance (see \cite{unitarity}) we should restrict to unitary CFT's \cite{unitarity}. This puts the further constraint $m=2n\pm 1$, from which we recover the Jain's series $\nu=\frac{n}{2n\pm 1}$. The operator algebra of the $(2n,m)$ minimal models is realized by the degenerate fields \cite{rif6}.

\begin{equation}
\Phi_{r,s}=\exp\left[ i(r-1)\frac{\phi}{\sqrt{\nu}}+i(s-1)\sqrt{\nu}\phi \right] 
\end{equation}

for $1\leq r < 2n, \ 1\leq s < m $, which satisfy the relations \cite{minimal}

\begin{equation}
\Phi_{r_{1},s_{1}} \times \Phi_{r_{2},s_{2}} = \sum_{ \substack{ k=1+\vert r_{1}-r_{2}\vert , k+r_{1}+r_{2}+1=0 \ \mathrm{mod} \ 2 \\  l=1+\vert r_{1}-r_{2}\vert , l+r_{1}+r_{2}+1=0 \ \mathrm{mod}\ 2 } }^{\substack{ k=r_{1}-r_{2}-1 \\  l=r_{1}-r_{2}-1 }} \Phi_{k,l}
\end{equation}

In particular, the $\Phi_{1,s}$ are identified with the quasi-holes operators, where $\Phi_{1,2}$ is the minimal excitation, and generate insertions of $(z_{i}-\zeta)^{s-1}$ factors in the wave function. This leads to the main difference between this model and the previous ones, where the quasi-holes of the states $\nu=\frac{n}{2n\pm 1}$ are predicted to be Abelian anyons. On the contrary, in this theory the quasi-holes should have the same fusion rules and non-Abelian braiding properties of the $(2n,2n\pm 1)$ minimal models. The embedding of minimal models in Liouville theory allows also to identify the bulk topological theory, which turns out to be a Chern-Simons theory based on an $SL(2,\mathbb{C})$ gauge group.

\subsection{The Vafa's Hamiltonian}

Among the several proposals of Vafa in \cite{rif1}, there is a connection between FQHE and supersymmetric $\mathcal{N}=2$ gauge theories in four dimensions, which arise from compactification of six dimensional $(2,0)$ theories on a punctured Riemann surface $\Sigma$, the so called Gaiotto curve \cite{rif2}. These are labelled by semisimple Lie algebras in the ADE classification. Picking an ADE group, the rank $r$ of the corresponding Dynkin graph is identified with the number of layers of FQHE. We associate to each node in the graph an integer $ N_{a}, a=1,...,r $ which denotes the number of electrons in the $a$-th layer. The worldvolume of FQHE is identified with $\Sigma\times \mathbb{R}$, where $\mathbb{R}$ is the time direction and the Riemann surface $\Sigma$ is the target space of FQHE. For each node we assign also a meromorphic $(1,0)$ form $W_{a}(z)^{\prime}dz$ on $\Sigma$. The meromorphic functions $W_{a}(z)^{\prime}$ encode the interaction between an electron of coordinate $z$ in the $a$-th layer and the quasi-holes in the quantum Hall fluid. We focus on the case of target space $\mathbb{P}^{1}$, where $W_{a}(z)^{\prime}$ are just rational functions on the complex plane. Finally, we introduce for each node of the graph the set of chiral superfields $z_{a,k_{a}}$, $k_{a}=1,...,N_{a}$, which play the role of electron coordinates for the different layers of FQHE. Hence, with the ingredients defined above, a Dynkin diagram of ADE type identifies a $4$-SQM Landau-Ginzburg theory with superpotential

\begin{equation}\label{potuno}
\begin{split}
\mathcal{W}(z_{a,k_{a}})= & \sum_{a=1}^{r} \sum_{k_{a}=1}^{N_{a}} W_{a}(z_{a,k_{a}}) + \beta \sum_{a=1}^{r} \ \sum_{1\leq k_{a}<h_{a}\leq N_{a}} \log (z_{a,k_{a}}-z_{a,h_{a}})^{2}  \\ & -\beta \sum_{a,b=1}^{r}\sum_{k_{a},h_{a}} I_{a,b} \log(z_{a,k_{a}}-z_{b,h_{a}}),
\end{split}
\end{equation}

where $C_{a,b}=2\delta_{a,b}-I_{a,b}$ is the Cartan matrix of the chosen ADE group and $\beta$ can be a generic complex coupling. We are mostly interested in the case of single layer FQHE. This corresponds to pick the $A_{1} \ (2,0)$ six dimensional theory. In the single layer case we have a unique rational function $W^{\prime}(z)$ containing the interaction between electrons and quasi-holes. It is natural to describe the quasi-holes with simple poles. Indeed, the function $W^{\prime}(z)$ has the physical interpretation of electrostatic field generated by the quasi-holes and the two dimensional Coulomb interaction between point-like particles decays with the inverse of the distance. This is also the most general case, since higher degree poles can be obtained by confluent limit of simple poles. Hence, denoting with $\zeta_{\ell}, \ell=1,...,M $ the positions of the punctures on $\mathbb{P}^{1}$, the holomorphic superpotential corresponding to this setting is

\begin{equation}\label{potdue}
\mathcal{W}(z_{i};\zeta_{\ell})= \sum_{i=1}^{N} W(z_{i};\zeta_{\ell}) + \beta\sum_{1\leq i<j \leq N} \log (z_{i}-z_{j})^{2},
\end{equation} 

where 

\begin{equation}
W(z;\zeta_{\ell})= \sum_{\ell=1}^{M} \alpha_{\ell}\log (z-\zeta_{\ell}).
\end{equation}

This model can be defined for general residues $\alpha_{\ell}$. However, as we are going to discuss in section \ref{wedisc}, if we want to identify the punctures at positions $\zeta_{\ell}$ either with the quasi-holes or the magnetic fluxes of FQHE one should set respectively $\alpha_{\ell}=\pm 1$. \\ In \cite{rif1} Vafa makes the following predictions:

\begin{itemize}\label{list}

\item The degeneracy of the lowest Landau level can be mapped to the degeneracy of ground states of a supersymmetric system.

\item The $\mathcal{N}=4$ Hamiltonian prescribed by the above superpotential is in the same universality class of the principal series of FQH states and capture the same topological order.

\item The minimal quasi-holes of the FQHE states have the same braiding properties of the primary field $\phi_{1,2}$ in minimal models, or equivalently of the spin $1/2$ Wilson line operator in $SU(2)$ Chern-Simons theory. Moreover, the Berry connection of the quantum mechanical model corresponds to the $SU(2)$ Knizhnik-Zamolodchikov connection. This provides a Hecke representation $H_{n}(q)$ of the braid group of $n$ quasi-holes, where the parameter $q=\pm e^{\frac{i\pi}{\nu}}$ is function of the filling fraction $\nu$ labelling the FQH states.

\item There is a distinguished state among the ground states of the supersymmetric Hamiltonian of Laughlin type. This canonical vacuum corresponds to the configuration in which the electrons are as spread out as possible among the lowest Landau levels. Moreover, it is the most symmetric one with respect to the permutation of the quasi-holes and magnetic fluxes.

\end{itemize}

The main goal of this thesis is to show that these predictions connecting FQHE and $\mathcal{N}=4$ supersymmetry are correct. The superpotentials \ref{potuno} and \ref{potdue} are motivated by the relation between topological string amplitudes, partition functions of ADE matrix models and chiral blocks of Toda theories \cite{rif7,rif8,rif9}. Focusing on the $A_{N-1}$ case, the topological string amplitudes are the objects which connect the Nekrasov partion functions of $\mathcal{N}=2 \ SU(N)$ gauge theories \cite{rif24} and the chiral conformal blocks of $A_{N-1}$ Toda theories in the context of AGT correspondence. In particular, one can specialize to the case of minimal models, which admit an embedding in the AGT set up with an appropriate refinement of the Nekrasov partition functions \cite{rif42,rif43,rif44}. In the case of single layer FQHE, the $\mathcal{N}=2$ four dimensional theories arising from the compactification on $\Sigma$ are $SU(2)$ gauge theories with a matter sector which depends on the structure of the punctures \cite{rif2}. According to the original formulation of AGT correspondence \cite{rif3}, the Nekrasov partition functions of $SU(2)$ gauge theories compute the chiral blocks of the Liouville CFT. In Appendix \ref{classS} we provide a short review about the Gaiotto theory and the relation between topological string amplitudes and Liouville chiral blocks which motivate the Vafa Hamiltonian. Once identified the effective theory for the principal series of FQHE, Vafa proposes also a microscopic Hamiltonian that in principle may support the topological phases described by the minimal models as ground states. This Hamiltonian is $\mathcal{N}=4$ supersymmetric and is motivated by string theoretical considerations. The key idea of this program is that with extended supersymmetry we have several techniques at our disposal, such as hyperKahler geometry, isomonodromic deformations, triangle and cluster categories, mirror symmetry, integrable and Hitchin systems, etc..., which we may use to understand the phenomenology of FQHE. These condensed matter systems seem to hide a rich mathematical structure beyond Chern-Simons theory and $2d$ CFT, which involves all the main recent developments in string theory and supersymmetric QFT.

\section{Supersymmetry and $tt^{*}$ Geometry}\label{reviewtt}

\subsection{Supersymmetric Quantum Mechanics}

Before exploring the connections between FQHE and supersymmetry, we need to recall some basic knoledge about supersymmetric quantum mechanics and related tools to study the geometry of vacua. We begin with the case of two supercharges.

\subsubsection{$\mathcal{N}=2$ SQM}

$\mathcal{N}=2$ supersymmetric quantum mechanics is essentially the Witten's reformulation of Morse theory \cite{rif19,rif22}. The theory has a graded Hilbert space $\mathcal{H}=\mathcal{H}^{+}\oplus \mathcal{H}^{-}$, where $\mathcal{H}^{\pm}$ are spaces of bosonic and fermionic states respectively. These are eigenspaces of the Fermi parity operator $I=(-1)^{F}$ which counts the fermion number $F$ of a state modulo two. This operator is Hermitian and satisfies $I^{2}=1$. By definition of $\mathcal{N}=2$ supersymmetry, the theory has an Hermitian operator $\mathcal{Q}$ (and its adjoint $\mathcal{Q}^{\dagger}$) which maps bosonic states to fermionic states and viceversa. The susy charges $\mathcal{Q},\mathcal{Q}^{\dagger}$ and the Fermi parity satisfy the algebraic relations

\begin{equation} 
\begin{split}
& \left\lbrace \mathcal{Q},I\right\rbrace = \left\lbrace \mathcal{Q}^{\dagger} , I \right\rbrace=0 \\ 
& \mathcal{Q}^{2}=\mathcal{Q}^{\dagger 2}= \left\lbrace \mathcal{Q},\mathcal{Q}^{\dagger} \right\rbrace =0
\end{split}
\end{equation} 

Since they are symmetry operators of the system, the susy charges commute with the Hamiltonian $H$. In particular, the Hamiltonian is the ``Laplacian'' 

\begin{equation}
H= \frac{1}{2}\left( \mathcal{Q}\mathcal{Q}^{\dagger}+ \mathcal{Q}^{\dagger}\mathcal{Q}\right) .
\end{equation}

The operators $\mathcal{Q},\mathcal{Q}^{\dagger},F,H$ generate the supersymmetry algebra of the $\mathcal{N}=2$ theory.\\ Since it is a nilpotent operator, one can define the cohomology of $\mathcal{Q}$ in the Hilbert space. The cohomology classes of $\mathcal{Q}$ are in correspondence with the ground states of the theory. The vacuum wave functions $\ket{\Psi}$ are eigenstates of the Hamiltonian with zero energy and coincide with the harmonic representatives of the cohomology classes

\begin{equation}
H\ket{\Psi}=0 \Leftrightarrow \mathcal{Q}\ket{\Psi}= \mathcal{Q}^{\dagger}\ket{\Psi}=0.
\end{equation}

We consider a smooth compact manifold $X$ of dimension $\mathrm{dim}_{\mathbb{R}} X=n$, with a local coordinate system $x_{i},i=1,...,n$. To each coordinate $x_{i}$ we associate a couple of creation-annihilation operator $\psi^{i},\psi_{i}$ which satisfy the Clifford algebra 

\begin{equation}
\begin{split}
& \left\lbrace  \psi^{i}, \psi^{j}   \right\rbrace= \left\lbrace  \psi_{i}, \psi_{j} \right\rbrace = 0  \\ & \left\lbrace  \psi^{i}, \psi_{j}   \right\rbrace= \delta^{i}_{j}.
\end{split}
\end{equation}

The Hilbert space is a representation of this algebra and a general wave function reads 

\begin{equation}
\Phi=\sum_{k} \phi(x)_{i_{1},...,i_{k}}\psi^{i_{1}}...\psi^{i_{k}}\ket{0},
\end{equation}

where $\ket{0}$ is the Clifford vacuum and the functions $\phi(x)_{i_{1},...,i_{k}}$ are totally antisymmetric in the indices. The commutation relation between creation and annihilation operators can be read as the pairing between forms and vectors $dx^{i}(\partial_{j})= \delta^{i}_{j}$. The identification $\psi^{i}\rightarrow dx^{i}$ allows to rewrite the above wave function as a differential form on $M$. Hence, the space of differential forms $\Omega^{*}(X)$ plays the role of Hilbert space, where the Hermitian scalar products between states coincides with the Hodge inner product of forms

\begin{equation}\label{inner}
\braket{\Phi}{\Psi}= \int_{X} \Psi \wedge \ast \Phi^{\ast}.
\end{equation}

In this representation the susy charges act as differential operators on the space of forms, while the degree $k$ of a wave form has the interpretation of Fermi number. Moreover, the Hermitian conjugation of operators is defined with respect to the inner product \ref{inner} and has the usual expression of Hodge theory. \\ The easiest realization of $\mathcal{N}=2$ supersymmetric quantum mechanics is a sigma model on a Riemannian manifold. The supersymmetric charges $\mathcal{Q}, \mathcal{Q}^{\dagger}$ are given by the exterior derivative $d$ and the adjoint $d^{\dagger}=(-1)^{k} \ast^{-1}d \ast$. The corresponding Hamiltonian is the usual Laplacian on a Riemannian manifold

\begin{equation}
H=\frac{1}{2}\Delta=\frac{1}{2}(d^{\dagger}d+dd^{\dagger}).
\end{equation}

The vacuum of this theory is isomorphic to the de Rham cohomology $H^{*}(X;\mathbb{C})$ and the Witten index coincides the Euler character of $X$ 

\begin{equation}
I_{W}=\mathrm{Tr}_{\mathcal{H}}(-1)^{F}=\sum_{k=0}^{n} (-1)^{k}b_{k}= \chi (X),
\end{equation}

where the Betti numbers $b_{k}$ count the number of vacua with degree $k$.\\ One can deform the sigma model by introducing a smooth real superpotential $V$ on $X$. The susy algebra generators of this theory are the generalized differential operators 

\begin{equation}
\begin{split}
& d_{V}= d+ dV\wedge, \hspace{1cm} d_{V}^{\dagger}=(-1)^{k} \ast^{-1} (d +dV\wedge)\ast \\ \\ 
& \frac{1}{2}\Delta_{V}=\frac{1}{2} \left( \Delta + g^{ij}\partial_{i}V\partial_{j}V +  \partial_{i}\partial_{j}V [\ast dx^{i}, dx^{j}]\right)  ,
\end{split} 
\end{equation}

where $g_{ij}$ is a Riemannian metric on the target manifold. The susy charge $d_{V}$ is conjugated to the exterior derivative by the relation $d_{V}= e^{-V}d e^{V}$. It follows that the cohomology classes $[\tilde{\Psi}]$ of $d_{V}$ are in one to one correspondence with de Rham classes $[\Psi]$ through the map

\begin{equation}\label{conjugacy}
\tilde{\Psi}= e^{-V}\Psi.
\end{equation}

The isomorphism $\mathcal{H}_{\mathrm{vacuum}}\simeq H^{*}(X;\mathbb{C})$ implies that on a compact manifold the dimension of the ground state depends only on the topology of the manifold and not on the choice of the superpotential.\\ One can study the ground states of the theory also with a perturbative approach. We introduce the planck constant $\hslash$ and rescale the superpotential as $V\rightarrow \frac{1}{\hslash}V$. In the semiclassical limit $\hslash\ll 1$ the potential energy $\frac{1}{\hslash^{2}}g^{ij}\partial_{i}V\partial_{j}V$  becomes very large except around the critical points of $V$. For small $\hslash$ the eigenfunctions of the Hamiltonian are concentrated around the classical vacua and can be computed with an asymptotic expansion in terms of local data. This is the essence of Morse theory: we reconstruct the cohomology from the local data of a smooth function around its critical points. We consider the case in which $V$ is a Morse function. This implies that the critical points of $V$ are isolated and non degenerate, namely $\mathrm{det} \ \partial_{i}\partial_{j}V\neq 0$. Since the set of Morse functions is open and dense in the space of functions, these properties are very general. One can associate to each perturbative vacuum a Morse index $I_{M}$ which counts the number of negative eigenvalues of the Hessian of $V$ at the critical point. The Morse index has the physical interpretation of fermion number. Since the Witten index is invariant under smooth deformation of the theory, one can perform its computation also in the classical limit. Denoting with $c_{k}$ the number of classical vacua with Morse index $k$, one has 

\begin{equation}
I_{W}=\sum_{k=0}^{n}(-1)^{k}c_{k}= \sum_{k=0}^{n}(-1)^{k}b_{k}.
\end{equation}

The classical computation of vacua is exact at every order in perturbation theory, but is in general not correct at the non-perturbative level. Indeed, instanton corrections can lift some of the classical vacua. Hence, the true quantum vacua are generically less than the classical ones and we have $c_{k}\geq b_{k}$. Since the Witten index must be preserved, bosonic and fermionic vacua can be lifted by instantons only in pairs.\\ One can extend $\mathcal{N}=2$ SQM to a non compact Riemannian manifold. In the non compact case the Hilbert space is the space of differential forms whose coefficents are $L^{2}$ functions on $X$. In absence of a superpotential, the vacua are harmonic representatives in cohomology classes of $L^{2}$ forms. These states are exact in the absolute cohomology of $X$ and one can write for a vacuum $\Psi$ the relation

\begin{equation}
\Psi=d\Phi.
\end{equation}

However, $\Phi$ is not square integrable on the target space and so the wave form $\Psi$ defines a non trivial cohomology class in the Hilbert space. \\ 
As before, we can introduce a real superpotential $V$ on the target space which generates a potential term in the Hamiltonian. Differently from the compact case, now the structure of vacua depends on the choice of the superpotential. Morse theory admits an extension for non compact manifolds and one can show that the space of vacua is isomorphic to the relative de Rham cohomology $H^{*}(X,X_{V}; \mathbb{C})$, where 

\begin{equation}\label{setv}
X_{V}=\lbrace     x \in X \big \vert  \  V<-\Lambda; \ \Lambda \rightarrow \infty \rbrace
\end{equation}

is the asymptotic region of $X$ where the superpotential tends to $-\infty$. In particular, the relative de Rham classes in $H^{*}(X,X_{V}; \mathbb{C})$ are mapped to $d_{V}$-cohomology classes in the space of $L^{2}$ forms by the map \ref{conjugacy}. We can consider as an example the polynomial superpotential on the real line 

\begin{equation}
V= a x^{2n}+ \mathrm{lower \ order \ terms}, \hspace{1cm} x \in \mathbb{R}.
\end{equation}

The structure of the ground state depends on the sign of the coefficient $a$ in fron of the highest power of the polynomial. If $a$ is positive the superpotential is bounded from below and so we have $X_{V}=\varnothing$. The vacuum space is isomorphic to the de Rham cohomology of the real line

\begin{equation}
H^{\ast}(\mathbb{R})= \begin{cases}
H^{0}(\mathbb{R})= \mathbb{C}\\ H^{1}(\mathbb{R})= 0
\end{cases}.
\end{equation}

One can easily check that the eigenfunction of the Hamiltonian is the zero form

\begin{equation}
\psi= ce^{-V},
\end{equation}

where $c$ is some normalization constant. If $a$ is negative the superpotential is bounded from above and tends to $-\infty$ as $x\rightarrow \pm \infty$. In this case the vacuum is isomorphic to the cohomology with compact support on $\mathbb{R}$. This is Poincar\'e dual to the de Rham cohomology and reads 

\begin{equation}
H^{\ast}_{\mathrm{c}}(\mathbb{R})= \begin{cases}
H^{0}_{\mathrm{c}}(\mathbb{R})= 0 \\ H^{1}_{\mathrm{c}}(\mathbb{R})= \mathbb{C}
\end{cases}.
\end{equation}

The corresponding vacuum is the fermionic state

\begin{equation}
\psi= c e^{V}dx.
\end{equation}

\subsubsection{$\mathcal{N}=4$ SQM}\label{sqm}  

Supersymmetric quantum mechanics with four supercharges has the same structure of Hodge theory on a Kahler manifold $X$ \cite{rif14,rif19}. For any neighborhood of the manifold one can choose a set of holomorphic coordinates $z_{i},i=1,...,n=\mathrm{dim}_{\mathbb{C}}X$ and a local Kahler potential $K$. This defines also the corresponding Kahler metric $g_{ij}=\partial_{i}\bar{\partial}_{j}K$ and Kahler form $\omega_{K}= i g_{i\bar{j}}dz^{i}d\bar{z}^{j}$. The fermionic partners of the bosonic degrees of freedom $z_{i},\bar{z}_{i}$ are four creation and annihilation operators $\psi^{i},\psi_{i},\bar{\psi}^{i},\bar{\psi}_{i}$ which satisfy the Clifford relations 

\begin{equation}
\begin{split}
& \left\lbrace  \psi^{i}, \psi^{j}   \right\rbrace= \left\lbrace  \psi_{i}, \psi_{j} \right\rbrace = 0 \hspace{1cm } \left\lbrace  \bar{\psi}^{i}, \bar{\psi}^{j}   \right\rbrace= \left\lbrace  \bar{\psi}_{i}, \bar{\psi}_{j} \right\rbrace = 0 \\ & \left\lbrace  \psi^{i}, \psi_{j}   \right\rbrace= \delta^{i}_{j} \hspace{3cm} \left\lbrace   \bar{\psi}^{i}, \bar{\psi}_{j}   \right\rbrace= \delta^{i}_{j}.
\end{split}
\end{equation}

Similarly to the case with two supercharges, one can identify the creation and annihilation operators with forms and vectors on the target space. The map $\psi^{i}\rightarrow dz^{i}, \bar{\psi}^{i}\rightarrow d\bar{z}^{i}$ realizes the isomorphism between the space of $L^{2}$ smooth forms on $X$, endowed with the inner product \ref{inner}, and the Hilbert space $\mathcal{H}$ of a $\mathcal{N}=4$ theory. Compared to the Riemannian case, the richer structure of Kahler geometry leads to a larger supersymmetry algebra. The Hilbert space can be decomposed in irreducible representations of the Lefschetz $SU(2)$ symmetry group, whose algebra is generated by the Lefschetz operators 

\begin{equation}
L=\omega_{K}  \wedge , \hspace{1cm} \Lambda= (-1)^{k}\ast^{-1}L \ast
\end{equation}

and the shifted Fermi number $\tilde{F}=k-n$, where $k=p+q$ is the degree of a form which decomposes in holomorphic and anti-holomorphic degree $p,q$. These operators satisfy the algebraic relations 

\begin{equation}
\begin{split}
& [\tilde{F}, L ]=2L, \\  
& [ \tilde{F}, \Lambda ]=-2L, \\ 
& [ L, \Lambda ]=\tilde{F}.
\end{split}
\end{equation}

The Lefschetz group has the physical interpretation of $R$-symmetry group of the supersymmetry algebra.\\ The easiest realization of $\mathcal{N}=4$ SQM is a sigma model with target space $X$. In this theory the supercharges are given by the Dolbeault operators on the Kahler manifold 

\begin{equation}
\begin{split}
& \mathcal{Q}=\partial, \hspace{1cm} \overline{\mathcal{Q}}= \bar{\partial}, \\ 
& \mathcal{Q}^{\dagger}= \bar{\delta}, \hspace{1cm} \overline{\mathcal{Q}}^{\dagger}= \delta.
\end{split}
\end{equation}

The susy charges are nilpotent and anticommuting operators

\begin{equation}
\begin{split}
& \partial^{2}=\bar{\partial}^{2}=0 \hspace{1cm} \partial\bar{\partial} + \partial\bar{\partial}=0 \\ 
& \delta^{2}=\bar{\delta}^{2}=0 \hspace{1cm} \bar{\delta}\delta + \delta\bar{\delta}=0
\end{split}
\end{equation}

and with the R-symmetry generators satisfy the Kahler identities 

\begin{equation}\label{kahlid}
\begin{split}
& \left[ \bar{\partial}, L \right]=\left[ \partial, L \right] =0 \hspace{1cm} \left[ \bar{\partial}^{\dagger}, \Lambda \right]=\left[ \partial^{\dagger}, \Lambda \right] =0 \\ 
& \left[ \bar{\partial}^{\dagger}, L \right]= i\partial \hspace{2.12cm} \left[ \partial^{\dagger}, L \right] =-i\bar{\partial} \\ & H=\bar{\partial}\bar{\partial}^{\dagger}+ \bar{\partial}^{\dagger}\bar{\partial}= \partial\partial^{\dagger}+ \partial^{\dagger} \partial= \frac{1}{2}(dd^{\dagger}+ d^{\dagger}d),
\end{split}
\end{equation}

where $H$ is the Hamiltonian. The four susy charges, the R-symmetry generators and the Hamiltonian provide a complete basis of generators for the supersymmetry algebra.\\ If the Kahler manifold is non compact we can add to the theory an holomorphic superpotential $W$. A $\mathcal{N}=4$ theory with a superpotential is called Landau-Ginzburg (LG) theory. In presence of $W$ the supercharges become the generalized Dolbeault operators 

\begin{equation}
\begin{split}
& \partial_{W}= \partial+\bar{dW} \wedge, \hspace{1cm} \bar{\partial}_{W}= \bar{\partial} + dW \wedge, \\ 
& \bar{\delta}_{W}= \delta_{W}^{\dagger}, \hspace{2.2cm} \delta_{W}=  \bar{\partial}_{W}^{\dagger}.
\end{split}
\end{equation}

One can check that these operators satisfy the same algebraic relations of $\partial,\bar{\partial}$. In particular, the Hamiltonian is the generalized Laplacian 

\begin{equation}
H= \partial_{W}\partial_{W}^{\dagger}+ \partial_{W}^{\dagger} \partial_{W}= \bar{\partial}_{W}\bar{\partial}_{W}^{\dagger}+ \bar{\partial}_{W}^{\dagger}\bar{\partial}_{W}= \frac{1}{2} (d_{V}d_{V}^{\dagger}+ d_{V}^{\dagger}d_{V}),
\end{equation}

where $d_{V}=\partial_{W}+ \bar{\partial}_{W}$ is the generalized exterior derivative of $V=W+\overline{W}$.\\ We are interested in the space of supersymmetric vacua of a $\mathcal{N}=4$ Landau-Ginzburg theory on a non compact space. By the Kahler identities \ref{kahlid}, we can study the cohomology in the Hilbert space of one of the two Dolbeault derivatives. For convenience, one typically considers the complex of $\overline{\mathcal{Q}}=\bar{\partial}_{W}$. Similarly to Hodge theory, the vacuum wave forms coincide with the harmonic representatives of the $\bar{\partial}_{W}$-cohomology classes. There exists a subclass of manifolds that are very common in the applications in which this problem simplifies considerably, namely Stein manifolds. There are several characterizations of this type of spaces, in particular capturing the property of having ``many'' holomorphic functions. Denoting with $O(X)$ the ring of holomorphic functions on $X$, the following two conditions are satisfied:

\begin{itemize}
\item $X$ is holomorphically convex, namely for every compact subset $C\subset X$, the so-called holomorphically convex hull

\begin{equation}
\bar{C}=\left\lbrace z \in X \big \vert \vert f(z)\vert \leq \mathrm{sup}_{w\in K} \vert f(w)\vert \ \forall f \in O(X) \right\rbrace 
\end{equation}

is also a compact subset of $X$.

\item $X$ is holomorphically separable, i.e. $\forall x,y \in X$ with $x\neq y$ there is an holomorphic function $f$ such that $f(x)\neq f(y)$.
\end{itemize}

Examples of Stein manifolds are $\mathbb{C}^{n}$, as well as its holomorphic domains, and all the non compact Riemann surfaces. Among the properties of a Stein space, we have that $X$ is holomorphically spreadable, i.e. for every point $x \in X$ there is a set of $n=\mathrm{dim}_{\mathbb{C}} X$ globally defined holomorphic functions which form a local coordinate system when restricted to a neighborhood of $x$. Another relevant property is that the Dolbeault cohomologies $H^{p,q}_{\bar{\partial}}(X)$ have a simple structure, i.e.

\begin{equation}\label{dolbcom}
H^{p,q}_{\bar{\partial}}(X)= \begin{cases} 0, \hspace{2cm} q\geq 1 \\ \Omega^{p}(X), \hspace{1cm} q=0 \end{cases} 
\end{equation}

where $ \Omega^{p}(X)$ is the space of holomorphic $p$-forms on $X$. The cohomologies $H^{p,q}_{\partial}(X)$ have a similar structure and can be obtained by complex conjugation. This result implies that a Stein space admits a globally defined Kahler potential $K$, since the corresponding Kahler form is exact in the Dolbeault cohomology and can be written globally as $\omega=i\partial\bar{\partial}K$. Moreover, a Stein space admits geodesically complete Kahler metrics.\\ In order to describe the ground states of a Landau-Ginzburg theory it is relevant to define the BRST cohomology of $\overline{\mathcal{Q}}$ in the space of operators. The supercharge acts on an operator $\phi$ as 

\begin{equation}
\left[ \overline{Q}, \phi \right],
\end{equation}

where the brackets denote respectively a commutator or anticommutator if the operator has even or odd fermion number. An operator is closed if the commutator with the susy charge is vanishing, while it is exact if there exist an operator $\varphi$ such that $\phi=\left[ \overline{Q}, \varphi \right]$. The BRST cohomology of $\overline{\mathcal{Q}}$ forms a ring under multiplication of operators, also known as chiral ring \cite{rif14,rif19}

\begin{equation}
\mathcal{R}=\frac{ \overline{\mathcal{Q}}-\mathrm{closed \ opers}}{ \overline{\mathcal{Q}}-\mathrm{exact \ opers}}.
\end{equation}

In particular, $\mathcal{R}$ is a unital associative algebra defined over $\mathbb{C}$. Let us consider an operator which multiplies a wave form by a smooth function $\chi$. This operator is closed if 

\begin{equation}
\left[ \overline{Q}, \chi \right]=\bar{\partial}\chi=0,
\end{equation}

namely if $\chi$ is holomorphic. Moreover, the operator is exact if there exists an holomorphic section $\upsilon$ of the tangent bundle on $X$ such that 

\begin{equation}
\left\lbrace \iota_{\upsilon}, \overline{\mathcal{Q}} \right\rbrace = \upsilon^{i}\partial_{i}W,
\end{equation}

where $\iota_{\upsilon}$ is the contraction of differential forms by the vector field $\upsilon$. It turns out that the chiral ring is isomorphic to the commutative quotient algebra

\begin{equation}
\mathcal{R}\simeq O(X)/J_{W},
\end{equation}

where $J_{W}= \langle \partial_{1}W,...,\partial_{n}W\rangle$ is the Jacobian ideal generated by the partial derivatives of $W$. It is know from singularity theory that if the critical point of $W$ are isolated, the quotient algebra is finite dimensional and localizes around the critical points. Denoting with $p_{i},i=1,...,N$ the set of critical points, we have the isomorphism of complex algebras

\begin{equation}
\mathcal{R}\simeq \prod_{i=1}^{N} \mathcal{R}_{i}, \hspace{1cm} \mathcal{R}_{i}= O_{p_{i}}/J_{W} \simeq \mathbb{C}^{\mu_{i}}
\end{equation}

where $O_{p_{i}}$ are the germs of holomorphic functions at $p_{i}$. The dimension of the local ring $\mu_{i}=\mathrm{dim}_{\mathbb{C}}  \mathcal{R}_{i}$ is called Milnor number of the critical point $p_{i}$ and is equal to the degeneracy of $p_{i}$. The dimension of the algebra is computed by the sum of the Milnor numbers $\mathrm{dim}_{\mathbb{C}}\mathcal{R}= \sum_{i}\mu_{i}$. If $W$ is non degenerate at the critical points, the chiral ring is a semisimple algebra and can be written as product of simple factors $\mathcal{R}\simeq\prod_{i=1}^{N} \mathbb{C}_{i} \simeq \mathbb{C}^{N}$. In particular, we have $\mathcal{R}_{i}\simeq \mathbb{C}_{i}$ and $\mu_{i}=1$ for each non degenerate $p_{i}$. One can introduce a basis of orthogonal idempotents $e_{i},i=1,...,N$ in which the chiral ring is completely diagonal. This is defined by the condition

\begin{equation}
e_{i}(p_{j})=\delta_{ij}
\end{equation}

and provides a canonical isomorphism between $\mathcal{R}$ and $\mathbb{C}^{N}$ as complex algebra. This basis can always be constructed if $X$ is a Stein manifold and the set of non degenerate critical points has no accumulation points. An holomorphic operator $\phi$ is identified by this map with its set of critical values

\begin{equation}
\phi\rightarrow \begin{pmatrix} \phi(p_{1}) \\  . \\ . \\ . \\ \phi(p_{N}) \end{pmatrix}.
\end{equation}

One can introduce a non degenerate symmetric pairing between chiral operators

\begin{equation}
\langle \phi_{i} \phi_{j}\rangle = \eta_{ij}
\end{equation}

which is defined by the Grothendieck formula \cite{rif10}

\begin{equation}\label{grotform}
\langle \phi \rangle= \frac{1}{(2\pi i)^{n}} \oint \frac{\phi \ dz_{1}...dz_{n}}{\partial_{1}W...\partial_{n}W},
\end{equation}

where the integration contour encircles the critical points of $W$. Endowed with the above pairing, the chiral ring is a symmetric Frobenius algebra.\\ The structure of the vacuum space $\mathcal{V}$ of $\mathcal{N}=4$ Landau-Ginzburg theory is described by the following theorem \cite{rif14,rif15}. Under the condition that $X$ is a Stein manifold, with $\mathrm{dim}_{\mathbb{C}}X=n$ and the critical points of $W$ are isolated 

\begin{itemize}

\item The supersymmetric vacua $\Psi_{i}$ are invariant $n$-forms under the Lefschetz symmetry group, namely 

\begin{equation}
L\Psi_{i}=\Lambda \Psi_{i}=0
\end{equation}

\item The vacuum wave forms are ``essentially independent of the Kahler metric''. Denoting with $\Psi_{0}$ a wave function for a Kahler potential $K_{0}$, given another potential $K$ the corresponding wave function is 

\begin{equation}
\Psi= \Psi_{0}+ \sum_{k=0}^{[n/2]} a_{k} L^{k}\Lambda^{k} \Psi_{0}
\end{equation}

for some coefficients $a_{k}$. In particular $\Psi^{n,0}= \Psi_{0}^{n,0}$ and $\Psi^{0,n}= \Psi_{0}^{0,n}$.

\item The Hamiltonian is compatible with complex conjugation: if $\Psi_{i}$ is a basis of vacuum wave functions, also the Hermitian conjugates 

\begin{equation}\label{realstructure}
\Psi^{\ast}_{i}= M_{\bar{i}}^{j} \Psi_{j}
\end{equation}

form a complete basis, where the complex conjugation is defined by the real structure $M$ of the ground state.

\item The space of vacua is isomorphic to the chiral ring as vector space 

\begin{equation}
\mathcal{V}\simeq \mathcal{R}.
\end{equation}

Denoting with $\phi_{i}$ a basis of $\mathcal{R}$, the vacuum wave function $\Psi_{i}$ representing the $\phi_{i}$-class reads

\begin{equation}
\Psi_{i}= \phi_{i}dz_{1}\wedge....\wedge dz_{n} + \overline{\mathcal{Q}}-\mathrm{exact \ term}. 
\end{equation}

\end{itemize}

We make a few comments about this result. The vacuum wave functions can be seen as representatives of the $\bar{\partial}_{W}$-cohomology in the Hilbert space. One can show that if the $\bar{\partial}$-cohomology has the structure \ref{dolbcom}, which is guaranteed by the fact that $X$ is a Stein manifold, the $\bar{\partial}_{W}$-cohomology is concentrated in the middle cohomology group $H^{n}_{\bar{\partial}_{W}}$, which is isomorphic to the vector space 

\begin{equation}
H^{n}_{\bar{\partial}_{W}} \simeq \tilde{\mathcal{R}}= \frac{\Omega^{n}}{dW \wedge \Omega^{n-1}},
\end{equation}

namely the space of holomorphic $n$-forms modulo those that can be written as $dW \wedge \alpha$ for some $n-1$ holomorphic form $\alpha$. The map $\phi(z)\rightarrow \phi(z)dz^{1}....dz^{n}$ makes explicit the isomorphism between $\mathcal{\tilde{R}}$ and the chiral ring

\begin{equation}\label{module}
\mathcal{\tilde{R}}\simeq \mathcal{R}
\end{equation}

as modules over $\mathcal{R}$. Moreover, $H^{n}_{\bar{\partial}_{W}}$ has the same local structure of the chiral ring. The fact that the map $dW \wedge :\Omega^{n-1}(M)\rightarrow \Omega^{n}(M)$ fails to be surjective only at the critical points $p_{i}$ of $W$, which are assumed to be isolated, implies that the cohomology is localized around the classical vacua. Indeed, one can prove that

\begin{equation}\label{localization}
\mathcal{\tilde{R}} \simeq \bigoplus_{i} \mathcal{\tilde{R}}_{i}, \hspace{1cm} \mathcal{\tilde{R}}_{i}= \frac{\Omega_{i}^{n}(M)}{dW \wedge \Omega_{i}^{n-1}(M)},
\end{equation}

where $\Omega_{i}^{n}(M)$ are the germs of holomorphic $n$-forms at $p_{i}$. Moreover, each space $\mathcal{\tilde{R}}_{i}$ is isomorphic to the local ring 

\begin{equation}
\mathcal{R}_{i}=\frac{\mathcal{O}_{p_{i}}}{I_{W}}
\end{equation}

as module over $ \mathcal{R}_{i}$. It is proved in \cite{rif14} that each $\bar{\partial}_{W}$-class has representatives with compact support, hence in the Hilbert space, and viceversa a class in the Hilbert space determines a class in the space of smooth forms $\Lambda^{*}(M)$. Therefore, the cohomology $H_{\bar{\partial}_{W}}(M)$ is equivalent to the cohomology in the Hilbert space and we have the isomorphism of vector spaces

\begin{equation}
 H_{\bar{\partial}_{W}}(M)\simeq H_{\bar{\partial}_{W}}(\mathcal{H})\simeq \mathcal{R}.
\end{equation}

The vacuum space has vanishing Lefschetz angular momentum $\tilde{F}$ and forms a trivial representation of the R-symmetry group. In particular, the Witten index $I_{W}=\mathrm{Tr}_{\mathcal{H}}(-1)^{\tilde{F}+n} $ counts the number of vacua up to the sign and is given by the sum of the Milnor numbers 

\begin{equation}
I_{W}= (-1)^{n}\sum_{i} \mu_{i}.
\end{equation}

If $W$ is a Morse function, all the critical points are non degenerate and the dimension of the ground state is equal to the number of classical vacua. Indeed, since they have equal fermion number, the classical vacua cannot be lifted by instantons corrections and, differently from the case with two supercharges, the perturbative computation is always exact.\\ By the Frobenius structure of the chiral ring, the space of vacua is naturally endowed with a symmetric pairing 

\begin{equation}\label{symmpair}
\eta_{ij}= \frac{1}{(2\pi i)^{n}}\int \Psi_{i} \wedge \ast \Psi_{j}= \oint \frac{\phi_{i}\phi_{j} \ dz_{1}...dz_{n}}{\partial_{1}W...\partial_{n}W}= \langle \phi_{i}\phi_{j} \rangle,
\end{equation}

where $\phi_{i},\phi_{j}$ are representative in the chiral ring of the vacua $\Psi_{i},\Psi_{j}$. Viceversa, the scalar product of the Hilbert space allows to define an Hermitian pairing between chiral and anti-chiral operators, also known as $tt^{*}$ metric \cite{rif10}

\begin{equation}
g_{j\bar{i}}=\braket{\overline{\phi}_{i}}{\phi_{j}}= \braket{\Psi_{i}}{\Psi_{j}}= \int \Psi_{j} \wedge \ast \Psi_{j}^{\ast}.
\end{equation}

Using the identities \ref{realstructure}, \ref{symmpair} one finds that the ground state metric $g$, the real structure $M$ and the symmetric pairing $\eta$ are related by 

\begin{equation}
 g_{j\bar{i}}=M_{\bar{i}}^{k} \int \Psi_{j} \wedge \ast \Psi_{k}= \eta_{jk} M_{\bar{i}}^{k}.
\end{equation}

Moreover, since the real structure satisfies $MM^{*}=1$, where $M^{*}$ denotes the complex (not Hermitian) conjugate matrix, one can write an identity between $\eta$ and $g$ 

\begin{equation}\label{realitycon}
\eta^{-1}g(\eta^{-1}g)^{*}=1
\end{equation}

also known as reality constraint.\\ 
A Landau-Ginzburg theory with four supercharges is in particular $\mathcal{N}=2$ supersymmetric. In the $\mathcal{N}=4$ supersymmetry algebra one can define a family of $\mathcal{N}=2$ subalgebras parametrized by an angle $\theta$ as 

\begin{equation}
d_{V_{\theta}}= d+( e^{i\theta}\partial W+e^{-i\theta} \overline{\partial W})\wedge, \hspace{1cm}   \delta_{h_{\theta}}=d_{h_{\theta}}^{\dagger},
\end{equation}

where the real superpotential $V_{\theta}$ is the harmonic function

\begin{equation}
V_{\theta}= 2Re( e^{i\theta}W). 
\end{equation}

On a Kahler space the complex structure is compatible with the Riemannian structure and one can see the vacuum space also as $L^{2}$-cohomology of $d_{V_{\theta}}$.\\ The cohomology classes which label the vacua of the supersymmetric system are defined over the complex numbers. However, it is possible to endow the vacua also with an integral structure. By Morse cobordism, we have an isomorphism between the vacuum space and the relative de Rham cohomology \cite{rif10}

\begin{equation}
\mathcal{V}=H^{*}(X,X_{\mathrm{Re} (e^{i\theta}W)};\mathbb{C}),
\end{equation}

where the set $X_{\mathrm{Re} (e^{i\theta}W)}\subset X$ is defined according to \ref{setv}.  The vacuum wave forms $\Psi_{i}$ are conjugated to some closed $n$-forms $\omega_{i}$ in the relative de Rham classes by 

\begin{equation} \label{relaz}
\Psi_{i}= e^{-2\mathrm{Re}( e^{i\theta}W)} \omega_{i}, \hspace{1cm} i=1,...,N.
\end{equation}

The classes in the dual homology group $H_{n}(X,X_{\mathrm{Re} (e^{i\theta}W)};\mathbb{C})$ are represented by non compact cycles with boundary in the region $X_{\mathrm{Re} (e^{i\theta}W)}$. This space has a natural integral structure given by the homology with integer coefficients 

\begin{equation}
\mathcal{V}^{\vee}\simeq H_{\ast}(X,X_{\mathrm{Re} (e^{i\theta}W)};\mathbb{Z})\otimes_{\mathbb{Z}} \mathbb{C}.
\end{equation}

In case of Morse superpotential a canonical integral basis is given by the Lefschetz thimbles $B_{a}(\theta),a=1,...,N$ describing the gradient flow of the superpotential $e^{i\theta} W$ \cite{rif11,rif19}. These are special Lagrangian middle-dimensional submanifolds of $X$ which arise from the critical point of $W$. The images of these cycles on the $W$-plane are straight lines stretched in the $e^{i\theta}$ direction starting at the critical values of the superpotential. The relative homology is also called the space of branes, because in $2d$ the corresponding objects have the physical interpretation of half-BPS branes \cite{rif11}. In particular, the angle $\theta$ specifies which linear combinations of the original $4$ supercharges leave the brane invariant. One can use the above formula and the standard pairing between cycles and forms to introduce the so called $tt^{*}$ brane amplitudes 

\begin{equation}
\braket{B_{a}(\theta)}{\Psi_{i}}=\Psi_{a,i}(\theta)= \int_{B_{a}(\theta)} e^{2\mathrm{Re}( e^{i\theta}W)} \Psi_{i},
\end{equation}

which define a non-degenerate $N \times N$ matrix whose components are not univalued as function of the couplings and the phase $e^{i\theta}$.

\subsection{$tt^{*}$ Geometry}

\subsubsection{$tt^{*}$ Geometry in SQM}

Once we solve the Schroedinger equation for the zero energy level, it is interesting to study the evolution of the ground states as we vary the parameters of the theory. We consider a family of Landau-Ginzburg models $W(z_{i},t_{a})$ parametrized by a set of holomorphic couplings $t_{a}$. These are local holomorphic coordinates on a complex manifold $\mathcal{P}$. We assume for this family of theories that the target manifold $X$ is a Stein space and the critical points of $W$ are isolated. One defines a complex vector bundle $\mathcal{H}\rightarrow \mathcal{P}$, also called Hilbert bundle, where the fiber is the Hilbert space $\mathcal{H}$. This is a trivial bundle and can be written globally as $\mathcal{H}\times \mathcal{P} \rightarrow \mathcal{P}$. The vacuum bundle $\mathcal{V}\rightarrow \mathcal{P} $ is the subbundle of the Hilbert bundle whose fiber is the space of vacua $\mathcal{H}_{\mathrm{vac}}$. Differently from the Hilbert bundle, the bundle of vacua is non trivial, since the ground state of the Hamiltonian vary non trivially as we change the couplings of the superpotential. The canonical trivial connection of the Hilbert bundle induces by projection $P:\mathcal{H}\rightarrow \mathcal{H}_{\mathrm{vac}} $ a non trivial connection on the vacuum bundle, which is by definition the Berry connection. The equations which prescribe the curvature of the Berry connection are known as $tt^{*}$ equations \cite{rif10,rif14}. We recall that a generic wave form $\Psi$ in the Hilbert space has a unique orthogonal Hodge decomposition 

\begin{equation}
\Psi=\Psi_{0}+\bar{ \partial}_{W}\alpha + \bar{\partial}_{W}^{\dagger} \beta
\end{equation}

where $\Psi_{0}$ is $W$-harmonic. The operator $P$ projects the state along the vacuum space, namely

\begin{equation}
P:\Psi\rightarrow \Psi_{0}.
\end{equation}

The vacua $\Psi_{k},k=1,...,N$, satisfy the equations 

\begin{equation}
\begin{split}
& (\bar{\partial}+ dW\wedge) \Psi_{k}=0, \\
& (\bar{\partial}+ dW\wedge)^{\dagger} \Psi_{k}=0.
\end{split}
\end{equation}

We take the derivative of these equations with respect to some coupling $t_{a}$. Denoting $W_{a}=\partial_{t_{a}}W$, we find 

\begin{equation}\label{connectiondef1}
\begin{split}
\partial_{t_{a}} \left[  (\bar{\partial}+ dW\wedge) \Psi_{k} \right] =&\  \bar{\partial}_{W}(\partial_{t_{a}} \Psi_{k})+ d(\partial_{a}W)\wedge \Psi_{k} \\ = & \ \bar{\partial}_{W}(\partial_{t_{a}} \Psi_{k} ) + \partial_{W}(W_{a}\Psi_{k})=0.
\end{split}
\end{equation}

and 

\begin{equation}\label{connectiondef2}
\partial_{t_{a}} \left[  (\bar{\partial}+ dW\wedge)^{\dagger} \Psi_{k} \right]= \bar{\partial}_{W}^{\dagger}(\partial_{t_{a}}\Psi_{k})=0.
\end{equation}

We introduce the forms $\lambda_{a,k}$ which satisfy 

\begin{equation}
\bar{\partial}_{W}\lambda_{a,k}= W_{a}\Psi_{k}-P(W_{a}\Psi_{k}), \hspace{1cm} \bar{\partial}_{W}^{\dagger}\lambda_{a,k}=0.
\end{equation}

In particular, the first equation is consistent with the fact that $W_{a}\Psi_{k}$ is $\bar{\partial}_{W}$-closed. One can rewrite the equations \ref{connectiondef1} and \ref{connectiondef2} as 

\begin{equation}
\begin{split}
& \bar{\partial}_{W}\left( \partial_{t_{a}}\Psi_{k}-\partial_{W}\lambda_{a,k}\right) =0, \\ 
& \bar{\partial}_{W}^{\dagger}\left( \partial_{t_{a}}\Psi_{k}-\partial_{W}\lambda_{a,k}\right) =0.
\end{split}
\end{equation}

So, we see that $\partial_{t_{a}}\Psi_{k}-\partial_{W}\lambda_{a,k}$ is in the ground state and can be written as linear combination of $\Psi_{k}$. The coefficients of the linear combination defines the Berry connection 

\begin{equation}
\partial_{t_{a}}\Psi_{k}-\partial_{W}\lambda_{a,k}= -(A_{a})_{k}^{l}\Psi_{l}.
\end{equation}

One can introduce the corresponding covariant derivative $D_{a}$ and rewrite this relation as 

\begin{equation}
D_{a}\Psi_{k}= \partial_{W}\lambda_{a,k}.
\end{equation}

We can repeat the same steps also for the other vacuum equations 

\begin{equation}
\begin{split}
& (\partial+ d\overline{W}\wedge) \Psi_{k}=0 ,\\
& (\partial+ d\overline{W}\wedge)^{\dagger} \Psi_{k}=0.
\end{split}
\end{equation}

We find the definition of the antiholomorphic part of the Berry connection 

\begin{equation}
\overline{D}_{a}\Psi_{k}=\bar{\partial}_{W}\tilde{\lambda}_{a,k}
\end{equation},

where $\tilde{\lambda}_{a,k}$ are wave forms satisfying 

\begin{equation}
\begin{split}
\partial_{W}\tilde{\lambda}_{a,k}= \overline{W}_{a}\Psi_{k}-P(\overline{W}_{a}\Psi_{k}), \hspace{1cm}
\partial_{W}^{\dagger} \tilde{\lambda}_{a,k}=0.
\end{split}
\end{equation}

The bundle of vacua has a natural structure of holomorphic vector bundle. A basis of holomorphic sections is given by the operators in the chiral ring $\mathcal{R}$, which are holomorphic representatives of the vacua in the $\bar{\partial}_{W}$-cohomology. The holomorphic operators provide also a basis for the tangent space of the couplings manifold. Indeed, the deformations of the superpotential $W_{a}$ are elements of the chiral ring. Denoting with $\phi_{k}$ a basis for $\mathcal{R}$, the chiral ring algebra is encoded in the relation

\begin{equation}
\phi_{i}\phi_{j}=C_{ij}^{k}\phi_{k},
\end{equation}

where the numbers $C_{ij}^{k}$ are the structure constants of the ring. These matrices describe the action of the chiral operators on the vacua. Projecting on the ground state the action of $W_{a}$ on a wave function $\Psi_{k}$ one finds

\begin{equation}
P(W_{a}\Psi_{k})= W_{a}\phi_{k} \ dz_{1}\wedge...\wedge dz^{n} + \bar{\partial}_{W}(...)= C_{ak}^{l}\phi_{l} \ dz_{1}\wedge...\wedge dz^{n}+ \bar{\partial}_{W}(...).
\end{equation}

The curvature of the Berry connection can be computed in terms of the chiral ring coefficients $(C_{a})_{k}^{l}$ with Hodge theoretical arguments. One gets the $tt^{*}$ equations \cite{rif10}

\begin{equation}
\begin{split}
& [\overline{D}_{a},D_{b}] =-[ \overline{C}_{a},C_{b}] \\ 
& [ D_{a},D_{b} ] = [ C_{a},C_{b}]=0 \\ 
& [ \overline{D}_{a},\overline{D}_{b}] = [ \overline{C}_{a},\overline{C}_{b}]=0 \\ 
\end{split}
\end{equation}

In addition, the chiral ring coefficients satisfy 

\begin{equation}
\overline{D}_{a}C_{b}=D_{a}\overline{C}_{b}=0 \hspace{1cm} D_{a}C_{b}=D_{b}C_{a} \hspace{1cm} \overline{D}_{a}\overline{C}_{b}=\overline{D}_{b}\overline{C}_{a}.
\end{equation}

The structure of the $tt^{*}$ equations depends uniquely on the holomorphic data of the chiral ring, while there is no dependence on the Kahler metric of the target space. In general, once the topology of the target manifold is fixed, the geometry of the vacuum bundle is completely determined by the superpotential. From the above equations one can see that the Berry connection preserves the holomorphic structure of the bundle and is compatible with the metric. In particular, only the $(1,1)$ component of the curvature is non vanishing. Hence, the Berry connection is the unique Chern connection of the bundle of vacua. In an holomorphic trivialization the coefficients of the connections read

\begin{equation}\label{ciao}
A_{ak}^{j}= g_{k\bar{l}}(\partial_{a}g^{-1})^{\bar{l} j}, \hspace{1cm} A_{\bar{a} k}^{j}=0,
\end{equation}

where $g_{k\bar{l}}= \braket{\overline{\phi}_{l}}{\phi_{k}}$ is the ground state metric. One can express $\overline{C}_{a}$ in this basis in terms of $C_{a}$ and $g$ as 

\begin{equation}
(\overline{C}_{a})^{k}_{l}= g_{k\bar{j}} (C^{\dagger})_{\bar{r}}^{\bar{j}}g^{\bar{r}l}= (gC^{\dagger}g^{-1})^{k}_{l}.
\end{equation}

Plugging the expression \ref{ciao} in the $tt^{*}$ equations one obtains a set of differential equations for $g$

\begin{equation}
\bar{\partial}_{i}(g\partial_{j}g^{-1})=\left[ C_{j},\overline{C}_{i}\right].
\end{equation}

Solving this equation with the appropriate boundary conditions and the reality contraint \ref{realitycon} allows to determine the geometry of the vacuum bundle.\\ We specialize the analysis to the class of Landau-Ginzburg models in which the critical points of the superpotential $p_{i},i=1,...,N$ are non degenerate. In this case, as we said in the previous section, the chiral ring is semisimple and factorizes as product of simple factors $\mathcal{R}\simeq \prod_{i=1}^{N}\mathbb{C}_{i}$. Moreover, one can introduce a basis of orthogonal idempotents $e_{i}$ which provide a canonical isomorphism between $\mathcal{R}$ and $\mathbb{C}^{N}$. This is also called `point basis ', since the operators satisfy the property $e_{i}(p_{j})=\delta_{ij}$. In this basis the chiral ring algebra diagonalizes completely 

\begin{equation}
e_{i}e_{j}=\delta_{ij}e_{j},
\end{equation}

and a generic operator $\Phi=\sum_{i=1}^{N}\Phi(p_{i})e_{i}$ can be identified with its set of critical values $\Phi(p_{i})$. It is convenient to rescale the generators $e_{i}$ in order to normalize the symmetric pairing $\eta_{ij}$ of the chiral ring to the identity. For a Morse superpotential the Grothendieck residue \ref{grotform} becomes 

\begin{equation}\label{residue}
\eta_{ij}=\langle \phi_{i}\phi_{j}\rangle= \sum_{p_{i}} \phi_{i}\phi_{j}\left( \mathrm{det} \, \partial_{k}\partial_{l}W\right)^{-1}.
\end{equation}

In the basis of orthogonal idempotent we have 

\begin{equation}
\eta_{ij}= \left( \mathrm{det} \, \partial_{k}\partial_{l}W(p_{i})\right)^{-1} \delta_{ij}, \hspace{1cm} \langle e_{i}\rangle= \left( \mathrm{det} \, \partial_{k}\partial_{l}W(p_{i})\right) ^{-1}.
\end{equation}

The symmetric pairing can be normalized to the identity with the rescaling $e_{i}\rightarrow\sqrt{\mathrm{det} \, \partial_{k}\partial_{l}W(p_{i})}e_{i}$. The so redefined basis is called canonical basis and is unique up to the sign. Since the Berry connection is compatible also with the symmetric pairing, in the canonical basis we have

\begin{equation}
A_{i}^{t}=-A_{i},
\end{equation}

namely the Berry connection takes value in the algebra of the orthogonal group.\\ The LG theories with a Morse superpotential posses a natural system of coordinates in the coupling constant space. The critical map $w:\mathcal{P}\rightarrow \mathbb{C}^{N}$ given by $t_{a}\rightarrow (w_{1}(t_{a}),...,w_{N}(t_{a}))$ is a local immersion and the set of critical values $w_{i}(t_{a})=W(p_{i};t_{a})$ form a local coordinate system on the Frobenius manifold of all couplings of the theory which contains the physical coupling space $\mathcal{P}$ as a submanifold. These are usually called canonical coordinates \cite{rif12,rif32,rif33}. In the canonical basis, the chiral ring coefficients associated to the variations of $w_{i}$ have the simple structure 

\begin{equation}\label{canonicalcoeff}
(C_{i})^{k}_{l}= \delta_{il}\delta^{k}_{l}.
\end{equation}

In these coordinates the $tt^{*}$ equations become universal and inequivalent physical systems are distinguished by different boundary conditions.\\ It is known that the vacuum bundle of a $\mathcal{N}=4$ Landau-Ginzburg theory and its two dimensional counterpart have the same $tt^{*}$ geometry \cite{rif10}. Even if the physical interpretation is different, the formulas are exactly the same and one can use the language of $2d$ $\mathcal{N}=(2,2)$ quantum field theory to study the $tt^{*}$ geometry of the model. The usual strategy to solve the $tt^{*}$ equations requires to rescale the superpotential $W\rightarrow \beta W$ and consider variations of the overall coupling $\beta$. In two dimensions this is equivalent to study the RG flow of the theory \cite{rif10}. Even though the superpotential is protected by non-renormalization theorems, the F-term picks up a factor due to the rescaling of the superspace coordinates. From $z\rightarrow \beta z$ and $\theta \rightarrow \beta^{-1/2} \theta $ we get 

\begin{equation}
\int d^{2}\theta d^{2}z \ W \longrightarrow \beta \int  d^{2}\theta d^{2}z \ W.
\end{equation}

Variations in the overall factor $\beta$ generate a flow which has UV limit for $\beta \rightarrow 0$ and IR limit for $\beta \rightarrow \infty$.\\ We denote with $\dot{\mathcal{P}} \in \mathcal{P}$ the dense open domain in the space of couplings in which $\mathcal{R}$ is semisimple and $w_{i}\neq w_{j}$ for $i\neq j$. In other words, $\dot{\mathcal{P}}$ is the domain in which $W$ is strictly Morse. A quantity of great interest is the Berry's connection in the direction of the RG flow \cite{rif12,rif20}

\begin{equation}
Q=\iota_{\varepsilon}A= \sum_{i}w^{i}A_{i},
\end{equation}

where $\varepsilon=\sum_{i}w^{i}\partial_{w_{i}}=\frac{1}{2}\beta\partial_{\beta}$ is the Euler vector in $\dot{\mathcal{P}}$. The equations $ D_{i}C_{j}= D_{j}C_{i}$ in the canonical coordinates become 

\begin{equation}
\left[ A_{i},C_{j}\right] = \left[ A_{j},C_{i}\right]. 
\end{equation}

From this equality and the definition of $Q$ we find 

\begin{equation}
\left[ A_{i}, w^{j}C_{j}\right] = \left[ A_{j}w^{j},C_{i}\right] =\left[ Q,C_{i} \right] .
\end{equation}

The $k,l$ component of the above expression is 

\begin{equation}
(A_{i})_{kl}w^{l}-w^{k}(A_{i})_{kl}= Q_{ki}-Q_{li},
\end{equation}

which allows to write the Berry's connection in terms of $Q$ only as \cite{rif32}

\begin{equation}\label{basicformula}
(g\partial g^{-1})_{kl}=Q_{kl}\frac{d(w_{k}-w_{l})}{w_{k}-w_{l}},
\end{equation}

where we used the fact that $Q$ is antisymmetric in the canonical basis. Hence, it is enough to know $Q$ in order to specify completely the solution of the $tt^{*}$ equations.\\ In $2d \ (2,2)$ models $Q_{ij}$ is a pseudo-index which plays two roles \cite{rif20}. First, it captures  the half-BPS solitons with boundary conditions given by the $i,j$-th vacuum repsectively at $x\rightarrow \pm \infty$. In the IR limit the $tt^{*}$ solution can be written in the form of soliton expansion, with the boundary condition represented by the soliton spectrum. The IR region in canonical coordinates is identified by large masses $m_{ij}= 2\vert w_{i}-w_{j} \vert$ of the kinks interpolating the $i$-th and $j$-th vacua. In the canonical basis the IR expansion $(\beta\rightarrow \infty)$ of the metric and $Q_{ij}$ are \cite{rif12}

\begin{equation}
\begin{split}
& g_{i\bar{j}}\simeq \delta_{ij}-\frac{i}{\pi}\mu_{ij} K_{0}(m_{ij}\beta) \\ 
& Q_{ij} \simeq -\frac{i}{2\pi} \mu_{ij}m_{ij}\beta K_{1}(m_{ij}\beta)
\end{split}
\end{equation}

where $K_{0},K_{1}$ are the modified Bessel function and $\mu_{ij}$ is the soliton matrix which counts with sign the number of soliton species in the $i,j$ sector. The existence of regular solutions requires $\mu_{ij}$ to be real, although they are integral in the physical case. Moreover, the CPT invariance imposes $\mu_{ji}=-\mu_{ij}$.\\
The second role of $Q_{ij}$ is that it captures the $U(1)_{R}$ charges of the Ramond vacua at the UV fixed point. As we said, the UV fixed point is reached when $\beta \rightarrow 0$. If there are no Landau poles along the RG flow and the UV theory exists, the model can flow to a SCFT or an asymptotically free theory. In this limit the operators in the chiral ring are the chiral primary fields of the conformal theory \cite{rif10,rif12}. It is known that a generic Landau-Ginzburg theory at the critical point gains the $U(1)_{V}$ R-symmetry, which is broken off-criticality by the superpotential. The associated charge is the scaling dimension of the conformal fields. One can choose a basis of chiral primaries for the chiral ring and order the R-charges $q_{i},i=1,...,N=\mathrm{dim}\mathcal{R}$ in a non-decreasing sequence

\begin{equation}
0=q_{1}\leq q_{2}\leq...\leq q_{N}=c/3=q_{\mathrm{max}},
\end{equation}

where $c$ is the central charge of the CFT. In particular, there is always an operator with $0$ charge in the spectrum which is given by the identity operator. The Berry connection in the direction of the RG flow has the same eigenvalues of the generator of the $U(1)_{R}$ symmetry $\boldsymbol{Q}$ of the $(2,2)$ SCFT \cite{rif10,rif12,rif20}. In this sense, the Berry's connection associated to the RG flow can be seen as an off-criticality definition of the $U(1)_{R}$ generator. The eigenvalues of $Q$ at the critical point are the Ramond charges of the vacua \cite{rif10}

\begin{equation}
q_{i}^{R}=q_{i}-\hat{c}/2,
\end{equation}

which are symmetrically distributed between $-\hat{c}/2$ and $\hat{c}/2$.

\subsubsection{The Integral Formulation of $tt^{*}$ Geometry}\label{integral} 

We previously introduced D-branes to provide the vacua with an integral structure. One can define a dual bundle whose fiber is the lattice of relative homology cycles which pair with the vacuum wave functions. This bundle is parametrized by a phase $\zeta \in \mathbb{P}^{1}, \vert \zeta \vert=1$ which defines which half of supersymmetry is preserved by the D-branes. The pairing between D-branes and vacua in the Hilbert space allows to define a basis of sections for the vacuum bundle 

\begin{equation}
\Psi_{i,a}= \braket{\phi_{i}}{B_{a}(\zeta)}.
\end{equation}

We remind that, because of the integral structure of D-branes, these amplitudes are locally constant in the couplings $w_{i}$ and the spectral parameter $\zeta$. We can define a family of flat connections parametrized by $\zeta$ \cite{rif12}

\begin{equation}
\begin{split}
& \nabla_{i}^{\zeta}= \partial_{i}+(g\partial_{i}g^{-1})-\zeta C_{i}, \\ \\ & \bar{\nabla}_{i}^{\zeta}= \bar{\partial}_{i}-\zeta^{-1}\overline{C}_{i},
\end{split}
\end{equation}

anso known as $tt^{*}$ Lax connection. The $tt^{*}$ equations for the Berry connection can be rephrased as the flatness conditions for the above connection, i.e.

\begin{equation}
(\nabla^{\zeta})^{2}= (\bar{\nabla}^{\zeta})^{2} =  \nabla^{\zeta} \bar{\nabla}^{\zeta} + \bar{\nabla}^{\zeta} \nabla^{\zeta}=0.
\end{equation}

The D-brane amplitudes are horizontal sections of the Lax connection 

\begin{equation}
\nabla_{i}^{\zeta}\Psi^{a}(\zeta)=\bar{\nabla}_{i}^{\zeta}\Psi^{a}(\zeta)=0.
\end{equation}

Differently from the Berry connection, the Lax connection is not compatible with the ground state metric and provides a non unitary representation of $\pi_{1}(\mathcal{P})$. Let $\Psi(w_{i},\zeta)$ be a $N\times N$ matrix whose columns are linearly independent solutions of the above linear equation. Taking the analytic continuation of the solutions along a non trivial loop $\gamma$ in the space of couplings we obtain the monodromy $\rho_{\zeta}(\gamma)$ defined by

\begin{equation}
\Psi(w_{i},\zeta)\rightarrow \Psi(w_{i},\zeta)\rho_{\zeta}(\gamma).
\end{equation}

Since the Lax connection is flat, the matrix $\rho_{\zeta}(\gamma)$ depends only on the homotopy class of $\gamma$. The $tt^{*}$ monodromy representation $\rho_{\zeta}$ in $GL(N,\mathbb{C})$ may be conjugated such that it lays in the arithmetic subgroup $SL(N,\mathbb{Z})$. Since the branes are representatives of integral homology classes, for each $\zeta \in \mathbb{P}^{1}$ they define a local system on $\mathcal{P}$ canonically equipped with a flat connection, the Gauss-Manin one. Dually, the brane amplitudes define a $\mathbb{P}^{1}$- family of flat connections on the vacuum bundle which is naturally identified with the $\mathbb{P}^{1}$-family of $tt^{*}$ Lax connections. Hence, modulo conjugation we have 

\begin{equation}
\rho_{\zeta}: \pi_{1}(\mathcal{P})\rightarrow SL(N,\mathbb{Z}).
\end{equation}

Since the entries of the matrix $\rho_{\zeta}(\gamma)$ and its inverse are integers, they are locally indipendent from the couplings and the spectral parameter $\zeta$. This implies that the $tt^{*}$ monodromy representations is indipendent from the point $t \in \mathcal{P}$ and $\zeta$. \\ The $tt^{*}$ geometry is a set of isomonodromic equations for the Lax linear system and admits a reformulation as Riemann-Hilbert problem \cite{rif12,rif33}. It is known that the solution is captured by the monodromy of the flat sections around the movable singularities. These are poles for the equations and have the $2d$ physical interpretation of UV fixed points in the parameter space. A condition we need to incorporate in this setup is the indipendence of the ground state metric from an overall rotations of $w_{i}$. Indeed, an overall phase can always be absorbed in the fermionic measure of the superspace. We can consider the dependence of $\Psi_{a}$ on the RG scale $\beta \in \mathbb{R}_{\geq 0}$ and a phase $e^{i\theta}$ by redefining the canonical coordinates as 

\begin{equation}
w_{i}\rightarrow \beta e^{i\theta} w_{i}.
\end{equation} 

After the identification $\zeta= e^{i\theta}$, the Lax equations become 

\begin{equation}\label{linearprobl}
\begin{split}
\zeta\partial_{\zeta} \Psi= \left( \beta\zeta C + Q-\beta \zeta^{-1}\overline{C} \right) \Psi ,\\ \\ 
\beta\partial_{\beta}\Psi= \left(  \beta\zeta C + Q+\beta \zeta^{-1}\overline{C}  \right) \Psi.
\end{split}
\end{equation}

The compatibility of these equations automatically implies the indipendence of the $tt^{*}$ solution from the angle $\theta$. Indeed, $Q,C, \overline{C}$ are consistently indipendent from $\zeta$.\\ The identification of the overall phase with the spectral parameter allows to extend its domain to whole complex plane. Thus, the equation above has two singular points in $\zeta=0,\infty$ and $\Psi(\zeta,w_{i})$ undergoes monodromy as $w_{i}\rightarrow e^{2\pi i} w_{i}$. In this case there are two Stokes sector which can be chosen to be the upper and lower half $\zeta$-plane. The solutions on these two half planes $\Psi_{+},\Psi_{-}$ overlap on the real line and are related by

\begin{equation}\label{RH}
\begin{split}
&\Psi_{-}(y)=\Psi_{+}(y)S \\ \\ 
&\Psi_{-}(-y)=\Psi_{+}(-y)S^{t}
\end{split}
\end{equation}

where $y>0$. The stokes matrix $S$ is given by the phase-ordered product of the Stokes jumps 

\begin{equation}\label{piclef}
M_{ij}=\delta_{ij}-A_{ij}
\end{equation}

which are generated when $\zeta$ crosses the BPS central charge $Z_{ij}=2(w_{i}-w_{j})$ of a primitive soliton connecting the $j$-th to the $i$-th vacuum. The matrix $A_{ij}$ has a unique non vanishing coefficient in the $ij$ entry. In the basis of Lefschetz thimbles this is equal to the BPS multiplicity $A_{ij}=\mu_{ij}$ of the $ij$ sector. When $\zeta$ spans the whole upper half plane in the anticlockwise sense we find the definition of the Stokes matrix

\begin{equation}
S= \prod_{0< \arg Z_{ij} < \pi} M_{ij},
\end{equation}

where we are assuming that there are no solitons with central charges aligned with the Stokes axes. This can always be achieved with a rotation of the axes.\\ One can solve the Riemann-Hilbert problem \ref{RH} imposing the correct boundary condition at the infinity of the $\zeta$-plane

\begin{equation}\label{conditmetric}
\lim_{\zeta\rightarrow \infty} \Psi(x)\exp \left[ \beta \left( xC+x^{-1}C^{\dagger} \right) \right] =1.
\end{equation}

Using this boundary condition and the known identity

\begin{equation}
\frac{1}{x-y \mp i\epsilon}= P\frac{1}{x-y} \pm i\pi \delta(x-y)
\end{equation}

one rewrites the Riemann-Hilbert problem as the integral equation

\begin{equation}
\begin{split}
\Phi(x)_{ij}= \delta_{ij}& +\frac{1}{2\pi i}\sum_{k} \int_{0}^{\infty} \frac{dy}{y-\zeta+i\epsilon} \Phi(y)_{ik} A_{kj}e^{-\beta(y\delta_{kj}+y^{-1}\bar{\delta}_{kj}} \\ & + \frac{1}{2\pi i}\sum_{k} \int_{-\infty}^{0} \frac{dy}{y-\zeta+i\epsilon} \Phi(y)_{ik} A^{\mathrm{t}}_{kj}e^{-\beta(y\delta_{kj}+y^{-1}\bar{\delta}_{kj}},
\end{split}
\end{equation}

where $\delta_{kj}=w_{k}-w_{j}$. The Riemann-Hilbert problem has a unique solution given by the piecewise constant function $\Psi = (\Psi_{+}, \Psi _{−})$, where   

\begin{equation}
\Psi_{+}(x)=\Phi(x)\exp\left[-\beta\left( \zeta C+\zeta^{-1}C^{\dagger} \right) \right] .
\end{equation}

The solution $\Psi = (\Psi_{+}, \Psi _{−})$ satisfy \cite{rif12,rif33} 

\begin{equation}
\begin{split}
& \Psi(x)\Psi^{t}(-x)=1 \\ 
& \overline\Psi(1/\bar{x})=g^{-1}\Psi(x),
\end{split}
\end{equation}

where the second relation means that in the canonical basis the complex conjugation acts on the vacuum wave functions as the ground state metric. Indeed, in this basis we have $\eta=1$ and so the ground state metric and the real structure coincide. From the above relations and the boundary condition \ref{conditmetric} one can extract the ground state metric 

\begin{equation}
g_{i\bar{j}}=\lim_{\zeta\rightarrow 0}\Phi(x)_{i\bar{j}}.
\end{equation}

The object which specifies the $tt^{*}$ solution is the monodromy of the D-brane states around the singular point $\zeta=0$, also known as the quantum monodromy \cite{rif12}

\begin{equation}
H= S(S^{t})^{-1}.
\end{equation}

According to the theory of isomonodromic deformations, the monodromy group of the Lax connection is invariant up to conjugacy under deformations of the parameters and can be computed in the limit that we prefer. The eigenvalues of $H$ are the phases $e^{2\pi i q^{R}_{j}}$ which encode the Ramond charges of the chiral primary fields at the UV conformal point. Given the dependence of the Stokes matrix on the BPS multiplicities, the quantum monodromy puts in relation the spectrum of solitons and chiral primaries, providing a map between IR and UV fixed points. The quantum monodromy is an element of $SL(N,\mathbb{Z})$ and may have non trivial Jordan blocks. It is known that a non trivial Jordan structure is related to logarithmic violations of scaling and reveals that the theory is asymptotocially free in the UV \cite{rif12}.

\subsubsection{Computing the $tt^{*}$ Monodromy Representation}\label{uvapproach}

Our main problem is to compute the $SL(N,\mathbb{Z})$ monodromy representation of $\pi_{1}(\mathcal{P})$ given by the $tt^{*}$ Lax connection. We are mostly interested in the case in which the chiral ring of the theory is semi-simple. As we mentioned before, since $tt^{*}$ is an isomonodromic problem, we have the freedom to deform continuously the model in the coupling constant space. The monodromy groups we find will be related by an overall conjugation. In particular, the Jordan block structure and the eigenvalues of the monodromy matrices  are invariant under finite continuous deformations. However, the useful limits in which the computations really simplify are typically singular limits in the coupling constant space. The monodromy representation that we find in these limits may be related to the original by a singular conjugation matrix. The eigenvalues of the matrices will not be modified, but the Jordan blocks may decompose in smaller ones. Indeed, having a certain spacetrum is a closed condition in the matrix space, while having Jordan blocks of dimension $>1$ is a closed condition. Therefore, the monodromy eigenvalues are typically easy to compute, while knowing the Jordan structure is subtler. However, in many situations we know a priori that the monodromy matrix is semi-simple and so we do not loose any information. For instance, as in the case relevant for the FQHE, when $\pi_{1}(\mathcal{P})$ is a complicated non-Abelian group the Jordan blocks are severely restricted by the group relations and so it is plausible that they can be recovered from the knowledge of the eigenvalues. The limits that are relevant for our purposes are the so-called asymmetric limit and the UV limit (in the $2d$ language). In the first approach we rescale the critical values 

\begin{equation}
w_{i}\rightarrow \beta w_{i}
\end{equation}

so that the $tt^{*}$ flat connection becomes 

\begin{equation}
\nabla_{\zeta}=D + \frac{\beta}{\zeta}C, \hspace{1cm} \overline{\nabla}_{\zeta}= \overline{D}+\zeta \beta \overline{C},
\end{equation}

while the Berry curvature changes as 

\begin{equation}
F=-\beta^{2}\left[ C, \overline{C}\right] .
\end{equation}

The asymmetric limit consists in taking the unphysical limit $\beta\rightarrow 0$ and $\zeta\rightarrow \infty$ with $\beta/\zeta$ fixed. The Berry curvature vanishes in the limit, so the metric connection $A$ is pure gauge. The $tt^*$ linear problem \ref{linearprobl} reduces formally to

\begin{equation}
\left( \partial_{i}+ A_{i} + \frac{\beta}{\zeta}C_{i} \right) \Psi= \bar{\partial}_{i}\Psi=0.
\end{equation}

A basis of solutions to this equation is given by the integrals

\begin{equation}
\Psi_{i}^{a}= \int_{\Gamma_{a}} \phi_{i}(z)e^{\frac{\beta}{\zeta} W(z;w)}dz^{1}\wedge...\wedge dz^{n},
\end{equation}

where the cycles $\Gamma_{a}$ are an integral basis of branes and $\phi_{i}(z)$ is a basis of the chiral ring. The homology classes of the branes with given $\zeta$ are locally constant in coupling constant space, but jump at loci where (in the 2d language) there are BPS solitons which preserve the same two supercharges as the branes \cite{rif12}. The jump in homology at such a locus are given by the Picard-Lefshetz transformations of the forms \ref{piclef}. In the canonical basis of branes, namely the Lefshetz thimbles, the Stokes jumps can be written in terms of the soliton multiplicities computed in the IR limit of the $2d$ theory. Taking into account all the jumps in homology one encounters along the path (controlled by the 2d BPS spectrum), one gets the monodromy matrix which is automatically integral of determinant $1$. The full monodromy representation is given by the combinatorics of the PL transformations. This approach is convenient from the practical point of view and the monodromy matrices computed is this limit are manifestly integral. On the other side, the fact that we consider a limit which do not correspond to any unitary quantum system tends to make the physics somewhat obscure. For our present purposes the UV approach seems more natural. As discussed before, this just requires to send $\beta\rightarrow 0$, while the spectral parameter $\zeta$ remain fixed and can be taken in the unitary locus $\vert \zeta \vert=1$. One can see that in this limit the Lax connection reduces to the Berry one 

\begin{equation}
\nabla_{\zeta} +\overline{\nabla}_{\zeta}  \xrightarrow{\beta \rightarrow 0} D + \overline{D}.
\end{equation}

Hence, the Berry connection becomes flat in this limit. Since the monodromy of the flat connection is independent of $\beta$, the flat UV Berry connection should have the same monodromy modulo the subtlety with the size of the Jordan blocks. This concept can be clarified by comparing the quantum monodromy computed in the UV and IR limit of the theory. The holonomy on the $\zeta$-plane of the UV Berry connection, which corresponds to the Lax connection in the same limit, is conjugated to the unitary matrix $e^{2\pi i \boldsymbol{Q}}$. On the other hand, as we discussed previously, the quantum monodromy is computed in the IR limit in terms of the Stokes matrices as $H=SS^{-t}$. The two matrices have the same unitary spectrum, but for asymptotically free theories the IR monodromy has non trivial Jordan blocks. It is clear in general that the monodromy representation computed in the UV limit is unitary, since the Berry connection is metric, but differently from the homological approach the integral structure of the matrices is not manifest. In order to give an explicit description of the holonomy representation of the UV Berry connection we need additional details about the $tt^{*}$ geometry that we want to study.

\section{The Vafa Model and the Microscopic Physics of FQHE}\label{wedisc}

\subsection{The Supersymmetric Structure of Landau Levels}

In this section we start to explore the connections between supersymmetry and the physics of quantum Hall effect. It turns out that the structure of the lowest Landau level naturally admits a $\mathcal{N}=4$ supersymmetric description. One can describe the Landau levels of a single electron on a Riemann surface $\Sigma$ of arbitrary genus by rephrasing the construction in section \ref{eleclandau} in a more geometrical language. We consider a complex line bundle $\mathcal{L}\rightarrow \Sigma $ with first Chern class $c_{1}(\mathcal{L})= \Phi/2\pi$, where $\Phi>0$ is the total magnetic flux through the surface $\Sigma$. The complex structure of the Riemann surface is irrelevant in the discussion and we can choose it according to convenience. A state of the system is represented by a smooth section $\psi: \Sigma\rightarrow \mathcal{L}$ of the line bundle. Every complex line bundle over a one dimensional complex manifold is holomorphic and so one can always find a local holomorphic trivialization. The line bundle is endowed with an hermitian metric $h$ which allows to define an inner product on the Hilbert space 

\begin{equation}
\braket{\psi_{1}}{\psi_{2}}= \int_{\Sigma} dzd\bar{z} h \ \psi_{1}^{*}\psi_{2}.
\end{equation}

The lowest Landau level is defined as the kernel of the Hamiltonian operator 

\begin{equation}
\tilde{H}=-D_{z}D_{\bar{z}} + E_{0}
\end{equation}

where $D$ is the connection with respect to the $U(1)$ structure group of the bundle. More precisely, $D$ is the Chern connection associated to $h$. In an holomorphic trivialization one has 

\begin{equation}
D_{z}=-h\partial_{z}h^{-1}, \hspace{1cm} D_{\bar{z}}=\partial_{\bar{z}}.
\end{equation}

The magnetic field is encoded in the curvature of the connection, which is a closed and real $(1,1)$ form 

\begin{equation}
F_{h}=\bar{\partial}\partial \log h,
\end{equation}

where $\partial, \bar{\partial}$ are the Dolbeault operators of the complex manifold. Choosing natural units, in the case of the complex plane one has $h= e^{-B\vert z \vert^{2}/2}$ and $F_{h}=\frac{B}{2}dzd\bar{z}$. The Riemann surface $\Sigma$ endowed with the curvature of the line bundle, which is positive for $B>0$, is naturally a Kahler manifold. The fact that the configuration space of an electron in a constant magnetic field is a Kahler manifold, in particular a Stein space with a globally defined Kahler potential $K=\log h$, provides a first important hint that quantum Hall effect may be related to $\mathcal{N}=4$ supersymmetry. Isomorphism classes of line bundles over a compact Riemann surface are in correspondence with the divisors of the surface modulo linear equivalence. These form a free abelian group on the points of the surface. We write a divisor $D$ as a finite linear combination of points 

\begin{equation}
D= \sum_{i} n_{i} p_{i},
\end{equation}

where $p_{i} \in \Sigma$ and $n_{i}$ are integers. We demand the divisor to be effective, namely with $n_{i}>0$. Since the electrons are particles with spin, in this discussion we have also to endow the line bundle with a spin structure. This is associated with a divisor $S$ on the Riemann surface such that $2S$ is in the class of canonical divisors. The divisor identifying the twisted line bundle is the sum $D+S$ and is unique up to linear equivalence. The vacuum wave functions satisfy $D_{\bar{z}} \phi=0$ and so we have the definition 

\begin{equation}
\mathcal{H}_{\mathrm{LLL}}= \Gamma (\Sigma, \mathcal{L}(D+S)),
\end{equation}

where $\mathcal{H}_{\mathrm{LLL}} $ is the lowest Landau level and $\Gamma (\Sigma, \mathcal{L}(D+S))$ is the space of holomorphic sections of the twisted line bundle $\mathcal{L}(D+S)$. The Riemann-Roch theorem for a compact genus-$g$ Riemann surface states that the dimension $\ell(D+S)$ of the vector space $\Gamma (\Sigma, \mathcal{L}(D+S))$ is 

\begin{equation}
\ell(D+S)= \mathrm{deg}(D+S)-g+1
\end{equation}

where $\mathrm{deg}(D+S)$ is the degree of the divisor. This result holds as long as $D+S$ has degree at least $2g-1$. For definiteness we put the system in a finite box with periodic boundary conditions. This is equivalent to choose an elliptic curve $E$ as target manifold for the electron coordinate. It is convenient to pick an even spin structure associated to a divisor $S=p_{0}-q$, where $p_{0},q$ are distinct points on the elliptic curve which satisfy $2p_{0}=2q$. The choice of $p_{0}$ does not affect the discussion and we can translate it as we prefer. In the case of a genus-$1$ surface the above relation becomes 

\begin{equation}
\ell(D+S)= \mathrm{deg}(D+S)= \sum_{i}n_{i}.
\end{equation}

Since $S$ has vanishing degree, the dimension of the lowest Landau level is entirely determined by $D$. This is consistent with the fact that the spin of the electrons in FQH systems is a frozen degree of freedom and is irrelevant for the physics of the ground state. From this equality we see that the divisor $D$ parametrizes topologically the magnetic flux of the system. In particular, the degree of the divisor is equal to the total magnetic flux $\Phi/2\pi$ which defines the degeneracy of the lowest Landau level. The divisor $D+S$ has a defining meromorphic section $\psi_{0}$ with zeros of order $n_{i}$ at $p_{i}$, a simple zero at $p_{0}$ and a single pole at $q$. The map $\psi\rightarrow \psi/\psi_{0}$ provides a canonical identification between holomorphic sections in  $\Gamma (E, \mathcal{L}(D+S))$ and meromorphic functions on the elliptic curve with poles at most given by $D_{\infty}=D+p_{0}$ and vanishing in $q$.\\ Taking the limit of infinite volume we recover a system of electrons moving on the complex plane. In this case the spin structure is associated to a divisor $S=-q$, where $q$ is a generic point on the plane. The contribution of $\mathrm{deg} \,S=-1$ cancels the $+1$ factor in the Riemann-Roch formula and we have again

\begin{equation}
\ell(D-q)= \mathrm{deg}D= \sum_{i}n_{i}.
\end{equation}

The states in the lowest Landau level are represented by elements $\psi \in \Gamma(\mathbb{P}^{1},\mathcal{L}(D-q))$, which are canonically associated by the map $\psi\rightarrow \psi/\psi_{0}$ to meromorphic functions on $\mathbb{P}^{1}$ with poles at most given by $D$ and vanishing at $q$. \\ In order to get the Hilbert space of the $N$ electrons system we simply need to take the antisymmetric tensor product of the single particle space

\begin{equation}
\mathcal{H}_{\Phi}= \bigwedge^{N} \Gamma (\Sigma, \mathcal{L}(D+S)),
\end{equation}

where the dimension is given by Fermi statistics $\begin{pmatrix} \Phi/2\pi \\ N \end{pmatrix}$.\\ We can show that the problem of studying the Landau levels of a charged particle in magnetic field is equivalent to find the vacua of a $\mathcal{N}=4$ supersymmetric Hamiltonian. In particular, the holomorphic structure of Landau levels require to have four supercharges. As we recalled in the previous section, a theory in supersymmetric quantum mechanics with four supercharges is specified by the choice of a Kahler potential $K$ and an holomorphic superpotential $W$. The Kahler potential prescribes an hermitian metric on the target space. Once the topology of the Riemann surface is fixed, the choice of the Kahler metric does not affect the structure of the vacua, which depends only on the superpotential. Hence, we can choose $K=\log h$ as globally defined Kahler potential on the Riemann surface. The number of vacua is given by the Witten index, which is equal to the number of zeros counted with multiplicitly of the $1$-form $dW$. In order to compare this description with the previous one we choose as target space the manifold $\mathcal{K}=E\setminus \mathrm{supp}\ F$, where $E$ is an elliptic curve and $F$ is an effective divisor on $E$. As we discussed above, a divisor identifies up to linear equivalence a line bundle with its set of holomorphic sections. However, for a linear combination of points on the Riemann surface one can associate also a $\mathcal{N}=4$ supersymmetric system. Given an effective divisor $D=\sum_{i}n_{i}p_{i}$, we assign a closed meromorphic $1$-form $dW$ on $E$ with zeros in $p_{i}$ of order $n_{i}$. We choose $W^{\prime}(z)$ such that its polar divisor is given by $F$ and in making a precise dictionary between the two models we can set $p_{0} \in \mathrm{supp}\ F$. The ground states of a $\mathcal{N}=4$ theory are in correspondence with cohomology classes of the susy charge $\overline{\mathcal{Q}}= \bar{\partial} + dW \wedge$ in the space of differential forms, which are labelled by holomorphic operators of the chiral ring $\mathcal{R}$. The meromorphic functions $\psi/\psi_{0}$ defines canonically a basis of wave forms for the supersymmetric vacuum space $\mathcal{V}$ through the map 

\begin{equation}
\frac{\psi}{\psi_{0}}\rightarrow \frac{\psi}{\psi_{0}} W^{\prime} dz + \overline{\mathcal{Q}}(....).
\end{equation}

Despite we can write $\frac{\psi}{\psi_{0}} W^{\prime}dz=\overline{\mathcal{Q}}(\frac{\psi}{\psi_{0}} )$, the $1$-forms above define non trivial representatives in the $\overline{\mathcal{Q}}$ cohomology, since the meromorphic functions $\frac{\psi}{\psi_{0}}$ are singular at the zeros of the superpotential and so are not elements of the chiral ring.\\ When the electrons move on the complex plane the corresponding $\mathcal{N}=4$ Landau-Ginzburg model is defined by a one-form $dW$ which is a rational differential on the plane with a pole of order $\geq 2$ at $\infty$ and zero-divisor $D=\sum_{i}n_{i}z_{i}$, namely 

\begin{equation}
dW(z)=\frac{\prod_{i}(z-z_{i})^{n_{i}}}{P(z)}, \hspace{0.5 cm} \mathrm{deg}P(z)\leq \sum_{i}n_{i}=\frac{\Phi}{2\pi}.
\end{equation}

The prescription about the behaviour of $dW$ at $\infty$ implies that the scalar potential $\vert W^{\prime}\vert^{2}$ in the supersymmetric hamiltonian is non vanishing at infinity for all the complete Kahler metrics on the complex plane. This condition ensures the normalizability of the vacuum wave functions and the existence of an energy gap between the ground state and the first excited level of the Hamiltonian. Moreover, it guarantees the absence of run-away vacua in the $2d$ $\mathcal{N}=(2,2)$ version of the theory. In the correspondence with the Landau description we choose the reference point of the spin structure such that $q \not\in \mathrm{Supp} \ D\cup \left\lbrace \infty \right\rbrace$. As in the case of periodic boundary conditions, the meromorphic functions $\frac{\psi}{\psi_{0}}W^{\prime}$ define a basis of chiral operators labelling the supersymmetric vacua of the model. In conclusion, we have the isomorphisms of vector spaces

\begin{equation}\label{isom}
U_{\Phi}: \mathcal{H}_{\mathrm{LLL}} \xrightarrow{\sim} \mathcal{V}, \hspace{0.5 cm} U_{\Phi}:\mathcal{H}_{\Phi} \xrightarrow{\sim} \mathcal{V}_{N}= \bigwedge^{N}\mathcal{V}
\end{equation}

which confirms the first prediction in the list \ref{list}.

\subsection{Comparison of the Hermitian Structures}

The correspondence between the lowest Landau level and the vacuum space of the LG model does not guarantee that the two systems have the same geometry of the vacuum bundle. One can conclude that the vacuum sector of a supersymmetric system and the lowest Landau level are two equivalent description of the same physical system if the above isomorphism is also an isometry between Hilbert spaces. If the norm of the Hilbert space is preserved by the above map, the Berry connection induced on the vacuum bundle and consequently the topological order of the ground state are the same in the two systems.\\
We focus on the case of the complex plane as target manifold and consider a system with a single electron. The generalization to the multi-particle case is straightforward. In the Landau description the probability of finding an electron at a position $z$ is given by

\begin{equation}
P(z)_{\mathrm{LLL}} = \vert f(z) \vert^{2} e^{-B \vert z \vert^{2}},
\end{equation}

where $f(z)$ is an holomorphic function and $B>0$. On a macroscopic volume the probability measure satisfies

\begin{equation}\label{plll}
\log P(z)_{\mathrm{LLL}} = -B\vert z \vert^{2} + \mathrm{subleading \ as \ } \vert z \vert \rightarrow \infty.
\end{equation}

An exact identification between the hermitian structures of the Hilbert spaces is a too strong requirement. What we can ask is an equivalence in measure, i.e. measurements of long range observables on small but macroscopic domains $U \subset \mathbb{C}$ give the same answers on the two sides. This is enough for our purposes, since the long range observables are the ones that characterize the quantum topological order. According to the previous discussion, the supersymmetric representation of the Landau levels requires to introduce a meromorphic $1$-form $dW$ on the plane, whose primitive plays the role of superpotential. We consider a differential form $dW(z)$ with $\Phi/2\pi$ zeroes and polar divisor of the type $F=F_{f}+ 2\infty$. This has the form 

\begin{equation}\label{modelledby}
dW(z)= \left( \mu + \sum_{i=1}^{\Phi/2\pi} \frac{a_{i}}{z-\zeta_{i}} \right) dz, \hspace{0.5 cm} F_{f}=\sum_{i} \zeta_{i}
\end{equation} 
 
where $\mu, a_{i} \in \mathbb{C}^{\times}$ and $\zeta_{i}$ are all distinct. Since we can always redefine the overall phase of the superpotential, we assume $a_{i} \in \mathbb{R}$ without loss of generality. In order to reproduce a macroscopically uniform magnetic field we should take the residues $a_{i}$ all equal and consider a uniform distribution of the flux sources $\zeta_{i}$ in $\mathbb{C}$. The typical separation of the punctures should be much smaller than the size of the macroscopic domain $U \subset \mathbb{C}$ on which we want to measure observables. In the present context being macroscopic means

\begin{equation}
\frac{1}{2\pi}\left( \mathrm{magnetic \ flux \ through \ U}\right) =\int_{U} \frac{B}{2\pi}= \# \left\lbrace \zeta_{i} \in U \right\rbrace  \gg 1,
\end{equation}

namely the domain $U$ contains a large number of fluxes. The supersymmetric wave functions have the form

\begin{equation}
\psi(z)_{\mathrm{SUSY}}= \Phi(z)dz + \tilde{\Phi}(z)d\overline{z}
\end{equation}

and the corresponding probability distribution is 

\begin{equation}
P(z)_{\mathrm{SUSY}}= \vert \Phi(z) \vert^{2} + \vert \tilde{\Phi}(z) \vert^{2}.
\end{equation}

We can choose a real basis of wave functions such that $\tilde{\Phi}(z)=\overline{\Phi}(z)$. Then, the Schroedinger equation for the zero energy levels is \cite{rif14} 

\begin{equation}
\left( -\frac{\partial^{2}}{\partial z \partial\bar{z}} + \bigg\vert \frac{dW}{dz} \bigg\vert^{2}\right) \frac{\Phi}{W^{\prime}}=0.
\end{equation}

An asymptotic behaviour of the solution in the macroscopic limit which is consistent with the Schroedinger equation is 

\begin{equation}
\Phi(z)= e^{\pm 2  \mathrm{Re}  W(z) + \mathrm{\ subleading \ as \ } \vert z \vert \rightarrow \infty},
\end{equation}

where the subleading terms are smooth and bounded functions in the domain $U$ which ensure that the wave function is single-valued and normalizable on the whole complex plane. We notice that the function 

\begin{equation}
2  \mathrm{Re} \, W= \mu z + \overline{\mu z}+  \sum_{i}a_{i}\log\vert z-\zeta_{i}\vert^{2}
\end{equation}

is the electrostatic potential of a system of point-like charges of size $a_{i}$ at positions $\zeta_{i}$ superimposed to a constant background electric field $\mu$. When averaged on a macroscopic region $U$, it looks like the potential for a continuous charge distribution with density $\sigma(z)$ such that

\begin{equation}
\int_{U}d^{2}z \ \sigma(z)= \sum_{\zeta_{i} \in U}a_{i} 
\end{equation}

for any $U \subset \mathbb{C}$. The conclusion is that for any macroscopic domain $U\subset \mathbb{C}$ we have\footnote{Recall that the volume form of $\mathbb{R}^{2}$ is $dx\wedge dy=\frac{i}{2}dz \wedge d\bar{z}$. }

\begin{equation}
\begin{split}
\mathrm{magnetic \ flux \ through \ U}  =& \,  \frac{i}{2}\int_{U}\bar{\partial}\partial \log P(z)_{\mathrm{SUSY}} \approx \pm i \int_{U} \bar{\partial} \partial (2 \mathrm{Re} \, W)= \\ \\ &  \mp 2\pi\sum_{\zeta_{i} \in U} a_{i},
\end{split}
\end{equation}

where in the last equality we used the Poisson equation of $2d$ electrostatics. It is clear that the background electric potential does not contribute to the magnetic flux of the system. We see that $\log P(z)_{\mathrm{SUSY}}$ matches the behaviour of $\log P(z)_{\mathrm{LLL}}$ in \ref{plll} when averaged on any macroscopic domain $U \subset \mathbb{C}$ iff we set respectively $a_{i}=-1$ or $a_{i}=+1$ for all $\zeta_{i}$. The two choices are related by a change of orientation. We fix the conventions so that the external magnetic field is modelled in the susy side by \ref{modelledby} with $a_{i} = -1$ for all $i$.

\subsection{Including Defects and Interactions}

It is very natural to introduce defects on the supersymmetric side. One just need to flip the sign of the residues $a_{i}$ for a bunch of punctures. Now there is a small mismatch between the number of vacua and the effective magnetic field  measured by the fall-off of the wave function at infinity: we have two extra vacua per defect. The extra vacua are localized near the position of the corresponding defect in the plane and may be interpreted as internal states of the defect. We identify these defects with the quasi-holes of FQHE.\\ We have shown in the previous sections that the low energy physics of a system of charged electrons in a uniform magnetic field is described at large $B$ by a quantum system with Hilbert space $\mathcal{H}_{\Phi}$. Moreover, this system admits two equivalent descriptions which are related by the isomorphism $U_{\Phi}$ in \ref{isom}. In particular, the original FQHE Hamiltonian $H_{\mathrm{FQHE}}=U_{\Phi}^{-1}\hat{H}U_{\Phi}$ is mapped to some Hamiltonian $\hat{H}$ which can be seen as a deformation of a $4$-SQM model. The free part of the Hamiltonian si supersymmetric and the corresponding superpotential is given by the sum of $N$ copies of the single particle superpotential describing the interaction between a single electron and the punctures. The interacting part of the Hamiltonian can be splitted in two groups: the interactions which preserve supersymmetry and the ones which are susy-breaking. Including the susy-preserving interaction we get a differential $d\mathcal{W}$ of the form 

\begin{equation}
d\mathcal{W}= \sum_{i}^{N}\left( dW(z_{i})+ \sum_{a=1}^{h}\frac{dz_{i}}{z-x_{a}}\right) + \sum_{i=1}^{N}U_{i}(z_{1},...,z_{N})dz_{i}.
\end{equation}

The term $dW(z)$ models the interaction between the electrons and the macroscopic magnetic field. The Vafa proposal for this term is 

\begin{equation}
dW_{\mathrm{Vafa}}(z)=-\sum_{k=1}^{\Phi/2\pi-h} \frac{dz}{z-\zeta_{k}},
\end{equation}

where $\zeta_{k}$ form a regular lattice. Working on the plane is convenient to add a constant to $dW_{\mathrm{Vafa}}(z)$ in order to regularize the double pole at infinity. This may be seen as integration constant for the electrostatic Poisson equation satisfied by $2\mathrm{Re} W$ and does not affect the magnetic flux of the system. Hence, one obtains

\begin{equation}
dW(z)= dW_{\mathrm{Vafa}}(z)+ \mu dz
\end{equation}

where $\mu\neq 0$. The meromorphic $1$-form $U_{i}dz_{i}$ describes the Coulomb interaction between an electron and the other ones. As a function of the position $z_{i}$ of the i-electron at fixed $z_{j\neq i}$, this form can have poles only when $z_{i} = z_{j}$. Generically $U_{i}dz_{i}$ has only simple poles and the residues must be entire bounded functions on the plane, namely they are constants. Since $\mathcal{W}$ must be symmetric under permutation of the electron coordinates, the most general interacting part of the superpotential reads

\begin{equation}
d\mathcal{W}_{\mathrm{int}}= 2\beta\sum_{1\leq i<j\leq N} \frac{d(z_{i}-z_{j})}{z_{i}-z_{j}},
\end{equation}

where $\beta$ is some complex coupling. We are mostly interested in FQHE on the plane and in such case $\beta$ can be left a generic coplex parameter. However, if we put the system on a finite box with periodic boundary conditions $\beta$ gets quantized to a rational number. In this case $d\mathcal{W}$ is a meromorphic $1$-form on $E^{N}$, where $E$ is an elliptic curve. If we project the model on the target space of a single electron $z_{i}$ by fixing the positions $z_{j},j\neq i$ of the other ones, we obtain a meromorphic differential on $E$ with poles at $\zeta_{k}$ with residue $-1$, at $x_{a}$ with residue $+1$ and at $z_{j}, j\neq i $ with residue $2\beta$. Since the sum of the residues of an elliptic function must vanish, we have the condition

\begin{equation}
0= -\left( \frac{\Phi}{2\pi}-h \right)  + h + 2\beta (N-1)\approx (2\beta \nu-1)\frac{\Phi}{2\pi},
\end{equation}

where in the last equality we have used $N\gg 1$ and the FQHE relation $N=\nu \Phi/2\pi$. Hence, we obtain the quantization condition 

\begin{equation}
2\beta= 1/\nu \in \mathbb{Q}_{>0},
\end{equation}

which is the value given in \cite{rif1}. So, the Vafa model on $E^{N}$ is described by the differential 

\begin{equation}
d\mathcal{W}=\sum_{i=1}^{N}\left( \sum_{a}U(z_{i},x_{a})-\sum_{k}U(z_{i},\zeta_{k}) + \frac{1}{\nu}\sum_{j\neq i}U(z_{i},z_{j})\right) ,
\end{equation}

where 

\begin{equation}
U(z,w)=\frac{\wp^{\prime}(w/2)dz}{\wp(z-w/2)-\wp(w/2)}.
\end{equation}

\subsection{Emergence of a Single Vacuum}

The isomorphism $U_{\Phi}$ maps the original Landau Hamiltonian to the new Hamiltonian 

\begin{equation}
\hat{H}=H_{\mathcal{W}}+ H_{\mathrm{su.br}},
\end{equation}

which provides an equivalent description of FQHE systems. The piece $H_{\mathcal{W}}$ is the supersymmetric Hamiltonian associated to the superpotential $\mathcal{W}$ discussed previously, while $H_{\mathrm{su.br}}$ is the susy-breaking  part. For large magnetic fields $H_{\mathcal{W}}$ is of order $O(B)$, while $H_{\mathrm{su.br}}$ is of order $O(1)$ and represents a small perturbation. However, this does not mean and we can neglect it when we study the topological order of FQHE. Indeed, the supersymmetric part alone cannot be in the same universality class of FQHE states. As we are going to discuss in the next section, the vacuum sector with Fermi statistics $\mathcal{V}_{\mathrm{Fer}}$ of the Vafa Hamiltonian has dimension $d=\begin{pmatrix} \Phi/2\pi \\ N \end{pmatrix}$. The vacuum bundle of the theory is an holomorphic vector bundle of rank $d$ which is endowed with the $tt^{*}$ flat connection $\nabla$ which extends holomorphically the Berry connection $D$. In particular, the topological order of $H_{\mathcal{W}}$ is captured by the monodromy representations of $\nabla$. According to the Laughlin argument in the original description of FQHE, the degeneracy of the ground state is lifted by the Coulomb interactions which select a single vacuum. In a similar way, the susy-breaking part of the Hamiltonian should select a unique ground state $\ket{\mathrm{vac}}$ in the Hilbert space $H_{\Phi}$ which encodes the topological quantum order of FQHE. The sub-bundle over the coupling constant space with fiber spanned by $\ket{\mathrm{vac}}$ is endowed with two canonical sub-bundle connections $\nabla^{\mathrm{vac}},D^{\mathrm{vac}}$ which are induced from $\nabla$ and $D$ respectively. In general the sub-bundle curvature is different from the curvatural of the original one and the monodromy of $\nabla^{\mathrm{vac}}$ is a priori neither well defined nor simply related to the one of $\nabla$. Hence, a priori there is no relation between the topological order captured by $H_{\mathcal{W}}$ and the FQHE one. In order to have such relation the following two conditions must be satisfied:

\begin{itemize}

\item  The monodromy representation of $\nabla$ must be reducible with an invariant sub-bundle of rank $1$. The fiber of this eigenbundle defines a unique preferred vacuum for the $\mathcal{N}=4$ Hamiltonian.

\item The physical vacuum $\ket{\mathrm{vac}}$ is mapped by the isomorphism $U_{\Phi}$ to the preferred vacuum of the susy Hamiltonian.

\end{itemize}

The first question is purely related to the supersymmetric model. It is suggested in $\cite{rif1}$ that such a preferred vacuum exists and should correspond to the identity operator. While this sounds as a natural guess, it is in general not true in $tt^{*}$ geometry that the identity operator spans an invariant subspace of the flat connection. This is an extremely non trivial fact that we have to check by studying the monodromy representation of $\nabla$. The validity of the second point is based on the fact that the preferred vacuum, if it exists, should be the most symmetric one under permutations of the quasi-holes. Then one may argue euristically that \cite{rif1}, indipendently from the details of the interactions between electrons\footnote{ A fundamental requirement the interactions between electrons should satisfy is to vanish in the limit $\vert z_{i}-z_{j}\vert\rightarrow \infty$. This ensures that the dynamics of the system at large $\vert z \vert$ is governed by the magnetic field $B$, so that on any macroscopic domain the probability of finding the electrons at positions $z_{i}$ satisfies $\log P(z_{i})\sim -B\sum_{i}\vert z_{i} \vert^{2}$. The interaction term may diverge when two particles become closer. This is not an issue, since the hyperplanes $z_{i}=z_{j}$ are not part of the configuration space of $N$ identical particles and the wave functions of the Hilbert space are vanishing there.}, as long as the susy-breaking part $H_{\mathrm{su.br}}$ respects the permutation symmetry of electrons and quasi-holes, the unique vacuum $\ket{\mathrm{vac}}$ will also be the maximally symmetric one. This state should correspond to the preferred vacuum of the susy Hamiltonian.\\ The conclusion is that, under our mild assumptions, the quantum order of the FQHE is captured by the 4-susy SQM model proposed in \cite{rif1}. In the present discussion we are actually arguing more than this. Since our considerations do not depend on the details of the interactions between the electrons, we claim that the supersymmetric model represents the correct universality class of any multi-particle system in a strong uniform magnetic field.

\section{Symmetry and Statistics}\label{chap5}

\subsection{$tt^{*}$ Functoriality}\label{functandcovering}

Supersymmetric quantum mechanics is functorial with respect to branched coverings \cite{rif10}. Let us consider a Landau-Ginzburg theory with target space $X_{1}$ and superpotential $W(x;\lambda)$ depending on a set of couplings $\lambda \in \mathcal{P}$. We denote with $f:X_{2}\rightarrow X_{1}$ an holomorphic but not globally invertible map between a covering space $X_{2}$ and $X_{1}$. We assume that the map is indipendent from the point of the coupling constant space. One can use $f$ to pull-back the LG model on $X_{2}$. Denoting with $x_{i}, y_{i}, \ i=1,...,n$ a set of coordinates on $X_{1},X_{2}$ respectively, the superpotential of the covering model can be obtained by making the substitution $x_{i}=f_{i}(y_{j})$, namely

\begin{equation}
f^{*}W(x_{i})= W(f_{i}(y_{j}))=W_{f}(y_{j}).
\end{equation}

For the susy charges we have

\begin{equation}
\mathcal{Q}= f^{*}\mathcal{Q}= \partial+d\overline{W}_{f} , \hspace{1cm} \overline{\mathcal{Q}}= f^{*}\overline{\mathcal{Q}}= \bar{\partial}+dW_{f},
\end{equation}

while the Lefschetz operators can be defined by pulling-back the Kahler form on the cover space. The algebraic relations satisfied by the generators of the supersymmetry algebra are preserved by the pull-back operation. The pull-back of the vacuum wave forms of the original theory are not in general ground states for the covering model. Given  a basis of vacuum wave functions $\Psi_{k}, \ k=1,...,N_{1}$, the pulled-back forms $f^{*}\Psi_{k}$ are cohomologous to the true vacua on $X_{2}$. Indeed, the conditions solved by the vacua which are compatible with the pull-back are 

\begin{equation}
\begin{split}
& \bar{\partial}_{f^{*}W} f^{*}\Psi_{k}= f^{*} (\bar{\partial}_{W}\Psi_{k})=0 \\ 
& \partial_{f^{*}W} f^{*}\Psi_{k}= f^{*} (\partial_{W}\Psi_{k})=0 \\
& f^{*}\omega \wedge f^{*}\psi_{k}= f^{*}( \omega \wedge \Psi_{k})=0,
\end{split}
\end{equation}

where $f^{*}\omega$ is the pulled-back Kahler form. The vacuum equations involving the charges $\mathcal{Q}^{\dagger},\overline{\mathcal{Q}}^{\dagger}$ depend D-term and in general are not preserved by the pull-back. The covering theory has generically more vacua than the original LG model. The number of vacua in the new theory $N_{2}$ is related to $N_{1}$ by

\begin{equation}
N_{2}= \mathrm{deg}f \cdot N_{1} + \# \mathrm{zeroes} \ \mathrm{of} \  J_{f},
\end{equation}

where $\mathrm{deg} f$ is the degree of the cover and $J_{f}$ is the Jacobian of $f$. From the pulled-back expression of the vacuum wave functions 

\begin{equation}
\begin{split}
f^{*}\Psi_{k}= & f^{*}\left( \phi_{k}(z_{i}) dz_{1}\wedge....\wedge dz_{n} +\bar{\partial}_{W}(....)\right)  \\ = & \phi_{k}(f_{i}(y_{j}))J_{f} \ dy_{1} \wedge....\wedge dy_{n} + \bar{\partial}_{W_{f}}(....)
\end{split}
\end{equation}

one can read how the chiral ring roperators of the vacua transform by pull-back

\begin{equation}
f^{\sharp}(\phi_{k})= f^{*}(\phi_{k})J_{f}.
\end{equation}

The linear map $f^{\sharp}: \mathcal{R}_{1}\rightarrow \mathcal{R}_{2}$ is an isometry for the topological metric

\begin{equation}\label{cond1}
\langle f^{\sharp}(\phi_{k}), f^{\sharp}(\phi_{j}) \rangle_{X_{2}} = \langle \phi_{k} ,\phi_{j} \rangle_{X_{1}}.
\end{equation}

and is compatible with the $\mathcal{R}_{1}$-module structure

\begin{equation}\label{cond2}
f^{\sharp}(\phi_{k}\cdot \phi_{j})= f^{*}(\phi_{k})\cdot f^{\sharp}(\phi_{j}) \in \mathcal{R}_{2}.
\end{equation}

The chiral ring $\mathcal{R}_{2}$ decomposes as direct sum

\begin{equation}
\mathcal{R}_{2}= f^{\sharp}(\mathcal{R}_{1}) \oplus f^{\sharp}(\mathcal{R}_{1})^{\perp},
\end{equation}

where $(\cdot)^{\perp}$ denotes the orthogonal complement with respect to the $tt^{*}$ metric. The $tt^{*}$ functoriality is the statement that $f^{\sharp}: \mathcal{R}_{1}\rightarrow \mathcal{R}_{2}$ is an isometry also for the $tt^{*}$ metric. In order to show this fact one has to check that the two $tt^{*}$ metrics solve the same equations and satisfy the same boundary conditions. Since the classes in $R_{2}$ of the operators $\partial_{\lambda}W$ belong to the subspace $f^{\sharp}(R_{1})$, the chiral ring coefficients $C_{\lambda}$ are functorial by \ref{cond2}. Since also the topological metric $\eta_{ij}$ is functorial by the \ref{cond1}, we conclude that the $tt^{*}$ equations are preserved by $f^{\sharp}$. The boundary conditions which select the correct solution to the $tt^{*}$ equations are encoded in the 2d BPS soliton multiplicities. The BPS solitons are the connected preimages of straight lines in the $W$-plane ending at critical points \cite{rif12,rif19}. Since the map $W_{f}:X_{2}\rightarrow \mathbb{C}$: factorizes through $W:X_{1}\rightarrow \mathbb{C}$, so do the counterimages of straight lines, and therefore the counting of solitons agrees in the two theories.\\ $tt^{*}$ functoriality preserves also the integral structure of D-branes. If $B\subset X_{2}$ is a D-brane for the LG model on $X_{2}$, then $f(B)$ is a D-brane for the model on $X_{1}$. In particular, $B$ is a special Lagrangian submanifold of $X_{2}$ and this property is invariant under pull-back and push-forward operations by smooth functions. The functorial property of SQM regards also the integral pairing between branes and vacua

\begin{equation}
\braket{f(B)}{\Psi}_{X_{1}}= \braket{B}{f^{*}\Psi}_{X_{2}}.
\end{equation}

One gives in general the following definition: a $tt^{*}$-duality between two $4$-susy theories is a Frobenius algebra isomorphism between their chiral rings $\mathcal{R}_{2}\rightarrow \mathcal{R}_{1}$ which is an isometry for the $tt^{*}$ metric and so for the brane amplitudes. $tt^{*}$ functoriality produces several interesting $tt^{*}$-dual pairs and for an appropriate choice of the respective D-terms implies the equaivalence of the full quantum theories.

\subsection{Covering Spaces}\label{form}

\subsubsection{Abelian Covers}

The functorial property of SQM is particularly important when we study Landau-Ginzburg models in which the target manifold $X$ is not simply connected and the superpotential differential $dW$ is a closed meromorphic $1$-form. Despite it is not possible to define a primitive $W$ on $X$, the model is still well defined, since the Hamiltonian and the susy charges depend only on the derivatives of the superpotential. However, in these systems one can consider deformations of the theory which are not described by operators in the chiral ring \cite{rif13}. For instance, the chiral operator associated to the RG flow deformation is precisely the superpotential, which is not an holomorphic function on $X$. In order to write and solve the $tt^{*}$ equations with respect to these variations we need to pull-back the model on the universal cover $\mathcal{K} $ of the target manifold. This can be defined in an abstract way as the space of curves 

\begin{equation}\label{defcov}
\begin{split}
& \hspace{3cm}\mathcal{K}= \left\lbrace p:[0,1]\longrightarrow  X, p(0)=p^{*} \in X \right\rbrace / \sim, \\ \\ 
& \sim: \hspace{2cm} 
p\sim q=
\begin{cases}
p(1)=q(1), \\ 
p \cdot q^{-1} =0 \ \mathrm{in \ the \ fundamental \ group} \  \pi_{1}(X,\mathbb{Z}).
\end{cases}
\end{split}
\end{equation}

We can pull-back $dW$ on this space and give a formal definition of superpotential:

\begin{equation}
W(p)= \int_{p} dW.
\end{equation}

The fundamental group $\pi_{1}(X)$ plays the role of Galois group of the cover and is a symmetry group of the model on $\mathcal{K}$. The action of a loop generator $\ell$ of $\pi_{1}(X)$ on $W$ is

\begin{equation}\label{homologygenerator}
\ell^{*}W(p)= \int_{\ell \cdot p} dW= W(p) + \int_{\ell} dW.
\end{equation}

Compatibily with the definition of symmetry, the superpotential is left invariant up to a constant factor. The Galois group of the universal cover is generically non abelian and may not have unitary representations. In the context of quantum mechanics, since the symmetry transformations must be unitary, one considers tipically the abelian universal cover of the target space. We note that the action of the fundamental group is abelianized at the level of superpotential and the model naturally descends on the abelian universal cover. This space is the minimal simply connected cover of the target space on which $W$ is single-valued. The Galois group of the abelian cover is obtained by dividing the fundamental group $\pi_{1}(X)$ by the commutator subgroup $\left[ \pi_{1}(X),\pi_{1}(X)\right] $. This subgroup is normal in $\pi_{1}(X)$ and so the quotient $\pi(X)^{\mathrm{Ab}}=\pi_{1}(X)/\left[ \pi_{1}(X),\pi_{1}(X)\right]$ is still a Galois group. More precisely, it is the first homology group $H_{1}(X;\mathbb{Z})$ of the target space. By replacing the fundamental group with the homology group in \ref{defcov}, one gets the definition of abelian universal cover $\mathcal{A}$ of $X$. The first homology of the target space is an abelian group with a freely generated part and a torsion part. It is clear by the above formula that the torsion subgroup has a trivial action on the superpotential. Hence, one can consider only the torsion-free part $ H_{1}(X;\mathbb{Z})/\mathrm{tor}\simeq\mathbb{Z}^{b_{1}}$, where $b_{1}$ is the first Betti number.\\ The vacuum space of the LG model on $\mathcal{A}$ decomposes in a direct sum of unitary irreducible representations of the homology group:

\begin{equation}
\mathcal{V}_{\mathcal{A}}= \bigoplus_{\chi \in \mathrm{Hom}\left( H_{1}(X;\mathbb{Z}),U(1)\right) } \mathcal{V}_{\chi}, \hspace{1cm} \mathrm{dim}\mathcal{V}_{\chi}=d,
\end{equation}

where $d=\mathrm{dim}\mathcal{V}_{X}$ is the dimension of the vacuum space of the theory on $X$. Identifying $H_{1}(X;\mathbb{Z})$ with $\mathbb{Z}^{b_{1}}$, the characters labelling the unitary representations $\mathcal{V}_{\chi}$ can be written as 

\begin{equation}
\chi_{\vec{\theta}}: \vec{n}\rightarrow e^{i \vec{n}\cdot\vec{\theta}}
\end{equation}

and we can call $\vec{\theta}$-vacua the states in the eigenspace $\mathcal{V}_{\chi}=\mathcal{V}_{\vec{\theta}}$.
Since it is metric, the orthogonal decomposition of the vacuum space in $\theta$-sectors is preserved by the parallel transport with the Berry connection. However, it is not generically left invariant by the $tt^{*}$ Lax connection. The subrepresentation of the $tt^{*}$ monodromy are associated with subgroups of the Galois group. Let $H\in \pi_{1}(X)^{\mathrm{Ab}}$ a subgroup and let $\mathcal{A}_{H}=\mathcal{A}/H$. We have an Abelian cover $\mathcal{A}_{H}\rightarrow X$ with Galois group $\pi_{1}(X)^{\mathrm{Ab}}/H$. This cover is not simply connected and the fundamental group $\pi_{1}(\mathcal{A}_{H})$ is the kernel of the surjective homomorphism

\begin{equation}
\beta: \pi_{1}(X)^{\mathrm{Ab}}\rightarrow \pi_{1}(X)^{\mathrm{Ab}}/H,
\end{equation}

which is precisely the normal subgroup $\mathrm{ker}  \beta= H$. One can consistently formulate the $4$-SQM model on the target space $\mathcal{A}_{H}$. The vacuum space of the LG theory on $\mathcal{A}_{H}$ is identified with the $H$-invariant subspace of the theory on the universal cover

\begin{equation}
\mathcal{V}_{H}=\bigoplus_{\chi: \chi\vert_{H}=\mathrm{trivial}} \mathcal{V}_{\chi}.
\end{equation}

The covering map $\mathcal{A}\rightarrow\mathcal{A}/H$ allows to pull-back on $\mathcal{A}$ also the branes and the $tt^{*}$ brane amplitudes of the $\mathcal{A}/H$-theory. Hence, we know by $tt^{*}$ functoriality that $\mathcal{V}_{H}$ must be preserved by the $tt^{*}$ monodromy. We conclude that for each subgroup $H$ of $\pi_{1}(X)^{\mathrm{Ab}}$ we have a monodromy subrepresentation $\mathrm{Mon}_{H}$ of the Lax connection. Moreover, to a sequence of subgroups 

\begin{equation}
...\subset H_{k} \subset H_{k-1} \subset ....\subset H_{1} \subset H_{0}= \pi_{1}(X)^{\mathrm{Ab}},
\end{equation}

there corresponds an inverse sequence of $tt^{*}$ monodromy representations

\begin{equation}
\mathrm{Mon}_{H_{0}} \subset \mathrm{Mon}_{H_{1}} \subset ....\subset \mathrm{Mon}_{H_{k-1}} \subset \mathrm{Mon}_{H_{k}} \subset .... 
\end{equation}

where $\mathrm{Mon}_{H_{0}}$ is the monodromy representation for the original model defined on $X$.\\ We can choose a subgroup $H$ of finite index in $\pi_{1}(X)^{\mathrm{Ab}}$ in such a way that $\pi_{1}(X)^{\mathrm{Ab}}/H$ is a finite Abelian torsion group. In this case the theory on $\mathcal{A}_{H}$ has finite Witten index

\begin{equation}
d_{H}=\left[ \pi(X)^{\mathrm{Ab}} : H \right]\cdot d.
\end{equation}

A periodic character $\chi$ of $\pi_{1}(X)^{\mathrm{Ab}}$ is identified by theta angles

\begin{equation}
\vec{\theta} \in (2\pi \mathbb{Q})^{b_{1}}.
\end{equation}

Denoting with $J_{\chi}$ the finite cyclic group generated by $\chi$ and $H_{\chi} = \mathrm{ker} \chi \in \pi_{1}(X)^{\mathrm{Ab}}$ the corresponding finite-index normal subgroup, we have

\begin{equation}
\pi_{1}(X)^{\mathrm{Ab}}/H_{\chi} \simeq J_{\chi}.
\end{equation}

In this case we may reduce from the infinite universal cover to a finite cover with Galois group $J_{\chi}$. From the physical viewpoint, torsion characters $\chi$ have the special property that they allow a consisten truncation of the chiral ring $\mathcal{R}_{\mathcal{A}}$ to a finite-dimensional ring $\mathcal{R_{\chi}}$. In this way the $\vec{\theta}$-vacua become normalizable, which is a basic requirement in quantum mechanics, while they are never normalizable for non-torsion $\chi$. 

\subsubsection{$tt^{*}$ Equations with $\theta$-Vacua}\label{cov}

On the universal cover we have more vacua than on the target space, but also a larger symmetry group to classify them. We denote with $z_{i},i=1,...,d$ the zeroes of $dW$ on $X$, which we assume to be non degenerate. Choosing a representative $p^{0}_{i}$ of $z_{i}$ on the abelian cover, all the critica points of $W$ can be obtained by composing $p^{0}_{i}$ with the generators $\ell_{1},...,\ell_{b_{1}}$ of the torsion-free part of the homology group $H_{1}(X,\mathbb{Z})/\mathrm{torsion}$. Hence, the points $z_{i}$ label equivalence classes of vacua which are isomorphic to the Galois group of the cover.\\ A basis for the chiral ring of the theory on $\mathcal{A}$ can be constructed as follows. Since $X$ is a Stein space we can find a basis of holomorphic one forms $\rho_{k},k=1,...,b_{1} \in \Omega(X)$ dual to $\ell_{k}$ whose classes generates $H^{1}(X,\mathbb{Z})/\mathrm{tor}$. Since the homology has an integer structure, we can choose the one forms $\rho_{k}$ such that

\begin{equation}
\int_{\ell_{k}}\rho_{k}= c_{k} \in \mathbb{Z}.
\end{equation}

Without loss of generality we can normalize the constants $c_{k}$ to $1$. On $\mathcal{A}$ we can find holomorphic functions $h_{k}$ such that $dh_{k}=\rho_{k}$. Let $ \lbrace \phi_{a} \rbrace \in \mathcal{R}_{X}$ be holomorphic functions on $X$ forming a basis for the chiral ring of the model, with $\phi_{0}=1_{X}$ and relations $\phi_{a}\phi_{b}=C_{ab}^{c}\phi_{c}$. Then, let $\varpi:\mathcal{A}\rightarrow X$ be the projection map from the abelian universal cover to the target space. The holomorphic functions on $\mathcal{A}$

\begin{equation}
\Phi_{a}(\vec{\theta})= \varpi^{*}\phi_{a} \cdot e^{i \vec{\theta}\cdot \vec{h}}, \hspace{1cm} \vec{\theta} \in [0,2\pi)^{b_{1}},
\end{equation}

form a basis for $\mathcal{R}_{\mathcal{A}}$ which is diagonal in the characters of the Galois group. Indeed, the action of a generator of the homology $\ell_{k}$ on the above state is 

\begin{equation}
\ell_{k}( \varpi^{*}\phi_{a} \cdot e^{i \vec{\theta}\cdot \vec{h}})= \varpi^{*}\phi_{a} \cdot e^{i \vec{\theta}\cdot ( \vec{h} + \int_{\ell_{k}}d\vec{h}) } =e^{i\theta_{k}} \varpi^{*}\phi_{a} \cdot e^{i \vec{\theta}\cdot \vec{h}}
\end{equation}

The product table  of $\mathcal{R}_{\mathcal{A}}$ is 

\begin{equation}
\Phi_{a}(\vec{\theta})\cdot \Phi_{a}(\vec{\varphi})= C_{ab}^{c}\Phi_{c}(\vec{\theta}+\vec{\varphi}).
\end{equation}

This equation implies that the Ramond charges of the chiral operators are piece-wise linear functions of the angles. The same applies to the UV Berry connection $A(\vec{\theta})^{\mathrm{UV}}$, which is function of the $U(1)_{R}$ charges. The discontinuous jumps of $A(\vec{\theta})^{UV}$ correspond to gauge transformations. On the contrary, the characters of the monodromy representation are continuous. For generic $\vec{\theta}$ the eigenvalues of the monodromy matrices are distinct, and hence no Jordan blocks are present; at characters where we have jumps typically non-trivial Jordan blocks appear.\\ Let $H$ a subgroup of $\pi_{1}(X)^{\mathrm{Ab}}$ and $J$ the subgroup of characters which are trivial on $H$. Then, the chiral ring $\mathcal{R}_{H}$ of the model on $\mathcal{A}_{H}$ is spanned by the chiral operators 

\begin{equation}
\lbrace\Phi_{a}(\vec{\theta})\rbrace_{\vec{\theta} \in J}.
\end{equation}

Using the fact that $ \ell_{k}$ are symmetries of the model, one finds that the ground state metric diagonalizes with respect to the angles:

\begin{equation}
\braket{\overline{\Phi_{j}(\vec{\theta})}}{\Phi_{k}(\vec{\theta^{\prime}})}= \delta(\vec{\theta}-\vec{\theta^{\prime}}) g_{k,\bar{j}}(\vec{\theta}), \hspace{1cm} g_{k,\bar{j}}(\vec{\theta})= \sum_{\vec{r} \in \mathbb{Z}^{b_{1}}} e^{i \vec{\theta}\cdot \vec{r}} g_{k,\bar{j}}.
\end{equation}

In presence of theta sectors, the complex conjugate operator $g^{*}(\vec{\theta})$ must be intended as the Fourier series of the complex conjugated coefficients, namely

\begin{equation}
g^{*}(\vec{\theta})=[g(-\vec{\theta})]^{*}.
\end{equation}

Hence, using the fact that $g(\vec{\theta})$ must be hermitian, the $tt^{*}$ reality constraint in the canonical basis becomes

\begin{equation}\label{odd}
g(-\vec{\theta})^{t}=g(\vec{\theta})^{-1}.
\end{equation}

We can use the action of the homology generators \ref{homologygenerator} to compute the critical values of the superpotential. These reads 

\begin{equation}
W( (\ell_{1}^{n_{1}}....\ \ell_{b_{1}}^{n_{b_{1}}})\cdot p_{0}^{i})= (\ell_{1}^{n_{1}} ....\ \ell_{b_{1}}^{n_{b_{1}}})^{*}W(p_{0}^{i})= W(p_{0}^{i}) + \sum_{i=1}^{b_{1}} n_{i}\omega_{i},
\end{equation}

where $\omega_{i}=\int_{\ell_{i}}dW$ are constants in the chiral fields. We choose $\omega_{i}$ as the first $b_{1}$ local coordinates on the coupling constant space $\mathcal{P}$ and denote with $t_{a} \in \mathcal{P}$ the remaining couplings such that $\partial_{t_{a}}W$ are well defined holomorphic functions representing elements of $\mathcal{R}_{X}$. The $tt^{*}$ metric can be thought as function of the variables 

\begin{equation}
(\omega_{i},\theta_{i}) \in (\mathbb{C}\times S^{1})^{b_{1}}
\end{equation}

at fixed $t_{a}$. One can consistently define the action of the chiral operator $C_{\omega_{i}}$ associated to the coupling $\omega_{i}$ on the theta vacua. This can be see as a $U(d)$ covariant derivative in the $\theta_{i}$-direction 

\begin{equation}\label{togeqn}
C_{\omega_{i}}= \overline{D}_{\overline{\vartheta}_{i}}= \frac{\partial}{\partial\theta_{i}}+ M_{\omega_{i}}, \hspace{1cm} \overline{C}_{\overline{\omega}_{i}}=-D_{\vartheta_{i}} =\frac{\partial}{\partial\theta_{i}}- g\partial_{\theta_{i}}g^{-1}-g M^{\dagger}_{\omega_{i}} g^{-1},
\end{equation}

where $M_{\omega_{i}}$ is a $d\times d$ matrix. At fixed $t_{a}$, the component of the $tt^{*}$ flat connection take the form 

\begin{equation}
D_{\omega_{i}}+ \frac{1}{\zeta}\overline{D}_{\overline{\vartheta}_{i}}= \boldsymbol{D}^{(\zeta)}_{1,i}, \hspace{1cm} \overline{D}_{\overline{\omega}_{i}}-\zeta D_{\vartheta_{i}}= \boldsymbol{D}^{(\zeta)}_{2,i}
\end{equation} 

We can see the $\vartheta_{i}$ as complex coordinates with real part $\theta_{i}$ and introduce the new complex coordinates $(\eta^{\zeta}_{i},\xi^{\zeta}_{i}), i = 1,...,b_{1}$ 

\begin{equation}
\eta^{\zeta}_{i} = \omega_{i} -\zeta \overline{\vartheta}_{i}, \hspace{1cm} \xi^{\zeta}_{i} = \overline{\omega}_{i} + \frac{1}{\zeta} \vartheta_{i}, 
\end{equation}

which defines a $\mathbb{P}^{1}$ family of complex structures parametrized by the twistor variable $\zeta$ and a flat hyperKhaler geometry with holomorphic symplectic structures $d\xi_{i}^{\zeta}\wedge d\eta_{i}^{\zeta}$. The $tt^{*}$ Lax connection annihilates with spectral parameter $\zeta$ all the holomorphic coordinates $(\eta_{i}^{\zeta},\xi_{i}^{\zeta})$

\begin{equation}
\boldsymbol{D}^{(\zeta)}_{\alpha,i}\eta_{j}^{\zeta}= \boldsymbol{D}^{(\zeta)}_{\alpha,i}\xi_{j}^{\zeta}=0, \hspace{1cm} \alpha=1,2
\end{equation}

which means that, in the complex structure $\zeta$, it is the $(0,1)$ part of a connection $\boldsymbol{A}$ on the hyperkahler space $(\mathbb{R}^{2}\times T^{2})^{b_{1}}$. The Lax equations

\begin{equation}
\boldsymbol{D}^{(\zeta)}_{\alpha,i}\Psi(\zeta)=0, \hspace{1cm}\alpha=1,2
\end{equation}

say that the brane amplitudes $\Psi(\zeta)$ are holomorphic in complex structure $\zeta$ and independent of $\mathrm{Im} \  \vartheta_{i}$ \cite{rif13}. The $tt^{*}$ equations then say that the curvature of the connection $\boldsymbol{D}(\zeta)$ on the flat hyperKahler manifold is of type $(1,1)$ in all complex structures, i.e. $\Psi(\zeta)$ is a section of a hyperholomorphic vector bundle \cite{rif13}. The hyperholomorphic condition, supplemented by the condition on translation invariance in $\mathrm{Im}(\vartheta_{i})$, is equivalent to the higher dimensional generalization of the Bogomolnji monopole equations on $(\mathbb{R}^{2} \times S^{1})^{b_{1}}$. \\ The $tt^{*}$ geometry decomposes into an Abelian $U(1)$ monopole and a non-Abelian $SU(d)$ monopole. Restricted to the Abelian part, the $tt^{*}$ equations become linear. Writing $L(\vec{\theta})=-\log(\mathrm{det}g(\vec{\theta}))$, they read

\begin{equation}
\begin{split}
& \left( \frac{\partial^{2}}{\partial\omega_{i}\partial\overline{\omega_{j}}} + \frac{\partial^{2}}{\partial\theta_{i}\partial\theta_{j}}\right) L(\vec{\theta})=0 \\ \\ & \frac{\partial^{2}}{\partial t_{a}\partial \overline{\omega}_{j}}L(\vec{\theta})= \frac{\partial^{2}}{\partial t_{a} \partial \overline{t}_{b}}L(\vec{\theta})=0.
\end{split}
\end{equation}

The Abelian part of the Berry connection is

\begin{equation}
A^{\mathrm{Ab}} = \partial L(\vec{\theta}) = \partial_{\omega_{i}} L(\vec{\theta})d\omega_{i}+ \partial_{t_{a}} L(\vec{\theta}) dt_{a}.
\end{equation}

The $tt^{*}$ relation $[A_{\omega_{i}}, C_{t_{a}}] = [A_{t_{a}} , C_{\omega_{i}} ]$, together with \ref{togeqn}, implies

\begin{equation}
\partial_{\theta_{i}} A_{t_{a}} =[A_{t_{a}},M_{\omega_{i}}]-[A_{\omega_{i}},C_{t_{a}}].
\end{equation}

Taking the trace gives $\partial_{\theta_{i}} A^{\mathrm{Ab}}_{t_{a}} = 0$. Since $A^{\mathrm{Ab}}$ is odd in $\vec{\theta}$ by \ref{odd}, we conclude that the $t_{a}$-components of the $U(1)$ connection vanish.\\ For the sake of comparison with the literature on representation of braid groups and the Knizhnik-Zamolodchikov equation \cite{rif61,rif62} we state the above result in a different way. We write $q_{i} = e^{i\theta_{i}}$ for $i = 1, . . . , b_{1}$. The Frobenious algebra $\mathcal{R}_{\mathcal{A}}$ is a module over the ring $\mathbb{C}[q^{\pm 1}]$ of Laurent polynomials in $q_{1}, ... , q_{b_{1}}$. The isomorphism with the space of branes 

\begin{equation}
\mathcal{R}_{\mathcal{A}}\simeq \mathcal{B}_{\mathcal{A}}(\zeta)= H_{*}(\mathcal{A},\mathcal{A}_{\mathrm{Re}(\zeta W)};\mathbb{Z})\otimes_{\mathbb{Z}} \mathbb{C}
\end{equation}

allows us to restrict the scalars to $\mathbb{Z}$. Thus, we have that $\mathcal{B}(\zeta)_{\mathcal{A}} \simeq \mathcal{R}_{\mathcal{A}} \simeq \mathcal{V}_{\mathcal{A}}$ is a free $\mathbb{Z}[q^{\pm 1}]$-module of rank $d$. Moreover, the $tt^{*}$ Lax connection provides the group homomorphism

\begin{equation}
\rho: \pi_{1}(\mathcal{P})\rightarrow GL(d,\mathbb{Z}[q^{\pm 1}]).
\end{equation}

\subsection{The Problem of Statistics}

The concept of statistics is crucial in understanding the physics of FQHE. We want now to discuss this problem in the context of $\mathcal{N}=4$ SQM. We consider for simplicitly the case in which the target space is $\mathbb{C}^{N}$. We suppose to have a superpotential $W(z_{1},...,z_{N})$ which is symmetric under arbitrary permutations of the coordinates $z_{i}\leftrightarrow z_{j}$. In other words, the particles of our theory are indistinguishable. Since it is a symmetry of the model, the Hilbert space and in particular its subspace of vacua $\mathcal{V}$ must decompose in irreducible representations of the permutation group of $N$ object $S_{N}$

\begin{equation}
\mathcal{V}=\bigoplus_{\eta \in \mathrm{irrep}(S_{N})} \mathcal{V}_{\eta}.
\end{equation}

The relevant representation for physical applications are the trivial representation $\mathcal{V}_{\mathrm{s}}$, which corresponds to symmetric wave functions, and the sign representation $\mathcal{V}_{\mathrm{a}}$, which corresponds to antisymmetric wave functions. The symmetric group acts also on the chiral ring $\mathcal{R}$, whose elements labels the zero energy solutions of the Schroedinger equation. Similarly to the Hilbert space, the chiral ring decomposes linearly as 

\begin{equation}
\mathcal{R}=\bigoplus_{\eta \in \mathrm{irrep}(S_{N})} \mathcal{R}_{\eta}.
\end{equation}

The space of symmetric chiral operators $\mathcal{R}_{\mathrm{s}}$ is a ring, while all the other components, as the space of antisymmetric chiral operators $\mathcal{R}_{\mathrm{a}}$, are $\mathcal{R}_{\mathrm{s}}$-modules. One can see that the map between chiral operators $\phi \in \mathcal{R}$ and vacuum wave functions 

\begin{equation}
\phi \rightarrow \phi \  dz_{1}\wedge....\wedge dz_{N} +\overline{\mathcal{Q}}(....)
\end{equation}

provides the isomorphism of $\mathcal{R}_{\mathrm{s}}$-modules

\begin{equation}
\mathcal{R}_{\eta}\simeq \mathcal{V}_{\mathrm{a\cdot\eta}}.
\end{equation}

This follows from the antisymmetric property of the fermionic operator $dz_{1}\wedge....\wedge dz_{N}$ in the vacuum wave form. Counter intuition, we claim that Fermi (Bose) statistics corresponds to have symmetric (antisymmetric) wave functions. To justify our definition let us count the number of ground states in an important special case. We consider the class of superpotentials which can be written as sum of single field superpotentials

\begin{equation}
\mathcal{W}(z_{1},...,z_{N})=\sum_{i=1}^{N}W(z_{i}),
\end{equation}

where each superpotential $W(z_{i})$ is defined on a target space $\mathbb{C}$. We denote with $\mathcal{R}_{1}$ the one-particle chiral ring and with $d=\mathrm{dim}\mathcal{R}_{1}$ the dimension of the single particle ground state. The chiral ring of the $N$-particle model is given by the tensor product 

\begin{equation}
\mathcal{R}= \bigotimes^{N} \mathcal{R}_{1}.
\end{equation}

The symmetric and antisymmetric subspaces of $\mathcal{R}$ are 

\begin{equation}
\mathcal{R}_{\mathrm{s}}=\bigodot^{N} \mathcal{R}_{1}, \hspace{1cm} \mathcal{R}_{\mathrm{a}}=\bigwedge^{N}  \mathcal{R}_{1},
\end{equation}

which have dimensions respectively 

\begin{equation}
\mathrm{dim} \mathcal{R}_{\mathrm{s}}= \begin{pmatrix} N+d-1 \\ N
\end{pmatrix}, \hspace{1cm} \mathrm{dim} \mathcal{R}_{\mathrm{a}}= \begin{pmatrix} d \\ N
\end{pmatrix},
\end{equation}

which correspond to Bose and Fermi statistics. The above relations remain true also if we add to the superpotential an arbitrary supersymmetric interactions which do not change the behaviour at infinity in field space, since the dimension of the chiral ring is the Witten index $I_{W}=d$. Using the isomorphism $\mathcal{R}\simeq \mathcal{V}$ we have

\begin{equation}
 \mathcal{V}_{B}= \bigodot^{N} \mathcal{V}_{1},    \hspace{1cm} \mathcal{V}_{F}= \bigwedge^{N} \mathcal{V}_{1}
\end{equation}

and the $tt^{*}$ metric, connection and brane amplitudes are induced by the single particle ones.\\ Coming back to the general case, one can see that any antisymmetric operator $\phi \in \mathcal{R}_{\mathrm{a}}$ can be written as   

\begin{equation}
\phi=\hat{\phi} \prod_{i<j}(z_{i}-z_{j}),
\end{equation}

where $\hat{\phi} \in \mathcal{R}_{\mathrm{s}}$. This provides the definition of chiral ring of the Fermi model 

\begin{equation}
\mathcal{R}_{F}= \mathcal{R}_{\mathrm{s}}/\mathcal{I}_{\mathrm{Van}},
\end{equation}

where $\mathcal{I}_{\mathrm{Van}}\subset \mathcal{R}_{\mathrm{s}}$ is the annihilator ideal of the Vandermonde determinant $\Delta(z_{i})=\prod_{i<j}(z_{i}-z_{j})$. We have the linear isomorphism of $\mathcal{R}_{\mathrm{s}}$-modules

\begin{equation}
\mathcal{R}_{F}\simeq \mathcal{V}_{\mathrm{s}}=\mathcal{V}_{F}, \hspace{1cm} \mathrm{dim} \mathcal{R}_{F}= \begin{pmatrix}
d \\ N \end{pmatrix}.
\end{equation}

In order to study the fermionic sector of the theory we can use the functoriality of $tt^{*}$ geometry. A generic superpotential $\mathcal{W}(z_{1},...,z_{N})$ which is invariant under permutation of the $z_{i}$ can be rewritten as function of the elementary symmetric polynomials 

\begin{equation}
e_{k}=\sum_{1\leq i_{1}<....<i_{N}\leq N} z_{i_{1}}z_{i_{2}}....z_{i_{k}}, \hspace{0.5cm} e_{0}=1.
\end{equation}

Hence, the superpotential $\mathcal{W}:\mathbb{C}^{N}\rightarrow \mathbb{C}$ factorizes through the branched covering map $E: z_{k}\rightarrow e_{k}$ of degree $N!$. One can pull-back the vacuum wave forms for the LG theory with superpotential $\mathcal{W}(e_{i})$ on the cover with coordinates $z_{i}$

\begin{equation}
\begin{split}
\Psi(e_{1},...,e_{N})= & \ \hat{\phi}( e_{1},...,e_{N} ) de_{1} \wedge...\wedge de_{N} + \overline{\mathcal{Q}}(....) \\ = & \ \hat{\phi}(e_{1},...,e_{N}) \mathrm{det}\left( \frac{\partial e_{i}}{\partial_{z_{j}}} \right) dz_{1} \wedge...\wedge dz_{N} + \overline{\mathcal{Q}}(....) \\ = & \ \hat{\phi}(e_{1},...,e_{N}) \prod_{i<j} (z_{i}-z_{j})dz_{1}\wedge...\wedge dz_{N} + \overline{\mathcal{Q}}(...),
\end{split}
\end{equation}

which defines non trivial $\overline{\mathcal{Q}}$-cohomology classes for the vacua of the theory with superpotential $\mathcal{W}(z_{i})$. The above pull-back leads to a correspondence between the chiral ring $\mathcal{R}_{e}$ of the $\mathcal{W}(e_{i})$-model and the Fermi chiral ring $\mathcal{R}_{F}$ of the $\mathcal{W}(z_{i})$-theory. We have 

\begin{equation}
E^{\sharp}: \ket{h}_{e}\rightarrow \ket{h\delta}_{z}, \hspace{1cm} h \in \mathcal{R}_{e}= \mathbb{C}[e_{1},...,e_{N}]/\left( \partial_{e_{1}}\mathcal{W},...,\partial_{e_{N}}\mathcal{W}\right) 
\end{equation}

which implies the isomorphism 

\begin{equation}
E^{\sharp}(\mathcal{R}_{e})\simeq \mathcal{R}_{F}.
\end{equation}

By $tt^{*}$ functoriality, the map $E^{\sharp}$ is also an isometry for the $tt^{*}$ metric and sets an equivalence between the Fermi sector of the original theory and the LG model with superpotential $\mathcal{W}(e_{i})$.

\subsection{$tt^{*}$ Geometry of the $\nu=1$ Phase }\label{nu1state}

We want to study the fermionic sector of the $N$ particles LG theory with superpotential 

\begin{equation}\label{initialtheory}
\mathcal{W}(z_{1},...,z_{N})= \sum_{i=1}^{N}\left( \mu z_{i}+ \sum_{\ell=1}^{N}\alpha_{\ell} \log (z_{i}+\zeta_{\ell}) \right)
\end{equation}

where we let $\alpha_{\ell}, \ \mu$ to be generic complex couplings and take $\zeta_{\ell}$ all distinct. The coupling constant manifold of the theory at fixed $\alpha_{\ell}, \ \mu$ is given by the space of $N$ distinct ordered points on $\mathbb{C}$

\begin{equation}
C_{N}=\left\lbrace (\zeta_{1},...,\zeta_{N}) \in \mathbb{C}^{N} \vert \zeta_{i}\neq \zeta_{j}, i\neq j \right\rbrace .
\end{equation}

If we demand the residues $\alpha_{\ell}$ to be equal, the manifold of couplings is naturally projected on $Y_{N}=C_{N}/S_{N}$, namely the space of $N$ identical particles on $\mathbb{C}$. In parituclar, it is clear from the discussion in section \ref{wedisc} that if we set the couplings $\alpha_{\ell}=-1$ we obtain the Vafa model for the $\nu=1$ phase of quantum Hall effect. Writing the superpotential as 

\begin{equation}
\mathcal{W}(z_{1},...,z_{N})= \sum_{i=1}^{N}\mu z_{i} + \sum_{\ell=1}^{N}\alpha_{\ell} \log \left( \prod_{i=1}^{N}(z_{i}+\zeta_{\ell}) \right)
\end{equation}

we note that the argument of the logarithm can be expanded as

\begin{equation}
\prod_{i=1}^{N}(z_{i}+\zeta_{\ell})= \sum_{k=0}^{N}e_{k}\zeta_{\ell}^{N-k}= P(\zeta_{\ell}).
\end{equation}

The polynomials $P(\zeta_{\ell})$ are linear in the dynamical fields and depend holomorphically on the couplings. Hence, one can do a linear redefinition of variables 

\begin{equation}
e_{\ell}\rightarrow P(\zeta_{\ell})=P_{\ell}.
\end{equation}

Rewriting the interaction $\mu \sum_{i} z_{i}$ as linear combination of $P_{\ell}$, the superpotential as function of the new variables reads

\begin{equation}
\mathcal{W}(P(\zeta_{1}),...,P(\zeta_{N}))= \sum_{\ell=1}^{N} \left( \mu_{\ell}P_{\ell} + \alpha_{\ell}\log P_{\ell} \right).
\end{equation}

We note that in the dynamical fields $u_{\ell}=P_{\ell}$ the superpotential of the theory can be written as sum of $N$ copies of the single field superpotential

\begin{equation}\label{singleparticlemodel}
W(u)= \mu u + \alpha \log u .
\end{equation}

From this example we see how much powerful $tt^{*}$ functoriality can be. In the starting theory \ref{initialtheory} we have a large number of punctures, equal to the dimension of the lowest Landau level, and an equal number of electrons which are correlated by the Fermi statistics. On the other hand, in the $tt^{*}$ dual we have just a unit of magnetic flux concentrated in a single puncture and the particles are not correlated by the statistics. The $tt^{*}$ functoriality ensures that we can study the Berry connection and its holonomy representations in the $tt^{*}$ dual of the model. Since the Hilbert space of the dual model is simply the tensor producs of single particle Hilbert spaces, it is enough to study the $tt^{*}$ geometry of the LG theory with superpotential \ref{singleparticlemodel}.\\
The equation for the vacua $\partial_{P_{k}}\mathcal{W}=0$ is given by a set of $N$ indipendent equations

\begin{equation}
\mu_{\ell} + \frac{\alpha_{\ell}}{P_{\ell}}=0,
\end{equation}

which has unique solution

\begin{equation}
P_{\ell}=-\frac{\alpha_{\ell}}{\mu_{\ell}}.
\end{equation}

As expected, the dimension of the Fermi chiral ring is $\mathrm{dim}\mathcal{R}_{F}=1$.\\ 
We want to study the $tt^{*}$ geometry of the vacua for this theory. The essential couplings are the positions of the punctures $\zeta_{\ell}$ which are coordinates on the space $C_{N}$. Since we have a unique vacuum, the Berry connection must be Abelian. The chiral operators which describe the deformation of the theory with respect to $\zeta_{\ell}$ are 

\begin{equation}
\partial_{\zeta_{\ell}}\mathcal{W}= \sum_{i=1}^{N}\frac{\alpha_{\ell}}{z_{i}+\zeta_{\ell}}.
\end{equation}

The chiral ring coefficients $C_{\ell}$ which represent the action of these operators on the vacua are $1\times 1$ matrices. Hence, all the commutators $\left[ C_{\ell}, C_{m}\right]$ are vanishing and the $tt^{*}$ equations read

\begin{equation}
\bar{\partial}_{\ell} (g \partial_{m} g^{-1})=0,
\end{equation}

where the ground state metric $g$ is just a real function of the couplings. This set of equations is solved by 

\begin{equation}
g=\vert f(\zeta_{1},...,\zeta_{N})\vert^{2},
\end{equation}

where $f(\zeta_{\ell})$ is an holomorphic function of the couplings $\zeta_{\ell}$. We also have to impose the reality constraint 

\begin{equation}
\eta^{-1}g(\eta^{-1}g)^{*}= 1.
\end{equation}

In the canonical basis the Frobenious pairing is the identity and we have $gg^{*}=1$. Since $g$ is a real function we conclude that 

\begin{equation}
g=1.
\end{equation}

So, as expected for the $\nu=1$ phase, the Berry connection on $C_{N}$ is vanishing and its monodromy representation in the UV limit is trivial.\\ We note that the target manifold is not simply connected and the theory admits non trivial theta sectors. The solution that we found above corresponds to the sector of the Hilbert space with trivial character. To study the dependence of $tt^{*}$ geometry on the theta angles we have to consider deformations of the theory which are not in the chiral ring. The corresponding couplings are the residues $\alpha_{\ell}$ of $d\mathcal{W}$ at the poles $\zeta_{\ell}$. The associated operators 

\begin{equation}
\partial_{\alpha_{\ell}}\mathcal{W}=\sum_{i=1}^{N} \log (z_{i}+\zeta_{\ell})
\end{equation}

are not univalued functions on the target space so are not elements of $\mathcal{R}$. By the functoriality of $tt^{*}$ geometry, the ground state metric of the full theory factorizes as tensor product of metrics of the single field models with superpotential \ref{singleparticlemodel}. This superpotential describes in $2d$ a free chiral superfield with a twisted mass. The corresponding $tt^{*}$ geometry has already been studied in \cite{rif13}. The model is defined on $\mathbb{C}\setminus\lbrace 0 \rbrace$ and the Galois group of the universal cover is generated by the loop which encircles the origin. The universal cover $\mathcal{A}$ is Abelian and is simply the complex plane. An explicit map between the covering space and the target manifold is given by the exponential $u=e^{Y}$. Pulling-back the model on the complex plane one has the superpotential 

\begin{equation}
W(Y)= e^{Y}+\alpha Y,
\end{equation}

where the action of the Galois group is given by the shift $T:Y\rightarrow Y + 2\pi i$. We can easily construct the unique theta-vacua of this theory. We denote with $\ket{0}$ the vacuum state corresponding to some idempotent element $e_{0}$ of $\mathcal{R}_{\mathcal{A}}$. Then, the vacuum space is spanned by

\begin{equation}
\ket{x}= \sum_{n\in \mathbb{Z}} e^{2\pi i n x} \ T^{n}\ket{0},
\end{equation}

where $x$ is the theta-angle normalized in the interval $[0,1]$. We see that the above state is an eigenstate of $T$ with eigenvalue the character $e^{-2\pi ix}$. Acting with the generator of the Galois group $T$ on the classical vacuum corresponding to $\ket{0}$ we obtain all the vacua of the covering model. We can set to $0$ the critical value corresponding to $\ket{0} $ by adding a constant to the superpotential. Then, the whole set of critical values is simply

\begin{equation}
W_{n}= 2\pi i \alpha n.
\end{equation}

We want to derive the $tt^{*}$ equation in the parameter $\alpha$ and study the critical limit of the solution. The chiral ring operator $C_{\alpha}$ acts on the theta-vacuum as differential operator in the angle

\begin{equation}
C_{\alpha}\ket{x}=  \sum_{n\in \mathbb{Z}} e^{2\pi i n x}2\pi i n T^{n}\ket{0}=
\frac{\partial}{\partial x} \ket{x}.
\end{equation}

We define the ground state metric on the vacuum bundle 

\begin{equation}
g(t,x)= \braket{\overline{x}}{x} = e^{L(t ,x)},
\end{equation}

where $L(t,x)$ is a real function of the angles and the RG scale $t=\vert \alpha\vert$. The $tt^{*}$ metric can be expanded in Fourier series as 

\begin{equation}
g(x) = \sum_{r \in \mathbb{Z}} e^{2\pi i r x} \braket{\overline{r}}{0}.
\end{equation}

In order to solve the $tt^{*}$ equation we have to impose the reality constraint of the metric. In presence of theta sectors the coplex conjugation of the metric must be intended as 

\begin{equation}
g^{*}(x)=[g(-x)]^{*}.
\end{equation}

Hence, using the fact that $g$ is real, the $tt^{*}$ reality constraint in the canonical basis becomes

\begin{equation}
g(-x)=g^{-1}(x),
\end{equation}

which implies

\begin{equation}
L(-x)=-L(x). 
\end{equation}

Thus, the $tt^{*}$ equation for ground state metric reads

\begin{equation}
\left( \partial_{\alpha}\partial_{\bar{\alpha}}+ \frac{\partial^{2}}{\partial x^{2}}\right)  L(t,x)=0.
\end{equation}

which is a $U(1)$ monopole equation on $\mathbb{R}^{2}\times S^{1}$ as expected from the general discussion in \ref{cov}. The solution can be expanded in terms of Bessel functions as

\begin{equation}
L(t, x)= \sum_{m=1}^{\infty} a_{m}\sin( 2\pi m x )K_{0}(4\pi m t )
\end{equation}

where the coefficients $a_{m}$ are determined by the boundary conditions. In the limit $t \rightarrow 0$ and for $x \neq 0$ we must have the asymptotics

\begin{equation}
L(t, x) \overset{t \rightarrow 0}{\sim} -2 (q(x) - \hat{c}/2 )\log t
\end{equation}

where $q(x)$ is the SCFT $U(1)$ charge of the chiral primary $e^{xY}$ at the UV fixed point. Since $e^{Y}$ has charge $1$, one has $q(x)= x \in [0,1]$ and $\hat{c}=q_{\mathrm{max}}=1$. Therefore, from the limit  

\begin{equation}
K_{0}(x) \overset{x \rightarrow 0}{\sim} - \log(x/2)
\end{equation} 

we get the relation

\begin{equation}
\frac{1}{2} \sum_{m=1}^{\infty} a_{m} \sin( 2\pi m x)= \left(\frac{\lambda}{2\pi}-\frac{1}{2} \right) .
\end{equation}

From this equality we learn that the coefficients $a_{m}$ are those of the Fourier expansion of the first periodic Bernoulli polynomial, i.e. 

\begin{equation}
a_{m}= -\frac{2}{\pi} \frac{1}{m}.
\end{equation}

Consistently with the discussion in \ref{cov}, also for non vanishing theta angles the components of the $U(1)$ Berry connection in the $\zeta_{\ell}$-directions are vanishing. In the case of trivial character we recover the constant solution $g=1$.

\subsection{The Vacuum Space of the Interacting Theory}\label{fermtrunc}

\subsubsection{The Heine-Stieltjes Problem}

We consider a LG model with $N$ chiral fields and superpotential differential 

\begin{equation}\label{vafa}
d\mathcal{W}=2\beta \sum_{1\leq i < j \leq N} \frac{d(z_{i}-z_{j})}{z_{i}-z_{j}} + \sum_{i=1}^{N} dW(z_{i}),
\end{equation}

where $dW(z_{i})$ is a rational differential with $d$ zeros and a pole of order $1\leq \ell \leq d+2$ at $\infty$. In the most general case $dW(z_{i})$ has $p=d+2-\ell$ simple poles at finite points $\left\lbrace y_{1},...,y_{p} \right\rbrace \in \mathbb{C}$ which are all distinct, namely 

\begin{equation}
dW(z)= \frac{B(z)}{A(z)}dz, \hspace{1cm} A(z)=\prod_{s=1}^{p}(z-y_{s}),
\end{equation}

for some polynomial $B(z)$ of degree $d$ coprime with $A(z)$. In the case of the Vafa model the residues are $\pm 1$ and $2\beta=1/\nu$. The fundamental degrees of freedom of this theory are not the fields $z_{i}$, but the symmetric polynomials

\begin{equation}
e_{k}= \sum_{1\leq i_{1}<....<i_{k}\leq N} z_{i_{1}}....z_{i_{k}}. 
\end{equation}

Indeed the superpotential is manifestly invariant under $S_{N}$ permutations of the $z_{i}$ and can be rewritten in these coordinates. The target manifold of the theory is the configuration space of $N$ identical particles on $\mathbb{C}\setminus\left\lbrace \zeta_{\ell} \right\rbrace $, i.e. 

\begin{equation}
\mathcal{M}_{N}=\left\lbrace (z_{1},...,z_{N}) \in (\mathbb{C}\setminus\left\lbrace \zeta_{\ell} \right\rbrace )^{N} \vert z_{i}\neq z_{j} \right\rbrace / S_{N}. 
\end{equation}

In order to find the classical vacua of the theory one has to solve the equations $\partial_{e_{k}}\mathcal{W}(e_{j})=0$. This algebraic problem can be mapped to a well known differential problem \cite{rif23}. One can encode a $S_{N}$-invariant vacuum configuration $\lbrace X_{i}\rbrace^{S_{N}}$ in a polynomial of degree $N$

\begin{equation}
P(z)=\prod_{i=1}^{N}(z-X_{i})=z^{N}+\sum_{k=1}^{N}(-1)^{k} e_{k} \ z^{N-k}.
\end{equation}

The equation for the configuration $\left\lbrace X_{i}\right\rbrace_{a} $ is equivalent to the differential equation for the polynomial $P_{a}(z)$

\begin{equation}\label{lamè}
2\beta A(z)P_{a}^{\prime\prime}(z)+B(z)P_{a}^{\prime}(z)= f_{a}(z)P_{a}(z)
\end{equation}

where $a=1,....,d$ labels the vacua of the theory and $f_{a}(z)$ is a polynomial of degree $d-1$ such that $P_{a}(z)$ solves the equation above. It is clear that counting the number of such solutions is equivalent to count the number of vacua up to permutation of the particles. The general problem of finding the couples $(f_{a}(z),P_{a}(z))$ which solve a second order differential equation of type \ref{lamè} is called in mathematical literature the Heine-Stieltjes problem, and the generalized eigenvalues $f_{a}(z)$ are called van Vleck polynomials. If $dW$ is a generic rational differential, with just simple poles in $\mathbb{P}^{1}$, the equation \ref{lamè} is a generalized $d$-Lam\'e equation. The $d$-Lam\'e equation \cite{rif56} corresponds to the special case in which we set $\beta = 1$ and $dW = d \log Q(z)$, where $Q(z)$ is a polynomial of degree d. Taking the same differential $dW$ , but choosing $\beta = −1$, the superpotential $\mathcal{W}$ becomes the Yang-Yang functional \cite{rif55,rif57} (and its exponential the master function \cite{rif58,rif59}) of the $sl(2)$ Gaudin integrable model on $V^{\bigotimes d}$, the Heine-Stieltjes equation is equivalent to the corresponding algebraic Bethe ansatz equations, and the roots of $P(z)$ are the Bethe roots \cite{rif55,rif60}. In the case of the Vafa model, solving the Heine-Stieltjes equation is equivalent by construction to solve the equation

\begin{equation}
\partial_{z_{i}}\mathcal{W}=\sum_{j\neq i}\frac{2\beta}{z_{i}-z_{j}} + \sum_{a}\frac{1}{z_{i}-x_{a}} - \sum_{\ell}\frac{1}{z_{i}-\zeta_{\ell}}=0
\end{equation}

where the solutions are counted modulo permutations of $S_{N}$. The above equation is a generalization of the Algebraic Bethe Anzatz equation for the Gaudin model, where the $z_{i}$ are the analog of the Bethe roots. The Gaudin model arises from the semi-classical limit of the solutions to the Knizhnik-Zamolodchikov, and it is natural to expect that the relation remains valid in the present slightly more general context. The most relevant case for us is actually when the poles at $\infty$ in $\mathbb{P}^{1}$ is double. As we alrady observed, this is a very convenient limit. In this case the Heine-Stieltjes equation is a confluent generalized $d$-Lam\'e equation. The ODE isequivalent to the Bethe ansatz equation for the Gaudin model with an irregular singularity \cite{rif55}. A well known results from Heine-Stieltjes theory is that for generic couplings the number of solutions $(P(z),f(z))$ is at most 

\begin{equation}
d= \begin{pmatrix}
N+d-1 \\ N
\end{pmatrix}.
\end{equation}

By definition, this is also the Witten index of the LG theory with superpotential $\mathcal{W}$ on the target space $\mathcal{M}_{N}$.

\subsubsection{Fermionic Truncation}

Following Vafa \cite{rif1}, we wish to interpret the SQM model defined by the superpotential \ref{vafa} on $\mathcal{M}_{N}$ as a theory of FQHE where $N$ electrons are coupled to $d$ units of magnetic flux and an external background electric field included in $dW(z)$, while the Vandermonde coupling describes the Coulomb interactions between the electrons. We observe that the multiplicity of vacua given by the Heine-Stieltjes problem corresponds to the Bose statistics and not the Fermi one. It is clear what happens when we take the limit of $\beta\rightarrow 0$: in a classical vacuum configuration for $\mathcal{W}$ the $z_{i}$ are close to the vacua of the single particle model and several of them may take different valures in the vicinity of the same one-field vacuum. Since these values differ by orders $O(\beta)$, in the $\beta\rightarrow 0$ limit the vacuum configurations are simply labelled by a set of positive integers $(N_{1},....,N_{d})$ which denote how many particles we put in the vacua of the single field model $z_{1},...,z_{d}$. A natural guess is that, in order to get the correct FQHE phenomenology, one should consider only the subspace

\begin{equation}
\mathcal{V}_{\mathrm{Fer}}\subset \mathcal{V}, \hspace{1cm} \mathrm{dim}\mathcal{V}_{\mathrm{Fer}}=\begin{pmatrix}
d \\ N
\end{pmatrix}
\end{equation}

which contains the vacua that survive in the $\beta\rightarrow 0$ limit. In this limit all other vacua $\in \mathcal{V}_{\mathrm{Fer}}^{\perp}$ escape at the infinite end of $\mathcal{M}_{N}$ where two or more $z_{i}$ coincide. The fermionic truncation from $\mathcal{V}$ to $\mathcal{V}_{\mathrm{Fer}}$ is geometrically consistent if and only if it is preserved under parallel transport by the $tt^{*}$ flat connection, i.e. if $\mathcal{V}_{\mathrm{Fer}}$ is a subrepresentation of the monodromy representation. Since the flat $tt^{*}$ connection is the Gauss-Manin connection of the local system on $\mathcal{P}$ provided by the BPS branes (for a fixed $\mathcal{\zeta} \in \mathbb{P}^{1}$), this is equivalent to the condition that the model has $\begin{pmatrix} d \\ N \end{pmatrix}$ preferred branes which remain regular as $\beta\rightarrow 0 $ spanning the dual space of $\mathcal{V}_{\mathrm{Fer}}$.\\ 
The fermionic truncation has already been studied by Gaiotto and Witten in a strictly related context \cite{rif55}. They show that preferred branes with the required monodromy properties do exist. We review their argument in our notation. We assume that the rational one-form $dW$ has a double pole at infinity of strength $\mu$ and $d$ simple poles in general positions. Then we can write 

\begin{equation}
B(z)= \mu A(z) + \mathrm{lower \ degree}.
\end{equation}

Hence, the Heine-Stieltjes equation becomes

\begin{equation}
2\beta A(z)P(z)^{\prime \prime}+ \left( \mu A(z)+... \right) P^{\prime} (z)=\mu\tilde{f}(z)P(z),
\end{equation}

where $ f(z)=\mu\tilde{f}(z)$. The monodromy representation is independent of $\mu$ as long as it is non-zero. Taking $\beta \sim O(1)$ and $\mu$ finite but very large (the reasonable regime for FQHE), the above equation up to a $O(1/\mu)$ correction becomes 

\begin{equation}
A(z)P^{\prime}(z)=\tilde{f}(z)P(z)
\end{equation}

which implies that up to $O(1/\mu)$ corrections the zeros of $P(z)$ coincide with the zeros of $A(z)$, namely the positions of the punctures. At large $\mu$ these approximate also the zeros of $B(z)$, i.e. the vacua of the single field model. The fermionic truncation amounts to require that their multiplicities are at most one, namely the polynomials $P(z)$ and $P^{\prime}(z)$ are coprime. In this regime, the product of $N$ one-particle Lefshetz thimbles starting at distinct zeros of $B(z)$ is approximatively a brane for the full interacting model. While the actual brane differs from the product of one-particle ones by some $O(1/\mu)$ correction, they are equivalent in homology and this is sufficient to study the $tt^{*}$ monodromy. Since it is dual to the space of branes of the $N$-particles Fermi model, the fermionic sector $\mathcal{V}_{\mathrm{Fer}} \in \mathcal{V}$ defines by construction a sub-representation of the $tt^{*}$ monodromy. So, differently from the case of non-interacting electrons, the fermionic sector is selected by the $tt^{*}$ solution and not by the statistics.

\section{Braid Group Representations in $tt^{*}$ Geometry}\label{chap6}

To compute the monodromy representation of the Vafa model in the UV approach we need a more in-depth understanding of $tt^{*}$ geometry. It turns out that for a special class of theories the UV Berry connection is a Kohno connection \cite{rif62,rif65}. This appears in the theory of the braid group representation \cite{rif61,rif64} as a solution of the Knizhnik-Zamolodchikov equations \cite{rif63}. In this section we go through the details of this beautiful relation.

\subsection{$tt^{*}$ Monodromy and the Universal Pure Braid Representation}

Following the strategy discussed in \ref{uvapproach} we rescale the critical values $w_{i}\rightarrow \beta w_{i}$. We consider $w_{i} \in \dot{\mathcal{P}}$, where $\dot{\mathcal{P}}$ is an open domain in the space of couplings such that the chiral ring is semi-simple. We note that if $w_{i} \in \dot{\mathcal{P}}$ also $\beta w_{i} \in \dot{\mathcal{P}}$ for all $\beta > 0$, so the limiting point indeed lays in the closure $\overline{\dot{\mathcal{P}}}$ of the semi-simple domain. As we approach the UV fixed point of the RG the $tt^{*}$ equations imply that $\bar{\partial}Q\rightarrow 0$. Since $Q$ is Hermitian, we have also $\partial Q\rightarrow 0$, so that $\lim_{\beta\rightarrow 0}Q$ is a constant matrix. Naively, to get the UV Berry connection we just replace this constant matrix in the the basic formula \ref{basicformula}. However, this is not the correct way to define the $\beta\rightarrow 0$ limit. Indeed, the formula \ref{basicformula} is derived in the canonical trivialization, which becomes too singular in the UV limit: the chiral ring $\mathcal{R}$ is believed to be regular (even as a Frobenius algebra) in the UV limit but, since the limit ring is no longer semi-simple, its generators are related to the canonical ones by a singular change of basis. A trivialization which is better behaved as $\beta\rightarrow 0$ is the natural one given by the orthogonal idempotents $e_{i}$. Starting from \ref{basicformula} and performing the diagonal gauge transformation, we get

\begin{equation}
A_{kl}=h_{k}Q_{kl}h_{l}^{-1}\frac{d(w_{k}-w_{l})}{w_{k}-w_{l}}-\delta_{kl}d \log h_{l},
\end{equation}

where $h_{l}=\sqrt{\langle e_{l} \rangle}$. By the residue formula \ref{residue}, the norm $\langle e_{l} \rangle$ should be a meromorphic function of the critical coordinates with poles at $w_{l}=w_{i}$ for $l\neq i $. Indeed, the superpotential becomes degenerate when two critical points coincide and, for a strictly Morse superpotential, the approach of two critical values imply the same for the corresponding critical points. In the $\beta\rightarrow 0$ limit the connection $A_{kl}$ becomes locally a meromorphic one form with simple poles at $w_{l}=w_{j}$ for $l\neq j$ and invariant under $w_{j}\rightarrow w_{j}+c$ and $w_{j}\rightarrow \lambda w_{j}$. In addition, its contraction with the Euler vector $\xi=\sum_{k}w_{k}\partial_{w_{k}}$ has no poles. Hence, in the UV limit the Berry connection $\mathcal{D}=D+\bar{\partial}$ can be written locally as

\begin{equation}\label{berkohno}
\mathcal{D}= d+ \sum_{1\leq i<j \leq d} B_{ij} \frac{d(w_{i}-w_{j})}{w_{i}-w_{j}},
\end{equation}

where $B_{ij}$ are holomorphic functions in $w_{j}-w_{k}$ homogeneous of degree $0$. Moreover, they should reproduce the correct quantum monodromy

\begin{equation}\label{corrqu}
\exp \left[ \sum_{i<j}B_{ij}\right] = (-1)^{r}\exp \left[ 2\pi i \boldsymbol{\mathcal{Q}}\right] , \hspace{0.5cm} \mathrm{up \ to \ conjugacy},
\end{equation}

where $\boldsymbol{\mathcal{Q}}$ is the $U(1)_{R}$ generator at the UV fixed point written in the natural basis which makes it symmetric traceless. The prefactor $(-1)^{r}$ is generated by the anomalous term 

\begin{equation}
\iota_{\xi}(-\delta_{kl}d \log h_{l})= \frac{r}{2}\delta_{kl}\frac{d\beta}{\beta},
\end{equation} 

where the norm $\langle e_{l} \rangle$ scales homogeneously as $\lambda^{-r}, \ r \in \mathbb{Q}$ under the rescaling $w_{i}\rightarrow \lambda w_{i}$. The matrices $B_{ij}$ are restricted to satisfy a set of equations which follow from the flatness condition

\begin{equation}
\mathcal{D}^{2}=0,
\end{equation}

as predicted by the $tt^{*}$ equations in the UV limit.

\subsection{Complete and Very Complete $tt^{*}$ Geometries}

We recall the definition of configuration space $C_{d}$ of $d$ ordered distinct points in the plane

\begin{equation}
\mathcal{C}_{d}=\left\lbrace  (w_{1},...,w_{d}) \in \mathbb{C}^{d} \vert w_{i}\neq w_{j} \ \mathrm{for} \ i\neq j \right\rbrace  .
\end{equation}

The cohomology ring $H^{*}(C_{d}, \mathbb{Z})$ is generated by the $\begin{pmatrix} d \\ 2\end{pmatrix}$ $1$-forms 

\begin{equation}
\omega_{ij}=\omega_{ji}= \frac{1}{2\pi i } \frac{d(w_{i}-w_{j})}{w_{i}-w_{j}}
\end{equation}

which satisfy the relations \cite{rif66}

\begin{equation}
\omega_{ij}\wedge\omega_{jk}+ \omega_{jk}\wedge\omega_{ki}+ \omega_{ki}\wedge\omega_{ij}=0.
\end{equation}

The fundamental group $P_{d}=\pi_{1}(C_{d})$ is called the pure braid group of $d$ strings. The configuration space of $d$ unordered points is the quotient space

\begin{equation}
Y_{d}=C_{d}/S_{d}
\end{equation}

and its fundamental group $B_{d}=\pi_{1}(Y_{d})$ is the Artin braid group of $d$ strings. It is an extension of the symmetric group $S_{d}$ by the pure braid group

\begin{equation}
1\rightarrow P_{d} \xrightarrow{\iota} B_{d}  \xrightarrow{\beta} S_{d}\rightarrow 1.
\end{equation}

$B_{d}$ has a presentation with $d-1$ generators $\sigma_{i}$ and the relations 

\begin{equation}
\sigma_{i}\sigma_{i+1}\sigma_{i}=\sigma_{i+1}\sigma_{i}\sigma_{i+1}= \sigma_{i+1}\sigma_{i}\sigma_{i+1}, \hspace{1cm} \sigma_{i}\sigma_{j}=\sigma_{j}\sigma_{i} \ \mathrm{for} \ \vert i-j \vert \geq 2.
\end{equation} 

We define the critical value map 

\begin{equation}
w: \dot{\mathcal{P}}\rightarrow Y_{d}, \hspace{1cm} t\rightarrow \lbrace w_{1},...,w_{d} \rbrace.
\end{equation}

For semi-simple chiral rings the critical values provides a set of local coordinates for the space of couplings, implying that the above map is an holomorphic immersion. We say that the $tt^{*}$ geometry is \textbf{complete} if, in addition, $w$ is also a submersion. This is equivalent to say that $w$ is a local isomorphism and so a covering map from $\dot{\mathcal{P}}$ to $Y_{d}$. The perturbations of the UV fixed point are generated by the chiral primary operators $\phi \in \mathcal{R}$. Not all these deformations can be included in the superpotential, since for some of them the couplings can be UV relevant and the theory can develop Landau poles. Hence, being complete means that all the chiral primary operators are IR relevant or marginal non-dangerous. In this case the dimension of the manifold of physical couplings is precisely $d$.\\ We also say that $tt^{*}$ geometry is \textbf{very complete} if the canonical projection $p:C_{d}\rightarrow Y_{d}$ factors through the critical value map $w$. This means that the manifold $C_{d}$ plays the role of cover space for $\dot{\mathcal{P}}$ and denoting with $ s: C_{d}\rightarrow \dot{\mathcal{P}}$ the cover map we have $p=w \circ s$. If $tt^{*}$ geometry is very complete we can pull-back the vacuum bundle $\mathcal{V} \rightarrow\dot{\mathcal{P}}$ to a bundle over $C_{d}$ and consider the $tt^{*}$ geometry on the configuration space $C_{d}$. In a very complete $tt^{*}$ geometry, pulled-back to $C_{d}$, the local expression \ref{berkohno} becomes global, since in this case the $w_{i}$ are global coordinates and the partials $\partial_{w_{i}} W$ define a global trivialization of the bundle $\mathcal{R} \rightarrow C_{d}$.
In the very complete case the entries of the matrices $B_{ij}$ are holomorphic functions on $C_{d}$, homogenous of degree zero and invariant under overall translation, which satisfy \ref{corrqu}. We conclude that the matrices $B_{ij}$ should be constant. In the general case the $B_{ij}$ are only locally constant and can have jumps under certain perturbations of the critical values $w_{k}$. Being very complete means essentially that the theory has no wall-crossing phenomena (see \cite{rif12}). The $B_{ij}$ are further constrained by the flatness condition $\mathcal{D}^{2} = 0$, which leads to the theory of Kohno connections \cite{rif65,rif67}. If a connection $\mathcal{D}=d+A$ of the form \ref{berkohno} satisfies $\mathcal{D}^{2}=0$, then the following relations hold

\begin{equation}
\begin{split}
\left[   B_{ij},B_{ik}+B_{jk}\right] = \left[ B_{ij}+B_{ik},B_{jk}\right] =0, \hspace{1cm} & \mathrm{for} \ i<j<k, \\ 
\left[ B_{ij},B_{kl}\right] =0, \hspace{5cm} & \mathrm{for \ distinct \ i,j,k,l.}
\end{split}
\end{equation}

The above equations are called the infinitesimal pure braid relations. A connection of the form \ref{berkohno} where the constant $d\times d$ matrices $B_{ij}$ satisfy the above relations is called a rank-$d$ Kohno connection. A connection of this type defines a representation of the pure braid group $P_{d}$ of $d$ strings 

\begin{equation}
\sigma: P_{d}\rightarrow GL(d,\mathbb{C})
\end{equation}

through the parallel transport on the configuration space $C_{d}$

\begin{equation}
\sigma: \gamma \in C_{d}\rightarrow P \exp (-\int_{\gamma} A) \in GL(d,\mathbb{C}).
\end{equation}

The family of representations $\sigma$ parametrized by the matrices $B_{ij}$ satisfying the infinitesimal pure braid relations is called the universal monodromy \cite{rif67}. We conclude that for a very complete $tt^{*}$ geometry the UV Berry connection is the universal monodromy representation of the pure braid group $P_{d}$ of $d$ strings specialized to Kohno matrices $B_{ij}$ computed in terms of the $U(1)_{R}$ spectrum of the UV chiral ring. A very important class of very complete $tt^{*}$ geometries are the symmetric ones. In this case the Kohno matrices satisfy 

\begin{equation}
(B_{\pi(i)\pi(j)})_{\pi(k)\pi(l)}=(B_{ij})_{kl}, \hspace{1cm} \mathrm{for \ all \ } \pi \in S_{d}
\end{equation}

In this case the connection $\mathcal{D}$ descends to a flat connection on a suitable bundle $\mathcal{V}\rightarrow Y_{d}$, providing a representation of the full Braid group \cite{rif62,rif65,rif67}. This happens when the critical map $w$ is an isomorphism.

\subsection{Ising Model and SQM}

It is known that $\mathcal{N}=4$ SQM is closely related to the Ising model in $2d$ \cite{rif32}. The ising model consists in a free massive Majorana field $\Psi(w)$ which satisfies the Euclidean Dirac equation 

\begin{equation}
(\slashed{\partial}-m)\Psi(w)=0,
\end{equation}

where $m$ is the mass of the field and 

\begin{equation}
\slashed{\partial}= \begin{pmatrix}
0 & \partial \\ \bar{\partial} & 0
\end{pmatrix}, \hspace{1cm} \Psi(w)= \begin{pmatrix} \Psi_{+}(w) \\ \Psi_{-}(w) \end{pmatrix}.
\end{equation}

The space of solutions to the Dirac equation is endowed with an inner product, which coincide with the hermitian norm of the first quantized Hilbert space 

\begin{equation}
\Vert \Psi \Vert^{2}= \int \left( \vert \Psi{+} \vert^{2} + \vert \Psi_{-} \vert^{2}\right) d^{2}w,
\end{equation}

where $\Psi_{+},\Psi_{-}$ must be $L^{2}$ functions on the plane. It is possible to rephrase the Dirac equation as cohomological problem. The Majorana field $\Psi(w)$ is not univalued on the $W$-plane, but has complicated branching properties because of the insertion of topological defect operators $O_{k}(w_{k})$ at points $w_{k}$. The OPE

\begin{equation}
\Psi(w)_{\pm}O_{k}(w_{k})
\end{equation}

is singular as $w\rightarrow w_{k}$. One considers two possible defect insertions at $w_{k}$, which we denote by $\sigma_{k},\mu_{k}$ \cite{rif32}. The OPEs of these operators with the Majorana field are 

\begin{equation}\label{OPEs}
\begin{split}
& \Psi_{+}(u)\sigma_{k}(w)\sim  \frac{i}{2} (u-w)^{-1/2}\mu_{k}(w), \hspace{0.5 cm} \Psi_{-}(u)\sigma_{k}(w)\sim  -\frac{i}{2} (\bar{u}-\bar{w})^{-1/2}\mu_{k}(w) \\ 
& \Psi_{+}(u)\mu_{k}(w)\sim  \frac{1}{2} (u-w)^{-1/2}\sigma_{k}(w), \hspace{0.5 cm} \Psi_{-}(u)\mu_{k}(w)\sim  \frac{1}{2} (\bar{u}-\bar{w})^{-1/2}\sigma_{k}(w)
\end{split}
\end{equation}

up to $O(\vert u-w\vert^{1/2})$ contributions. Although the defect operators $\mu_{k}, \sigma_{k}$ have the same OPE singularities with the fermion field $\Psi(w)$ as the Ising order/disorder operators, they are not in general mere Ising order/disorder operators since globally they have different topological  properties. The fermion field $\Psi(w)$ is univalued on a suitable connected cover of the plane punctured at the insertion points. By Riemann existence theorem \cite{rif68}, this can be extended to a cover of the Riemann sphere 

\begin{equation}
W: \Sigma\rightarrow \mathbb{P}^{1}
\end{equation}

branched at $\lbrace w_{1},...,w_{d}, \infty \rbrace $. In case of a cover of finite degree $m$, $f$ is specified by its Hurwitz data at the $(d + 1)$ branching points \cite{rif68}. The Hurwitz data consist of an element $\pi_{k} \in S_{m}$ for each finite branching point $w_{k}$, while $\pi_{\infty} = (\pi_{1}\pi_{2}.... \pi_{d})^{-1}$. The monodromy group of the cover is the subgroup of the permutation group $S_{m}$ generated by the $\pi_{k}$'s. When $m$ is infinite the monodromy group is an infinite group and the geometry is a bit more involved. Hence, the topological defect operator $O_{k}(w_{k})$ inserted at the $k$-th branching point is specified by the choice between $\sigma$-type and $\mu$-type and the monodromy element $\pi_{k} \in S_{m}$.\\ 
One can replace the Majorana field $\Psi(w)$ with the $1$-form on the complex plane

\begin{equation}
\psi(w) = \Psi_{+}dw+\Psi_{-}d\overline{w}.
\end{equation}

The map $\Psi\rightarrow \psi$ is consistent with the structure of the Hilbert space of the Dirac equation. Indeed, the hermitian product of the wave functions is identified with the inner product of forms 

\begin{equation}\label{isingprod}
\Vert \Psi \Vert^{2} = \int \psi \wedge \ast \psi^{\ast}.
\end{equation}

It is natural to introduce the operators $\overline{\mathcal{D}},\mathcal{D}$ which act on forms as

\begin{equation}
\overline{\mathcal{D}}\psi=\bar{\partial}\psi + dw\wedge \psi, \hspace{1cm} \mathcal{D}\psi=\partial\psi + d\overline{w}\wedge \psi.
\end{equation}

These operators are nilpotent

\begin{equation}
\mathcal{D}^{2}= \overline{\mathcal{D}}^{2}= \overline{\mathcal{D}}\mathcal{D} + \mathcal{D}\overline{\mathcal{D}}=0,
\end{equation}

and satisfy the Kahler-Hodge identites \ref{kahlid} with the Lefschetz operators $L,\Lambda$ on the plane. In particular, we have 

\begin{equation}
\overline{\mathcal{D}}^{\dagger}=i \left[ \Lambda, \mathcal{D}\right] .
\end{equation}

It is easy to see that the Dirac equation for $m=1$ can be written as the cohomological relations 

\begin{equation}\label{compar}
\mathcal{D}\psi= \overline{\mathcal{D}}\psi=0,
\end{equation}

and using the above Kahler identity we also have 

\begin{equation}
 \overline{\mathcal{D}}^{\dagger}\psi=0.
\end{equation}

Hence, by standard Hodge theory, the normalizable solutions to the Dirac equation are the harmonic representatives of $\overline{\mathcal{D}}$-cohomology classes in the space of $L^{2}$ forms on the complex plane. One considers the following system of solutions \cite{rif32}

\begin{equation}\label{isingcorr}
\begin{split}
\psi_{i}(z;\zeta)= & \langle \Psi_{+}(W(z))\mu_{1}(w_{1})...\sigma_{j}(w_{j})...\mu_{d}(w_{d})\rangle \frac{W^{\prime}(z)dz}{\zeta \tau(w_{j})}  + \\ & \langle \Psi_{-}(W(z))\mu_{1}(w_{1})...\sigma_{j}(w_{j})...\mu_{d}(w_{d})\rangle  \frac{\overline{W}^{\prime}(\bar{z})\zeta d\bar{z}}{ \tau(w_{j})} 
\end{split}
\end{equation}

where $z$ is a local coordinate on the cover $\Sigma$, $\zeta \in \mathbb{P}^{1}$ is an arbitrary parameter such that $\vert \zeta \vert =1$ and the normalization constant is the Sato-Miwa-Jimbo $\tau$-function 

\begin{equation}
\tau=\langle \mu(w_{1})...\mu(w_{n})\rangle.
\end{equation}

The wave functions $\psi_{i}(w,\zeta)$ are singular at the branching points $w_{i}$. These singularities can be computed using the OPEs \ref{OPEs} and encode the cohomology class of the $j$-th vacuum. We can use the hermitian scalar product \ref{isingprod} to define a ground state metric on the Hilbert space spanned by the above Ising one-forms. Studying variations of the metric in the branching points $w_{k}$, one finds that the bundle with fiber spanned by the local system of solutions \ref{isingcorr} has the same $tt^{*}$ geometry of the vacuum bundle of a $4$-SQM model with semi-simple $d$-dimensional chiral ring \cite{rif32}. In this formal correspondence the insertion points $w_{k}$ are identified with the critical coordinates on the coupling constant space of the SQM model. Hence, $4$-SQM and the off-critical Ising model provide two equivalent description of the same geometry.\\ 
The relation between the Ising model on the $W$-plane and $4$-SQM is especially simple when the superpotential depends on a single chiral field $z$. Let us consider a $\mathcal{N}=4$ Landau-Ginzburg model on a Stein space $\Sigma$ and a Morse superpotential $W$. Comparing the expression of $\mathcal{D},\overline{\mathcal{D}}$ with the susy charges 

\begin{equation}
\mathcal{Q}= \partial + d\overline{W}\wedge, \hspace{1 cm} \overline{\mathcal{Q}}= \partial + dW\wedge,
\end{equation}

we see that the holomorphic map 

\begin{equation}
W: \Sigma\rightarrow \mathbb{C}
\end{equation}

provides the relations

\begin{equation}
\mathcal{Q}=W^{*}\mathcal{D},\hspace{1cm} \overline{\mathcal{Q}}=W^{*}\overline{\mathcal{D}}.
\end{equation}

Namely, one can map the Schroedinger equation for the vacua of the LG model on $\Sigma$ to the Euclidean Dirac equation on the $W$–plane. In this correspondence, the one-form $\psi_{i}(z,\zeta)$ is identified with the Schroedinger wave-function of the $i$-th vacuum corresponding to the idempotent $e_{i} \in \mathcal{R}_{w}$.\\ We see that the formulation in critical coordinates is universal. The informations which distinguish different models and make precise the dictionary between SQM and Ising model are encoded in the Hurwitz data. The wave functions are multi-valued on the $W$-plane with branching points just given by $w_{k}$. The Hurwitz data should be chosen such that $\psi_{i}(z,\zeta)$ are univalued in the coordinate $z$ of the target manifold $\Sigma$. This space can be seen as a cover of the $W$-plane with cover map $W : \Sigma \rightarrow \mathbb{C}$ branched at the critical points $w_{k}$. By the Riemann existence theorem, the Hurwitz data specify uniquely up to isomorphism the cover $(W,\Sigma)$ of the punctured plane. \\ A similar formula for the vacuum wave functions in the $N$-fields case can be constructed as follows. We consider the inverse image of a point $w$ on the $W$-plane. This has the homotopy type of a bouquet of $N-1$ spheres \cite{rif12,rif19}. We set a basis of cycles $S_{\alpha}(w), \alpha=1,...,d$ for the middle dimensional homology of the fiber. A vacuum wave function $\Psi$ is a $N$-form on the target space $X$. Pulling-back $\Psi$ on the $w$-level curve of the superpotential and integrating it over the cycles $S_{\alpha}(w)$, we get a set of $d$ one-forms on the $W$-plane

\begin{equation}\label{singlebutmulti}
\psi_{\alpha}(w)= \int_{S_{\alpha}(w)}\Psi.
\end{equation}

If we transport the homology cycles along a closed loop along in the $W$-plane (punctured at the critical values) we come back with a different (integral) basis of $(N-1)$-cycles $S_{\alpha}^{\prime} (w) = M_{\alpha\beta}S_{\beta}(w)$. The integral matrix $M_{\alpha \beta}$ is described by the Picard-Lefshetz theory \cite{rif12}. Thus, the \ref{singlebutmulti} is best interpreted as a single but multivalued wave-function $\psi(w)$ on the $W$-plane branched at the critical points, whose monodromy representation is determined by the Picard-Lefshetz formula. Let $\Sigma$ be the minimal branched cover of the $W$-plane such that $\psi_{\alpha}(w)$ is single valued. The map $W:X\rightarrow \mathbb{C}$ factorizes through $\Sigma$. Pulling-back the wave forms on $\Sigma$ by the covering map $w:\Sigma\rightarrow \mathbb{C}$, one can replace the original LG model with $N$ chiral fields and Morse superpotential $W$ with an LG model with target space $\Sigma$ and superpotential $w$ given by the factorization of $W$ thorugh $\Sigma$. Hence, we can use the formula for the single field case and write 

\begin{equation}
\begin{split}
\psi_{i}(w;\zeta)= & \langle \Psi_{+}(w)\mu_{1}(w_{1})...\sigma_{i}(w_{i})...\mu_{d}(w_{d})\rangle\frac{dw}{\zeta \tau(w_{j})}+ \\ & \langle\Psi_{-}(w)\mu_{1}(w_{1})...\sigma_{i}(w_{i})...\mu_{d}(w_{d})\rangle \frac{\zeta d\bar{w}}{ \tau(w_{j})}.
\end{split}
\end{equation}

The branes $B_{k}(\zeta)$ of the model on the $W$-plane are straight lines starting at the critical points $w_{k}$ and strethced towards the $\mathrm{Re} \ w/\zeta=-\infty$ direction. Denoting with $\Gamma_{k}(\zeta)$ the support of $B_{k}$ on $\Sigma$, we can define the brane amplitudes 

\begin{equation}
\braket{\Gamma_{k}(\zeta)}{\Psi_{i}}= \int_{\Gamma_{k}(\zeta)} e^{\beta w/\zeta + \beta \zeta \bar{w} }\psi_{i}(w;\zeta).
\end{equation}

In the physical $2d$ (2,2) LG model, the UV limit consists in sending to zero the overall coupling $\beta$. But, as one can see comparing the Dirac equation with the vacuum equation \ref{compar}, $\beta$ is also the mass $m$ of the Majorana fermion in the context of the QFT on the $W$-plane. Hence, the physical UV limit of the $2d $ LG model coincides with the UV limit of the fermion theory on the $W$-plane. As $\beta\rightarrow 0$  the left and right modes of the fermion $\Psi$ decouple and the multi-valued correlation functions become sums of products of left/right conformal blocks. The statement holds (roughly) for all $tt^{*}$ quantities: in the UV they become some complicate combination of conformal blocks. Then the differential equations they satisfy, the $tt^{*}$ equations, should be related in a simple way to the PDEs for the conformal blocks. From this viewpoint the fact that in the UV limit the Berry connection, coinciding with the $tt^{*}$ Lax one in the $\beta\rightarrow 0$ limit, has the Kohno form, which is typical of the monodromy action on conformal blocks, is not surprising. In connecting the $tt^{*}$ monodromy with the braid representation of conformal blocks, we need to use the precise disctionary between the two. From \ref{isingcorr} we see that the wave functions, being normalized, are to be seen as ratios of $n$-point functions in the $W$-plane CFT

\begin{equation}
\frac{\langle \Psi_{\pm}(w)\mu_{1}(w_{1})...\sigma_{i}(w_{i})...\mu_{d}(w_{d})\rangle}{\langle \mu_{1}(w_{1})...\mu_{n}(w_{n})\rangle}
\end{equation}

rather than correlators. Hence, the actual braid representation on the CFT operators is the $tt^{*}$ one twisted by the one defined by the $\tau$-function.

\subsection{Hecke Algebra Representations of the $tt^{*}$ Connection}\label{hecke}

We want to determine the matrices $B_{ij}$ for a very complete $tt^{*}$ geometry. $B_{ij}$ is the residue of the pole of the UV Berry connection as $w_{i}\rightarrow w_{j}$. In order to compute $B_{ij}$ it is enough to consider this limit from the point of view of the $2d$ model. Without loss of generality, we may deforme the $D$-terms so that the only light degrees of freedom are the BPS solitons interpolating between vacua $i$ and $j$ of mass $2\vert w_{i}-w_{j} \vert$. We may integrate out all other degrees of freedom and we end up with an effective IR description with just these two susy vacua. The theory that we get may be not UV complete, but this is not an issue, since it is just an effective description valid up to some non-zero energy scale. From the viewpoint of SQM, the $2d$ BPS solitons are BPS instantons. The effect of these BPS instantons is to split the two vacua not in energy as it happens in non-susy QM, but in the charge $q$ of the $U(1)_{R}$ symmetry which emerges in the $w_{i}-w_{j} \rightarrow 0$ limit. In this limit there is also an emergent $\mathbb{Z}_{2}$ symmetry which interchanges the classical vacua. The $U(1)_{R}$ symmetry is broken off-criticality to $\mathbb{Z}_{2}$ and the eigenstates are the symmetric and antisymmetric combination of the classical vacua \cite{rif10,rif12}. The corresponding charges must be opposite by PCT symmetry. We renumber the critical values $w_{k}$ such that $w_{i},w_{j}$ are $w_{1},w_{2}$. With a convenient choice of the relative phases of the two states, the upper-left $2 \times 2$ block of the $U(1)_{R}$ generator $\boldsymbol{\mathcal{Q}}$ takes the form

\begin{equation}
\boldsymbol{\mathcal{Q}}_{\mathrm{upper-left}}= -\lambda \begin{pmatrix}0 & 1 \\ 1 & 0
\end{pmatrix}= -\lambda\sigma_{1},
\end{equation}

for some constant $\lambda$. We have to remember that in the trivialization given by the idempotent basis the UV Berry connection gains the extra piece $\delta_{kl}d\log h_{l}$ which generates an additional term in the above formula proportional to the identity. To be general we consider a shift by a constant $\mu$

\begin{equation}
-\lambda \begin{pmatrix}0 & 1 \\ 1 & 0
\end{pmatrix} \rightarrow -\lambda \begin{pmatrix}0 & 1 \\ 1 & 0
\end{pmatrix}+ \mu \begin{pmatrix}
1 & 0 \\ 0 & 1
\end{pmatrix}
\end{equation}

From the Ising model point of view, at each of the two critical points $w_{1,2}$ we may insert either a $\sigma$-like defect or a $\mu$-like one and different choices correspond to different vacua of the original LG theory. The matrix $\sigma_{1}$ has the effect of flipping the two-vacua system, which corresponds to the exchange $ \sigma\leftrightarrow \mu$ in the correlators. It is therefore convenient to introduce a two-component notation

\begin{equation}
\Sigma_{k,\alpha}(w)= \begin{pmatrix}
\sigma_{k}(w) \\ \mu_{k}(w)
\end{pmatrix}.
\end{equation}

The UV conformal blocks for the effective theory with two vacua have the form 

\begin{equation}
\langle \Psi_{\pm}(w) \Sigma_{1,\alpha}(w_{1})\Sigma_{2,\alpha}(w_{2}) \rangle \in V_{1}\otimes V_{2}
\end{equation}

where $V_{a} \simeq \mathbb{C}^{2},a=1,2$, are two copies of the fundamental representation space of $sl(2, \mathbb{C})$. We note that the amplitudes representing the vacua span only a two dimensional subspace of $V_{1} \otimes V_{2}$ of dimension $2$. The action of $B_{12}$ can be extended on the full four dimensional space of conformal blocks as 

\begin{equation}
B_{12}=-\lambda\left( \sigma_{+}^{1}\otimes \sigma_{-}^{2} + \sigma_{-}^{1}\otimes \sigma_{+}^{2} \right) -\mu\ \sigma_{3}^{1}\otimes \sigma_{3}^{2},
\end{equation}

where $\sigma^{\ell}_{a}$ is the Pauli matrix acting on the $a$-th copy $V_{a}$ of $\mathbb{C}^{2}$. We conclude that the UV Berry connection  $\mathcal{D}$ of a very complete $tt^{*}$ geometry with $d$ vacua has the general form 

\begin{equation}\label{genkohno}
\mathcal{D}=d -2\sum_{i<j}\left( \lambda_{ij}s_{\ell}^{i}s_{\ell}^{j}+ \mu_{ij}s_{3}^{i}s_{3}^{j} \right) \frac{d(w_{i}-w_{j})}{w_{i}-w_{j}}
\end{equation}

and act on the sections of a bundle $\boldsymbol{V}\rightarrow C_{d}$ whose fibers are modelled on the vector space 

\begin{equation}
V^{\otimes d}= V_{1}\otimes V_{2} \otimes......\otimes V_{d},
\end{equation}

spanned by the conformal blocks of Ising type. The matrices $s_{\ell}^{a}, \ell=1,2,3$ are the $su(2)$ generators acting on the $V_{a}\simeq \mathbb{C}^{2}$ factor, namely 

\begin{equation}
s_{\ell}^{a}= 1 \otimes....\otimes 1 \otimes \frac{1}{2}\sigma_{\ell} \otimes 1 \otimes ....\otimes 1.
\end{equation}

The natural connection on the CFT conformal blocks may differ from $\mathcal{D}$ by a line bundle twisting; Indeed, the $d$-point functions representing the vacua are rescaled by the normalization factor $\tau(w_{j})={\langle \mu_{1}(w_{1})...\mu_{n}(w_{n})\rangle}$. This corresponds to replacing $\mathcal{D} \rightarrow \mathcal{D} + 1 \cdot d \log f$ for some multivalued holomorphic function $f$.\\ The actual brane amplitudes live in a rank $d$ sub-bundle $\mathcal{V}$ of the rank $2^{d}$ bundle $\boldsymbol{V}$. The $tt^{*}$ Lax equations requires this sub-bundle to be preserved by parallel transport with the connection $\mathcal{D}$. To see this, consider the total angular momentum

\begin{equation}
L_{\ell}= \sum_{a} s_{\ell}^{a}, \hspace{0.5cm} \ell=1,2,3.
\end{equation}

One can easily check that $L_{3}$ commutes with all the infinitesimal braiding $B_{ij}$ and so the eigenbundles $\mathcal{V}_{m}$ of $L_{3}$ are preserved by the parallel transport of $\mathcal{D}$. In particular, since the correlators representing the vacua has $d-1$ insertions of $\mu$-type and a single insertion of $\sigma$-type, the vacuum bundle corresponds to the $L_{3}$-eigenbundle with eigenvalue $m=1-d/2$. So, we have 

\begin{equation}
\mathcal{V}=\mathcal{V}_{1-d/2}.
\end{equation}

The constants $\lambda_{ij},\mu_{ij}$ are restricted by further conditions:

\begin{itemize}

\item $\mathcal{D}$ is flat acting on $\mathcal{V}$

\item Since $\mathcal{D}$ coincide with the Lax connection in the UV limit, the monodromy representation must be arithmetic.

\item If the very complete $tt^{*}$ geometry is symmetric, the constant $\lambda_{ij},\mu_{ij}$ must be indipendent from $i,j$, i.e. $\lambda_{ij}=\lambda, \mu_{ij}=\mu$.

\end{itemize}

A well known solution to the Knizhnik-Zamolodchikov equations is \cite{rif65,rif67}

\begin{equation}
\lambda_{ij}=\frac{\lambda}{2}, \hspace{0.5cm} \mu_{ij}=0,
\end{equation}

which gives 

\begin{equation}\label{symmkohno}
\mathcal{D}=d+ \lambda\sum_{i<j} \frac{s_{\ell}^{i}s_{\ell}^{j}}{w_{i}-w_{j}}d(w_{i}-w_{j}).
\end{equation}

This connection is automatically flat for all $\lambda$ when acting on sections of the big bundle $\boldsymbol{V}$ and is also symmetric. On the other hand, it is easy to check that the only symmetric solution to the flatness condition for a connection of the form \ref{genkohno} is given by \ref{symmkohno}. We conclude that a symmetric very complete $tt^{*}$ geometry
has a UV Berry connection of the above form.\\ Since $\mathcal{D}$ is flat on the larger bundle $\boldsymbol{V}$, the UV limit of the $tt^{*}$ linear problem, $\mathcal{D}\Psi = 0$ with $\Psi \in \Gamma(C_{d} , \mathcal{V} )$, may be extended to the big bundle

\begin{equation}
\mathcal{D}\boldsymbol{\Psi} = 0, \hspace{1cm} \boldsymbol{\Psi} \in \Gamma(C_{d} , \boldsymbol{V} ).
\end{equation}

in which we recognize the $sl(2)$ Knizhnik-Zamolodchikov equation for the $d$-point functions in
the $2d$ WZW model with group $SU(2)$ \cite{rif63}. In that context $\lambda$ is quantized in discrete values 

\begin{equation}
\lambda=\frac{2}{k+2}
\end{equation}

for the $2d$ $SU(2)$ current algebra at level $k \in \mathbb{Z}$. Since the connection \ref{symmkohno} is invariant under the symmetric group $S_{d}$, the representation of the pure braid group $P_{d}$ extends to a representation of the full braid group of $d$ strands $B_{d}$. A representation of this type is a Hecke algebra representation \cite{rif67}, which we briefly recall. Given $q \in \mathbb{C}^{\times}$, the Hecke algebra of the symmetric group $S_{d}$, $H_{d}(q)$ is the $\mathbb{C}$-algebra \cite{rif64} with generators $1,g_{1},...,g_{d-1}$ and relations 

\begin{equation}
g_{i}g_{i+1}g_{i}=g_{i+1}g_{i}g_{i+1}, \hspace{0.5cm} g_{i}g_{j}=g_{j}g_{i}, \ \vert i-j \vert\ \geq 2, \hspace{0.5cm} (g_{i}+1)(g_{i}.-q)=0.
\end{equation}

For $q=1$ we just recover the group algebra $\mathbb{C}[S_{d}]$ of the symmetric group.
Comparing the relations of the braid group $B_{d}$ with the above ones, we see that the map $\sigma_{i}\rightarrow g_{i}$ provides
an algebra homomorphism 

\begin{equation}
\varpi: \mathbb{C}[B_{d}]\rightarrow H_{d}(q).
\end{equation}

A linear representation $\rho$ of the braid group $B_{d}$ is called a Hecke algebra representation if it factorizes through $\varpi$. In such a representation the generators $\rho(g_{i})$ have at most two distinct eigenvalues: $−1$ and $q$.\\ The Hecke algebra may be rewritten in terms of generators $e_{i}=(q-g_{i})/(1+q)$ satisfying the relations 

\begin{equation}\label{first}
e_{i}^{2}=e_{i}, \hspace{1cm} e_{i}e_{j}=e_{j}e_{i}, \ \vert j-1 \vert \geq 2, 
\end{equation}

\begin{equation}\label{second}
e_{i}e_{i+1}e_{i}-\beta^{-1}e_{i}=e_{i+1}e_{i}e_{i+1}-\beta^{-1}e_{i+1}
\end{equation}

with $\beta=2+q+q^{-1}$. The Temperley-Lieb algebra $A_{d}(q)$ \cite{rif69} is the $\mathbb{C}$-algebra over the generators $1, e_{1},...,e_{d-1}$ satisfying the relation \ref{first}, while the relation \ref{second} is replaced by the stronger condition

\begin{equation}
e_{i}e_{i+1}e_{i} - \beta^{-1}e_{i} = 0.
\end{equation}

A special class of Hecke algebra representations of $B_{d}$ are the ones which factorize through the Temperley-Lieb algebra $A_{d}(q)$. It is shown in \cite{rif65} that the monodromy representation of the flat connection \ref{symmkohno} is a Hecke algebra representation of the braid group $B_{d}$ which factorizes through the Temperley-Lieb algebra $A_{d}(q)$ with 

\begin{equation}
q= \exp (\pi i \lambda)
\end{equation}

given by the correspondence 

\begin{equation}
\sigma_{i}\rightarrow q^{-3/4}(q-(1+q)e_{i}), \hspace{0.5cm} i=1,....,d-1.
\end{equation}

The eigenvalues of the monodromies $\sigma_{i}$ belong to the set 

\begin{equation}
\left\lbrace  q^{1/4}, -q^{-3/4} \right\rbrace .
\end{equation}

By arithmeticity of the $tt^{*}$ monodromy representation, in the present context we have to require the eigenvalues of the braiding matrices to be roots of unity. Hence, the Hecke algebra representations which appear in $tt^{*}$ are the ones with $\lambda \in \mathbb{Q}$ and $q$ a root of $1$. Moreover, the identification of $\mathcal{D}_{\mathcal{V}}$ with the UV Berry connection of a very complete $tt^{*}$ geometry entails that its monodromy representation is unitary. As we shall see momentarily, the monodromy representation of the braid group $B_{d}$ defined by restricting the Knizhnik-Zamolodchikov connection to the $tt^{*}$ sub-bundle is isomorphic to the Burau one, which is known to be unitary \cite{rif61,rif70,rif71}. We note further that the Knizhnik-Zamolodchikov connection preserves not only the $tt^{*}$ vacuum bundle $\mathcal{V}$, but also all the eigenbundles $\mathcal{V}_{l,m} \in \mathcal{V}$ of given total angular momentum, i.e.

\begin{equation}
\psi \in \mathcal{V}_{l,m} \Leftrightarrow (L^{2}-l(l+1))\psi=(L_{3}-m)\psi=0
\end{equation}

with $m=l,l-1,l-2,....,-l,$ for $l \in \frac{1}{2}\mathbb{N}$.\\ The Braid group representations which factorize through the Temperley-Lieb algebra appear in many areas of mathematical physics. In particular, they describe the braiding properties of the $(p,q)$ Virasoro minimal models \cite{rif72}. In Virasoro minimal models the primary operators have dimension 

\begin{equation}
h_{r,s}= \frac{(pr-qr)^{2}-(p-q)^{2}}{4pq}, \hspace{1cm} 1\leq r \leq q-1, \ \ \ \ 1\leq s \leq p-1.
\end{equation}

The braiding the operator $\phi_{1,2}$ correspond to the Temperley-Lieb algebra with 

\begin{equation}
q = e^{\pi i \lambda} = e^{2\pi i q/p}
\end{equation}

The fusion rule for $\phi_{1,2}$ is 

\begin{equation}
\phi_{1,2}\cdot\phi_{1,2}= 1 +\phi_{1,3}.
\end{equation}

Transporting a field $\phi_{1,2}$ around another one we get a matrix with eigenvalues 

\begin{equation}
\left\lbrace  e^{2\pi i(h_{1,3}-2h_{1,2})}, e^{2\pi i(h_{1,1}-2h_{1,2})} \right\rbrace = \left\lbrace    e^{2\pi i q/2p}  , e^{-6\pi i q/2p} \right\rbrace = \left\lbrace    q^{1/2} , q^{-3/2} \right\rbrace
\end{equation}

where one has 

\begin{equation}
h_{1,1}-2h_{1,2}=1-3\frac{q}{2p}, \hspace{1cm} h_{1,3}-2h_{1,2}=\frac{q}{2p}.
\end{equation}
. 

\section{$tt^{*}$ Geometry of the Vafa model}\label{chap7}

\subsection{Generalities}

In this last section we want to study the $tt^{*}$ geometry of the Vafa model for FQHE. We consider $N$ electrons moving on the plane $\mathbb{C}$. We denote with $z_{i}, i=1,...,N$ and $x_{a}, a=1,...,n$ the positions of electrons and quasi-holes respectively, and with $\zeta_{\alpha},\alpha=1,...,M$ the support of the polar divisor $D_{\infty}$ parametrizing the magnetic flux. We take the $d=M+n$ points $\zeta_{\alpha},x_{a}$ all distinct. Hence, the target space of the SQM Vafa model of FQHE is 

\begin{equation}\label{space}
X_{d,N}=\left\lbrace (z_{1},...,z_{N}) \in \left( \mathbb{C}\setminus \lbrace x_{a},\zeta_{\alpha}\rbrace \right)^{N} \big\vert z_{i}\neq z_{j}, \  \mathrm{for} \ i\neq j \right\rbrace \big/ S_{N}.
\end{equation}

In the experimental set-up $N$ is very large, while $N/d = \nu$ and $n$ are fixed. Despite this, we shall keep $N$ arbitrary\footnote{Because of Fermi statistics we should demand at least $N\leq m+M$.} as our arguments apply to any $N$. The superpotential of the LG model is 

\begin{equation}\label{formby}
\mathcal{W}= \beta \sum_{i<j}\log (z_{i}-z_{j})^{2} + \sum_{i}\left( \mu z_{i} + \sum_{a=1}^{n}\log (z_{i}-x_{a}) - \sum_{\alpha=1}^{M} \log(z_{i}-\zeta_{\alpha})\right) ,
\end{equation}

which can be rewritten in terms of the elementary symmetric polynomials as 

\begin{equation}
\mathcal{W}= \beta \log \mathrm{discr}(P)+ \mu e_{1} + \sum_{a=1}^{n}\log P(x_{a})- \sum_{\alpha=1}^{M}\log P(\zeta_{\alpha}),
\end{equation}

where $P(z)= \sum_{k=0}^{N} (-1)^{k}e_{k}z^{N-k} $. As explained in section \ref{wedisc}, we have introduced the coupling $\mu$ to make the problem better behaved. As long it is non zero, it can be fixed to the value that we prefer, since the $tt^{*}$ monodromy is indipendent from the couplings. We note that the superpotential $\mathcal{W}$ is not univalued in $X$. In this model we have two kinds of couplings: the residues of $d\mathcal{W}$ at its poles and the positions $x_{a}, \zeta_{\alpha}$. The residues of the poles at $x_{a}$ and $\zeta_{\alpha}$ are frozen to the values $±1$ by the arguments in \ref{wedisc} and do not play any role in the discussion. The only relevant coupling is $\beta$. Working in a periodic box $\beta$ is frozen to the rational number $1/2\nu$. On the other hand, when the electrons move on the plane the model is well defined for arbitrary values of $\beta$. Since the monodromy representation is independent of $\beta$ we are free to deform it away from its physical value $1/2\nu$ to simplify the analysis. The non-frozen couplings are the $x_{a}$ and the $\zeta_{\alpha}$. These form a set of $d$ distinct points in $\mathbb{C}$ in which are identified modulo permutation the ones with equal charge. The manifold of essential couplings
is therefore 

\begin{equation}
\mathcal{P}=C_{n+M}/S_{n}\times S_{M}.
\end{equation}

The $\zeta_{\alpha}$ are homogeneously distributed in order to reproduce the macroscopic magnetic field. The details of the distribution are not important for our purposes and we can focus on the projection on $Y_{n}$. The fundamental group $\pi_{1}(\mathcal{P})$ contains $B_{n}$. The UV Berry connection yields a family of unitary arithmetic representations of $\pi_{1}(\mathcal{P})$. Restricting to $B_{n}$ we get a monodromy representation 

\begin{equation}
\rho(\vec{\theta}):B_{n}\rightarrow GL(\mathcal{V}_{\theta})
\end{equation}

which is parametrized by the character $\vec{\theta}$ of the Galois group of the universal cover $\pi_{1}(X_{d,N})$. The dimension of the vacuum sector at fixed $\vec{\theta}$ is $d_{d,N}= \begin{pmatrix} N+d-1 \\ N \end{pmatrix}$, which reduces to $d_{d,N}^{f}= \begin{pmatrix} d \\  N \end{pmatrix}$ after the fermionic truncation discussed in \ref{fermtrunc}.\\ The fundamental group of the target manifold $X_{d,N}$ is 

\begin{equation}
\pi_{1}(X_{d,N})= \mathcal{B}(N,S_{0,d+1}),
\end{equation}

where $\mathcal{B}(N,S_{g,p}) $ is the braid group of $N$ strings on the surface $S_{g,p}$ of genus $g$ with $p$ punctures. This has the following  presentation 

\begin{equation}
\begin{split}
& \mathrm{generators}:  \sigma_{1},\sigma_{2},...,\sigma_{N-1},z_{1},z_{2},...,z_{p-1} \\ 
& \mathrm{relations}: \begin{cases} \sigma_{i}\sigma_{i+1}\sigma_{i}= \sigma_{i+1}\sigma_{i}\sigma_{i+1}, \hspace{2cm} \sigma_{i}\sigma_{j}=\sigma_{j}\sigma_{i} \ \ \mathrm{for} \vert i-j \vert \geq 2, \\ 
z_{j}\sigma_{i}=\sigma_{i}z_{j} \ \mathrm{for}\ i\neq 1, \hspace{2.3 cm} \sigma_{1}^{-1}z_{j}\sigma_{1}^{-1}z_{j}= z_{j}\sigma_{1}^{-1}z_{j}\sigma_{1}^{-1}, \\ 
\sigma_{1}^{-1}z_{j}\sigma_{1}^{-1}z_{l}= z_{l}\sigma_{1}^{-1}z_{j}\sigma_{1}^{-1} \ \ \mathrm{for} \ j<l.
\end{cases}
\end{split}
\end{equation}

The $\sigma_{i}$ generates a subgroup of $\mathcal{B}(N,S_{0,d+1})$ which is isomorphic to the Artin Braid group $\mathcal{B}_{N}$. The abelianized Galois group $\mathcal{B}(N,S_{0,d+1})^{\mathrm{Ab}}\simeq \mathbb{Z}^{d+1}$ is the homology group of the target space generated by $\sigma,z_{1},...,z_{d}$. The dual cohomology group $H^{1}(X_{d,N},\mathbb{Z})$ has the generators 

\begin{equation}
\frac{1}{2\pi i} d \log \mathrm{discr}(P), \hspace{1cm} \frac{1}{2\pi i }d\log P(x_{a}), \hspace{1cm} \frac{1}{2\pi i} d\log P(\zeta_{\alpha}).
\end{equation}

For each of this generator we have an angle 

\begin{equation}
\theta, \hspace{1cm} \phi_{a}, \hspace{1cm} \varphi_{\alpha}.
\end{equation}

It is natural to consider the quasi-holes and the magnetic-flux units to be indistinguishable and take the corresponding angles to be all equal $\phi_{a}=\phi$ and $\varphi_{\alpha} = \varphi$. This is equivalent to take the quotient group of $H_{1}(X_{d,N},\mathbb{Z})$ dual to the subgroup of $H^{1}(X_{d,N},\mathbb{Z})$ generated by the differentials 

\begin{equation}
\frac{1}{2\pi i} d \log \mathrm{discr}(P), \hspace{1cm} \frac{1}{2\pi i }d\prod_{a} \log P(x_{a}), \hspace{1cm} \frac{1}{2\pi i} d\log \prod_{\alpha} P(\zeta_{\alpha}),
\end{equation}

and define the LG model on the Abelial cover $\mathcal{A}_{H}=\mathcal{A}/H$, where $H$ is the kernel of the group homomorphism $H_{1}(X_{d,N},\mathbb{Z})\rightarrow \mathrm{Gal}(\mathcal{A}_{H})$. In particular, the Galois group of the cover is $\mathrm{Gal}(\mathcal{A}_{H})\simeq \mathbb{Z}^{3}$ and so we have three angles $\theta,\phi$ and $\varphi$. Setting
$q=e^{i\theta},t=e^{i\phi}$,and $y=e^{i \varphi}$ we conclude that in the LG model with indistinguishable defects the BPS branes span a free module of rank $d_{d,N}$ over the ring $\mathbb{Z}[q^{\pm 1}, t^{\pm 1}, y^{\pm 1}]$. The normalizability of the ground states requires to $q, t $ and $y$ to be roots of unity. Moreover, as we are going to see, the physical FQHE requires further truncations.

\subsection{The Physical FQHE}

The FQHE quantum system is a particular version of the 4-susy LG model with superpotential $\mathcal{W}$ given previously. Quasi-holes and magnetic-flux units are indistinguishable, but we have still to fix the characters $\theta, \phi$ and $\varphi$.
The roles of the punctures $\zeta_{\alpha}$ is to reproduce the uniform constant magnetic field $B$ through the isomorphism discuss in section \ref{wedisc}, which holds for the trivial character $\phi=0 \ \mathrm{mod} \ 2\pi$. Since the quasi-holes are just insertion of mangetic fluxes with the opposite sign, it is natural to take also $\varphi=0 \ \mathrm{mod} \ 2\pi$.
We remain with just one non-trivial angle $\theta$ associated to the Vandermonde coupling $\beta$. In order to have normalizable states the angle $\theta$ should be a rational multiple of $2\pi$

\begin{equation}
\theta= \pi \left( 1+\frac{a}{b}\right) , \hspace{1cm} a \in \mathbb{Z}, b \in \mathbb{N}, \hspace{0.5cm} \big\vert \mathrm{gcd}(a,b)=1 \hspace{0.5cm} \mathrm{and} \ -b\leq a \leq b.
\end{equation}

The $tt^{*}$ reality structure relates $\theta$ with $-\theta$ and so $a$ with $-a$. So, if a phase with pair $(a,b)$ exists, it is natural to expect also a dual phase associated to $(-a,b)$. These corresponds to the same representation of the Galois group and we can write $(\pm a,b)$ with $1\leq a \leq b$ to cover both phases at once. The LG model which we obtain with these restrictions is defined on the abelian cover of $X$ associated to $\theta$

\begin{equation}
\mathcal{A}_{\theta}=\left\lbrace p :[0,1] \rightarrow X, p_{0}=p_{*} \right\rbrace / \sim_{b},
\end{equation}

where 

\begin{equation}
\sim_{b}= \begin{cases}  p(1)=p^{\prime}(1) \\ 
p^{\prime}=p\cdot \sigma^{2bk} \ \mathrm{in} \ H^{1}(X_{d,N},\mathbb{Z}), \ k \in \mathbb{Z}.
\end{cases}
\end{equation}

The dimension of the generic $\theta$-sector of the ground state $V_{\theta}$ corresponds to the bosonic statistics $d_{d,N}$. To get the correct physical counting of states we need to consider its fermionic truncation, i.e. to keep only the states which are stable as $\beta\rightarrow 0$. Since the monodromy is independent of $\beta$ (as long as it is not zero), and its limit as $\beta\rightarrow 0$ is smooth after the fermionic truncation, we may set $\beta=0$ while keeping track of the non-trivial topology of the target space through the character $\theta$. In the language of \cite{rif23} this is called $\theta$-limit. According to the discussion in \ref{cov}, switching a non zero $\theta$ is equivalent to pull-back the operators $\phi_{a}$ of the fermionic chiral ring $\mathcal{R}_{F}$ on $\mathcal{A}_{H}$ and define the basis 

\begin{equation}
\Phi_{a}(\theta)= \varpi^{*}\phi_{a}e^{i\theta h},
\end{equation}

where $\varpi: \mathcal{A}_{\theta}\rightarrow X$ and the chiral field $e^{ih}$ is proportional to 

\begin{equation}
\mathrm{discr}(P)^{\theta/2\pi}= \prod_{i<j}(z_{i}-z_{j})^{\theta/\pi}.
\end{equation}

It follows that the entries of the $tt^{*}$ monodromy matrices will take values in $\mathbb{Z}[q^{\pm1}]$, with $q=e^{i\theta}$. Keeping into account the Jacobian factor arising from the map $\lbrace z_{i}\rbrace \rightarrow \lbrace e_{k} \rbrace $, the vacuum wave functions in terms of $z_{i}$ contains the factor 

\begin{equation}
\prod_{i<j}(z_{i}-z_{j})^{1+\theta/\pi}, \hspace{0.5cm} 0\leq \theta \leq 2\pi .
\end{equation}

The comparison with the Laughlin wave function \ref{Laugh} leads to the identification 

\begin{equation}\label{fillingab}
\frac{1}{\nu}= 1 + \frac{\theta}{\pi}= 2 \pm \frac{a}{b}
\end{equation}

which gives $1\leq 1/\nu \leq 3$. In particular, the minimal $b$-torsion character, $a=1$ yields the FQHE principal series 

\begin{equation}
\nu=\frac{b}{2b \pm 1}, \ b \in \mathbb{N}.
\end{equation}

Although this series is the most natural LG quantum systems of the form \ref{formby}, it is by no means the only possibility in the present framework. We are going to discuss the allowed values of the filling fraction in \ref{detlam}. The above analysis recalls in some sense the idea of Jain's composite fermion theory \cite{ab4}. If we consider the character $e^{2\pi i (1+\theta/\pi)} $ as the Aharonov-Bohm phase generated by taking an electron around another one, we can interpret the degrees of freedom of the Vafa model in the $\theta$-limit as free fermions bounded to $\frac{\theta}{\pi}=1/\nu-1$ units of magnetic flux. Similarly to the Jain's proposal, the picture emerging from the $4$-SQM model is that we can treat the fractional quantum Hall effect as an integer one for composite fermions.

\subsection{Fermi Statistics and Hecke Algebra Representations}

The fermionic truncation allows us to effectively put the coupling $\beta$ to zero and consider the LG Fermi model of $N$ chiral fields with superpotential 

\begin{equation}
\mathcal{W}(e_{i})= \sum_{i=1}^{N} W(z_{i})
\end{equation}

where the one-field model is 

\begin{equation}
W(z)= \mu z +\sum _{a=1}^{n}\log(z-x_{a})-\sum_{\alpha=1}^{M}\log(z-\zeta_{\alpha}).
\end{equation}

Following \cite{rif55} we can prove that the $tt^{*}$ geometry of the single particle model is very complete. Since the monodromy representation is indipendent from $\mu$, we can take $\mu\gg 1$. The equations for the classical vacua

\begin{equation}
\mu= \sum_{\alpha}=\frac{1}{z-\zeta_{\alpha}}-\sum_{a}\frac{1}{z-x_{a}},
\end{equation}

have $d=n+M$ solutions of the form $z=x_{a}+O(1/\mu)$ or $z=\zeta_{\alpha}+ O(1/\mu)$. Up to an irrelevant rescaling by a factor $\mu^{-1}$, the set of critical values is

\begin{equation}
\left\lbrace w_{1},...,w_{n+M}\right\rbrace = \left\lbrace x_{1},...,x_{n},\zeta_{1},...,\zeta_{M}\right\rbrace + O\left( \frac{1}{\mu}\log \mu \right). 
\end{equation}

The cover of the coupling constant space $\mathcal{P}=C_{n+M}/S_{n}\times S_{M}$ is the configuration space of $n+M$ distinct points $C_{n+M}$, which can be seen also as the cover of the space $Y_{n+M}=C_{n+M}/S_{n+M}$ where the critical map $w:\mathcal{P}\rightarrow Y_{n+M}$ take values. By the above equality we see that the canonical projection $p:C_{n+M}\rightarrow Y_{n+M}$ factorizes thorugh the critical map. This shows that the $tt^{*}$ geometry of the one-field model is very complete and so the UV Berry connection is a flat $sl(2)$ Kohno connection 

\begin{equation}
\mathcal{D}=d+ \lambda \sum_{i<j}s_{\ell}^{i}s_{\ell}^{j} \frac{d(w_{i}-w_{j})}{w_{i}-w_{j}},
\end{equation}

which provides a monodromy representation of $\pi_{1}(Y_{n+M})=B_{n+M}$ acting on the space $V^{n+M}=\bigotimes_{i} V_{i}$ with $V_{i}\simeq \mathbb{C}^{2},i=1,...,n+M$, restricted to the subspace $\mathcal{V}_{1}$ with total angular momentum

\begin{equation}
L_{3}=1-(n+M)/2.
\end{equation}

Taking the projection $p:Y_{n+M}\rightarrow Y_{n}$ we obtain a monodromy representation of the subgroup $B_{n}$ which factors through a Temperley-Lieb algebra with $q=e^{\pi i \lambda}$. \\ The vacuum bundle of the $N$ particle model is 

\begin{equation}
\mathcal{V}=\bigwedge^{N} \mathcal{V}_{1}\rightarrow \mathcal{P}
\end{equation}

and the UV Berry connection is the one induced on the $N$-th rank antisymmetric representation by the single field UV Berry connection. It is convenient to introduce the ``Grand-canonical '' bundle 

\begin{equation}
\mathcal{W}= \bigoplus_{N=0}^{d}\mathcal{V}_{N}\rightarrow \mathcal{P}, \hspace{1cm} \mathrm{rank} \ \mathcal{W}=\sum_{N=0}^{d} \begin{pmatrix} d \\ N \end{pmatrix}= 2^{d},
\end{equation}

where the total number of states is $2^{d}$ since each of the $d=n+M$ one-particle (vacuum) states may be either empty or occupied. In the above direct sum of bundles we included $\mathcal{V}_{0}$ which does not correspond to any LG model, since the number of fields is zero. This can be done without any harm by the isomorphism $\mathcal{V}_{0}\simeq \mathcal{V}_{d}$. The bundle $\mathcal{V}_{d}$ corresponds to the trivial $\nu=1$ phase and should have a trivial monodromy representation. It is easy to see that 

\begin{equation}
\mathrm{fiber}(\mathcal{W})\simeq V^{\otimes d},
\end{equation}

where $V^{\otimes d}$ is the fiber of the big bundle $\boldsymbol{V}\rightarrow \mathcal{P}$. According to the discussion in section \ref{hecke}, one associates to each vacuum of the single particle model a spin degree of freedom $s_{\ell}^{j}$. A spin down (up) corresponds to the insertion in the Ising-like correlators of a $\mu \ (\sigma)$-like operator at the critical value $w_{j}$ of the $j$-th vacuum. We can define a correspondence with the fiber of the Grand bundle as follows: a spin down (up) means that the $j$-th vacuum is empty (occupied). Hence, a vacuum with occupied state $\left\lbrace j_{1},....,j_{N}\right\rbrace $ corresponds linearly to the element of the $N$-particles Fermi chiral ring 

\begin{equation}\label{frame}
\sum_{\sigma \in S_{N}}\mathrm{sign}( \sigma)e_{j_{1}}(z_{\sigma(1)})e_{j_{2}}(z_{\sigma(2)}).......e_{j_{N}}(z_{\sigma(N)}) \in \left( \mathcal{R}_{N} \right)_{a},
\end{equation}

where $e_{j}(z) ,j=1,...,d$ are the orthogonal idempotents of the one-particle chiral ring $\mathcal{R}_{1}$. So we conclude that 

\begin{equation}
\mathcal{W}\simeq \boldsymbol{V},
\end{equation}

and the linear PDEs satisfied by the brane amplitudes are just the $sl(2)$ Knizhnik-Zamolodchikov equations up to a twist by  normalization factors. In this correspondence the angular momentum $L_{3}=\sum_{j=1}^{d}s_{3}^{j}$ is related to the operator number of particles by 

\begin{equation}
\hat{N}=L_{3} + \frac{d}{2},
\end{equation}

where  $L_{\ell}=\sum_{j=1}^{d}s^{j}_{\ell}$ are the generators of $sl(2)_{\mathrm{diag}}$. The underlying one-particle model, having a very complete symmetric $tt^{*}$ geometry, defines a Kohno connection acting on $V^{\otimes d}$ that we argued has the $sl(2)$ KZ form up to an overall twist, i.e.

\begin{equation}\label{twisting}
\mathcal{D}=d + \lambda(\theta)\sum_{i<j}s_{\ell}^{i}s_{\ell}^{j} \frac{d(w_{i}-w_{j})}{w_{i}-w_{j}} + \xi(\theta) \sum_{i<j}d\log(w_{i}-w_{j}).
\end{equation}

As we shall see, it turns out that the constant $\xi(\theta)$ is related to $\lambda(\theta)$ by consistency conditions. The factor $\lambda(\theta)$ should be some piece-wise linear function of $\theta$ and in the context of physical FQHE is related to the filling fraction labelling the different quantum phases. We are going to determine it in the next section. The eigenbundles $\mathcal{V}_{N}$ are preserved by parallel transport with $\mathcal{D}$, and hence define a monodromy representation of $B_{n}$ which is the one associated to the $N$-field Fermi theory. These representations are generically reducible, since also the eigenbundles of the total angular momentum $L^{2}=\sum_{\ell}L_{\ell}L_{\ell}$ are preserved by parallel transport. We have the monodromy invariant decomposition 

\begin{equation}
\mathcal{W}=\bigoplus_{l=0, \ l=\frac{1}{2}\mathbb{N}}^{d/2} \ \bigoplus_{m=-l}^{l}\mathcal{V}_{l,m}, \hspace{1cm}\mathcal{V}_{l,m}= \mathrm{ker} \left( L^{2}-l(l+1)\right) \bigcap \mathrm{ker} \left( L_{3}-m\right) 
\end{equation}

and 

\begin{equation}\label{decomp}
\mathcal{V}_{N}=\bigoplus_{l= \vert N-d/2 \vert }^{d/2} \mathcal{V}_{l,m=N-d/2}.
\end{equation}

Since the monodromy representation centralizes with respect to $sl(2)_{\mathrm{diag}}$ we have 

\begin{equation}
\mathcal{V}_{l,m} \simeq \mathcal{V}_{l,m^{\prime}}, \hspace{0.5cm} -l \leq m,m^{\prime} \leq l,
\end{equation}

from which it follows that the monodromy representations $\mathcal{V}_{N},\mathcal{V}_{d-N}$ are isomorphic. The rank of $\mathcal{V}_{l,m}$ is given by the Catalan triangle 

\begin{equation}
\mathrm{rank} \ \mathcal{V}_{l,m}= \begin{pmatrix} d \\ \frac{d}{2}-l \end{pmatrix} - \begin{pmatrix}
d \\ \frac{d}{2}-l -1 \end{pmatrix} \hspace{0.5cm} \mathrm{for} \ 0\leq l \leq \frac{d}{2}, \ -l\leq m \leq l.
\end{equation}

In the decomposition \ref{decomp} of $\mathcal{V}_{N}$ there is a unique eigenbundle of rank $1$ which corresponds to $\mathcal{V}_{d/2,N-d/2}$, namely the one which maximizes $L^{2}$. This is spanned by the unique monodromy invariant vacuum $\ket{d/2,N-d/2}$ of the Fermi model with $N$ electrons. It is tempting to identify this state with the preferred vacuum discussed in \ref{wedisc}. For a given number of particles $N$, the determinant of the brane amplitude matrix spans the fiber of a rank $1$ eigenbundle of the UV Berry connection. Hence, we have the isomorphism of bundles 

\begin{equation}
\mathrm{det}: \bigwedge^{N}\mathcal{V}_{1}\rightarrow \mathcal{V}_{d/2,N-d/2}\simeq \mathcal{V}_{d/2,d/2}.
\end{equation}

The isomorphism between $\mathcal{V}_{d/2,N-d/2}$ and the vacuum bundle of the trivial phase $\mathcal{V}_{N=d}$ implies that the preferred vacuum should have trivial monodromy. This allows to fix the constant $\xi(\theta)$. The restriction of the Kohno connection to the subbundle $\mathcal{V}_{d/2,d/2}$ is 

\begin{equation}
\mathcal{D}_{\mathcal{V}_{d/2,d/2}}= d+ \frac{\lambda(\theta)}{4}\sum_{i<j}d\log(w_{i}-w_{j}).
\end{equation}

In order to have a trivial UV Berry connection we have to demand 

\begin{equation}
\xi(\theta)=-\frac{\lambda(\theta)}{4}.
\end{equation}

In other words, the normalized amplitudes $\Psi_{\mathrm{norm}}$ of the Grand-bundle are related to the KZ ones $\Psi$ by 

\begin{equation}
\Psi_{\mathrm{norm}}= \frac{\Psi}{\Psi_{\mathrm{priv}}}
\end{equation}

where $\Psi_{\mathrm{priv}}$ is a parallel section of the line-bundle $\mathcal{V}_{d/2,d/2}$. In particular, the normalized monodromy is trivial for the $\nu = 1 $ phase.\\ Since it left invariant by the UV Berry connection, the preferred vacuum is the most symmetric one under permutations of the spin degrees of freedom. This state should correspond to some linear combinations of the idempotents $e_{i}$ of the chiral ring $\mathcal{R}_{N}$ such that the electrons are as spread out as possible among the single particle vacua.  As discussed at the end of section \ref{wedisc}, as long as the susy-breaking part of the interacting Hamiltonian preserves the symmetry between the holes and the units of fluxes, it lifts the degeneracy keeping the most symmetric state as the ground state. So it is natural to identify this state with the true FQHE ground state under the isomorphism developed in \ref{wedisc}.

\subsection{Determining $\lambda(\theta)$ and the allowed filling fractions}\label{detlam}

A direct computation of $\lambda(\theta)$ is quite hard and subtle. We can fix it by imposing consistency conditions. An intrinsecally defined quantity which is indipendent from the bundle trivialization is $q(\theta)^{2}$, namely the ratio of the two eigenvalues of $\sigma_{i}^{2} \in P_{n+M}$ which correspond to the operation of transporting a quasi-hole around another one and getting back to the original position. We have the relation 

\begin{equation}
q(\theta)^{2}= e^{2\pi i \lambda(\theta)}.
\end{equation}

By the functorial property of the Hecke algebra representations, the parameter $\lambda(\theta)$ should be a universal function indipendent from $n$ and $M$. Moreover, since it is related to the Ramond charges of the LG model, it should be a piece-wise linear function of the angle defined mod $1$, i.e.

\begin{equation}
\lambda(\theta)= C_{1}+C_{2}\frac{\theta}{\pi} \ \mathrm{mod} \ 1,
\end{equation}

for some real constants $C_{1},C_{2}$. We can assume $C_{2}>0$ by changing the sign of $\theta$. The periodicity in $\theta$ and the $tt^{*}$ reality structure which relates $\pm \theta$ imply 

\begin{equation}
q(\theta+2\pi)^{2}=q(\theta)^{2}, \hspace{1cm} q(-\theta)^{2}= q(\theta)^{-2}.
\end{equation}

These conditions require $2C_{1}=0 \ \mathrm{mod } \ 1$ and $2C_{2}=0 \ \mathrm{mod } \ 1$. Imposing the same condition on the ratio $q(\theta)$ of the two eigenvalues of $\sigma_{i} \in B_{n+M}$ one obtains the stronger conditions $C_{1}=0 \ \mathrm{mod } \ 1$ and $C_{2}=0 \ \mathrm{mod } \ 1$. Demanding $q(\theta)$ to be strictly periodic of $2\pi$ (and not of a fraction $2\pi/C_{2}$), we find the solution

\begin{equation}
\lambda(\theta)= \frac{\theta}{\pi} \ \mathrm{mod} \ 1.
\end{equation}

This identification is quite natural, since the $tt^{*}$ monodromy is defined over the ring $\mathbb{Z}[e^{\pm i \theta}]$, while the KZ monodromy is defined over $\mathbb{Z}[e^{\pm i \pi \lambda(\theta)}]$. Hence, the above solution leads to the identification of the two Laurent polynomial rings. By the relation \ref{fillingab} we have

\begin{equation}\label{piùsol}
q(\theta)^{2}= e^{2\pi i/\nu}.
\end{equation}

We assume that the UV Berry connection is a Knizhnik-Zamolodchikov connection for $SU(2)$ current algebra with level $k$ quantized in integral units, namely 

\begin{equation}
\lambda=\pm \frac{2}{k+2}, \hspace{1 cm} k \in \mathbb{Z}.
\end{equation}

From the point of view of $tt^{*}$ geometry we do not know any compelling argument to require $k+2 \in \mathbb{Z}$. However, it is believed that $2d$ WZW models describe the modes which live at the edge of a non-Abelian Chern-Simons theory with boundary. Further, it turns out that the braiding of their conformal blocks coincides with the braiding of Wilson lines in the Chern-Simons theory. Hence, if we want our $\mathcal{N}=4$ theory to admit an effective IR description in terms of a $3d$ topological field theory, it is natural to consider a subclass of LG models compatible with the above condition. By the $tt^{*}$ reality structure we have to consider the solutions with both signs. We first study the class of solutions with $C_{1}$ an odd integer. This is the most natural case, since we have 

\begin{equation}
e^{\pm 2\pi i/(k+2)}= e^{i\pi /\nu}.
\end{equation}

The equation 

\begin{equation}\label{secondsol}
\pm \frac{2}{k+2}= 1 +\frac{\theta}{\pi} \ \mathrm{mod} \ 2= \pm \frac{a}{b} \ \mathrm{mod} \ 2
\end{equation}

has solutions $a=1$ with $k$ even and $a=2$ with $b$ and $k$ odd. In agreement with the prediction in \cite{rif1}, the first case corresponds to the principal series of FQHE

\begin{equation}
\nu= \frac{b}{2b \pm 1}, \hspace{0.5cm} b \in \mathbb{N}.
\end{equation}

From point of view of \cite{rif1}, the element $\sigma_{i}^{2}$ of the pure braid group for the principal series has two distinct eigenvalues, in correspondence with the two different fusion channels of the $\phi_{1,2}$ operator in the minimal $(2n, 2n\pm1)$ Virasoro model. The ratio of the two eigenvalues is

\begin{equation}
q^{2}=\frac{\exp [ 2\pi i (h_{1,3}-2h_{1,2} ) ] }{\exp [ 2\pi i (h_{1,1}-2h_{1,2})] }=\exp(2\pi i/\nu),
\end{equation}

which is in agreement with the $tt^{*}$ result. The model admits also another list of filling fractions with denominatior divisible by $4$, namely 

\begin{equation}
\nu= \frac{b}{2(b \pm 1)}, \hspace{0.5cm} b \ \mathrm{odd}
\end{equation}

which correspond to the second solution of \ref{secondsol}.\\ We may consider also the less natural class of solutions in which $C_{1}$ is an even integer. In this case we have

\begin{equation}
q(\theta)=-e^{i\pi/\nu}= e^{\pm 2\pi i /(k+2)}
\end{equation}

Demanding $1\leq 1/\nu \leq 3$ we find two possible series of filling fractins. One corresponds to 

\begin{equation}
1+ \frac{2}{k+2}= 1 + \frac{\theta}{\pi} \ \mathrm{mod} \ 2= \frac{1}{\nu} \ \mathrm{mod} \ 2
\end{equation}

which implies 

\begin{equation}
\nu= \frac{m}{m+2}, \hspace{1cm} m=k+2 \in \mathbb{N}\geq 2.
\end{equation}

This series contains the values of $\nu$ corresponding to the Moore-Read \cite{nab2} and Read-Rezayi models \cite{nab3}. The second possibility is 

\begin{equation}
3-\frac{2}{k+2}= \frac{1}{\nu} \ \mathrm{mod} \ 2
\end{equation}

which gives 

\begin{equation}
\nu= \frac{m}{3m-2}, \hspace{1cm} m=k+2\geq 2.
\end{equation}

\subsection{Comparison with the Homological Approach}

We can compare the $tt^{*}$ monodromy representation computed in the UV limit with the one obtained in the asymmetric limit. In general the two are different unless the UV limit is regular. However, for a very complete $tt^{*}$ geometry, where $\pi_{1}(\mathcal{P})$ = $B_{d}$, the asymmetric limit monodromy yields the so-called homology representation of the braid group, which is essentially equivalent to the monodromy of the Knizhnik-Zamolodchikov connection. Following again \cite{rif55}, we can write an explicit integral representation of solutions to the $sl(2)$ Knizhnik-Zamolodchikov equation

\begin{equation}\label{integralez}
\Psi_{\mathrm{KZ}}= f(\zeta_{1},...,\zeta_{d}) \int_{\Gamma} \prod_{i,\alpha}(z_{i}-\zeta_{\alpha})^{\frac{1}{b^{2}}}e^{\tilde{\mu} \sum_{i} z_{i}}\prod_{i<j}(z_{i}-z_{j})^{-\frac{2}{b^{2}}} dz_{1} \wedge \cdot \cdot \cdot \wedge dz_{N}
\end{equation}

where $\tilde{\mu} \in \mathbb{C}^{\times} $ and 

\begin{equation}\label{overallfactor}
f(\zeta_{\alpha})=\prod_{\alpha <\beta}(\zeta_{\alpha}-\zeta_{\beta})^{-\frac{1}{2b^{2}}}e^{-\frac{\tilde{\mu}}{2}\sum_{\alpha}\zeta_{\alpha}}.
\end{equation}

The amplitudes \ref{integralez} are known to compute the Virasoro conformal blocks \cite{rif55}

\begin{equation}
\langle \phi_{1,2}(\zeta_{1})...\phi_{1,2}(\zeta_{d})V_{\tilde{\mu}}(\infty)\rangle
\end{equation} 

containing insertions of the primary field $\phi_{1,2}$ of the Virasoro algebra and an irregular vertex operator $V_{\tilde{\mu}}$ \cite{rif29,rif30,rif31}. The central charge $c$ is related to the Virasoro parameter $b$ by 

\begin{equation}
c=1+6 Q^{2}, \hspace{1cm} Q=b+\frac{1}{b} .
\end{equation}

The integration contour $\Gamma$ denotes a basis of reltative cycles which ensure the convergence of the integral. One denotes with $C_{a}, \ a=1,...,d$ a basis of Lefschetz thimbles for the homorphic potential

\begin{equation}
W(z;\zeta_{\alpha})=  \tilde{\mu} z  + \frac{1}{b^{2}}\sum_{\alpha=1}^{d}\log(z-\zeta_{\alpha}) ,
\end{equation}

which has the same functional form of the Vafa superpotential for a single electron and $d$ quasi-holes, but with `renormalized' couplings. Then, $\Gamma$ can be constructed by taking products of cycles

\begin{equation}
C_{j_{1},...,j_{N}}= C_{j_{1}} \times C_{j_{2}} \times ..... \times C_{j_{N}},
\end{equation}

in which $j_{i}\neq j_{k}$ for $i\neq k$. Hence, for a fixed $N$ we have $\begin{pmatrix}d \\ N \end{pmatrix}$ integration cycles. By parallel transport with the Knizhnik-Zamolodchikov connection, the space of Virasoro conformal blocks has the structure of a Hecke algebra representation of $B_{d}$ which factorizes through the Temperley-Lieb algebra $A_{d}(q)$ with  \cite{rif55}

\begin{equation}
q=e^{-2\pi i/b^{2}}.
\end{equation}

As suggested in \cite{rif1}, it is natural to identify the above integrals with the asymmetric limit of the $tt^{*}$ brane amplitudes given by the pairing between branes and the topological vacuum, i.e. the one corresponding to the identity operator. After the rescaling $w_{j} \rightarrow \beta w_{j}$ we have to send $\beta$ and the spectral parameter $\zeta$ to zero in such a way that the ratio $\beta/\zeta$ remains finite. Hence, we write

\begin{equation}\label{correspasymm}
\Psi^{\mathrm{asym}}_{\Gamma}= \int_{\Gamma} e^{\frac{\beta}{\zeta} \mathcal{W}} \prod_{i<j}(z_{i}-z_{j})^{1+\frac{\theta}{\pi}} dz_{1} \wedge \cdot \cdot \cdot \wedge dz_{N},
\end{equation}

where 

\begin{equation*}
\mathcal{W}= \sum_{i=1}^{N}\left(  \mu z_{i}  + \sum_{\alpha=1}^{d} \log(z_{i}-\zeta_{\alpha}) \right) .
\end{equation*}

The identity operator inside the brane amplitudes gets corrected by the factor $\prod_{i<j}(z_{i}-z_{j})^{1+\frac{\theta}{\pi} }$, which is generated by the projection on the fermionic sector and pull-back on the abelian cover $\mathcal{A}_{\theta}$. The correspondence with the Virasoro conformal blocks requires $\beta/\zeta=1/b^{2}$ and $\tilde{\mu}=\beta\mu/\zeta$. From the equality \ref{fillingab} we have

\begin{equation}
b=i\sqrt{2\nu}.
\end{equation}

The relative cycles $\Gamma$ provide a basis for the space of branes $B_{N}(\zeta)$ of the $N$ particle Fermi model, which has the structure of a $\mathbb{Z}[e^{\pm i \theta}]$-module of rank $\begin{pmatrix}d \\ N \end{pmatrix}$. The monodromy representation of the Knizhnik-Zamolodchikov connection of the $N$ particle sector coincides with the monodromy representation of the Gauss-Manin connection acting on $B_{N}(\zeta)$. We observe that the $tt^{*}$ amplitudes differ from the Virasoro conformal blocks by the overall normalization factor $f(\zeta_{\alpha})$, which is a multi-valued function of the $\zeta_{\alpha}$. Hence, as the UV Berry connection \ref{twisting}, the $tt^{*}$ connection in the asymmetric limit coincides with the Knizhnik-Zamolodchikov connection up to a line bundle twisting

\begin{equation}
\mathcal{D}\rightarrow \mathcal{D} + \xi\sum_{\alpha < \beta}d\log(\zeta_{\alpha}-\zeta_{\beta})
\end{equation}

where 

\begin{equation}
\xi=\frac{1}{2b^{2}}.
\end{equation}

The comparison between \ref{integralez} and \ref{correspasymm} leads to fix the parameter $\lambda$ of the Kohno connection in terms of $\nu$. One obtains

\begin{equation}
q=e^{i\pi \lambda(\theta)}=e^{i\pi/\nu}
\end{equation}

which is the solution to \ref{piùsol} with positive sign. However, there are other relative normalizations which are compatible with $tt^{*}$ periodicity and reality structure. According to the discussion in the previous section, some of these may lead to the solution with negative sign.

\appendix

\section{The Long-Range Limit of the Vafa Hamiltonian}

\subsection{One Electron in Uniform Magnetic Field}

We discussed in section \ref{wedisc} the equivalence between the Landau description of an electron in a uniform magnetic field and   a certain class of $4$-SQM Landau-Ginzburg theories. In the case of a particle moving on the complex plane, the Vafa superpotential is given by 

\begin{equation} \label{wlattice}
W= \sum_{i} \Phi_{i}\log(z-\zeta_{i}),
\end{equation}

where $\Phi_{i} \in \mathbb{C}^{\times}$ and $\zeta_{i}$ are all distinct. In order to reproduce a large macroscopically uniform magnetic field we have to take the residues $\Phi_{i}$ all equal and consider a uniform distribution of the flux sources $\zeta_{i}$ in $\mathbb{C}$. We may proceed by analogy with the CFT description of FQHE and consider a lattice of fundamental units of magnetic flux $\Phi(a)=a^{2}\frac{B}{2\pi}$ at positions $\zeta_{m,n}(a)=am+ian$, $\ m,n \in \mathbb{Z}$, where $a$ is a real parameter. The lattice $\Lambda(a)= a\mathbb{Z}+ia\mathbb{Z}$ is contained in a disk $D(R)$ of radius $R\gg a $ which defines the size of the sample. As we already discussed in section \ref{theoryofedgestates}, one can recover the uniform constant $B$ field by taking the continuous limit of the lattice and then sending $R$ to infinity. The immaginary part of the logarithmic superpotential oscillates very rapidly for $a\ll 1$ and the average of the oscillations vanishes in the limit $a\rightarrow 0 $. Instead, the real part of $W$ coincides with the $2$-dimensional electrostatic potential of a discrete set of charges and tends in the same limit to the potential of a continuous charge distribution. Thus, we remain with a real superpotential 

\begin{equation}\label{limitsuperpotential}
\Phi(a)\sum_{\Lambda(a)\cap D(R)} \log (z-\zeta_{m,n}(a)) \xrightarrow{ a\rightarrow 0, \ R\rightarrow \infty} B \ \vert z \vert ^{2}/4.
\end{equation}

This convergence is natural from the point of view of electrostatics, but it is non-trivial in the context of SQM. The model is based on $\mathcal{N}=4$ supersymmetry which requires the superpotential to be holomorphic. So, the fact that only the real part of the superpotential survives in the limit seems to be inconsistent with supersymmetry. Moreover, the very different analytical properties of the final limit compared to the finite series suggest that this convergence cannot be intended in the strong topology. In the present section we want to clarify the interpretation of this limit from the point of view of supersymmetry. The function 
\begin{equation}
W(z;a,R)= \Phi(a)\sum_{\Lambda(a)\cap D(R)} \log (z-\zeta_{m,n}(a))
\end{equation}

is multivalued and cannot be really considered as the superpotential of the theory. Conversely, the derivative 

\begin{equation}\label{derw}
\partial_{z}W(z;a,R)= \Phi(a)\sum_{\Lambda(a)\cap D(R)} \frac{1}{z-\zeta_{m,n}(a)}
\end{equation}

is a well defined meromorphic function on the plane. The dimension of the degenerate vacuum space is given by the number of classical vacua, i.e. the solutions to the equation $\partial_{z} W=0$. These are counted by the Witten index of the theory which grows with the number of lattice points

\begin{equation}
I_{W}=\mathrm{Tr}(-1)^{F}=  -(\#(\Lambda(a) \cap D(R))-1 ),
\end{equation}

where the mignus sign keeps into account that the vacua have fermion number $F=1$. In the limit $a\rightarrow 0, R\rightarrow \infty$ the Witten index goes like $\vert I_{W} \vert \simeq \pi R^{2}/a^{2}$, which is an extremely large number of vacua. The meaning of the $a\rightarrow 0$ limit is that we are taking measurements of observables on a scale much larger than the lattice parameter $a$. As physically expected, it turns out that for this specific class of measures the outcomes are correctly reproduced by an effective model with a single vacuum.\\ Let us begin with understanding the physics of the system in the continuous limit. It is clear that the wave functions of the theory must vanish at the points of the lattice $\Lambda(a)$ and, in the limit of $a\rightarrow 0$, they oscillate rapidly on a length scale comparable with the lattice parameter $a$. If we consider measures of observables on a scale much larger than $a$, we are not able to detect anymore the fluctuation of the wave functions and see the discrete structure of the set of charges. One can check that long-range measurements do not see the many vacua also by the correlators of topological obervables in the two dimensional $\mathcal{N}=(2,2)$ version of the theory. In a generic $\mathcal{N}=(2,2)$ Landau-Ginzburg model with fundamental chiral fields $X_{i}$ and superpotential $W$, the topological sector is described by operators $\Phi_{k}$ in the chiral ring $\mathcal{R}=\frac{\mathbb{C}\left[ X_{i}\right] }{\partial_{j} W}$, which are in one to one correspondence with the vacua of the model. We recall that, if the critical points $p_{\alpha}$ of $W$ are non degenerate, the chiral ring is semisimple and we have $\mathcal{R}\simeq \mathbb{C}^{\#\mathrm{classical \ vacua}}$. The canonical isomorphism between these $\mathbb{C}$-algebras is provided by a basis of minimal orthogonal idempotents defined by 

\begin{equation}
\Phi_{\beta}(z_{\alpha})= \delta_{\alpha\beta}.
\end{equation}

In this basis each operator of the chiral ring is represented by its set of values at the critical points.\\ Coming back to our theory, we see that for $a\rightarrow 0$ the classical vacua becomes very close to each other and to the lattice singularities. The holomorphic functions representing $\Phi_{\alpha}$ on the target space must vanish on the critical points of $W$ except $z_{\alpha}$ and so become rapidly oscillating for $a\ll 1$. Hence, it is clear that these operators are irrelevant for long-range measurements and one should consider a basis of generators for the chiral ring which are smooth in this limit. A suited basis is given by the monomials $z^{k},\ k \in \mathbb{N}$, which are defined indipendently from the vacua structure. Let us consider the correlators of these observables in the limit $a\rightarrow 0$. Setting $\Phi_{k}=z^{k},\Phi_{j}=1$ in the Grothendieck formula \ref{grotform} for the topological two points function we get 

\begin{equation}
\langle z^{k} \rangle= \frac{1}{(2\pi i)} \int_{\Gamma} \frac{z^{k}dz}{\Phi(a)\displaystyle\sum_{\Lambda(a)\cap D(R)} \frac{1}{z-\zeta_{m,n}(a)}}.
\end{equation}

In this computation one has to choose a contour $\Gamma$ which encircles all the critical points of $W$. It is clear that these will be located in the region of the disk containg the holes and so one must take a contour encircling $D(R)$. In the limit of $a\rightarrow 0$ at fixed $R$, the discrete set of charges approaches a continuous distribution on $D(R)$. If the coordinate $z$ is outside the disk, the derivative of the superpotential tends to  

\begin{equation}
\Phi(a)\sum_{\Lambda(a)\cap D(R)} \frac{1}{z-\zeta_{m,n}(a)}\sim \frac{B R^{2}}{2z} + O(a),
\end{equation}

in which we recognize the $2$-dimensional electrostatic field generated by a radially symmetric charge distribution. Hence, we find

\begin{equation}
\langle z^{k} \rangle \propto \int_{\Gamma} dz z^{k+1} +O(a) \xrightarrow{a\rightarrow 0} 0.
\end{equation}

The triviality of the topological correlators reflect the fact that measurements at a scale $z$ much larger than $a$ do not detect the many vacua of the theory.\\ It is already expected from \ref{limitsuperpotential} which should be the effective model which captures the relevant physics of the FQHE. The fact that in the limit we obtain a real superpotential implies a passage from the initial $\mathcal{N}=4$ to a $\mathcal{N}=2$ theory in which the holomorphicity is not required. Now we are able to give a precise physical and mathematical sense to the transition between the two formalisms.\\
The considerations above justify the introduction of an effective Hilbert space, different from the previous one, on which we can make long-range measurements of observables without seeing the lattice structure. This is the Schwartz space $S(\mathbb{C})$ of rapidly decreasing functions with bounded derivatives. The concept is that expectation values of observables in this Hilbert space coincide with the expectation values in the original Hilbert space in the long-wave regime. It is clear now that the limit of the superpotential and all the other operators must be intended in the context of the weak topology induced on the space of operators by the Hermitian scalar product of $S(\mathbb{C})$. Hence, the correct limit we have to study regards the matrix element

\begin{equation}\label{matrixelement}
\begin{split}
& \lim_{R\rightarrow \infty} \ \lim_{a \rightarrow 0} \ \Phi(a)\sum_{\Lambda(a)\cap D(R)} \Big  \langle \phi \Big\vert \ \frac{1}{z-\zeta_{m,n}(a)} \Big\vert   \psi \Big \rangle  =\\ \\ &  \lim_{R\rightarrow \infty} \ \lim_{a \rightarrow 0} \ \int _{\mathbb{C}} d^{2}z \ \Phi(a)\sum_{\Lambda(a)\cap D(R)} \frac{\phi^{*}(z)\psi(z)}{z-\zeta_{m,n}(a)}.
\end{split}
\end{equation}

with $\phi,\psi \in S(\mathbb{C})$. The details of the computation are given in \ref{computation}. As one can expect we find

\begin{equation}
\begin{split}
\lim_{R\rightarrow \infty} \ \lim_{a \rightarrow 0} \ \langle \phi \vert \ \partial_{z}W(z;a,R) \vert   \psi  \rangle= 
\Big  \langle \phi \Big\vert  \frac{B}{2}\bar{z}  \Big\vert   \psi \Big \rangle
\end{split}
\end{equation}

which gives the same limit discussed in the previous section, (actually the derivative)

\begin{equation}\label{weak}
\partial_{z}W(z;a,R)\rightarrow \frac{B}{2}\bar{z}.
\end{equation}

Hovewer, the mathematical structure that we have introduced here gives to the above formula a precise mathematical meaning. It is evident that this convergence must be intended only in the weak sense of quantum measure. A convergence in norm $L^{p}$ is excluded for evident reasons of integrability, as well as the uniform one, given the complitely opposite behavour between the initial function \ref{derw} and the final limit. Indeed, the former is a meromorphic function with simple poles at the lattice points and vanishing at infinity, while the latter is a pure antiholomorphic object with no singularities at finite points. This is consistent with the previous considerations: long-distance experiments cannot detect the lattice structure.\\ Contrary to $\partial_{z}W$, we cannot define the weak limit of $\vert \partial W \vert ^{2}$ and $\partial ^{2}W$ because the corresponding matrix elements are divergent. Indeed, the weak convergence does not preserve in general products and derivative of operators (otherwise it would be a strong convergence). This fact, together with the lost of holomorphicity of the superpotential, implies that part of the supersymmetry algebra is not preserved by the weak limit (as expected), included the Hamiltonian.\\ Adopting the formalism of section \ref{sqm} we can write the susy charges as the generalized Dolbeault operators

\begin{equation}
\begin{split}
& \bar{\partial}_{W}= \bar{\partial} + \partial W \wedge ,\hspace{2cm} \delta_{W}= \bar{\partial}_{W}^{\dagger}, \\ \\ 
& \partial_{W}= \partial + \overline{\partial W} \wedge \hspace{2cm} \bar{\delta}_{W}= \partial_{W}^{\dagger}.
\end{split}
\end{equation}

The limit \ref{weak} is not compatible with the structure of the susy charges because it exchanges holomorphic with antiholomorphic derivatives of $W$. Instead, there is no ambiguity if we consider real combinations of the generators. Only the real part of the superpotential has a limit and so only the $N=2$ subalgebra generated by 

\begin{equation}
d_{h}= \bar{\partial}_{W} + \partial_{W}= d+ d h \ \wedge, \hspace{2cm}   \delta_{h}= \delta_{W}+ \bar{\delta}_{W}  =d_{h}^{\dagger}, 
\end{equation}

where $ h= \mathrm{Re} \ W $, is compatible with the weak limit. This corresponds to the $\theta=0$ algebra in the family of $\mathcal{N}=2$ subalgebras. Learning from \ref{weak} that 

\begin{equation}
dh \rightarrow \frac{B}{2}(z+\bar{z})
\end{equation}

we conclude that the initial $\mathcal{N}=4$ theory is equivalent for long-range measurements to a $\mathcal{N}=2 $ one with real superpotential

\begin{equation}\label{effectivesuper}
h(z,\bar{z})= \frac{B}{2}\vert z \vert^{2}.
\end{equation}

Depending on the sign of $B$ we obtain the superpotential of a two dimensional harmonic oscillator or repulsor. Both the models are studied in \cite{rif19}. In this section we assume $B$ to be negative. One can count and classify the vacua of this theory by exploiting the similarity relation between $d_{h}$ and the exterior derivative. The $d_{h}$-cohomology classes in the Hilbert space are conjugated by $e^{-h}$ to relative de Rham classes of smooth forms which decrease rapidly at infinity. These classes are in the cohomology with compact support $H^{*}_{\mathrm{c}}(\mathbb{C})$. Since only $H^{2}_{\mathrm{c}}(\mathbb{C})$ is non trivial, the theory has a unique ground state represented by the two form

\begin{equation}\label{bos}
\Phi(z,\bar{z})= e^{ -\frac{\vert B \vert }{2} \vert z \vert^{2}}dzd\bar{z},
\end{equation}

in which we recognize the typical $B$ field factor which appears in the wave functions of the Landau levels.\\
We know that the Witten index of a supersymmetric theory is robust under continuous deformations and, the fact that we started from a model with a huge number of vacua and we found in the end a theory with just one vacuum, may lead to some doubts. However, this paradox does not arise because the $\mathcal{N}=2$ model is not the limit of the initial one in a strict sense. Indeed, the new Hamiltonian is not directly related to the previous one and acts on a different Hilbert space. The meaning of the weak limit is that for long distance measurements we can replace the original $\mathcal{N}=4$ theory with an effective $\mathcal{N}=2$ one with superpotential \ref{effectivesuper}.

\subsection{The Single Field Model with the Quasi-Holes}

Now that we have consistently recovered in this mathematical framework the standard non-holomorphic description of the magnetic field, we can include the quasi-holes in the system. Let us consider a generic set $S$ of quasi-holes at position $p(s)$. Correspondingly, the holomorphic superpotential \ref{wlattice} gains the new piece 

\begin{equation}
W\rightarrow W+ \sum_{s \in S} e(s) \log (z-p(s)),
\end{equation} 

where we assume $e(s) \in \mathbb{R}$. The target manifold of the theory is now $\mathbb{C}\setminus S$ and, taking the weak limit as before, we find a $\mathcal{N}=2 $ model with superpotential

\begin{equation}\label{potentialreal}
h_{S}(z,\bar{z})= \frac{B}{2} \vert z \vert^{2} +\sum_{s \in S} e(s) \log \vert z-p(s)\vert.
\end{equation}

The logarithmic terms in the superpotential do not spoil the polynomial divergence at infinity, but now the relative cohomology classes representing the vacua depend also on the behaviour of $ h_{S}(z,\bar{z})$ near the quasi-holes positions $p(s)$. If we demand the $e(s)$ to be all positive, the superpotential is bounded from above also in the region of the quasi-holes and tends to $-\infty$ as $z\rightarrow p(s)$. The map $e^{-h_{S}}$ provides a correspondence between states in the Hilbert space and smooth forms with a sufficiently rapid decay on the region $B^{-}_{h_{S}}$ of $\mathbb{C}$ where $h_{S}\rightarrow -\infty$. Compared to the previous case, now the forms have to vanish also approaching the quasi-holes positions. Hence, it follows that the $L^{2}$ representatives of the $d_{h_{S}}$-classes in the Hilbert space are conjugated to closed forms in the relative cohomology $H^{*}(\mathbb{C},B_{h_{S}}^{-})=H^{*}_{\mathrm{c}}(\mathbb{C}\setminus S)$. One can easily see that the vacuum space contains again a two form 

\begin{equation}\label{bosonicvacuum}
\Phi_{S}(z,\bar{z})= e^{ -\frac{\vert B \vert}{2} \vert z \vert^{2}}\prod_{s \in S} \vert z-p(s)\vert ^{e(s)}dzd\bar{z},
\end{equation}

which is a deformation of \ref{bos} vanishing at the positions of the quasi-holes. In the case of FQHE one should set all the charges to $1$. In presence of the quasi-holes the vacuum space gains new fermionic states in addition to the bosonic one above. These correspond to cohomology classes which are dual to the $1$-homology cycles with boundary on $B^{-}_{h_{S}}$. By Poincar\'e duality with the de Rham cohomology we find that there are as many homology classes as the number of quasi-holes.\\ One can count the fermionic states of the vacuum space also with the following argument. We note that $h_{S}$ contains an harmonic part which can be written as real part of an homolomorphic superpotential $W$. The definition of this function is ambigous because we have the freedom to shift the electron coordinate. So, to be more general one should perform the translation $z\rightarrow z+\Delta$. In this way the holomorphic superpotential depends also on this extra parameter and has the expression

\begin{equation}\label{holomorphicpot}
W(z)=\frac{ \overline{\Delta} B}{2} z +\sum_{s \in S} e(s) \log ( z-p^{\prime}(s)) ,
\end{equation}

where we setted $ p^{\prime}(s)= p(s)-\Delta$. This function defines a $\mathcal{N}=4$ theory in which the vacuum wave forms are in correspondence with the solutions to the equation $\partial W=0$. For $m$ quasi-holes and generic parameters the superpotential has $m$ distinct critical points and so the vacuum space has dimension $m$.  Moreover, as discussed in section \ref{sqm}, the vacua of a $\mathcal{N}=4$ theory with Morse superpotential on $\mathbb{C}^{n}$ are in correspondence with the relative cohomology classes of $H^{n}(\mathbb{C}^{n},B^{-}_{ReW})$, where $B^{-}_{ReW}$ is the region of $\mathbb{C}^{n}$ where $\mathrm{Re} \ W\rightarrow -\infty$. In the present case, denoting with $\Phi_{\alpha}, \alpha=1,...,m$ a basis of $d_{ReW}$-closed representatives in the Hilbert space, one can use the \ref{relaz} and write

\begin{equation}\label{omegaal}
\Phi_{\alpha}=e^{-Re W}\omega_{\alpha},
\end{equation}

where $\omega_{\alpha}$ are $d$-closed forms in $H^{1}(\mathbb{C},B^{-}_{ReW})$. In the same way the relative de Rham classes are trivial in the absolute cohomology, also the states $\Phi_{\alpha}$ become exact in the $d_{ReW}$-cohomology with smooth coefficients and we can find $0$-forms $\chi_{\alpha}$ such that 

\begin{equation}
\Phi_{\alpha}=d_{ReW} \chi_{\alpha}.
\end{equation}

However, since $\chi_{\alpha}$ are not $L^{2}$ forms, $\Phi_{\alpha}$ represent non trivial classes in the Hilbert space cohomology. The forms $\omega_{\alpha}$ are dual to homology cycles in $H_{1}(\mathbb{C},B^{-}_{ReW})$. We recall that for a Morse superpotential one can define a canonical basis of cycles $D_{\alpha}$ which begin from the critical points of $W$ and have boundary on $B^{-}_{ReW}$ (see \ref{sqm}). The real part of $W$ preserves the behaviour of $h_{S}$ near the quasi-holes and partially also at infinity. The boundary set for the non compact cycles contains small neighborhoods around $p(s)$ and a connected region at infinity where $\mathrm{Re} (-\overline{\Delta} z)\rightarrow -\infty$. Hence, one can easily see that the relative classes $\left[ D_{\alpha}\right] $ provide a complete basis also for $H_{1}(\mathbb{C},B^{-}_{h_{S}})$, implying that the two homology groups are isomorphic. We note further that the susy generators of the full theory are related by the similarity transformations 

\begin{equation}
d_{h_{S}}= e^{-B\vert z \vert^{2}/2}d_{Re W} \ e^{B\vert z \vert^{2}/2} \hspace{1cm} e^{B\vert z \vert^{2}/2}\delta_{Re W} \ e^{-B\vert z \vert^{2}/2}
\end{equation}

to the charges of the $\mathcal{N}=2$ subalgebra with $\theta=0$ of the $W$-theory. By these conjugation relations we can map representatives $\Phi_{\alpha}$ of $d_{Re W}$-classes to $d_{h_{S}}$-closed forms 

\begin{equation}\label{psial}
\Psi_{\alpha}=e^{-B\vert z \vert^{2}/2} \Phi_{\alpha}.
\end{equation}

The multiplication by $e^{-B\vert z \vert ^{2}/2}$ is not an isometry between Hilbert spaces and could spoil in general the integrability of the wave functions. However, one can always find in each $d_{Re W}$-cohomology class a representative $\Phi_{\alpha}$ with compact support, or with sufficiently rapid decay on $\mathbb{C}\setminus S$, such that the corresponding $\Psi_{\alpha}$ is an $L^{2}$-wave function of the Hilbert space. It is clear that these states remain closed but not exact in the $d_{h_{S}}$-cohomology because the multiplication by $e^{-B\vert z \vert^{2}/2}$ cannot make the primitive $\chi_{\alpha}$ normalizable. Therefore, cohomology classes of $d_{ReW}$ defines cohomology classes of $d_{h_{S}}$ and both are dual to the Lefschetz thimbles $D_{\alpha}$. In this sense the fermionic subspace of the vacuum space represent the $\mathcal{N}=4$ sector of the theory. We note that $\Delta=0$ represents a critical limit in which the superpotential has a different behaviour at infinity. In this case we find a degenerate model in which one of the critical point of $W$ is sent to infinity and decouples from the other ones.\\ 
We said that the great advantage of the Vafa model of FQHE compared to the previous ones is supersymmetry. In particular, this shoud provide more tools to study the monodromy of the ground states induced by the quasi-holes braiding. However, in order to make this problem resonable to solve, we need the tools of $tt^{*}$ geometry ensured by $\mathcal{N}=4$ supersymmetry that we lost in the weak limit. A strategy one can follow at this point is to consider some region of the parameter space in which we have an enhancement of supersymmetry from $\mathcal{N}=2$ to $\mathcal{N}=4$. In such regime the bosonic state should decouple from the system in such a way that one could see only the fermionic vacua.  We note that, since it is the unique $2$-form in the vacuum space, the wavefunction \ref{bosonicvacuum} must be an eigenstate of the Berry's holonomy. Moreover, since this form is real for generic holes positions, the corresponding Berry's phase is trivial. The fact that the non trivial part of the holonomy regards only the fermionic sector of the vacuum space provides a further justification to consider this limit. \\ It is already clear from the discussion above that the effective superpotential we are looking for is given by the harmonic part of $h_{S}(z,\bar{z})$. This IR description corresponds to the region of the parameter space where the quasi-holes are placed in a neighborhood of a point $\Delta \in \mathbb{C}\setminus\left\lbrace 0 \right\rbrace$ of radious $r \ll \ell_{B}=1/\sqrt{\vert B \vert }$. Denoting with $\zeta(s)=(p(s)-\Delta)/r$ the quasi-holes positions in $r$ units and redefining the electron coordinate $z\rightarrow r z + \Delta$, one has to expand $h_{S}\left( rz+\Delta,r\bar{z}+\overline{\Delta} \right) $ for $r\ll \ell_{B}$ and finite $\zeta(s)$. At the leading order one obtains (up to constant terms)

\begin{equation}
 h_{S}(z,\bar{z}) \sim -z\frac{\bar{\Delta}r}{2\ell_{B}^{2}}-\bar{z}\frac{\Delta r}{2\ell_{B}^{2}} +\sum_{s \in S}\log \vert z-\zeta(s)\vert,
\end{equation}

where we setted the quasi-holes charges to $1$. The holomorphic superpotential that we get is the \ref{holomorphicpot} with rescaled coordinates in $1/r$ unit

\begin{equation}\label{consid}
W(z)= \mu z +\sum_{s \in S}\log ( z-\zeta(s)),
\end{equation}

where we introduced the notation $\mu=-\frac{\overline{\Delta}r}{2\ell_{B}^{2}}$. This theory allows to capture exclusively the physics of the fermionic vacuum space, restoring the initial amount of supersymmetry. As we already mentioned, the case of $\mu=0$ represents a critical point in the parameter space where also one of the fermionic vacua decouples from the system. In the language of the $(2,2)$ version of the model, this turns out to be an IR fixed point of the RG flow, while by deforming $\mu$ away from the origin the number of vacua computed by the Witten index is consistently stable. So, we should keep $\mu\neq 0$ if we want to study the physics of the full fermionic sector of the theory. From the point of view of FQHE, the linear interaction in the effective superpotential above describes the electric field induced by the magnetic field. Taking the limit $\mu\rightarrow 0$ is equivalent to turn off the magnetic field. \\ At the end of this chain of effective limits we ended up with a $\mathcal{N}=4$ theory which is definitely easier to study with respect to the initial one. The virtue of the simplifications we have considered is to make the problem resonable to approach analitically, remaining at the same time in a realistic setting for the physics of FQHE.

\section{Weak Limit of $\partial W$}\label{computation}

Let us take $\phi,\psi$ in the Schwartz space $S(\mathbb{C})$ of smooth functions with bounded derivatives. We want to compute the limit \ref{matrixelement}

\begin{equation}
\begin{split}
& \lim_{R\rightarrow \infty} \ \lim_{a \rightarrow 0} \ \Phi(a)\sum_{\Lambda(a)\cap D(R)} \Big  \langle \phi \Big\vert \ \frac{1}{z-\zeta_{m,n}(a)} \Big\vert   \psi \Big \rangle  =\\ \\ &  \lim_{R\rightarrow \infty} \ \lim_{a \rightarrow 0} \  \int _{\mathbb{C}} d^{2}z \ \Phi(a)\sum_{\Lambda(a)\cap D(R)} \frac{\phi^{*}(z)\psi(z)}{z-\zeta_{m,n}(a)},
\end{split}
\end{equation}

where $\Lambda(a)=a\mathbb{Z}+ia \mathbb{Z} $, $D(R)=\lbrace z \in \mathbb{C}, \vert z \vert \leq R\rbrace$ and $\Phi(a)= \frac{a^{2}B}{2\pi}.$ Let us first take $a\rightarrow 0$ with fixed $R$. Performing the change of variable $z\rightarrow z+\zeta_{m,n}(a)$, we get 

\begin{equation}
\begin{split}
 & \lim_{a \rightarrow 0} \ \int _{\mathbb{C}} \frac{d^{2}z}{z} \ \Phi(a)\sum_{\Lambda(a)\cap D(R)} \phi^{*}(z+\zeta_{m,n}(a))\psi(z+\zeta_{m,n}(a))= \\ \\ &\lim_{a \rightarrow 0} \  \Big  \langle \frac{1}{z} 
 \Big\vert \Phi(a) \phi^{*}(z+\zeta_{m,n}(a))\psi(z+\zeta_{m,n}(a))  \Big \rangle.
 \end{split}
\end{equation}

Given that $\frac{1}{z}$ acts on the space of test functions as a linear continuous functional\footnote{It is straighforward to see that $1/z$ is a tempered distribution of function type.}, one can take the limit under the sign of integration. In the expression 

\begin{equation}
a^{2}\sum_{\Lambda(a)\cap D(R)} \phi^{*}(z+\zeta_{m,n}(a))\psi(z+\zeta_{m,n}(a))
\end{equation}

we recognize the Riemann sum of $ \phi^{*}(z+\zeta)\psi(z+\zeta)$ with partition of $D(R)$ given by squares of area $a^{2}$. The smoothness and boundedness of the Schwartz functions allow to write 

\begin{equation}
\begin{split}
 \lim_{a \rightarrow 0} a^{2} &\sum_{\Lambda(a)\cap D(R)} \phi^{*}(z+\zeta_{m,n}(a))\psi(z+\zeta_{m,n}(a))=  \\ \\ & \int_{D(R)} d^{2}\zeta \ \phi^{*}(z+\zeta)\psi(z+\zeta).
\end{split}
\end{equation}

Then, after another shift of $z$ we remain with

\begin{equation}\label{secondl}
\lim_{R\rightarrow \infty} \int _{\mathbb{C}} d^{2}z \ \phi^{*}(z)\psi(z) \ \frac{B}{2\pi} \int_{D(R)} d^{2}\zeta\ \frac{1}{z-\zeta}.
\end{equation}

Let us focus on the integral in $\zeta,\bar{\zeta}$. Since we are going to take the limit of infinite volume, we expand the integral for $\vert z \vert /R \ll 1$. Introducing the new variable $w=\zeta-z$ and denoting with $R_{z}(\theta)=\vert \zeta_{\partial D}(\theta)-z \vert= \sqrt{\vert  z \vert ^{2}+R^{2}-z\overline{\zeta_{\partial D}(\theta)} -\bar{z} \zeta_{\partial D}(\theta)}, $ the distance between $z$ and the point $ \zeta_{\partial  D}(\theta) \in \partial D(R)$ with phase $\theta \in [0,2\pi]$, we find 

\begin{equation}
\begin{split}
& \int_{D(R)} d^{2}\zeta\ \frac{1}{z-\zeta}=-\int_{0}^{2\pi} d\theta R_{z}(\theta) e^{-i\theta}= \\ \\ &-\int_{0}^{2\pi} d\theta \left( R-\frac{z}{2R}\overline{\zeta_{\partial D}(\theta)} -\frac{\bar{z}}{2R} \zeta_{\partial D}(\theta)+ O(1/R) \right)  e^{-i\theta}.
\end{split}
\end{equation}

Using $ \zeta_{\partial  D}(\theta)=R e^{i\theta} + O(1)$ and sending $R$ to infinity one gets 

\begin{equation}
\int_{D(R)} d^{2}\zeta\ \frac{1}{\zeta-z}=\pi\bar{z} + O(1/R) .
\end{equation}

Plugging this large $R$ expansion in the integral \ref{secondl} one obtains a convergent series of integrals weighted by powers of $1/R$. In the limit of infinite $R$ we keep only the leading term and write finally

\begin{equation}
\begin{split}
\lim_{R\rightarrow \infty} \ \lim_{a \rightarrow 0} \ \langle \phi \vert \ \partial_{z}W(z;a,R) \vert   \psi  \rangle= 
\Big  \langle \phi \Big\vert \ \frac{B}{2} \bar{z} \Big\vert   \psi \Big \rangle .
\end{split}
\end{equation}

\section{FQHE and Gauge Theories}\label{classS}

In this section we review some well known facts about class $\mathcal{S}$ theories and the $2d/4d$ correspondence. We also recall the correspondence between matrix models and gauge theories. These are the basic ingredients which motivates the Vafa proposal and allow to formulate a precise dictionary between FQHE and supersymmetric gauge theories in $4d$.

\subsection{$\mathcal{T}_{n,g}$ Gauge Theories}

We recall some known facts about class $\mathcal{S}$ theories, namely four dimensional $\mathcal{N}=2$ gauge theories which arise from the compactification of six dimensional theories. In particular, we focus on the subclass which according to Vafa are related to FQHE. A theory in this class is denoted with $\mathcal{T}_{n+3,g}\left[A_{1} \right] $ and is associated to a Riemann surface $\mathcal{C}_{n+3,g}$ of genus $g$ and $n+3$ equivalent punctures. These models are identified \cite{rif2} with the twisted compactification of the $A_{1} (2,0)$ six-dimensional SCFT on the $g$-Riemann surface in the presence of $n$ defect operators. In the context of FQHE the surface $\mathcal{C}_{n+3,g}$ has also the interpretation of target space for the electrons, hence we are interested in the case of $g=0$. The first model of the series correspond to the three-punctured sphere. This theory simply contains four free hypermultiplets and the gauge dynamics is absent. Much more interesting are the $\mathcal{T}_{n+3,0}\left[A_{1} \right] $ models with $n>0$ which identify $\mathcal{N}=2 \ SU(2)$ gauge theories with massive deformations. The gauge group contains $n$ $SU(2)$ factors, each of them coupled to $N_{f}=4$ flavours. These theories are therefore conformal and admit a space of exactly marginal gauge couplings. The parameter space of gauge couplings turns out to coincide with the moduli space $\mathcal{M}_{n+3,0}$ of complex structures of the Riemann sphere with $n+3$ punctures $\mathcal{C}_{n+3,0}$. The space of couplings has boundaries where one of the $SU(2)$ gauge groups become weakly coupled. To each puncture corresponds also a flavour subgroup $SU(2)$. So, in total we have an $SU(2)^{n+3}$ flavour group. Each punture is also associated to a mass parameter $m_{a},a=1,...,n+3$ of the corresponding flavour subgroup $SU(2)_{a}$. These theories enjoy $S$-duality, which has the geometric interpretation of fundamental group $\pi_{1}$ of $\mathcal{M}_{n,0}$. The action of S-duality on the flavour symmetry groups is a permutation action and coincides with the permutation of the $n+3$ punctures on the sphere and the corresponding mass parameters. The $S$-duality can rearrange in different ways the matter content of the theory, but in each possible $S$-frame we remain always with $n$ $SU(2)$ gauge groups coupled to four hypermultiplet doublets and $n+3$ $SU(2)$ flavour groups. The various weakly coupled $S$-dual frames of the theory coincide with the different ways a sphere with $n+3$ points can degenerate completely to a set of $n + 1$ three-punctured spheres attached together at $n$ nodes. Labelling with $SU(2)_{i}, i=1,..n$ the gauge groups, in a possible weakly coupled $S$-duality frame we have two hypermultiplet doublets in the fundamental of $SU(2)_{1}$ with masses $m_{1}\pm m_{2}$, $n-1$ hypermultiplets in the bifundamental of $SU(2)_{i}\times SU(2)_{i+1}$ with masses $m_{i+2}$ and other two hypermultiplets in the fundamental of $SU(2)_{n}$ with masses $m_{n+2}\pm m_{n+3} $. The linear gauge quiver describing this frame is \\
 
 \begin{tikzpicture}\label{superconfquiver}
 \node (NM) [rectangle,draw,scale=1.5] at (0,0) {$2$};
\node (NP) [rectangle,draw,scale=1.5] at (13,0) {$2$};
 \node(NL)[circle,draw] at (2,0) {$SU(2)$};
  \node(NR) [circle,draw] at (4,0){$SU(2)$};
  \draw (NL.east)--(NR.west);
  \draw [dotted] (5,0) -- (8,0);
   \node(NC)[circle,draw] at (9,0) {$SU(2)$};
  \node(ND) [circle,draw] at (11,0){$SU(2)$};
   \draw (NC.east)--(ND.west);
  \draw (0.38,0) -- (1.21,0);
\draw (11.78,0) -- (12.65,0);
\end{tikzpicture}

where the rectangular nodes on the two sides represent the two couples of fundamental hypermultiplets, the circular nodes are the gauge groups and the lines between the gauge nodes are the bifundamental hypermultiplets.\\ A canonical example is the $\mathcal{N}=2$ $SU(2)$ gauge theory with $N_{f}=4$ fundamental flavours. The theory has an exact marginal coupling $\tau= \frac{\theta}{\pi} + \frac{4\pi i}{g^{2}}$, since the number of flavours is twice the number of colours. The flavour group $SU(2)^{4}$ is enhanced to $SO(8)$ and the four hypermultiplet doublets transform in the eight dimensional vector representation. In this case the $S$-duality group acts by fractional linear transformation of $SL(2,\mathbb{Z})$ on $\tau$ and by triality on $SO(8)$. Hence, the space of marginal couplings parametrized by $\tau $ is $\mathbb{H}/SL(2,\mathbb{Z})$, which is the complex structure moduli space $\mathcal{M}_{4,0}$ of a sphere with four equivalent punctures. If we quotient the upper half plane only by the subgroup $\Gamma(2)$ we get the moduli space of a sphere with four marked punctures, i.e. the modular curve $\mathbb{H}/\Gamma(2)\simeq \mathbb{P}^{1}\setminus\lbrace 0,1,\infty\rbrace $. A natural parametrization of this space is given by the cross-ratio $q$ of the positions of the four punctures. This can be seen as a coordinates on $\mathbb{P}^{1}\setminus{0,1,\infty}$ and is related to $\tau$ by the $\Gamma(2)$-invariant modular lambda function $\lambda(\tau)=\frac{\theta_{2}^{4}(0,\tau)}{\theta_{3}^{4}(0,\tau)}$ which realizes the isomorphism $\mathbb{H}/\Gamma(2)\simeq \mathbb{P}^{1}\setminus\left\lbrace 0,1,\infty\right\rbrace $. The action of $\Gamma(2)$ does not permute the punctures of the sphere and the moduli space has three cusps at $\tau=0,1,\infty$ corresponding to the three weakly coupled frames of the theory. The full $S$-duality group permutes the punctures among themselves and simultaneously the associated mass parameters, mapping between each other also the three weakly coupled description of the theory which are therefore physically equivalent.\\
The compactification of $A_{1}(2,0)$ theories on a puctured Riemann surface provides also the construction of a canonical Seiberg-Witten curve encoding the physics of the Coulomb phase of the theory. This curve is a ramified double cover of $\mathbb{C}_{n+3,g}$ defined by the equation

\begin{equation}
y^{2} = \phi_{2}(z),
\end{equation}

where $(y,z)$ are local coordinates in the cotangent bundle of $C_{n+3,g}$. The SW differential is the canonical one form $d\lambda=ydz$, while $d\lambda^{2}=\phi_{2}(z)dz^{2}$ is the associated Jerkin-Strebel quadratic differential with appropriate poles at the punctures. In the case of the $\mathcal{T}_{n+3,0}$ the quadratic differential on the $n+3$-punctured sphere has double poles at the punctures $\zeta_{a}$ with coefficients given by the square of the corresponding mass parameters.\\ More general theories in the $A_{1}$ class are associated to a quadratic differential with poles of higher order \cite{rif29}. These theories are obtained from superconformal theories by tuning some mass parameter $m_{a}$ to be very large, adjusting at the same time the coupled marginal gauge coupling in the UV so that the running coupling in the IR remains finite. From a six-dimensional perspective, the limiting procedure brings two or more standard punctures together to produce a single puncture with a larger degree of divergence. It is relevant for FQHE the case in which two punctures in the $\mathcal{T}_{n+3,0}$ theory collide to generate an irregular puncture. This appears as a quartic pole in the quadratic differential. In the $n=0$ case one obtains an Argyres-Douglas system of $D_{2}$ type, namely a theory of a free hypermutliplet doublet and $SU(2)^{2}$ flavour group. For $n>0$ we find an $SU(2)^{n}$ gauge theory in which $n-1$ $SU(2)$ factors are coupled to $N_{f}=4$ flavour and one $SU(2)$ gauge group is coupled to three fundamentals. Hence, one of the couplings has negative beta function and a Yang-Mills scale $\Lambda_{ \mathrm{YM}}$ is generated at one loop. This coupling vanishes in the UV limit, while the other ones are marginal. So, the UV limit of the theory is decribed by a $\mathcal{C}_{n+2,0}$ punctured sphere and corresponds to the superconformal $\mathcal{T}_{n+2,0}$ theory. The space of couplings is identified with the complex structure moduli space of a sphere with $n+1$ equivalent punctures and a marked puncture. Each puncture contributes to the flavour group with an $SU(2)$ factor, giving in total a $SU(2)^{n+2}$ flavour group. The $S$-duality group which permute the equivalent punctures in the conformal case now is broken to a subgroup which act only on the regular ones, acting simultaneously on the punctures and the corresponding mass parameters. This residual group plays the role of fundamental group of the space of couplings. The theory admits different weakly coupled description which correspond to the possible ways in which the punctured sphere degenerates to $n$ spheres with three regular punctures, corresponding to $\mathcal{T}_{3,0}$ models, and a sphere with a regular and an irregular puncture which is identified with a $D_{2}$ system.
Labelling with $SU(2)_{i}, i=1,..n$ the gauge groups, in a possible weakly coupled $S$-frame we have one hypermultiplet doublet in the fundamental of $SU(2)_{1}$ with masses $m_{0}$, $n-1$ hypermultiplets in the bifundamental of $SU(2)_{i}\times SU(2)_{i+1}$ with masses $m_{i}$ and other two hypermultiplets in the fundamental of $SU(2)_{n}$ with masses $m_{n}\pm m_{n+1}$. The linear gauge quiver describing this frame is \\
 
 \begin{tikzpicture}\label{newquiver}
 \node (NM) [rectangle,draw,scale=1.5] at (0,0) {$1$};
\node (NP) [rectangle,draw,scale=1.5] at (13,0) {$2$};
 \node(NL)[circle,draw] at (2,0) {$SU(2)$};
  \node(NR) [circle,draw] at (4,0){$SU(2)$};
  \draw (NL.east)--(NR.west);
  \draw [dotted] (5,0) -- (8,0);
   \node(NC)[circle,draw] at (9,0) {$SU(2)$};
  \node(ND) [circle,draw] at (11,0){$SU(2)$};
   \draw (NC.east)--(ND.west);
  \draw (0.38,0) -- (1.21,0);
\draw (11.78,0) -- (12.65,0);
\end{tikzpicture}

Compared to the superconformal quiver \ref{superconfquiver}, now attached to the the first $SU(2)$ node we have a rectangular node representing a single hypermultiplet doublet, implying that $SU(2)_{1}$ is coupled only to three flavours. A simple example is given by  the $\mathcal{N}=2$ $SU(2)$ gauge theory with $N_{f}=3$ fundamental flavours. Since $N_{f}<2N_{c}$ the theory is asymptotically free in the UV. The flavour group $SU(2)^{3}$ is enhanced to $SO(6)$ and the three hypermultiplet doublets transform in the sixth dimensional vector representation. The fundamental group of the complex structure moduli space of a sphere with two equivalent and a marked puncture has a unique generator. This acts by permuting the two regular punctures and  corresponds to the residual $S$-duality transformation that we have in all the asymptotically free $SU(2)$ gauge theories with $N_{f}<4$. In the UV limit the theory flows to the $\mathcal{T}_{3,0}$ theory of four free hypermultiplets described by the three punctured sphere. 

\subsection{Matrix Models and $2d/4d$ Correspondence } \label{largen}

One of the key ingredients to connect FQHE to the gauge theories introduced above is the relation between matrix models and four dimensional $\mathcal{N}=2$ gauge theories \cite{rif8,rif9,rif23}. This connection arises from string theory as follows. One considers the type IIB superstring on a background $CY \times \mathbb{R}^{4}$, where $CY$ denotes a fixed Calabi-Yau threefold with a non compact holomorphic curve $Y\subset CY$, with $N$ $D3$-branes on a subspace $Y \times \mathbb{R}^{2}$. The theory living on $\mathbb{R}^{4}$ admits $\Omega$-deformation \cite{rif24} with parameters $\epsilon_{1}$ for a rotation transverse to the brane and $\epsilon_{2}$ along the brane. The complex structure moduli of $CY$ is controlled by a set of parameters $t_{k}$ which play the role of couplings in the four dimensional theory. In this set up one considers a topological string on $CY$ with $N$ branes on $Y$. It has been found \cite{rif9,rif25,rif26} that the Nekrasov partition function $Z(t_{k},\epsilon_{1},\epsilon_{2})$ of the $4d$ gauge theory coincides with the open topological string partition function $Z(t_{k},g_{s})$ with $\epsilon_{2}=-\epsilon_{1}=g_{s}$. This admits the representation of matrix model partition function. In the case of $SU(2)$ gauge theories only a single $N\times N$ matrix $\Phi$ is required and we have 

\begin{equation}\label{partition}
\begin{split}
Z(t_{k},\epsilon_{1},\epsilon_{2})= & \int_{N\times N}d\Phi \exp\left( -\frac{1}{g_{s}}\mathrm{Tr} W(t_{k},\Phi)\right) \\ \\= &\int \prod_{i=1}^{N}dz_{i} \prod_{1\leq i< j \leqslant N} (z_{i}-z_{j})^{\frac{\epsilon_{1}}{\epsilon_{2}}} \exp\left( -\frac{1}{\epsilon_{2}} \sum_{i} W(t_{k};z_{i})   \right) ,
\end{split}
\end{equation}

where $W(t_{k};z_{i})$ is the potential governing the dynamics of the matrix eigenvalues $z_{i}$ and the $\beta$-ensemble is specified by the parameter $\beta=\epsilon_{1}/\epsilon_{2}$ to which the Van der Monde determinant $\Delta(x_{i})= \prod_{1\leq i< j \leqslant N} (z_{i}-z_{j})$ is raised. The string partition function can be expanded perturbatively in the string coupling as 

\begin{equation}
Z(t_{k},g_{s})= \exp \left( \sum_{g \geq 0}g_{s}^{2g-2} \mathcal{F}_{g}(t_{k})  \right) ,
\end{equation}

where $\mathcal{F}_{g}(t_{k})$ denotes the genus $g$ amplitude of topological strings. In the double-scaling limit $g_{s}\rightarrow 0, N\rightarrow\infty$ with $g_{s}N$ fixed, the string partition function reproduces the prepotential as determined by the Seiberg-Witten solution 

\begin{equation}
\mathcal{F}_{0}(t_{k},a_{j})= \mathrm{Lim}_{g_{s}\rightarrow 0, N\rightarrow \infty} \ g_{s}^{2}\log Z(t_{k},g_{s})
\end{equation}

where $a_{j}$ are the electric central charges of the Gauge theory. The geometry of the Seiberg-Witten curve emerge naturally in the large $N$ limit of the matrix model, providing the definition of the Coulomb branch coordinates in terms of the matrix model parameters. The expectation value of operators are computed in this regime with the saddle-point method, which requires to know the critical configurations of the eigenvalues. Taking $N\rightarrow\infty$ and $g_{s}\rightarrow 0$ with the 't Hooft coupling $g_{s}N$ fixed, the set of eigenvalues becomes continuum and it is distributed according to a density function $\rho(z)=\frac{1}{N}\sum_{i=1}^{N}\delta(z-X_{i})$, where the set $X_{i}$ denotes a vacuum configuration of the eigenvalues. Once exponentiated the Van der Monde determinant in the path integral one obtains the effective action 

\begin{equation}\label{mathcalpot}
\frac{1}{g_{s}}\mathcal{W}(z_{i})= \sum_{i=1}^{N}\frac{1}{g_{s}}W(z_{i})+\sum_{i<j}\log(z_{i}-z_{j}).
\end{equation}

A certain critical configuration $X_{i}$ must be a solution of the classical equation of motions

\begin{equation}\label{eqmot}
\frac{1}{g_{s}}W^{\prime}(X_{i})+\sum_{j\neq i}\frac{1}{X_{i}-X_{j}}=0.
\end{equation}

It can be shown that the saddle-point solution is captured in the continuum limit by the so-called spectral curve

\begin{equation}\label{ris}
y^{2}-W^{\prime}(z)^{2} - f(z)=0
\end{equation}

where $f(z)$ is the so-called quantum correction  

\begin{equation}\label{spectr}
f(z)= 2g_{s}\sum_{i=1}^{N}\frac{W^{\prime}(z)-W^{\prime}(X_{i})}{z-X_{i}}.
\end{equation}

One can define in terms of $X_{i}$ the function

\begin{equation}\label{seiberwitten}
y(z)=g_{s}\sum_{i}\frac{1}{z-X_{i}} + W^{\prime}(z).
\end{equation}

which has simple poles at $X_{i}$. This function solves the $\ref{ris}$ in the large $N$ limit and encodes the data about the density of eigenvalues. In this regime the domain filled by the eigenvalues consists of $m$ disconnected intervals $A_{j}$ called cuts. One can show that the number of branch cuts is equal to the number of critical points of the single field potential $W(z)$. So, a saddle-point configuration corresponds to a distribution $N = N_{1} + N_{2} + . . .N_{m}$ over the cuts $A_{1},A_{2},...,A_{m}$. The density of eigenvalues on a certain $A_{j}$ is given by the jump of $y(z)$ accross the interval 

\begin{equation}
\rho(z)=\frac{g_{s}N}{2\pi i}\left(y( z+i0)-y(z-i0) \right). 
\end{equation}

One can compute the filling fractions $a_{j}=g_{s}N_{j}$ of a specific cut by doing a countour integral around the cut 

\begin{equation}
a_{j}= \frac{1}{2\pi i} \oint_{A_{j}} y(z)dz.
\end{equation}

In the correspondence with the gauge theory the $1$-form $y(z)dz$ plays the role of Seiberg-Witten differential and the spectral curve arising in the double scaling limit is naturally identified with the Seiberg-Witten curve. The filling fractions $a_{j},j=1,...,m$ parametrizing the vacua of the theory have the natural interpretation of Coulomb parameters and the cycles encircling the branch cuts are the A-cycles of the gauge theory. These are dualized by B-cycles which connects the branch cuts to a cut-off point $\Lambda$. These cycles move eigenvalues from the branch cuts to infinity and generate a variation of the filling fraction $a_{j}$. The open string amplitude allows to identify the `free energy' of the matrix model with the Seiberg-Witten prepotential $\mathcal{F}_{0}(a_{1},...,a_{m})$. This functional can be used to express the B-cycles in terms of their duals by 

\begin{equation}\label{magnchem}
a_{D}^{j}=\frac{\partial \mathcal{F}_{0}}{\partial a_{j}}= \int_{B_{j}} y(z)dz,
\end{equation}

hence the magnetic central charge $a_{D}^{j}$ has the interpretation of `energy cost' to bring an eigenvalue from infinity to the $j$-th branch cut.\\ 
The matrix models which correspond to $\mathcal{T}_{n+3,0}[A_{1}]$ are the multi-Penner models with logarithmic superpotential \cite{rif3,rif7,rif8,rif9}

\begin{equation}
W(\Phi)= \sum_{a=1}^{n+2} m_{a}\log \left( \Phi-\zeta\right).
\end{equation}

The target space of the matrix eigenvalues is the Riemann sphere with $n+3$ punctures at the positions $\zeta_{a}, a=1,...,n+2$ and $\zeta_{n+3}=\infty$. In this case the equation 

\begin{equation}
W^{\prime}(z)=\sum_{a=1}^{n+2}\frac{1}{z-\zeta_{a}}=0
\end{equation}

has $n+1$ solutions and so a saddle-point configurations is determined by $n+1$ filling fractions $a_{j}=g_{s}N_{j}$, $j=1,...,n+1$. Using the expression of $W^{\prime}(z)$ one can easily see that $f(z)$ in \ref{spectr} takes the form 

\begin{equation}
f(z)=\sum_{a}^{n+2} \frac{c_{a}}{z-\zeta_{a}}
\end{equation}

where the coefficients $c_{a}$ are determined by the filling fractions. Moreover, they satisfy the constraint $\sum_{a}c_{a}=0$ which follows from the relation $\sum_{i=1}^{N}W^{\prime}(X_{i})=0$. Using the expressions for $f(z)$ and $W^{\prime}(z)$ in \ref{ris} we find for the spectral curve the equation 

\begin{equation}\label{quadraticdiff}   
y^{2}=\frac{P_{2n+2}(z)}{\Delta_{n+2}(z)^{2}},
\end{equation}

where

\begin{equation}
\Delta_{n+2}(z)= \prod_{a=1}^{n+2}(z-\zeta_{a})
\end{equation}

and $P_{2n+2}(z)$ is a polynomial of degree $2n+2$. While in the denominator we find the positions of the punctures, the $2n+3$ coefficients of $P_{2n+2}(z)$ encode two types of data. There are the $n+2$ coefficients $m_{a}^{2}$ of the double poles of the quadratic differential $y^{2}dz^{2}$ at $z=\zeta_{a}$, to which one should include the residue $m_{0}^{2}$ of the double pole at $\infty$. As we said in the previous section, the residue of a certain double pole is the squared of the mass parameter of the flavour subgroup $SU(2)$ associated to the corresponding puncture. One can read the mass parameters associated to the punctures from the residues of $y(z)$ at the simple poles using the general solution in \ref{seiberwitten}. It is immediate to see that the masses satisfy the relation

\begin{equation}\label{masses}
\sum_{a=0}^{n+2} m_{a}=\sum_{j=1}^{n+1}a_{j}=-g_{s}N,
\end{equation}

from which we learn that the t'Hooft coupling corresponds (up to the sign) to the mass parameter of diagonal subgroup of the flavour group. The remaining parameters are combinations of the coefficients $c_{i}$, or equivalently of the filling fractions $a_{j}$. Since the sum $\sum_{j=1}^{n+1}a_{j}$ is already fixed in terms of the masses, we have $n$ indipendent moduli 

\begin{equation}
S_{j}=a_{j+1}-a_{j}, \hspace{1cm} j=1,...,n,
\end{equation}

which define a basis of $A$-cycles encicrcling appropriately the $A_{j}$ and $A_{j+1}$ branch cuts. The period $S_{j}$ is the Coulomb parameter associated to the $j$-th $SU(2)$ factor of the gauge group $SU(2)^{n}$.\\
An important implication of the double scaling limit is that the number of eigenvalues $N$ becomes naturally a continuum variable and one can consider variations of the total number of fields. This is equivalent to take a many-particle system in the grand canonical ensemble where the number of particles in not fixed. One can use the B-cycles in \ref{magnchem} to bring particles from infinity to the branch cuts and vary the free energy of the system. Given that $N$ is not fixed anymore in this regime, the sum of the filling fractions $\sum_{j}a_{j}=g_{s}N$ becomes dynamical and we gain an extra modulus to parametrize the vacua in the Coulomb phase. By the relation \ref{masses} we learn that the diagonal combination of the masses is promoted to a vev of a scalar field, implying that taking the continuum limit $N\rightarrow \infty$ has the effect of gauging the $U(1)$ diagonal subgroup of the flavour group.\\ The class of matrix models which corresponds to the sphere with $n+1$ regular punctures and an irregular puncture is a generalization of the Penner model involving a linear interaction. The single eigenvalue potential reads

\begin{equation}
W(z)= \mu z + \sum_{a=1}^{n+1}\frac{m_{a}}{z-\zeta_{a}}.
\end{equation}

Using the \ref{spectr} one can easily check that $f(z)$ takes again the form 

\begin{equation}
f(z)=\sum_{a=1}^{n+1}\frac{c_{a}}{z-\zeta_{a}}
\end{equation}

for some coefficients $c_{a}$ depending on the vacuum configuration $X_{i}$ we are considering. The relation $\sum_{i}W^{\prime}(X_{i})=0$ implies in this case the constraint $\sum_{a=1}^{n+1}c_{a}=2g_{s}N\mu $. The single field superpotential has $n+1$ vacua and so the spectral curve has $n+1$ branch cuts where the particles are distributed. The solution \ref{seiberwitten} for $y(z)$ has simple poles at the regular punctures and the corresponding residues are the flavour mass parameters $m_{a}$. At infinity, where the irregular puncture is located, we find a simple pole, whose residue is the mass $m_{0}$, and a double pole with coefficient the coupling $\mu$. As in the previous case the sum of the filling fractions $\sum_{j}a_{j}=g_{s}N$ is fixed in term of the masses according to the relation \ref{masses} and the 't Hooft coupling corresponds to the mass associated to the diagonal subgroup of the flavour group. The meromorphic quadratic differential associated to the spectral curve can be recasted in the form \ref{quadraticdiff}. As in the Penner model the denominator contains double poles at the regular punctures $\zeta_{a}$. The polynomial $P_{2n+2}(z)$ of degree $2n+2$ at the numerator is specified by $2n+3$ parameters. Among these we have the $n+1$ masses $m_{a}^{2},a=1,...,n+1$ corresponding to the regular punctures and the mass $m_{0}^{2}$ of the irregular puncture at infinity. The remaining $n+1$ degrees of freedom are the $n$ moduli of the Coulomb branch and the coupling $\mu$. This is identified in the large $N$ correspondence with the Yang-Mills scale $\Lambda_{\mathrm{YM}}$. As we discussed before, in the large $N$ limit the mass parameter $g_{s}N$ is promoted to an electric central charge and the vacua of the theory are parametrized by $n+1$ Coulomb parameters.\\
The open string amplitude introduced in \ref{partition} specifies a perturbative expansion, but does not give an unambiguous non-perturbative answer. To get the non perturbative contributions to the partition function one has to specify a contour for the matrix integral. There is not a unique choice and one has many non perturbative complitions of the partition function. A natural basis of integration domains is given by the Lefschetz thimbles $D_{\alpha}$ of the potential $\mathcal{W}(z_{1},...,z_{N})/\epsilon_{2}$ in \ref{mathcalpot}. These cycles are attached to the critical points of $\mathcal{W}$ and are labelled by the saddle point configurations $\alpha=(a_{1},...,a_{m})$. Moreover, they have boundary where $\mathrm{Re}\left(  \mathcal{W}/\epsilon_{2}\right) \rightarrow + \infty$ and ensure the convergence of the integral. So, one can define the non perturbative topological string amplitudes  \cite{rif7,rif8,rif9,rif25,rif26}

\begin{equation}\label{topstring}
\begin{split}
Z_{\alpha}(\epsilon_{1},\epsilon_{2}) & =\int_{D_{\alpha}} \prod_{i=1}^{N}dz_{i} \prod_{1\leq i< j \leqslant N} (z_{i}-z_{j})^{\frac{\epsilon_{1}}{\epsilon_{2}}} \exp\left( -\frac{1}{\epsilon_{2}} \sum_{i} W(t_{k};z_{i})   \right) \\ \\ & =
\int_{D_{\alpha}} \exp\left( -\frac{1}{\epsilon_{2}} \sum_{i} \mathcal{W}(t_{k};z_{i})   \right),
\end{split}
\end{equation}

where 

\begin{equation}
\mathcal{W}(t_{k};z_{i})= \sum_{i} W(t_{k};z_{i}) - \epsilon_{1} \sum_{i<j}\log (z_{i}-z_{j}).
\end{equation}

The matrix model representation of the Nekrasov partition functions allows to connect $\mathcal{N}=2$ gauge theories in $4d$ and two dimensional $\mathcal{N}=(2,2)$ Landau-Ginzburg models with superpotential $\mathcal{W}(t_{k};z_{i})$. The key point is that the open string amplitudes can be identified with the flat sections of the $tt^{*}$ Lax connection in the asymmetric limit \cite{rif23}. This can be realized by compactifying the theory on a spatial circle of radious $R$ in such a way that the superpotential gets rescaled by $\mathcal{W}\rightarrow R\mathcal{W}$. With a rotation of the fermionic measure of the superspace one can also introduce an overall phase $\zeta$ multiplying the superpotential $\mathcal{W}\rightarrow \mathcal{W}/\zeta, \overline{\mathcal{W}}\rightarrow \zeta\overline{\mathcal{W}} $. Then one has to consider an analytic continuation of the phase $\zeta$ away from the locus $\vert \zeta \vert=1$ and take the limit $R\rightarrow 0, \zeta\rightarrow 0$ with $\epsilon_{2}=\zeta/R$ finite. This is equivalent to a non unitary deformation of the theory in which we set $\overline{\mathcal{W}}=0$ and rescale $W\rightarrow W/\epsilon_{2}$. Among the $tt^{*}$ brane amplitudes, a set of distinguished elements is given by the pairing of the D-brane states $\ket{D_{\alpha}}$ with the topological vacuum $\ket{0}$ corresponding to the identity operator. We denote such wave functions as

\begin{equation}
\psi_{\alpha}= \braket{0}{D_{\alpha}}.
\end{equation}

In the asymmetric limit one has an explicit formula for the above overlap which reads

\begin{equation}
\lim_{ \mathrm{asym} }\psi_{\alpha}(\epsilon_{1},\epsilon_{2})= Z_{\alpha}(\epsilon_{1},\epsilon_{2}),
\end{equation}

where one has to redefine the Nekrasov deformation $\epsilon_{2}$ with respect to the open string convention by setting $\epsilon_{2}=\zeta/\tilde{R}$. As one can see from the expression above the Nekrasov parameter $\epsilon_{1}$ plays the role of Van der Monde coupling and has mass dimension $1$. One can naturally extend the $2d/4d$ correspondence also to SQM. The one dimensional $\mathcal{N}=4$ version of the theory has the same structure of vacua and BPS spectrum, which implies that one can study the geometry of the vacuum bundle either in one or two dimensions \cite{rif10}. The superpotential in SQM is dimensionless and arises from the compactification of the $2d$ theory. The matching scale is given by the radious $R$ of the $tt^{*}$ circle and we have $\mathcal{W}_{\mathrm{SQM}}=-\mathcal{W}_{2d}/\epsilon_{2}$. According to the AGT correspondence, the amplitudes of the Penner matrix models compute the conformal blocks of the Liouville CFT and hence the FQHE wave functions in the Vafa's model. To match the normalization of the Liouville correlators given in \cite{rif1} one has to identify $-m_{a}/\epsilon_{2}=1$ and $1/\nu=\epsilon_{1}/\epsilon_{2}$. The AGT correspondence has been generalized in \cite{rif29} also to class $\mathcal{S}$ theories with irregular punctures in the Gaiotto surface. It turns out that the Nekrasov partition functions of these models reproduce irregular conformal blocks of the Liouville CFT. The presence of higher degree poles in the quadratic differential corresponds to insertions in the Liouville correlators of the so called irregular vertex operators \cite{rif30,rif31}. As for the mass parameters, the relevant coupling $\mu_{2d}$ of the $2d$ theory is related to its dimensionless counterpart $\mu_{1d}$ in SQM by $\mu_{1d}=-\mu_{2d}/\epsilon_{2}$.

\section{$tt^{*}$ Geometry of the One Electron Model with Two Quasi-Holes}

We consider a class of $2d$ $\mathcal{N}=(2,2)$ Landau-Ginzburg theory described by the superpotential

\begin{equation}
W(z)= \mu z + \log z + \log(z-\rho).
\end{equation}

The $4$-SQM version of this theory corresponds to the Vafa model for a single electron on the plane with two quasi-holes at distinct positions $0,\rho \in \mathbb{C}$ and a background electrostatic field $\mu \in \mathbb{C}^{\times}$. The above function is multi-valued and cannot be strictly considered as the superpotential of the model. We want to define this theory on the Abelian universal cover of the target space and study the correspoding $tt^{*}$ geometry. As discussed in section \ref{integral}, the solution to the $tt^{*}$ equations is captured by the Stokes matrices which describe the jumps of the brane amplitudes $\Psi(\zeta)$ on the $\zeta$-plane. The monodromy data and the leading IR behaviour of the $tt^{*}$ metric and connection are determined by the BPS spectrum of the model. The inital step is to find the classical vacua of the theory on the universal cover. The solutions $z_{\pm}$ to the equation $\partial_{z} W=0$ denote two equivalence classes of vacua which are isomorphic to the homology group $H_{1}(\mathbb{C}\setminus(0,\rho);\mathbb{Z})$. A natural basis for this group is given by the two loop generators $\ell_{0},\ell_{\rho}$ encircling the holes positions. The vacuum space of the theory on the Abelian cover $\mathcal{A}$ decomposes in a direct sum of irreducible representations of the homology group:

\begin{equation}
\mathcal{V}_{\mathcal{A}} \simeq \mathrm{L}^{2}\left(  \mathrm{Hom} \left(   \mathbb{Z}^{2}, U(1) \right) \right) \otimes \mathbb{C}^{2},
\end{equation}

where $ H_{1}(\mathbb{C}\setminus(0,\rho);\mathbb{Z})=\mathbb{Z}^{2}$. We denote the orthogonal idempotents of $\mathcal{R}_{\mathcal{A}}$ corresponding to the classical vacua on the universal cover with

\begin{equation}
\ket{\pm;m,n}= \ell_{0}^{m}\ell_{\rho}^{n} \ket{\pm;0,0}
\end{equation}

where $\ket{\pm;0,0}$ corresponds to some representative of $z_{\pm}$ on the universal cover. In terms of these states one can construct eigenstates of the loop generators

\begin{equation}\label{point}
\begin{split}
\ket{\pm;\phi,\varphi}= \sum_{m,n} e^{-i(m\phi + n\varphi)} \ket{\pm;m,n}
\end{split}
\end{equation}

where the angles $\phi,\varphi \in [0,2\pi]$ label representations of $H_{1}(\mathbb{C}\setminus(0,\rho),\mathbb{Z})$. By the fact that the homology is an abelian symmetry of the model, the $tt^{*}$ geometry diagonalizes completely with respect to the angles $\phi,\varphi$. In particular, introducing the labels $k,j=\pm$, the ground state metric in the point basis \ref{point} can be expanded in Fourier series as 

\begin{equation}\label{fourier} 
\begin{split}
& g_{k,\bar{j}}( \phi,\varphi) =  \sum_{r,s} e^{i( r\phi + s\varphi)} g_{k,\bar{j}}(r,s), \\ \\ 
& g_{k,\bar{j}}(r,s)= \braket{\overline{j;r,s}}{k;0,0}.
\end{split}
\end{equation}

From the IR expansion 

\begin{equation}\label{expansion}
\braket{\overline{j;r,s}}{k;0,0} \sim \delta_{k,j}\delta_{r,0}\delta_{s,0}-\frac{i}{\pi} \mu_{k,0,0;j,r,s} K_{0}(2\vert  w_{k,0,0}-w_{j,r,s}  \vert)
\end{equation}

we see that the one-soliton multiplicities $\mu_{k,0,0;j,r,s} $ are weighted by the Fourier phase factors $e^{i(r\phi+s\varphi)}$. The number of solitons saturating the Bogomonlyi bound can be obtained by solving the BPS equation \cite{rif12}

\begin{equation}
\partial_{\sigma}z= \alpha \overline{\partial_{z}W}
\end{equation}

where $\sigma$ is the spatial variable of the field $z$ and $\alpha= \triangle W/ \vert \triangle W \vert$ identifies the BPS sector we are considering. At spatial infinity $\sigma= \pm \infty$ one has also to impose the corresponding critical points as boundary condition. One can count the number of solutions to the BPS equations by plotting with a program the flow of the vector field 

\begin{equation}\label{flow}
V= \begin{pmatrix} \mathrm{Re} z \\ \mathrm{Im} z \end{pmatrix}= \begin{pmatrix} \mathrm{Re} \ \bar{\alpha}\partial W \\ -\mathrm{Im} \ \bar{\alpha}\partial W \end{pmatrix} .
\end{equation}

Let us start with the case of $\alpha=\pm i$. The solitons in this sector projected on the target manifold are respectively anti-clockwise and clockwise loops based at $z_{\pm}$. The spectrum that we find is the same of the models

\begin{equation}\label{IRtheories}
\begin{split}
& W_{-}(z)= \mu z + 2\log z, \\ \\ 
& W_{+}(z)= \log z + \log (z-\rho),
\end{split}
\end{equation} 

which describe the dynamics of the two vacua at $z_{\pm}$ respectively in the IR limits $\rho\rightarrow 0$ and $\mu\rightarrow 0$. In the case of $\alpha=i$ one finds

\begin{equation}\label{primitive}
\begin{split}
& \vert \mu_{-,m,n;-,r,s} \vert = \delta_{m,r+1}\delta_{n,s+1}, \\ \\ 
& \vert \mu_{+,m,n;+,r,s} \vert = \delta_{m,r+1}\delta_{n,s}+\delta_{m,r}\delta_{n,s+1},
\end{split}
\end{equation}

where the sign of the soliton multiplicities will be fixed by the computation of the Stokes matrices. \\ 
Let us consider now the solitons connecting $z_{-}$ to $z_{+}$. Given that the quantum monodromy is not affected by the choice of the parameters (up to conjugation) we can count the soliton solutions for $\mu=\rho=1$. The soliton equation for these BPS states must be solved with the central charges 

\begin{equation}\label{solboundcond}
\Delta W^{+-}_{k}= z_{+}-z_{-} +\log \bigg \vert \frac{z_{+}(z_{+}-1)}{z_{-}(z_{-}-1)}\bigg\vert +2\pi i(k+1/2) , \ \ \ \ \ k\in \mathbb{Z}
\end{equation}

and boundary conditions $x(-\infty)= z_{-}, \ x(+\infty)=z_{+}$, where $z_{\pm}= (1\pm \sqrt{5})/2$. Let us fix a representative $z_{-;0,0}$ of $z_{-}$ on the spectral cover and denote with $\ket{-;0,0}$ the corresponding state in the Hilbert space.

\begin{figure}[!h]
\centering
\begin{tikzpicture}
\coordinate [label=above:$z_{-}$] (A)  at (0,0);
\coordinate [label=above:$z_{+}$] (B)  at (8.1,0.2);
\coordinate[label= x] (x) at (6,0);
\coordinate[label= x ] (x) at (10,0);
\draw  ((0,0) to [out=-55, in=250] (8,0);
\end{tikzpicture}
\caption{Definition of $\gamma^{+}_{0,0}$}\label{curve}
\end{figure}
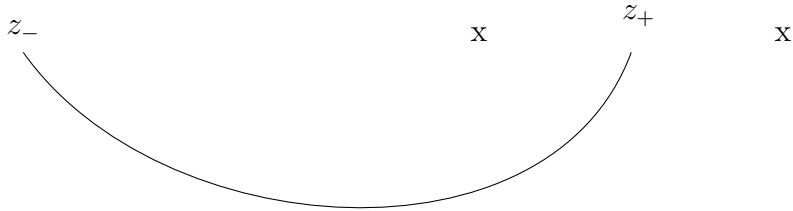

It is convenient to label the representatives of $z_{\pm}$ on the spectral cover with homotopy classes of curves which connect a fixed point $z_{\ast} \in \mathbb{C} \setminus \left\lbrace 0,\rho \right\rbrace $ to $z_{\pm}$. Let us choose $z_{\ast}=z_{-}$. In this way the curves $\gamma^{-}_{r,s}$ corresponding to the points $\ket{-;r,s}$ are simply the closed cycles $\ell_{0}^{r}\cdot\ell_{\rho}^{s}$ based at $z_{-}$. Then, we identify the curve $\gamma^{+}_{0,0}$ in figure \ref{curve} which connects $z_{-}$ to $z_{+}$ with the state $\ket{+;0,0}$. All the other curves $\gamma^{+}_{r,s}$ going from $z_{-}$ to $z_{+}$ which label the representatives $\ket{+;r,s}$ of $z_{+}$  are in the same homotopy class of $\gamma^{+}_{0,0}\cdot\ell_{0}^{r}\cdot\ell_{\rho}^{s}$. The solutions to the BPS equation which interpolate $z_{-;00}$ and $z_{+;r,s}$ belong to homotopy classes $\gamma^{+}_{r,s}$ for certain couples of integers $(r,s)$. The multiplicity matrix that we obtain by studying the flow of \ref{flow} for $\alpha= \Delta W^{+-}_{k}/\vert \Delta W^{+-}_{k}\vert $ is 

\begin{equation}\label{spec}
\vert \mu_{-,0,0;+,r,s} \vert = \delta_{r,s}+\delta_{r+1,s}
\end{equation}

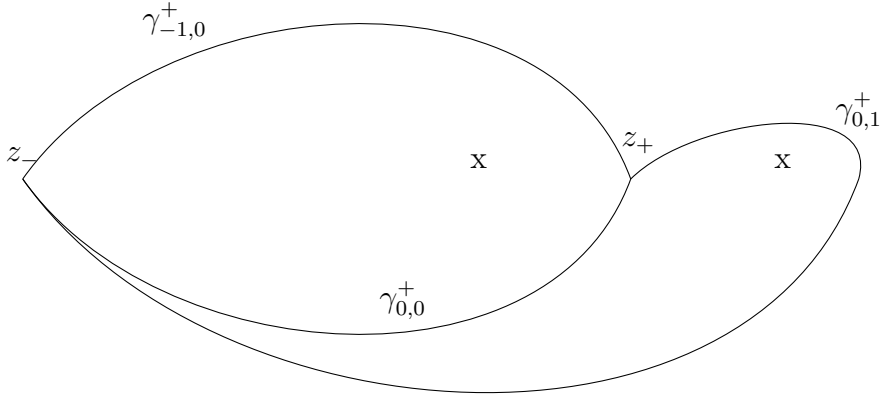
\begin{figure}[!h]
\centering
\begin{tikzpicture}
\coordinate [label=above:$z_{-}$] (A)  at (0,0);
\coordinate [label=above:$z_{+}$] (B)  at (8.1,0.2);
\coordinate[label= x] (x) at (6,0);
\coordinate[label= x ] (x) at (10,0);
\coordinate[label= $\gamma^{+}_{0,1}$ ] (x) at (11,0.5);
\coordinate[label= $\gamma^{+}_{0,0}$ ] (x) at (5,-2);
\coordinate[label= $\gamma^{+}_{-1,0}$ ] (x) at (2,1.7);
\draw  ((0,0) to [out=-55, in=250] (8,0);
\draw  ((0,0) to [out=55, in=-250] (8,0);
\draw ((0,0) to [out=-55, in=250] (11,0);
\draw ((11,0) to [out=75, in=45] (8,0);
\end{tikzpicture}
\caption{Some soliton solutions}\label{somesolitonsolutions}
\end{figure}

and the central charge of the BPS particle in the $(+,r,s;-,0,0)$ sector turns out to be 

\begin{equation}\label{centralcharge}
\Delta W^{+-}_{r,s}= z_{+}-z_{-} +\log \bigg \vert \frac{z_{+}(z_{+}-1)}{z_{-}(z_{-}-1)}\bigg\vert +2\pi i(r+s+1/2) .
\end{equation}

One can easily see that the braiding of the quasi-holes generates a spectral flow of the BPS spectrum. As we exchange the two singularities in the anticlockwise sense the curves $\gamma^{\pm}_{r,s}$ transform respectively as 

\begin{equation}\label{rule}
\gamma^{-}_{r,s}\rightarrow \gamma^{-}_{s,r}, \hspace{1cm} \gamma^{+}_{r,s}\rightarrow \gamma^{+}_{s,r+1},
\end{equation}

which implies that the entire BPS spectrum can be generated by the fundamental soliton $\gamma^{+}_{0,0}$ by iterating the braiding of the quasi-holes. At the same time the states $\ket{\pm;r,s}$ transform as

\begin{equation}
\ket{-;r,s}\rightarrow \ket{-;s,r}, \hspace{1cm} \ket{+;r,s}\rightarrow \ket{+;s,r+1},
\end{equation}

which implies

\begin{equation}
\braket{\overline{+;r,s}}{-;0,0} \rightarrow \braket{\overline{+;s,r+1}}{-;0,0}.
\end{equation}

On the other hand, from the IR expansion of the metric one gets also 

\begin{equation}
\begin{split}
\braket{\overline{+;r,s}}{-;0,0} & \sim -\frac{i}{\pi}\mu_{-,0,0;+,r,s} \ K_{0}(2\vert  w_{k,0,0}-w_{j,r,s}  \vert) \\ \\ & \rightarrow -\frac{i}{\pi}\mu_{-,0,0;+,r,s} \ K_{0}(2\vert  w_{k,0,0}-w_{j,s,r+1}  \vert).
\end{split}
\end{equation}

The compatibility of these two expressions puts the condition on the soliton matrix

\begin{equation}
\mu_{-,0,0;+,r,s} = \mu_{-,0,0;+,s,r+1}
\end{equation}

which implies that the sign of the BPS multiplicity in the $(-,0,0;+,r,s)$ sector is the same for each couple $(r,s)$. This sign has not an intrinsic meaning, since it can always be absorbed by changing the sign of one of the theta vacua in \ref{point} without affecting the diagonal elements $g_{k,\bar{k}}( \phi,\varphi) $.\\ We are able now to compute the Stokes operators of the system. In the basis of theta vacua $\ket{-;\phi,\varphi},\ket{+;\phi,\varphi}$ these are $2\times2$ matrices where the entries are functions of the angles $\phi,\varphi$. We choose the real axis of the $\zeta$-plane as Stokes axis. The Stokes matrix can be written as ordered product of three matrices 

\begin{equation}
S=S_{3}S_{2}S_{1}
\end{equation}

which we are going to describe. The first contribution $S_{1}$ is generated by the phase-ordered product of the solitons with positive real and immaginary part. For $\mu=\rho=1$ we have $\mathrm{Re} \ W(z_{-})> \mathrm{Re} \ W(z_{+})$. So, these BPS states have central charges $-\Delta W^{+-}_{r,s} $ and connect the points $z_{+;r,s}$ to $z_{-;0,0}$ with $r<0,s\leq 0 $. Each of them generates a Stokes jump 

\begin{equation}\label{stokesgenerated}
M_{r,s}= \begin{pmatrix}
1 & \mu_{-,0,0;+,r,s} \ e^{i(r\phi+s\varphi)} \\ 0 & 1
\end{pmatrix}
\end{equation}

where the phase factor comes from the Fourier expansion of the metric and we take the soliton multiplicities $\mu_{-,0,0;+,r,s}=-( \delta_{r,s}+\delta_{r+1,s})$ with the mignus sign. Taking the product of these factors for $r<0,s\leq 0 $ we get 

\begin{equation}
\begin{split}
S_{1}=& \prod_{r<0,s\leq 0}M_{r,s}= \begin{pmatrix}
1 & -\sum_{r<0,s\leq 0} e^{i(r\phi+s\varphi)} (\delta_{r,s}+\delta_{r+1,s})\\ 0 & 1
\end{pmatrix} \\ \\ =&\begin{pmatrix}
1 & \frac{1+e^{i\varphi}}{1-e^{i(\phi+\varphi)}}\\ 0 & 1
\end{pmatrix}.
\end{split}
\end{equation}

When $\zeta$ crosses the positive immaginary axis the Stokes matrix picks up the contribution $S_{2}$ of the closed solitons bases at $z_{\pm}$. This matrix has a diagonal structure and the entries $S_{\pm}$ correspond to the Stokes jumps of the abelian models \ref{IRtheories}. In presence of aligned critical values there are subtleties in the computation of the Stokes factors. Let us suppose to have $N$ vacua and a set of critical values $w_{k},k=1,....,m\leq N$ on a straight line in the W-plane with one physical soliton (and the time-reversed) connecting only the consecutive ones. As discussed in \cite{rif37}, one finds in the leading IR behavour of the ground state metric terms looking like there were solitons in the $k,k+n$ sector with multiplicity $\pm 1/n$. This could seem strange, since the number of BPS particles should be an integer. However, what is really required to be integral is the Stokes matrix $S\in SL(N,\mathbb{Z})$ and not the soliton matrix $\mu_{ij}$. In this case one has to define a matrix $F$ belonging to the Lie algebra $sl(N,\mathbb{Q})$ that encodes also the BPS particles with rational multiplicities. In order to have an integral Stokes operator we have to exponentiate the algebra element  

\begin{equation}\label{stokes}
S=e^{F}.
\end{equation}

Coming back to our model, one has to deal with an infinite number of vacua on the spectral cover with pure immaginary critical values. The correct Stokes jumps as function of the angles can be extracted by the asymmetric limit of the $tt^{*}$ brane amplitudes for the Landau-Ginzburg models $W_{\pm}(z)$ in \ref{IRtheories}. The computation in \cite{rif23} provides the following functions

\begin{equation}\label{magnfun}
\begin{split}
& F_{-}(\phi,\varphi)= \log\left( 1-e^{-i(\phi+\varphi)}\right)    \\ \\
 & F_{+}(\phi,\varphi)=   \log\left( 1-e^{-i\phi}\right)  +  \log\left( 1-e^{-i\varphi}\right)  - \log\left( 1-e^{-i(\phi+\varphi)}\right)      
\end{split}
\end{equation}

from which one can obtain the Stokes jump 

\begin{equation}
S_{2}= \begin{pmatrix}
e^{F_{-}} & 0 \\ 0 & e^{F_{+}} 
\end{pmatrix}.
\end{equation}

The matrix $S_{3}$ includes the contribution of the solitons with negative real part and positive immaginary part. These BPS particles connect the points $z_{-,r,s}$ with $r,s\leq 0$ to $z_{+,0,0}$ and have central charges $\Delta W^{+-}_{-r,-s}$. The corresponding multiplicities are $\mu_{+;0,0;-,r,s}=- \mu_{-;r,s;+,0,0}=-\mu_{-;0,0;+,-r,-s}.$ Each of these particle generates a contribution to the Stokes matrix which reads 

\begin{equation}
M_{r,s}= \begin{pmatrix}
1 & 0 \\ -\mu_{-,0,0;+,-r,-s} \ e^{i(r\phi+s\varphi)} & 1
\end{pmatrix}.
\end{equation}

Multiplying the Stokes jumps of these solitons for $r,s\leq 0$ one finds 

\begin{equation}
\begin{split}
S_{3}=& \prod_{r,s\leq 0}M_{r,s}= \begin{pmatrix}
1 & 0 \\  \sum_{r,s\leq 0} e^{i(r\phi+s\varphi)} (\delta_{-r,-s}+\delta_{-r+1,-s}) & 1
\end{pmatrix} \\ \\ =&\begin{pmatrix}
1 & 0 \\ \frac{1+e^{-i\varphi}}{1-e^{-i(\phi+\varphi)}} & 1
\end{pmatrix}.
\end{split}
\end{equation}

Putting all the pieces together we get 

\begin{equation}
S(\phi,\varphi)= \begin{pmatrix}
e^{F_{-}(\phi,\varphi)} & \frac{1+e^{i\varphi}}{1-e^{i(\phi+\varphi)}}e^{F_{-}(\phi,\varphi)} \\ \frac{1+e^{-i\varphi}}{1-e^{-i(\phi+\varphi)}}e^{F_{-}(\phi,\varphi)} & e^{F_{+}(\phi,\varphi)} + \frac{2(1+\cos\varphi)}{(1-e^{i(\phi+\varphi)})(1-e^{-i(\phi+\varphi)})}e^{F_{-}(\phi,\varphi)}
\end{pmatrix}.
\end{equation}

The Stokes matrix encoding the contribution of the BPS states with central charges in the lower part of the $\zeta$-plane corresponds to the operator $S^{-t}(\phi,\varphi)$. This has not to be intended as the inverse transpose in the $2\times2$ matrix sense. The correct definition is

\begin{equation}
S^{-t}(\phi,\varphi)=\left[ S(-\phi,-\varphi)\right] ^{-t},
\end{equation}

which requires also to change the sign of the angles. \\ In the present case we obtain 

\begin{equation}
S^{-t}(\phi,\varphi)= \begin{pmatrix}
 e^{-F_{-}(-\phi,-\varphi)} + \frac{2(1+\cos\varphi)e^{-F_{+}(-\phi,-\varphi)}}{(1-e^{i(\phi+\varphi)})(1-e^{-i(\phi+\varphi)})} & -\frac{(1+e^{i\varphi})e^{-F_{+}(-\phi,-\varphi)} }{1-e^{i(\phi+\varphi)}}\\ -\frac{(1+e^{-i\varphi})e^{-F_{+}(-\phi,-\varphi)}}{1-e^{-i(\phi+\varphi)}} & e^{-F_{+}(-\phi,-\varphi)}
\end{pmatrix}.
\end{equation}

So, the quantum monodromy $H=S^{-t}S$ of this system is 

\begin{equation}
\begin{split}
H(\phi,\varphi)= & \begin{pmatrix} e^{F_{-}(\phi,\varphi)-F_{-}(-\phi,-\varphi)} & \frac{1+e^{i\varphi}}{1-e^{i(\phi+\varphi)}}\left(  e^{F_{-}(\phi,\varphi)-F_{-}(-\phi,-\varphi)} -e^{F_{+}(\phi,\varphi)-F_{+}(-\phi,-\varphi)}\right)  \\ 0 & e^{F_{+}(\phi,\varphi)-F_{+}(-\phi,-\varphi)} 
\end{pmatrix} \\ \\  = & \begin{pmatrix}
-e^{-i(\phi+\varphi)} & -(1+e^{i\varphi})e^{-i(\phi + \varphi)} \\ 0 & -1
\end{pmatrix}
\end{split}
\end{equation}

where in the second equality we have used the expression for $F_{\pm}$ in \ref{magnfun}. We see that despite the Stokes matrices are singular for some values of the angles, the quantum monodromy has smooth coefficients. This is equivalent to the statement that the phases of the quantum monodromy $e^{2\pi i q^{R}_{\pm}(\phi,\varphi)}$ are smooth in the angles, while the Ramond charges $q^{R}_{\pm}(\phi,\varphi)$ are allowed to have integral jumps. The Ramond charges of the theta-vacua reads 

\begin{equation}
q^{R}_{-}=-B_{1}((\phi+\varphi)/2\pi), \hspace{1cm} q^{R}_{+}= B_{1}((\phi+\varphi)/2\pi) -B_{1}(\phi/2\pi) - B_{1}(\varphi/2\pi),
\end{equation}

where $B_{1}(x)=x-1/2$ is the first periodic Bernoulli polynomial of variable $x \in [0,1]$.

\section{$tt^{*}$ Geometry of Modular Curves}\label{papermodjhep}

\subsection{Quantum Hall Effect and Modular Curves}

In this section we are going to study a special class of models which reveal a beautiful connection between the physics of quantum Hall effect and the geometry of modular curves \cite{rif54}. Despite it is not relevant for phenomenological purposes, this class of theories has remarkable properties which enlarge further the rich mathematical structure of FQHE. We recall that the prototype of one-particle supersymmetric model which is relevant for the FQHE physics is given by the $\mathcal{N}=4$ Landau-Ginzburg theory with superpotential

\begin{equation}\label{class}   
W(z)= \sum_{\zeta \in L} e(\zeta)\log(z-\zeta),
\end{equation}

where $L$ is a discrete set of $\mathbb{C} $ and $e(\zeta)$ are real numbers. The variable $z$ is interpreted as the electron coordinate and the term $e(\zeta)\log(z-\zeta)$ is the two dimensional coulombic potential which describes the interaction between the electron and an external charge. In the standard setting of the FQH systems the source of electrostatic interaction is taken to be $L= \Lambda \cup S$, with $\Lambda$ a lattice and $S$ a set of positions of quasi-holes. The effect of the lattice is to reproduce the constant macroscopic magnetic field with $e(\lambda)=1$ units of magnetic flux at a point $\lambda \in \Lambda$. At this level the expression of the superpotential is just symbolic, since the sum is taken over an infinite set of points and the function is multi-valued. Therefore it requires a more precise definition according to the class of models that one is considering.\\ For a generic choice of the parameters defining $W(z)$, the classical vacua are isolated, and the elements in the chiral ring are identified with their set of values at the critical points. However, finding the classical vacua and studying the $tt^{*}$ geometry of these models is rather complicated, unless one arranges the set of quasi-holes in some special configuration to have an enhancement of symmetry. In this way we can construct degenerate models of FQHE which are not realistic for phenomenology, but at least analitically treatable. We want to consider a particular class of theories of this type which have an abelian subgroup of symmetry acting transitively on the set of vacua. This is the most convenient limit, since in this case the Berry's connection can be completely diagonalized in a basis of eigenstates of such symmetry and the $tt^{*}$ equations can be derived. It turns out that these models are parametrized modulo isomorphisms by the family of Riemann surfaces $Y_{1}(N)= \mathbb{H}/\Gamma_{1}(N)$, labelled by an integer $N\geq 2$, also known as modular curves for the congruence subgroup $\Gamma_{1}(N)$ of the modular group $SL(2,\mathbb{Z})$ \cite{rif45}. Each point of the curve of level $N$ identifies a theory where $\partial_{z}W(z)$ is an ellitpic function with a $\mathbb{Z}_{N}$ symmetry generated by a torsion point of the elliptic curve $\mathbb{C}/\Lambda$. For each connected component of the space of models one can define its spectral cover as the complex manifold whose points identify a model and a vacuum \cite{rif13}. These are the modular curves $Y(N)= \mathbb{H}/\Gamma(N)$ for the principal congruence subgroup $ \Gamma(N)$. More precisley, $Y(N)$ is a cover of $Y_{1}(N)$ of degree $N$, which is the number of vacua in the fundamental cell of the torus. The $tt^{*}$ equations simplify considerably on the spectral curve and can be normalized in the form of $\hat{A}_{N-1}$ Toda equations \cite{rif10}. These appear in all the models with a $\mathbb{Z}_{N}$ symmetry group which is transitive on the vacua. The modular curves are manifolds with cusps, which represent physically the RG flow fixed points of the theory. These are in correspondence with the equivalence classes of rationals with respect to the congruence subgroups. An outstandig fact is that for a given $N$ all the $\hat{A}_{Q-1}$ models with $Q \vert N$ are embedded in this class of theories as critical limits, providing the regularity conditions for the solutions to the equations. An exception is the case of $N=4$, where only $\hat{A}_{3}$ models appear. The beauty of the modular curves is that they possess various geometrical structures. For instance, in the modular curves $Y(N)$ of level $N=3,4,5 $ the cusps are located at the vertices of platonic solids inscribed in the Riemann sphere.\\ The theory of modular curves is also strictly related to number theory. These surfaces can be seen as projective algebraic curves defined over the real cyclotomic extension of the rationals $\mathbb{Q}(\zeta_{N}+\zeta_{-N})$, with $\zeta_{N}=e^{2\pi i/N}$. At the level of superpotential, the action of the Galois group is reproduced by a third congruence subgroup which enters in this theory, i.e. $\Gamma_{0}(N)$. An interesting implication is that the solutions of the $N$-Toda equations are related by the action of the Galois group, since the UV cusps described by the $\hat{A}_{N-1}$ models are all in the same orbit of $\Gamma_{0}(N)$.\\ Another interesting phenomenon appearing in this class of theories is that, despite the covariance of the $tt^{*}$ equations, neither the superpotential nor the ground state metric, and therefore the Berry's connection, are invariant under the action of $\Gamma(N)$. This apparent contraddiction finds a consistent explanation in the context of the abelian universal cover of the model. \\ \\ 
The rest of the chapter is organized as follows: 
In section \ref{sec2} we classify up to isomorphisms all the models of the type \ref{class} with an abelian subgroup of symmetry acting transitively on the vacua. The underlying structure of the modular curves and the relation between geometry and number theory arise naturally in the derivation. In section \ref{sec3} we provide an explicit description of these models. The target manifold is not simply connected and one needs to pull-back the model on the universal cover in order to define the Hilbert space and write the $tt^{*}$ equations. On this space the symmetry group contains also the generators of loops around the poles in the fundamental cell. The symmetry algebra is non-abelian on the universal cover and the abelian physics of the punctured plane can be recovered at the level of quantum states by considering trivial representations of the loop generators. We show that this can be done consistently with the $tt^{*}$ equations in \ref{trunc}. In section \ref{sec4} we study the modular properties of these systems. First we consider the transformation of $\partial_{z}W(z)$ under the congruence subgroups and then of the superpotential. In particular we focus on the critical value of one the vacua which we use to write the $tt^{*}$ equations in the Toda form. This can be defined as holomorphic function only on the upper half plane and shifts by a costant under a transformation of $\Gamma(N)$. We connect this phenomenon to the geometry of the modular curves in the simple cases of genus $0$, i.e. with $2\leq N \leq 5$. The details of the computation of the constant for a generic $N$ are instead given in \ref{modulartrnasofrmationmass}. The modular transformations have also the effect of changing the basis of the symmetry generators and act on the states by modifying the representation of the symmetry group. We study this action and how the components of the ground state metric are transformed. In \ref{final} we provide a classification of the cusps. In particular we study the behavour of the superpotentials around these points and distinguish between UV and IR critical regions. Finally, we discuss the boundary conditions of the solutions and how they are related by the Galois group.

\subsection{Classification of the Models}\label{sec2}

\subsubsection{Derivation}

Our first aim is to classify (up to isomorphisms) all the models in the class of FQHE theories \ref{class} with an abelian subgroup of symmetry acting transitively on the vacua. Since the punctured plane is not a simply connected space, one cannot define for these models a superpotential on the target manifold. So, we have to start the classification from the derivative

\begin{equation*}
 \partial_{z} W(z)= \sum_{\zeta \in \Lambda} \frac{1}{z-\zeta} + \sum_{s \in S} \frac{e(s)}{z-s},
 \end{equation*}
 
where

\begin{equation*}
\begin{split}
 & \Lambda= 2 \pi \mathbb{Z}  \oplus 2\pi \tau \mathbb{Z}, \ \tau \in \mathbb{H}, \\ \\ & e: S \longrightarrow \mathbb{C}.
\end{split} 
\end{equation*}

We stress again that the expression above is just formal and represents a meromorphic function with a simple pole at each point of $L=\Lambda \cup S$. Moreover, in this classification we allow the charges of the quasi-holes to be complex. This is mathematically consistent, since the superpotential is a complex function. \\
The action of the abelian group is transitive on the vacua and, in the case of a non trivial kernel, can always be made faithful. The transitivity implies that the set of vacua is a copy of the abelian group. In particular, given that the zeroes of $\partial_{z} W(z)$ cannot have accumulation points, it must be finitely generated. The abelian subgroups of $\mathbb{C}$ satisfying this property are lattices. Since the group acts freely also on the set of poles and the principal divisor of $\partial_{z} W(z)$ has degree $0$\footnote{This is true for a compact Riemann surface, as it turns out to be the target manifold.}, it is immediate to conclude that also $L$ is a lattice, as well as (the faithful representation of) the abelian subgroup of symmetries of our model.\\ If we want $L$ to be at least a pseudosymmetry for $\partial_{z} W(z)$, the function $e(s)$ must be extended to a multiplicative periodic character:

\begin{equation*}
\begin{split}
\partial _{z}W(z+\zeta)=e(\zeta) \partial_{z} W(z),\ \ \ \ \ \ \ \ \ \  e(\zeta+\lambda)= e(\zeta),
\end{split}
\end{equation*}

for each $\zeta \in L$, $\lambda \in \Lambda$. By definition of homomorphism, the kernel of $e$ must be a subgroup of $L$. Up to a redefinition of the initial set of holes, this is represented by the sublattice $\Lambda$, which is a symmetry for $\partial_{z} W(z)$ in a strict sense. \\ Up to this point, we have a model for each lattice $L$ and a character $e$ which is periodic of a sublattice $\Lambda \subset L$.\footnote{We are going to show that $e$ is non trivial.}\\ Given the periodicity of $e$, we can equivalently restrict the analysis to primitive characters, i.e. with trivial kernel:

\begin{equation*}
e: L/ \Lambda \longrightarrow \mathbb{C}, 
\end{equation*}

where $ L/ \Lambda \simeq \mathbb{Z}_{N_{1}} \oplus \mathbb{Z}_{N_{2}}$ for two positive integers $N_{1}, N_{2}$ such that $N_{1} \mid N_{2}$. The fact that $e$ is primitive implies the isomorphism $ \mathbb{Z}_{N_{1}} \oplus \mathbb{Z}_{N_{2}} \simeq \mathbb{Z}_{N_{1}N_{2}}$. But, according to the chinese remainder theorem, this can be true only if the two integers are coprime. The consistency between the two conditions on the integers $N_{1},N_{2}$ requires that $N_{1}=1$ and $N_{2}=N\geq 2$, with 

\begin{equation*}
L/ \Lambda \simeq \mathbb{Z}_{N}, \ \ \ \ \ \ N \geq 2.
\end{equation*}

If we set $N=1$ we obtain the trivial case in which there are no quasi-holes. \\ In conclusion, the models are classified by couples $( E_{\Lambda},Q)$, where

\begin{itemize}
\item $E_{\Lambda}$ is the elliptic curve $ \mathbb{C}/ \Lambda$,  
\item $ Q \in E[N]= \left\lbrace P \in \mathbb{C}/ \Lambda \mid NP \in \Lambda \right\rbrace $ such that $e(Q)= e^{\frac{2\pi i}{N}}$.
\end{itemize}

The set $E[N]$ is called the N-torsion subgroup of the additive torus group $\mathbb{C}/ \Lambda$. \\ Once $\Lambda$ and the level $N$ are fixed, the choice of the torsion point specifies an embedding of the cyclic subgroup $L/ \Lambda \simeq \mathbb{Z}_{N}$ in the elliptic curve. In particular, the torsion point must be of order $N$, i.e. such that $NQ \in \Lambda$ but $nQ \not \in \Lambda$ for $ 1<n<N$.

\subsubsection{Modular Curves}\label{modular}

Since we are classifying models up to strict equivalence, we have to identify those which are related by an isomorphism.  It is known that there is a bijection between the set of equivalence classes of elliptic curves endowed with a $N$-torsion point and the space 

\begin{equation*}
Y_{1}(N)= \mathbb{H}/\Gamma_{1}(N),
\end{equation*}

where $\Gamma_{1}(N)$ is a subgroup of $SL(2,\mathbb{Z})$ defined by the congruence condition

\begin{equation*}
\Gamma_{1}(N)= \left\lbrace \gamma \in SL(2,\mathbb{Z}) : \gamma=\begin{pmatrix} a & b \\ c & d

\end{pmatrix}= \begin{pmatrix} 1 & * \\ 0 & 1 \end{pmatrix} \mathrm{mod} \ N \right\rbrace .
\end{equation*}

A complete proof of this result can be found in \cite{rif45}. It is worth to recall that a matrix $\gamma \in SL(2,\mathbb{Z})$ acts on the upper half plane by the usual fractional linear transformation

\begin{equation}
\tau^{\prime}=\begin{pmatrix} a & b \\ c & d \end{pmatrix}\tau=  \frac{a\tau+b}{c\tau+d},
\end{equation}

and induces on the points of the elliptic curve the isogeny $z + \Lambda_{\tau} \longrightarrow m z + \Lambda_{\tau^{\prime}} $, for some $m \in \mathbb{C}$ such that $m\Lambda_{\tau}=\Lambda_{\tau^{\prime}}$. These maps are the only bijections which preserve the group structure of the elliptic curve. With these definitions it is immediate to show that the enhanced elliptic curve $(E_{\Lambda_{\tau^{\prime}}},Q)$ is isogenous to $\left( E_{\Lambda_{\tau}}, 2\pi/N +\Lambda_{\tau}\right)$, where $\tau^{\prime}=\gamma(\tau)$ for some $\gamma \in SL(2,\mathbb{Z})$, and that transformations of $\Gamma_{1}(N)$ are the only ones which preserve the choice of the torsion point. So, we can define the moduli space for $\Gamma_{1}(N)$ as

\begin{equation*}
 S_{1}(N)= \lbrace \left( E_{\Lambda_{\tau}}, 2\pi/N +\Lambda_{\tau}\right) , \tau \in \mathbb{H}  \rbrace / \sim 
\end{equation*}

where $\tau \sim \tau^{\prime}$ if and only if $\Gamma_{1}(N)\tau=\Gamma_{1}(N)\tau^{\prime}$, and state the bijection 

\begin{equation*}
 S_{1}(N) \overset{\sim}{\longrightarrow} Y_{1}(N).
\end{equation*}

The space $Y_{1}(N)$ is topologically a complex manifold with cusps and can be compactified. One first has to extend the action of $SL(2,\mathbb{Z})$ to the rational projective line $\mathbb{P}^{1}(\mathbb{Q})= \mathbb{Q} \cup \lbrace \infty  \rbrace$. Given  $\begin{pmatrix} p & q \\ r & t \end{pmatrix} \in SL(2,\mathbb{Z})$, we have 

\begin{equation*}
\begin{split}
& \mathbb{H} \longrightarrow \mathbb{\overline{H}}= \mathbb{H} \cup \mathbb{P}^{1}(\mathbb{Q}),  \\ \\ 
 \begin{pmatrix} p & q \\ r & t \end{pmatrix} \frac{a}{c}=&  \frac{pa+qc}{ra+tc},  \hspace{2cm } \begin{pmatrix} p & q \\ r & t \end{pmatrix} \infty = \frac{p}{r}
\end{split}
\end{equation*}

with $a,c \in \mathbb{Z}$ such that $\mathrm{gcd}(a,c)=1$.\\ The set of cusps is given by $C_{\Gamma_{1}(N)}=\mathbb{P}^{1}(\mathbb{Q})/ \Gamma_{1}(N)$ and is finite. By adding these points to $Y_{1}(N)$ one obtains

\begin{equation*}
Y_{1}(N)\longrightarrow X_{1}(N)= \overline{\mathbb{H}}/\Gamma_{1}(N).
\end{equation*}

The space $X_{1}(N)$ is called modular curve for $\Gamma_{1}(N)$ and can be shown to have the structure of a compact Riemann surface. Since the two dimensional lattice becomes degenerate when $\tau$ approaches the rational projective line, the cusps cannot be strictly considered as members of this class of theories, but rather as critical limits\footnote{It will be clear later that the cusps can be interpreted as RG flow fixed points.}.\\ So, we learn that the whole space of models can be written as union of connected components labelled by the integer $N$:

\begin{equation*}
\mathcal{A}= \bigcup _{N \geq 2} X_{1}(N),
\end{equation*}

where $X_{1}(N)$ parametrizes the subclass of theories of level $N$. It follows from the derivation that each point on $\mathcal{A}$ identifies a model up to isomorphisms. \\ The next step is to classify the vacua for this family of theories. Let us consider the modular curve of level $N$. The derivative of the superpotential is an elliptic function on the elliptic curve $\mathbb{C}/ \Lambda_{\tau}$ which has a simple pole at each point $kQ + \Lambda_{\tau},\  k=0,...,N-1$ and a simple zero at each point $P+kQ+ \Lambda_{\tau}, \ k=0,...,N-1$ for some $P \in C/ \Lambda_{\tau}$ such that $\partial_{z} W(P)=0$. Given that the principal divisor is vanishing by Abel's theorem, we get

\begin{equation*}
\mathrm{div}(\partial_{z} W(z))= \sum_{0\leq k < N} ([P]+k[Q])-\sum_{0\leq k < N} (k[Q])= N[P]=0.
\end{equation*}

So we deduce that also $P$ must be a torsion point and, once known, we can construct the whole set of vacua with the action of the symmetry generators. In order to find $P$, one can define the Weil pairing of order $N$ between torsion points:

\begin{equation*}
\begin{split}
& e_{N}: E[N] \ \mathrm{x} \ E[N]\longrightarrow \mu_{N}, \ \ \ \ \ \ \ \ \ \ \mu_{N}=\lbrace z \in \mathbb{C}\mid z^{N}=1\rbrace, \\ \\ \ \ \ \ & \hspace{3cm}e_{N}(P,Q)= \frac{F_{P}(z+Q)}{F_{P}(z)},
\end{split}
\end{equation*}

where $F_{P}(z)$ is any elliptic function with simple poles at $kQ+\Lambda_{\tau}$ and simple zeroes at $P+kQ+\Lambda_{\tau}$. Since $F_{P}(z)$ is unique up to multiplicative constants, there is no ambiguity in the definition. The Weil pairing is a sort of inner product on $E[N]$ and can be shown to be alternating, bilinear and non degenerate \cite{rif2,rif3}. These properties imply that, once the torsion point of the poles $Q$ is picked, a torsion point of the vacua $P$ is coupled by the condition $e_{N}(P,Q)= e^{\frac{2\pi i}{N}}$. Moreover, another point $P^{\prime}$ which has the same pairing with $Q$ must be of the form $P^{\prime}=P+kQ$ for some integer $k$. We can also give an explicit expression for $e_{N}$ \cite{rif2}. Chosen $\left( 2\pi/N + \Lambda_{\tau}, 2\pi \tau /N+ \Lambda_{\tau}\right) $ as basis of generators for the torsion group, the formula for the Weil pairing of $P,Q \in E[N]$ is 

\begin{equation*}
e_{N}(P,Q)= e^{2\pi i \ \mathrm{ det} \gamma /N},
\end{equation*}

where 

\begin{equation*}
\begin{bmatrix}
P \\ Q
\end{bmatrix}= \gamma \begin{bmatrix}
2\pi/N + \Lambda_{\tau} \\ 2\pi \tau/N + \Lambda_{\tau}
\end{bmatrix}, \ \ \ \ \ \mathrm{for} \ \gamma \in M_{2}(\mathbb{Z}).
\end{equation*}

This definition is actually independent of how the basis of the torsion group is chosen. One can see that, if $P$ and $Q$ generate $E[N]$, the matrix $\gamma$ is invertible and $e_{N}(P,Q)$ is a primitive complex $N$-th root of unity. Moreover, such expression is preserved under isogenies.\\
Now we can classify models and vacua with triplets  $(E_{\Lambda_{\tau}}, P, Q)$ which denote elliptic curves with associated two torsion data. If we set $Q=2\pi/N +\Lambda_{\tau}$ with a modular transformation, a torsion point satisfying the Weil condition is $P=-2\pi \tau/N+\Lambda_{\tau}$. It is straightforward to see that the congruence subgroup of $SL(2,\mathbb{Z})$ which preserves both the torsion points is:

\begin{equation*}
\Gamma(N)= \left\lbrace \gamma \in SL(2,\mathbb{Z}) : \gamma= \begin{pmatrix}
a & b \\ c & d \end{pmatrix} = \begin{pmatrix} 1 & 0 \\ 0 & 1 \end{pmatrix} \mathrm{mod} \ N\right\rbrace,
\end{equation*}

also called principal congruence subgroup. Similarly to the case of $S_{1}(N)$, one shows that the moduli space for the enhanced elliptic curves of principal type is 

\begin{equation*}
 S(N)= \lbrace \left( E_{\Lambda_{\tau}}, -2\pi\tau/N+\Lambda_{\tau}, 2\pi/N +\Lambda_{\tau}\right) , \tau \in \mathbb{H}  \rbrace / \sim 
\end{equation*}

where $\tau \sim \tau^{\prime}$ if and only $\Gamma(N)\tau=\Gamma(N)\tau^{\prime}$. This set of equivalence classes is therefore isomorphic to the complex manifold

\begin{equation*}
Y(N)= \mathbb{H}/ \Gamma(N).
\end{equation*}

As for $Y_{1}(N)$, we can compactify such space by adding the set of cusps $C_{\Gamma(N)}= \mathbb{P}^{1}(\mathbb{Q})/\Gamma(N)$ and obtain the compact Riemann surface

\begin{equation*}
Y(N)\longrightarrow X(N)= \overline{\mathbb{H}}/\Gamma(N),
\end{equation*}

which is known as the N-modular curve of principal type.\\ This curve represents the spectral cover of the space of models of level $N$ \cite{rif12}. Given that $\Gamma(N)$ is a normal subgroup of $\Gamma_{1}(N)$ with coset group $\Gamma_{1}(N)/\Gamma(N)\simeq \mathbb{Z}_{N}$, the degree of the cover is $[\Gamma_{1}(N):\Gamma(N)]=N$, which is exactly the number of vacua up to periodic identification.\\ A series of results which allows to count the cusps and describe the set $C_{\Gamma(N)}$ can be found in \cite{rif2}. In particular, denoting with $\begin{bmatrix} a \\ c \end{bmatrix}, \begin{bmatrix} a^{\prime} \\ c^{\prime} \end{bmatrix}$ two vectors of $\mathbb{Z}^{2}$, one can show that

\begin{equation}\label{width1}    
\begin{bmatrix} a^{\prime} \\ c^{\prime} \end{bmatrix}= \gamma \begin{bmatrix} a \\ c \end{bmatrix}, \ \mathrm{for} \ \gamma \in \Gamma(N) \Longleftrightarrow \begin{bmatrix} a^{\prime} \\ c^{\prime} \end{bmatrix} =  
\begin{bmatrix} a \\ c \end{bmatrix} \ \mathrm{mod} \ N.
\end{equation}

From this result, letting $s=a/c$ and $s^{\prime}= a^{\prime}/c^{\prime}$ two elements of $\mathbb{Q}\cup\left\lbrace \infty\right\rbrace $ such that $\mathrm{gcd}(a,c)= \mathrm{gcd}(a^{\prime},c^{\prime})=1$, it follows the equivalence relation

\begin{equation}\label{width2}   
\Gamma(N)s^{\prime}= \Gamma(N)s \Longleftrightarrow \begin{bmatrix} a^{\prime} \\ c^{\prime} \end{bmatrix} = \pm 
\begin{bmatrix} a \\ c \end{bmatrix} \ \mathrm{mod} \ N,
\end{equation}

where the sign `$-$'  keeps into account the projective action of the modular group on $\mathbb{P}^{1}(\mathbb{Q})$. Once collected the rationals in $\Gamma(N)$-classes according to this theorem, one can go futher and find also the cusps of $\Gamma_{1}(N)$. It is enough for this purpose to identify the cusps of $\Gamma(N)$ which belong to the same orbit of $T:\tau\rightarrow \tau +1$, i.e. the generator of $\Gamma_{1}(N)/\Gamma(N)$. Indeed, these points represent vacua of the same theory and are projected on the same cusp of $X_{1}(N)$.\\ From a geometrical point of view the cusps are the points for which the covering map $X(N)\rightarrow X_{1}(N)$ degenerates and the dimension of the orbit of $T$ is generically less than $N$. In physical terms, these represent the fixed points of the RG flow in the space of couplings. Being more precise, the cusps of $\Gamma(N)$ which are stabilized by $T$ are expected to be UV fixed points of the RG flow. Indeed, in this limits the zeroes of $\partial_{z}W(z)$ tend to a unique vacuum of order $N$ and the BPS states of the theory become consequently massless, implying that we are approaching a conformal field theory \cite{rif12}. On the other hand, in the IR limits the vacua become infinitely separated from each other and decouple at the leading order. If they are not simple zeroes of $\partial_{z}W(z)$, i.e. the corresponding orbit of $T$ has dimension between $1$ and $N$, these cusps represent again conformal fixed points on the spectral cover. Instead, if the vacuum is a simple zero and the orbit of the cusp under $T$ has dimension exactly $N$, we are dealing with a free massive field theory. We will provide a more precise description of the physics around the cusps in section $5$.

\subsubsection{The Role of Number Theory}

Another congruence subgroup of $SL(2,\mathbb{Z})$ which plays an important role in this classification is 

\begin{equation*}
\Gamma_{0}(N)= \left\lbrace \gamma \in SL(2,\mathbb{Z}): \gamma= \begin{pmatrix}
a & b \\ c & d \end{pmatrix} = \begin{pmatrix} * & * \\ 0 & * \end{pmatrix} \mathrm{mod} \ N \right\rbrace.
\end{equation*}

The three congruence subgroups that we have defined satisfy the chain of inclusions $\Gamma(N)\subset \Gamma_{1}(N) \subset \Gamma_{0}(N) \subset SL(2,\mathbb{Z})$. Moreover, $\Gamma_{1}(N)$ is a normal subgroup of $\Gamma_{0}(N)$. The action of $\Gamma_{0}(N)$ on the space of models is more clear when we choose $Q= \frac{2\pi}{N}$. It is evident that $\gamma \in \Gamma_{0}(N)$ does not leave the model invariant and maps the torsion point into an inequivalent one $\gamma^{*}Q= 2\pi \frac{(c\tau+d)}{N} \sim \frac{2\pi d}{N}$, for an integer $d$ coprime with $N$. The effect on the model corresponds to modify the character by the formula $\sigma_{a}(e(Q))= e(Q)^{a}=e^{\frac{2\pi i a}{N}}$, where $a$ is the inverse of $d$ in $\mathbb{Z}_{N}$. Therefore, it is manifest that $\Gamma_{0}(N)$ reproduces the action of the Galois group of the cyclotomic extension $[\mathbb{Q}(\zeta_{N}):\mathbb{Q}]$. This number field is obtained by adjoining a primitive $N$-th root of unity $\zeta_{N}$ to the rational numbers. Such remarkable connection reveals the algebraic nature of the modular curves. The above formula suggests to define a character $e_{l}(Q)=e^{\frac{2\pi i l}{N}}$ depending on an integer $l$ coprime with $N$, which we call co-level of the modular curve. It is clear that, for a fixed lattice $\Lambda_{\tau}$, a different choice of the co-level corresponds to pick a different point on the space of models. With this definition, a transformation of $\Gamma_{0}(N)$ can be seen as a permutation of the co-levels.\\ We can be more precise about these statements by keeping into account that $\Gamma_{0}(N)$ contains the matrix $-I$. We know that a point on $X_{1}(N)$ parametrizes an elliptic curve and a specific embedding of $\mathbb{Z}_{N}$. However, a cyclic group has always two generators which are one the inverse of the other. It is clear that $-I$ acts on the model as a parity transformation, since it does not change the point on $X_{1}(N)$ but inverts the sign of the torsion point. This means that the co-levels $l$ and $N-l$ are actually two descriptions of the same model. So, except for the trivial case of $N=2$, since $-I$ is in $ \Gamma_{0}(N)$ but not in $\Gamma_{1}(N)$ the degree of the cover $\mathbb{H}/ \Gamma_{0}(N)\rightarrow \mathbb{H}/ \Gamma_{1}(N)$ is $\left[ \Gamma_{0}(N):\Gamma_{1}(N) \right]/2  = \phi(N)/2$,  where $\phi(N)$ is the Euler totient function which count the elements of $\left\lbrace 0,...,N-1\right\rbrace$ coprime with $N$. As one can expect, this is also the degree of the number field defining this class of theories, which is actually the real cyclotomic extension $\mathbb{Q}(\zeta_{N}+\zeta_{-N})$. The cusps of $\Gamma_{1}(N)$ fall in equivalence classes of the Galois group described by the set $C_{\Gamma_{0}(N)}= \mathbb{P}^{1}(\mathbb{Q})/\Gamma_{0}(N)$. We will discuss the orbits of $\Gamma_{0}(N)$ in the set of critical theories in section $5$.

\subsection{Geometry of the Models}\label{sec3}

\subsubsection{The Model on the Target Manifold}

Now we want to translate our abstract classification into an explicit description of these models. A fundamental property of the elliptic functions is that they are uniquely specified (up to multiplicative constants) by the positions and orders of their zeroes and poles. Let us focus on the modular curve of level $N$ and co-level $l$ and choose $(P,Q)=(-\frac{2\pi l \tau}{N}+\Lambda_{\tau},\frac{2\pi }{N}+\Lambda_{\tau})$ as torsion points. For convenience we invert the lattices of poles and vacua with respect to the previous derivation. This can be done with the translation $z \rightarrow z- \frac{2\pi l \tau}{N}$. The derivative of the superpotential for this class of theories is

\begin{equation}
\partial_{z} W^{(N,l)}(z;\tau)= \sum_{k=0}^{N-1} e^{\frac{2\pi i l k}{N}} \left[ \zeta \left( z-\frac{2\pi}{N}(l\tau + k);\tau\right) +\frac{2 \eta_{1}k}{N}\right], 
\end{equation}

where  $  z \in \mathcal{S}= \mathbb{C} \setminus \left\lbrace \frac{2\pi}{N}(l\tau+k)+\Lambda_{\tau}, \ k=0,....,N-1; \ \Lambda_{\tau}= 2\pi\mathbb{Z} \oplus 2\pi \tau \mathbb{Z}\right\rbrace $. \\ The Weierstrass zeta function $\zeta(z;\tau)$ is defined with the conventions $\eta_{1}=\zeta(\pi;\tau)$, $ \eta_{2}=\zeta(\pi \tau;\tau)$ \cite{rif47}. This function is meromorphic on $\mathcal{S}$ with simple poles and simple zeroes respectively in $\frac{2\pi(l\tau+k)}{N}+\Lambda_{\tau}$ and $\frac{2\pi k}{N}+\Lambda_{\tau}$, $k=0,...,N-1$. By definition of elliptic function, the sum of the residue inside the fundamental cell is vanishing. From a physical point of view this means that the flux of the magnetic field is cancelled by that of the quasi-holes charges.\\ The algebra of the abelian symmetry group of this model is generated by three operators $\sigma,A,B$, defined by the actions

\begin{equation}\label{transform}
\begin{split}
& \sigma: z\longrightarrow z+ 2\pi/N ;\hspace{2cm} \partial_{z} W^{(N,l)}\left( z+2\pi/N;\tau\right) = e^{\frac{2\pi i l}{N}}\partial _{z}W^{(N,l)}\left( z;\tau\right),  \\ \\ 
& A: z\longrightarrow z+2\pi; \hspace{2.4cm} \partial_{z} W^{(N,l)}\left( z+2\pi;\tau\right) = \partial_{z} W^{(N,l)}\left( z;\tau\right),  \\ \\
& B: z\longrightarrow z +2\pi \tau ; \hspace{2.1cm} \partial_{z} W^{(N,l)}\left( z+2\pi \tau;\tau\right) = \partial_{z} W^{(N,l)}\left( z;\tau\right) ,
\end{split}
\end{equation}

with the additional relation $ \sigma^{N}= A$. These transformations follows from the quasi-periodicity properties of the Weierstrass function. The double periodicity of $\partial _{z}W^{(N,l)}\left( z;\tau\right) $ allows to identity points wich differ by an element of $\Lambda_{\tau}$, so that the model is naturally projected on the torus $\mathcal{K}=\mathcal{S}/\Lambda_{\tau} $. With this identification we can work with just $N$ critical points, denoted by the equivalence classes $\left[ \frac{2\pi k}{N}\right], k=0,...,N-1 $. From the property $\zeta(-z;\tau)=-\zeta(z;\tau)$, we find that the superpotentials of co-levels $l$ and $N-l$ are actually related by a parity transformation

\begin{equation}\label{colevel}
\partial_{z} W^{(N,N-l)}(-z;\tau)=-\partial_{z} W^{(N,l)}(z;\tau)
\end{equation}

and therefore, as pointed out in the previous section, they describe the same model with inverse torsion points generating the $\mathbb{Z}_{N}$ symmetry.\\ We can also write for this class of theories a symbolic `primitive'

\begin{equation}\label{symb}
W^{(N,l)}(z;\tau)= ``\sum_{k=0}^{N-1} e^{\frac{2\pi i l k}{N}} \log  \Theta \begin{bmatrix} \frac{1}{2}-\frac{l}{N} \\ \frac{1}{2}-\frac{k}{N}\end{bmatrix}\left( \frac{z}{2\pi};\tau \right) ",
\end{equation}

where $\Theta\begin{bmatrix} \alpha \\ \beta \end{bmatrix}(z;\tau)$ is the theta function of character $(\alpha,\beta) \in \mathbb{R}$ and quasi-periods $1,\tau$. This function is multi-valued and cannot be really considered as superpotential of the model. Indeed, the target manifold is not simply connected and we cannot find a true primitive on this space. 

\subsubsection{Abelian Universal Cover}

In order to define a superpotential, as well as the Hilbert space of the theory, we need to pull-back the model on the abelian universal cover of $\mathcal{S}$. We remind the definition of universal cover as space of curves

\begin{equation}\label{def}
\begin{split}
& \hspace{3cm}\mathcal{H}= \left\lbrace p:[0,1]\longrightarrow  \mathcal{S}, p(0)=p^{*} \in \mathcal{S} \right\rbrace / \sim, \\ \\ 
& \sim: \hspace{2cm} 
p\sim q=
\begin{cases}
p(1)=q(1), \\ 
p \cdot q^{-1} =0 \ \mathrm{in \ the \ homology \ group} \  H_{1}(\mathcal{S},\mathbb{Z}).
\end{cases}
\end{split}
\end{equation}

By the pull-back of $ \partial_{z} W^{(N,l)}(z;\tau)$ on this space we can define the superpotential of the theory:

\begin{equation}
W^{(N,l)}(p;\tau)= \int_{p} \partial_{z} W^{(N,l)}(z;\tau) dz .
\end{equation}

\begin{figure}[!h]
\centering
\begin{tikzpicture}
\coordinate [label=below left:$\gamma_{a}$] (A)  at (7,0);
\coordinate [label=above left:$\gamma_{b}$] (B)  at (1,1.5);
\coordinate [label=below left: $z$] (Z) at (0,0);
\coordinate[label= x] (x) at (1.5,1);
\coordinate[label= x ] (x) at (3,1);
\coordinate[label= x ] (x) at (6.8,1);
\coordinate[label= x ] (x) at (11,1);
\coordinate[label= x ] (x) at (13,1);
\coordinate[label=  . . . . . . . . . .] (x) at (4.87,1);
\coordinate[label=   . . . . . . . . . . . ] (x) at (8.88,1);
\coordinate [label= above: $\ell_{1} $]    (L)     at (2,2);
\coordinate [label= above: $\ell_{2} $]    (L)     at (4,2);
\coordinate [label= above: $\ell_{k} $]    (L)     at (7.5,1.7);
\coordinate [label= above: $\ell_{N-1} $]    (L)     at (11.5,1.47);

\draw  [->](0,0)--(13,0);
\draw  (13,0)--(15,3);
\draw [->](0,0)--(2,3);
\draw  (2,3)--(15,3);
\draw  ((0,0) to [out=35, in=180] (2,2);
\draw [->](0,0) to [out=0, in=0] (2,2);
\draw  ((0,0) to [out=2, in=180] (4,2);
\draw [->](0,0) to [out=0, in=0] (4,2);
\draw  ((0,0) to [out=0, in=180] (7.5,1.7);
\draw [->](0,0) to [out=0, in=0] (7.5,1.7);
\draw  ((0,0) to [out=0, in=180] (11.5,1.47);
\draw [->](0,0) to [out=0, in=0] (11.5,1.47);

\end{tikzpicture}
\caption{Homology generators}\label{homology}
\end{figure}
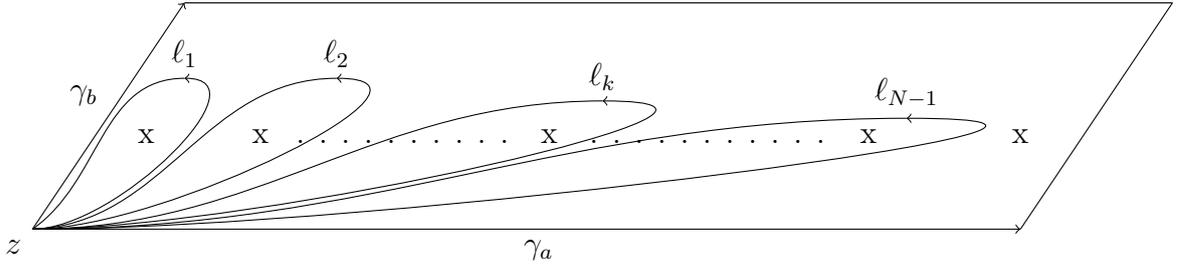

Contrary to \ref{symb}, this is a well defined function which assigns a single value to each equivalence class of paths in $\mathcal{H}$.\\
In the covering model we have many more vacua than before, but at the same time a larger symmetry group to classify them. An evident subgroup is the deck group of the cover, i.e. the homology. Let us fix a point $z \in \mathcal{S}$. We choose as local basis of $H_{1}(\mathcal{S};\mathbb{Z})$ the one given by the curves $\gamma_{a},\gamma_{b}$, obtained by applying respectively the lattice vectors $a=2\pi,b=2\pi \tau$ to the base point $z$, and the anticlockwise loops $\ell_{0},...,\ell_{N-2}$ encircling the first $N-1$ poles contained in the cell with sides $\gamma_{a},\gamma_{b}$ (Figure \ref{homology}). The action on the superpotential of the corresponding operators is

\begin{equation}\label{integrals}
\begin{split}
& A^{*}W^{(N,l)}(p;\tau)= \int_{\gamma_{a}\cdot p} \partial_{z} W^{(N,l)}(z;\tau)dz = W^{(N,l)}(p;\tau) + \int_{\gamma_{a}} \partial_{z} W^{(N,l)}(z;\tau)dz, \\ \\ 
& B^{*}W^{(N,l)}(p;\tau)= \int_{\gamma_{b}\cdot p} \partial_{z} W^{(N,l)}(z;\tau)dz = W^{(N,l)}(p;\tau) + \int_{\gamma_{b}} \partial_{z} W^{(N,l)}(z;\tau)dz, \\ \\ 
& L_{k}^{*}W^{(N,l)}(p;\tau)=  \int_{\ell _{k}\cdot p} \partial_{z} W^{(N,l)}(z;\tau)dz = W^{(N,l)}(p;\tau) + \oint_{\ell_{k}} \partial_{z} W^{(N,l)}(z;\tau)dz.
\end{split}
\end{equation}

Compatibly with the definition of symmetry, the superpotential shifts by a constant under the action of the homology generators. \\ \\ 

\begin{figure}[h!]
\centering
\begin{tikzpicture}
\coordinate [label=below left:$\gamma^{*}$] (A)  at (2.5,0);
\coordinate [label=above left:$\gamma_{\sigma}$] (B)  at (4.5,2);
\coordinate [label=below left: $p^{*}$] (P) at (0,0);
\coordinate [label=right:$\sigma (p)$] (A)  at (6,1.3);
\coordinate [label=left:$p$] (A)  at (0.7,1.3);
\coordinate [label=below right:$ \sigma(p^{*})$] (A)  at (5,0);
\coordinate [label=above left:$p(1)$] (A)  at (2,2);
\coordinate [label=above right:$\sigma(p_{1})$] (A)  at (7,2);

\draw  [->](0,0)--(5,0);
\draw  [->] ((0,0) to [out=60, in=200] (2,2);
\draw [->](2,2) to [out=-10, in=190] (7,2);
\draw  [->] ((5,0) to [out=60, in=200] (7,2);

\end{tikzpicture}
\caption{Definition of $\gamma_{\sigma}$}\label{sigma}
\end{figure}
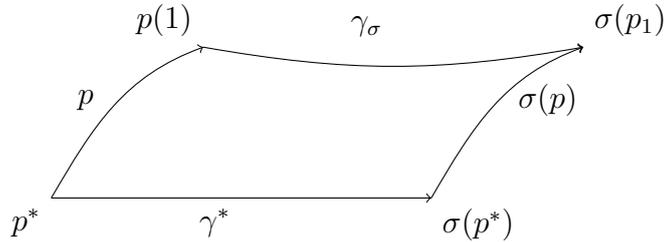

In order to define the action of $\sigma$ on $\mathcal{H}$ we need to specify a path $\gamma_{\sigma}$ connecting $z$ to $z+\frac{2\pi}{N}$. Let us fix the point $p^{*}$ in the definition \ref{def}. We choose a path $\gamma^{*}$ connecting $p^{*}$ to $p^{*}+\frac{2\pi}{N}$. If the point is not orizontally aligned with the poles we can take for instance a straight line. Now, given a curve $p$ in $\mathcal{H}$, we define $\gamma_{\sigma}$ such that $\gamma_{\sigma} \cdot p \ $= $ \sigma (p) \cdot \gamma^{*}$ in $H_{1}(\mathcal{S};\mathbb{Z})$, where $\sigma(p)$ denotes the curve $p$ shifted by $\frac{2\pi}{N}$ (Figure \ref{sigma}). Choosing the integration constant so that $W^{(N,l)}(\gamma^{*};\tau)=0$, we get the transformation of the superpotential

\begin{equation}\label{exact}
\begin{split}
\sigma^{*}W^{(N,l)}(p;\tau) = & \int_{\gamma_{\sigma}\cdot p} \partial_{z} W^{(N,l)}(z;\tau)dz=
\int_{\sigma (p)\cdot \gamma_{*}} \partial_{z} W^{(N,l)}(z;\tau)dz= \\ \\ & e^{\frac{2\pi i l}{N}} W^{(N,l)}(p;\tau) + W^{(N,l)}(\gamma^{*};\tau)= e^{\frac{2\pi i l}{N}} W^{(N,l)}(p;\tau) . 
\end{split}
\end{equation}

Inequivalent definitions of $\gamma_{\sigma}$ differ by compositions with the loops $\ell_{k} $ and generate a different constant $W^{(N,l)}(\gamma^{*};\tau)$ in the transformation. Indipendently from the choice, by composing $N$ times $\gamma_{\sigma}$ we obtain an curve in $H_{1}(\mathcal{S};\mathbb{Z})$ homologous to $\gamma_{a}$, consistently with the operatorial relation $\sigma^{N}=A$. \\ \\ 

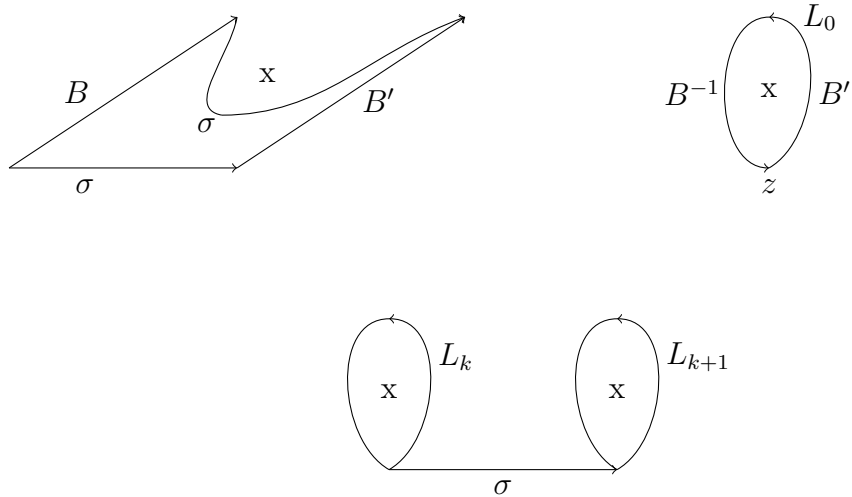
\begin{figure}[h!]
\centering
\begin{tikzpicture}
\coordinate [label=left:$B$] (A)  at (1.2,1);
\coordinate [label=right:$B^{\prime}$] (B)  at (4.5,0.9);
\coordinate [label=left:$B^{-1}$] (A)  at (9.5,1);
\coordinate [label=right:$B^{\prime}$] (B)  at (10.5,1);
\coordinate [label= x] (X) at (3.4,1);
\coordinate [label= x] (X) at (10,0.8);
\coordinate [label= x] (X) at (5,-3.2);
\coordinate [label= x] (X) at (8,-3.2);
\coordinate [label= below:$\sigma$] (s) at (2.6,0.8);
\coordinate [label= below:$z$] (X) at (10,0);
\coordinate [label= below:$\sigma$] (s) at (1,0);
\coordinate [label= below:$\sigma$] (s) at (6.5,-4);
\coordinate [label= right:$L_{k}$] (s) at (5.5,-2.5);
\coordinate [label= right:$L_{k+1}$] (s) at (8.5,-2.5);
\coordinate [label= right:$L_{0}$] (s) at (10.3,2);

\draw  [->](0,0)--(3,0);
\draw  [->](0,0)--(3,2);
\draw  [->](3,0)--(6,2);
\draw  [->](5,-4)--(8,-4);
\draw  ((3,2) to [out=-100, in=170] (2.8,0.7);
\draw  [->] ((2.8,0.7) to [out=00, in=200] (6,2);
\draw  [->] ((10,2) to [out=180, in=180] (10,0);
\draw  [->] ((10,0) to [out=30, in=00] (10,2);
\draw   ((5,-4) to [out=150, in=180] (5,-2);
\draw  [->] ((5,-4) to [out=30, in=00] (5,-2);
\draw   ((8,-4) to [out=150, in=180] (8,-2);
\draw  [->] ((8,-4) to [out=30, in=00] (8,-2);

\end{tikzpicture}
\caption{Algebra of curves}\label{algebr}
\end{figure}

The generators $\sigma, A,B,(L_{k})_{k=0,...,N-2}$ represent a complete basis for the symmetry group of the covering model. It is clear from Figure \ref{algebr} that the algebra is non-abelian on the space of curves. Indeed, the generator $\sigma$ do not commute with $B$ and $L_{k}$ according to the algebraic relations

\begin{equation}\label{identity}    
\sigma B= L_{0} B \sigma, \hspace{2cm} \sigma L_{k}= L_{k+1} \sigma, \ \ \ k=0,...,N-2,
\end{equation}

where we have $L_{N-1}= \prod\limits_{k=0}^{N-2}L_{k}^{-1}$. We see that the obstruction to abelianity on the universal cover is represented by the generators of loops. In the faithful representation on the plane these act trivially and $A,B,\sigma$ commute. This is not inconsistent, since the physics must be abelian on the target manifold but not necessarily on the covering model. The abelianity which we have asked in the classification can be recoverd at the quantum level by projecting the Hilbert space on the trivial representation of $L_{k}$. We will discuss this point in the next section. There are not problems instead in the commutation between $\sigma$ and $A$ and clearly the subgroup given by the homology is abelian. 

Now we can classify the vacua of the model defined on $\mathcal{H}$. The vacua on the target manifold are represented by the points $z_{k}= \frac{2\pi k}{N}, k=0,...,N-1,$ up to periodic identification of $\Lambda_{\tau}$.
For these points there is an ambiguity in the definition of the cycle $\gamma_{b}$, since there is a pole along the side of the fundamental cell. We choose conventionally the homology path which bypasses the pole on the left. Now that the homology based on the critical points is well defined, we fix $p^{*}$ away from the set of vacua and pick a curve $p_{0}$ connecting $p^{*}$ to $z_{0}$ as representative of $z_{0}$ in $\mathcal{H}$.  All the other representatives in the corresponding fiber can be obtained by composing $p_{0}$ with the cycles of $H_{1}(\mathcal{S},\mathbb{Z})$ based on $z_{0}$. Moreover, we can map fibers over different points to each other with the action of $\sigma$. Therefore, the vacua on $\mathcal{H}$ are labelled by the curves

\begin{equation}\label{vacua}
p_{k;m,n,n_{0},...,n_{N-2}}= \gamma_{a}^{m} \cdot \gamma_{b}^{n} \cdot \prod_{r=0}^{N-2} \ell_{r}^{n_{r}} \cdot \gamma_{\sigma}^{k} \cdot p_{0}
\end{equation}

where $k=0,...,N-1,$ and $m,n,n_{0},...,n_{N-2} \ \in \mathbb{Z}$.  By definition of abelian universal cover, each fiber is a copy of the homology group.\\ We can also provide the corresponding critical values by computing the integrals in  \ref{integrals}. Let us take again the curve $p_{0}$. By the relation $A=\sigma^{N}$ and the residue formula we have 

\begin{equation}
\begin{split}
& A^{*}W^{(N,l)}(p_{0};\tau)= W^{(N,l)}(p_{0};\tau),\\ \\ 
&  L_{k}^{*}W^{(N,l)}(p_{0};\tau)= W^{(N,l)}(p_{0};\tau) + 2\pi i \ e^{\frac{2\pi i l k}{N}}.
 \end{split}
\end{equation}

Moreover, from the first algebraic identity in \ref{identity}, we easily get

\begin{equation}
B^{*}W^{(N,l)}(p_{0};\tau)= W^{(N,l)}(p_{0};\tau) + \frac{2 \pi i}{e^{\frac{2\pi i l}{N}}-1}.
\end{equation}

Finally, by acting with the generators $\sigma, A,B,(L_{k})_{k=0,...,N-2}$ on $W^{(N,l)}(p_{0};\tau)$, we obtain the whole set of critical values

\begin{equation}\label{critical}
\begin{split}
W^{(N,l)} & (p_{k;m,n,n_{0},...,n_{N-2}};\tau)  = A^{* m} B^{*  n} \prod_{0\leq r \leq N-2} L_{r}^{*  n_{r}} \sigma^{*  k} \ W^{(N,l)}(p_{0};\tau)= \\ \\
  &  e^{ \frac{2 \pi i l k}{N}} W^{(N,l)}(p_{0};\tau)+ n \frac{2 \pi i}{e^{\frac{2\pi i l}{N}}-1} + \sum_{r=0}^{N-2} n_{r} \ 2\pi i \ e^{\frac{2\pi i l r}{N}} .
  \end{split}
\end{equation}

\subsection{$tt^{*}$ Equations}\label{trunc}

\subsubsection{General Case}

Our next aim is to derive the $tt^{*}$ equations for this class of models. We first introduce the point basis $\Phi_{k;m,n,n_{0},...,n_{N-2}}$ defined by

\begin{equation}\label{rep}
\Phi_{k;m,n,n_{0},...,n_{N-2}} ( p_{k^{\prime};m^{\prime},n^{\prime},n_{0}^{\prime},...,n_{N-2}^{\prime}};\tau) =\delta_{k,k^{\prime}}\delta_{m,m^{\prime}} \delta_{n,n^{\prime}} \prod_{r=0}^{N-2} \delta_{n_{r},n_{r}^{\prime}}.
\end{equation}

The isomorphism between $\mathcal{R}$ and $\mathbb{C}^{\# \mathrm{ classical \ vacua}}$ is realized by the identification

\begin{equation*}
\Phi_{k;m,n,n_{0},...,n_{N-2}}\longrightarrow \ket{p_{k;m,n,n_{0},...,n_{N-2}}}= A^{m} B^{ n} \prod_{r=0}^{N-2} L_{r}^{  n_{r}} \sigma^{ k}\ket{p_{0}},
\end{equation*}

where the operatorial action of the symmetry algebra is naturally transported on the Hilbert space. The vacuum space of the theory on $\mathcal{H}$ decomposes in a direct sum of irreducible representations of the homology group:

\begin{equation*}
\mathcal{V}_{\mathcal{H}} \simeq \mathrm{L}^{2}\left(  \mathrm{Hom} \left(   \mathbb{Z}^{N+1}, U(1) \right) \right) \otimes \mathbb{C}^{N},
\end{equation*}

provided that $N+1$ is the rank of $H_{1}(\mathcal{S},\mathbb{Z})$ and $N$ the number of vacua on $\mathcal{S}$. A point basis of `theta-vacua' for $\mathcal{V}_{\mathcal{H}}$ is 

\begin{equation}\label{pointb}
\begin{split}
 \ket{k;\alpha, \beta, \lambda_{0},...,\lambda_{N-2}}=& \ e^{\frac{-i \alpha k}{N}} \sum_{m,n,n_{r} \in \mathbb{Z}} e^{-i( m\alpha + n\beta+\sum_{r=0}^{N-2}n_{r}\lambda_{r} )} \\ \\ & A^{m} B^{n} \prod_{r=0}^{N-2} L_{r}^{ n_{r}} \sigma^{ k}\ket{p_{0}},
\end{split}
\end{equation}

where the angles $\alpha, \beta, \lambda_{0},...,\lambda_{N-2} \in [0,2\pi]$ label representations of $H_{1}(\mathcal{S},\mathbb{Z})$. Since the homology is abelian, the angles are defined simultaneously in this common basis of eigenstates:

\begin{equation*}
\begin{split}
& A\ket{k;\alpha, \beta, \lambda_{0},...,\lambda_{N-2}}=e^{i\alpha} \ket{k;\alpha, \beta, \lambda_{0},...,\lambda_{N-2}}, \\ \\ & B\ket{k;\alpha, \beta, \lambda_{0},...,\lambda_{N-2}}=e^{i\beta} \ket{k;\alpha, \beta, \lambda_{0},...,\lambda_{N-2}}, 
\\ \\ & L_{r}\ket{k;\alpha, \beta, \lambda_{0},...,\lambda_{N-2}}=e^{i\lambda_{r}} \ket{k;\alpha, \beta, \lambda_{0},...,\lambda_{N-2}}.
\end{split}
\end{equation*}

Using the symmetries $A,B,L_{r}$, one finds that in the point basis the ground state metric diagonalizes with respect to the angles:

\begin{equation*}
\begin{split}
& \braket{\overline{j;\alpha^{\prime}, \beta^{\prime}, \lambda_{0}^{\prime},...,\lambda_{N-2}^{\prime}}}{k;\alpha, \beta, \lambda_{0},...,\lambda_{N-2}}= \\ \\ & \delta(\alpha-\alpha^{\prime})\delta(\beta-\beta^{\prime}) \prod_{r=0}^{N-2} \delta( \lambda_{r}-\lambda_{r}^{\prime}) \  g_{k,\bar{j}}( \alpha, \beta, \lambda_{0},...,\lambda_{N-2}), 
\end{split} 
\end{equation*}

where the matrix coefficients $g_{k,\bar{j}}( \alpha, \beta, \lambda_{0},...,\lambda_{N-2})$ have the Fourier expansion:

\begin{equation*}
\begin{split}
& g_{k,\bar{j}}( \alpha, \beta, \lambda_{0},...,\lambda_{N-2}) =  e^{\frac{i}{N}\alpha (j-k)}\sum_{r,s,t_{q} \in \mathbb{Z}} e^{i( r\alpha + s\beta+\sum_{q=0}^{N-2}t_{q}\lambda_{q} )} g_{k,\bar{j}}(r,s,t_{0},...,t_{N-2}), \\ \\ 
& g_{k,\bar{j}}(r,s,t_{0},...,t_{N-2})= \braket{\overline{p_{j;m,n,n_{0},...,n_{N-2}}}}{p_{k;m^{\prime},n^{\prime},n_{0}^{\prime},...,n_{N-2}^{\prime}}} \vert_{m-m^{\prime}=r, n-n^{\prime}=s, n_{q}-n_{q}^{\prime}=t_{q}} .
\end{split}
\end{equation*}

The metric cannot be diagonalized also as $N \times N$ matrix, since $\sigma$ do not commute with $B$ and $L_{r}$. The pull-back of the Landau-Ginzburg model on the universal cover is a necessary operation to write the $tt^{*}$ equations, but the price to pay is to have more vacua and a larger space of solutions. We are not interested in the most general solution to the $tt^{*}$ equations, but in the subclass reproducing the abelian physics of the punctured plane. Moreover, the importance of an abelian symmetry group with a transitive action on the vacua is that we can completely diagonalize the ground state metric. Without this property, even writing the $tt^{*}$ equations becomes an hopeless problem. Therefore, we have to truncate the $tt^{*}$ equations in such a way that the solutions are compatible with the abelian representation of the symmetry group. We cannot just set the loop angles to be $0$, since these are differential variables which enter in the equations. So, we have to check that the ansatz of a solution with constant vanishing $\lambda_{r}$ is consistent with all the equations. From the equality 

\begin{equation*}
\frac{2 \pi i}{e^{\frac{2\pi i l}{N}}-1}= \frac{2\pi i}{N} \sum_{k=0}^{N-1} k e^{\frac{2\pi i l k}{N}}= \frac{2\pi i}{N} \sum_{k=0}^{N-2} (k-N+1) e^{\frac{2\pi i l k}{N}},
\end{equation*}

we learn that the combination $\tilde{B}= B^{N}\prod\limits_{k=0}^{N-2} L_{k}^{N-k-1}$ leaves invariant the superpotential. Moreover, one can see that it commutes with $\sigma$
\begin{equation*}
\begin{split}
&\sigma \tilde{B}= \sigma B^{N}\prod_{k=0}^{N-2} L_{k}^{N-k-1}= B^{N}L_{0}^{N}\prod_{k=0}^{N-3} L_{k+1}^{N-k-1} \prod_{k=0}^{N-2} L_{k}^{-1} \sigma = \\ & B L_{0}^{N-1} \prod_{k=0}^{N-3} L_{k+1}^{N-(k+1)-1}\sigma = B^{N}\prod\limits_{k=0}^{N-2} L_{k}^{N-k-1}\sigma = \tilde{B}\sigma
\end{split}
\end{equation*}

where we have used the commutation relations in \ref{identity}. Thus, if we choose $A,\tilde{B},L_{r}$ as generators of the homology, we can find a basis in which the eigenvalues of $\sigma$ are simultaneously defined with the two angles $\alpha,\tilde{\beta}= N\beta + \sum_{k=0}^{N-2}(N-k-1)\lambda_{k}$ corresponding to $A,\tilde{B}$.\\ The couplings we can vary in these models are the lattice parameter $\tau$ and the usual overall scale multiplying the superpotential. To avoid clash of notations, we denote with $\mu$ this coupling. The corresponding chiral ring coefficients in the point basis of the covering model read

\begin{equation}
\begin{split}
\left( C_{\tau}\right)_{ k;m,n,n_{0},...,n_{N-2}}^{k^{\prime};m^{\prime},n^{\prime},n_{0}^{\prime},...,n_{N-2}^{\prime}}= & \
 \mu \partial_{\tau} W^{(N,l)}\Phi_{k^{\prime};m^{\prime},n^{\prime},n_{0}^{\prime},...,n_{N-2}^{\prime}} (p_{k;m,n,n_{0},...,n_{N-2}};\tau) \\  = &  \ \delta_{k,k^{\prime}}\delta_{m,m^{\prime}} \delta_{n,n^{\prime}} \prod_{r=0}^{N-2} \delta_{n_{r},n_{r}^{\prime}} \ \mu \partial_{\tau} W^{(N,l)}(p_{0};\tau)e^{\frac{2\pi i l k}{N}}, \\ \\ 
\left( C_{\mu}\right)_{ k;m,n,n_{0},...,n_{N-2}}^{k^{\prime};m^{\prime},n^{\prime},n_{0}^{\prime},...,n_{N-2}^{\prime}}= & 
 \ W^{(N,l)}\Phi_{k^{\prime};m^{\prime},n^{\prime},n_{0}^{\prime},...,n_{N-2}^{\prime}}   (p_{k;m,n,n_{0},...,n_{N-2}};\tau) \\ = & \ \delta_{k,k^{\prime}}\delta_{m,m^{\prime}} \delta_{n,n^{\prime}} \prod_{r=0}^{N-2} \delta_{n_{r},n_{r}^{\prime}} \\ &  \mathrm{x} \ \left( W^{(N,l)}(p_{0};\tau)e^{\frac{2\pi i l k}{N}} +\sum_{r=0}^{N-2} n_{r} \ 2\pi i \ e^{\frac{2\pi i l r}{N}} \right), 
\end{split}
\end{equation}

where we have replaced $B$ with $\tilde{B}$ in the computation of the critical values in \ref{critical}.\\ The action of $C_{\tau}$ and $C_{\mu}$ is naturally projected on the theta-sectors of $\mathcal{V}_{\mathcal{H}}$:

\begin{equation}
\begin{split}
 C_{\tau}^{\mathrm{point \ basis}}(\alpha,\tilde{\beta} , \lambda_{0},...,\lambda_{N-2})_{j,k}=& \ \mu \partial_{\tau} W^{(N,l)}(p_{0};\tau) e^{\frac{2\pi i l k}{N}} \delta_{j,k}, \\ \\ 
 C_{\mu}^{\mathrm{point \ basis}}(\alpha, \tilde{\beta}, \lambda_{0},...,\lambda_{N-2})_{j,k}=& \left( W^{(N,l)}(p_{0};\tau)e^{\frac{2\pi i l k}{N}}- 2\pi \sum_{r=0}^{N-2}e^{\frac{2\pi i l r}{N}} \frac{\partial}{\partial \lambda_{r}} \right) \delta_{j,k}.
\end{split}
\end{equation}

The ground state metric satisfy the following set of equations 

\begin{equation}\label{set}
\begin{split}
& \partial_{\bar{\tau}}( g \partial_{\tau} g^{-1})= \left[C_{\tau}  ,g C_{\tau}^{\dagger}g^{-1}\right], \\ \\ 
& \partial_{\bar{\mu}}( g \partial_{\mu} g^{-1})= \left[C_{\mu}  ,g C_{\mu}^{\dagger}g^{-1}\right], \\ \\ 
& \partial_{\bar{\mu}}( g \partial_{\tau} g^{-1})= \left[C_{\tau}  ,g C_{\mu}^{\dagger}g^{-1}\right],
\end{split}
\end{equation}

as well as the complex conjugates. If we demand that $\lambda_{0}=...=\lambda_{N-2}=0$ and $\frac{\partial g}{\partial \lambda_{r}}=0$ for each $r$, the derivatives $\frac{\partial}{\partial \lambda_{r}}$ do not contribute to the $tt^{*}$ equations and the F-term deformations can be truncated at the non trivial part 

\begin{equation}
\begin{split}
 C_{\tau}^{\mathrm{point \ basis}}(\alpha, \tilde{\beta})_{j,k}=& \mu \partial_{\tau} W^{(N,l)}(p_{0};\tau) e^{\frac{2\pi i l k}{N}} \delta_{j,k}, \\ \\ 
 C_{\mu}^{\mathrm{point \ basis}}(\alpha, \tilde{\beta})_{j,k}=&  W^{(N,l)}(p_{0};\tau)e^{\frac{2\pi i l k}{N}} \delta_{j,k},
\end{split}
\end{equation}

with $\tilde{\beta}=N\beta$. It is clear that with this ansatz the equations given above can be written universally as a unique equation with respect to the critical value $w=W^{(N,l)}(p_{0};\tau)$:

\begin{equation}\label{canonical}   
\partial_{\bar{w}}( g \partial_{w} g^{-1})= \left[C_{w}  , g C_{w}^{\dagger}g^{-1}\right],
\end{equation}

where 

\begin{equation}
C_{w}^{\mathrm{point \ basis}}(\alpha, \tilde{\beta})_{j,k}= e^{\frac{2\pi i l k}{N}} \delta_{j,k}.
\end{equation}

The equations in the parameters $\tau$ and $\mu$ can be obtained from this one by specifying the variation of $w$ and $\bar{w}$. If the equation in the canonical variable is solved with the boundary conditions given by the cusps, the solution automatically satisfies all the equations in \ref{set}. Hence, it is convenient to pull-back the equation on the spectral cover of the model where one can use $w= W^{(N,l)}(p_{0};\tau)$ as unique complex coordinate to parametrize models and vacua. By consistency, the cusps should reproduce the correct regularity conditions for the truncated equations. We are going to discuss the boundary conditions for the $tt^{*}$ equations in section \ref{final} and we will see that this is the case.\\ Now that we have recovered the abelian representation of the symmetry group, we can construct a common basis of eigenstates for $\sigma,A,B$ in which the ground state metric diagonalizes completely. Given the action of $\sigma$ on the point basis 

\begin{equation*}
\sigma\ket{k;\alpha, \beta}=e^{\frac{i \alpha }{N}} \ket{k+1;\alpha, \beta},
\end{equation*}

we can define a set of $\sigma$-eigenstates as 

\begin{equation}\label{eigen}
\begin{split}
& \ket{j;\alpha, \beta}= \sum_{k=0}^{N-1}  e^{- \frac{2 \pi i l k j}{N}}  \ket{k;\alpha, \beta}, 
\\ \\ & \sigma \ket{j;\alpha, \beta}= e^{\frac{ i}{N} (\alpha + 2\pi l j)} \ket{j;\alpha, \beta}.
\end{split}
\end{equation}

With this normalization we have the periodicity 

\begin{equation}
\ket{k+N;\alpha, \beta}=\ket{k;\alpha, \beta}
\end{equation}

and we note that the given definitions are consistent with the relation $\sigma^{N}=A$. \\ In the $\sigma$-basis the metric becomes

\begin{equation}
 g_{k,\bar{j}}( \alpha, \beta) = \delta_{k,j} \ e^{\varphi_{k}( \alpha, \beta)},
\end{equation}

where the $\varphi_{k}( \alpha, \beta), k=0,..,N-1,$ are real functions of the angles. It is straightforward to derive the $tt^{*}$ equations satisfied by these functions. Using

\begin{equation*}
C_{w}^{\sigma-\mathrm{basis}}(\alpha, \beta)= 
\begin{pmatrix}  
0 & 1 & 0 & . . . & 0 & 0 \\ 
0 & 0 & 1 &. . . & 0 & 0 \\
. & . &  . &     & . & . \\ 
. & . &  . &     & . & . \\ 
 0 & 0 & 0 & . . . & 0 & 1 \\ 
 1 & 0 & 0 & . . . & 0 & 0 \\
\end{pmatrix},
\end{equation*}

one finds the well known $\hat{A}_{N-1}$ Toda equations:

\begin{equation}\label{toda} 
\begin{split}
& \partial_{\bar{w}}\partial_{w} \varphi_{0} + e^{\varphi_{1}-\varphi_{0}}-e^{\varphi_{0}-\varphi_{N-1}}=0 ,\\ \\
& \partial_{\bar{w}}\partial_{w} \varphi_{i} + e^{\varphi_{i+1}-\varphi_{i}}-e^{\varphi_{i}-\varphi_{i-1}}=0, \hspace{2cm} i=1,...,N-2 \\ \\
& \partial_{\bar{w}}\partial_{w} \varphi_{N-1} + e^{\varphi_{0}-\varphi_{N-1}}-e^{\varphi_{N-1}-\varphi_{N-2}}=0,
\end{split}
\end{equation}

where the dependence on the angles and $w$ is understood. These equations appear tipically in the models with a $\mathbb{Z}_{N}$ symmetry group acting transitively on the vacua. We are going to discuss the solution to these equations in section \ref{final}. In particular, we will see that the RG fixed points reproduce the appropriate boundary conditions for the Toda equations, providing a consistency check for the abelian truncation.\\ We can also provide an expression for the symmetric pairing of the topological theory. The operators in the chiral ring corresponding to the basis in \ref{pointb} with $\lambda_{r}=0$ and \ref{eigen} are respectively

\begin{equation*}
\begin{split}
& \chi_{k}(\alpha, \beta)= e^{\frac{-i \alpha k}{N}} \sum_{m,n \in \mathbb{Z}} e^{-i( m\alpha + n\beta)} \Phi_{k;m,n}, \\ \\ 
& \Psi_{j}(\alpha, \beta)= \sum_{k=0}^{N-1}  e^{- \frac{2 \pi i l k j}{N}}  \chi_{k}(\alpha, \beta).
\end{split}
\end{equation*}

The topological metric can be computed with the formula \ref{residue} for one chiral superfield. In the point basis the expression is

\begin{equation}
\begin{split}
\mathrm{Res}_{W} \left[ \chi_{k}(\alpha, \beta), \chi_{m}(\alpha^{\prime}, \beta^{\prime}) \right] = \delta(\alpha-\alpha^{\prime})\delta(\beta-\beta^{\prime}) \eta_{k,m}^{\mathrm{point \ basis}}(\alpha, \beta)  
\end{split}   
\end{equation}

where 

\begin{equation}
\eta_{k,m}^{\mathrm{point \ basis}}(\alpha, \beta) = \left( \partial^{2} _{z} W^{(N,l)}(0;\tau)\right) ^{-1} e^{-\frac{ i k}{N} (2\alpha + 2\pi l )}\delta_{k,m} .
\end{equation}

In the basis of $\sigma$-eigenstates one gets

\begin{equation}
\begin{split}
& \mathrm{Res}_{W} \left[ \Psi_{n}(\alpha, \beta), \Psi_{m}(\alpha^{\prime}, \beta^{\prime}) \right] =  \delta(\alpha-\alpha^{\prime})\delta(\beta-\beta^{\prime})\ \eta_{n,m}^{\sigma-\mathrm{basis}}(\alpha, \beta), 
\end{split}
\end{equation}

where

\begin{equation}
\eta_{n,m}^{\sigma-\mathrm{basis}}(\alpha, \beta)=\left( \partial^{2} _{z} W^{(N,l)}(0;\tau)\right) ^{-1}  \sum_{k=0}^{ N-1} e^{-\frac{ i k}{N} (2\pi l (m+n+1) + 2\alpha )} .
\end{equation}

We can choose a basis and use one of these expressions to impose the $tt^{*}$ reality constraint,
where the complex conjugates $g^{*}, \eta^{*}$ in terms of the matrices $g(\alpha,\beta)$, $\eta(\alpha,\beta)$ are respectively 

\begin{equation*}
\begin{split}
& g^{*}(\alpha, \beta)=\left[  g(-\alpha,- \beta) \right]  ^{*}, \\ & \eta^{*}(\alpha, \beta)=\left[  \eta(-\alpha, -\beta) \right]^{*}.
\end{split}
\end{equation*}

This condition gives $g(-\alpha, -\beta)$ in terms of $g(\alpha, \beta)$, but an explicit computation is hard in general. In the case of a trivial representation of the homology, the reality constraint simplifies, becoming the same of the $A_{N}$ models \cite{rif10}.

\subsubsection{Peculiarities of the N=2 Level}

We briefly discuss some peculiar aspects of the class of models of level $2$. The derivative of the superpotential is 

\begin{equation}
\partial_{z} W(z;\tau)= \zeta ( z-\pi \tau;\tau)-\zeta(z-\pi \tau -\pi ;\tau) -\eta_{1},
\end{equation}

where  $  z \in \mathcal{S}= \mathbb{C} \setminus \left\lbrace \pi \tau, \pi \tau +\pi +\Lambda_{\tau}; \  \Lambda_{\tau}= 2\pi\mathbb{Z} \oplus 2\pi \tau \mathbb{Z}\right\rbrace $.  The simple poles and simple zeroes are located respectively in $\pi \tau, \pi \tau +\pi +\Lambda_{\tau}$ and $0, \pi+\Lambda_{\tau}$. In this case the Galois group acts trivially and we have only one co-level.\\ In addition to the generators $\sigma, A, B,L$ already discussed, the symmetry group of the models of level $2$ contains also the parity transformation 

\begin{equation*}
\iota : z\longrightarrow -z; \hspace{2cm} \partial_{z} W(-z;\tau)=- \partial_{z} W(z;\tau),
\end{equation*}

which follows from \ref{colevel} for $N=2$ and $l=1$. \\ This operator satisfies the following commutation relations with the generators of the abelian group:

\begin{equation*}
\begin{split}
& \iota A=  A^{-1} \iota, \\ 
& \iota B=  B^{-1} \iota, \\ 
& \iota L=  L \iota, \\ 
& \iota \sigma=  \sigma^{-1} \iota .
\end{split}
\end{equation*}

The action of $ \iota$, as well as its algebraic relations, can be extended to the abelian universal cover. There are inequivalent choices which differ by compositions with the loop generator $L$. Indipendently from the definition, the set of curves with $p(1)=0$ is left invariant by the action of $\iota$.\\ The presence of an extra symmetry, as well as the possibility of working with $2 \times 2$ matrices, guarantees some semplifications in the derivation of the $tt^{*}$ equations. A point basis of theta-vacua for this family of theories is
 
\begin{equation*}
\begin{split}
& \ket{0;\alpha, \beta, \lambda}= \sum_{m,n,j \in \mathbb{Z}} e^{-i( m\alpha + n\beta+j\lambda)} A^{m} B^{  n} L^{j} \ket{p_{0}}, 
\\ &
\ket{1;\alpha, \beta, \lambda}= \sum_{m,n,j \in \mathbb{Z}} e^{-i( m\alpha + n\beta+j\lambda)} A^{m} B^{  n} L^{j} \sigma\ket{p_{0}}.
\end{split}
\end{equation*}

Setting $\lambda=0$, the operators $\sigma$ and $\iota$ act on these states as:

\begin{equation*}
\begin{split}
& \sigma \ket{0;\alpha, \beta}= \ket{1;\alpha, \beta}, \hspace{2cm} \sigma \ket{1;\alpha, \beta} = e^{i \alpha} \ket{1;\alpha, \beta}, \\ 
& \iota \ket{0;\alpha, \beta}= \ket{0;-\alpha, -\beta}, \hspace{1.5cm} \iota \ket{1;\alpha, \beta} = e^{-i \alpha} \ket{1;-\alpha, -\beta}.
\end{split}
\end{equation*}

The ground state metric in this basis is represented by the $2$ x $2$ hermitian matrix

\begin{equation*}
g(\alpha, \beta)= \begin{pmatrix}
g_{0\bar{0}}(\alpha, \beta) & g_{0\bar{1}}(\alpha, \beta) \\ g_{1\bar{0}}(\alpha, \beta) & g_{1\bar{1}}(\alpha, \beta)
\end{pmatrix}.
\end{equation*}

The symmetries $\sigma$ and $\iota $ imply respectively

\begin{equation*}
g_{0\bar{0}}(\alpha, \beta)= g_{1\bar{1}}(\alpha, \beta), \hspace{2cm} 
g_{0\bar{1}}(\alpha, \beta)= e^{i\alpha} g_{1\bar{0}}(\alpha, \beta),
\end{equation*}

and

\begin{equation*}
g_{0\bar{0}}(\alpha, \beta)= g_{0\bar{0}}(-\alpha, -\beta), \hspace{2cm} 
g_{1\bar{0}}(\alpha, \beta)= e^{-i\alpha} g_{1\bar{0}}(-\alpha, -\beta).
\end{equation*}

Therefore the metric can be written as 

\begin{equation*}
g(\alpha, \beta)= \begin{pmatrix}
A(\alpha, \beta) & e^{\frac{i\alpha}{2}}B(\alpha, \beta) \\ e^{\frac{-i\alpha}{2}}B(\alpha, \beta) & A(\alpha, \beta)
\end{pmatrix},
\end{equation*}

where $A(-\alpha, -\beta)=A(\alpha, \beta)$ and $B(-\alpha, -\beta)=B(\alpha, \beta)$.  The transpose and complex conjugate of $g$ read  

\begin{equation*}
\begin{split}
& g^{\mathrm{T}}(\alpha, \beta)=\left[  g(-\alpha,- \beta) \right]  ^{\mathrm{T}}, \\ & g^{*}(\alpha, \beta)=\left[  g(-\alpha, -\beta) \right]^{*},
\end{split}
\end{equation*}

and therefore 

\begin{equation*}
g^{\mathrm{\dagger}}(\alpha, \beta)= \left[  g(-\alpha,- \beta) \right]  ^{\mathrm{\dagger}}.
\end{equation*}

The hermiticity of $g$ implies that $A(\alpha, \beta)$ and $B(\alpha, \beta)$ are real functions, while the positivity requires that $A(\alpha, \beta) > 0$. The matrix $C_{w}$ and the topological metric in this basis are 

\begin{equation}
C_{w}(\alpha, \beta)= \begin{pmatrix}
1 & 0 \\ 0 & -1
\end{pmatrix}, \hspace{2cm }
\eta(\alpha, \beta)= \left( \partial^{2} _{z} W(0;\tau)\right) ^{-1} \begin{pmatrix}
1 & 0 \\ 0 & -1
\end{pmatrix}.
\end{equation}

We see that, by the parity properties of $A(\alpha, \beta)$ and $B(\alpha, \beta)$ implied by $ \iota$,  the reality constraint \ref{realityconstraint} reduces to 

\begin{equation}
\vert \partial^{2} _{z} W(0;\tau) \vert ^{2} ( A(\alpha, \beta)^{2}-B(\alpha, \beta)^{2} )=1.
\end{equation}

As a conquence, the metric can be parametrized in term of a single function of the angles

\begin{equation}
\begin{split}
& A(\alpha, \beta)= \frac{1}{\vert \partial^{2} _{z} W(0;\tau) \vert} \cosh \left[ L(\alpha, \beta)\right], \\ 
& B(\alpha, \beta)= \frac{1}{\vert \partial^{2} _{z} W(0;\tau) \vert} \sinh \left[ L(\alpha, \beta)\right] .
\end{split}
\end{equation}

Finally, by plugging the given expressions of $g(\alpha, \beta)$ and $C_{w}$ in \ref{canonical} we find the Sinh-Gordon equation

\begin{equation}
\partial_{\bar{w}}\partial_{w}  L(\alpha, \beta) = 2 \sinh \left[ 2 L(\alpha, \beta)\right].
\end{equation}

\subsection{Modular Properties of the Models}\label{sec4}

\subsubsection{Modular Transformations of $ \partial_{z} W^{(N,l)}(z;\tau)$}

The underlying structure of this class of models is the theory of the modular curves. In particular, as discussed in \ref{modular}, the space of models and its spectral cover are described respectively by the modular curves for $\Gamma_{1}(N)$ and $\Gamma(N)$. Now that we have provided an explicit description of these systems, we want to investigate the modular properties of the superpotential, as well as the ground state metric, with respect to the relevant congruence subgroups of $SL(2,\mathbb{Z})$. Let us first consider $\partial_{z} W^{(N,l)}(z;\tau)$. Given a modular transformation $\gamma = \begin{pmatrix} a & b \\ c & d \end{pmatrix} \in \Gamma_{1}(N)$, the derivative of the superpotential transforms as 

\begin{equation}
 \partial_{z} W^{(N,l)}\left( \frac{z}{c\tau + d};\frac{a\tau +b}{c\tau + d}\right) = 
\end{equation}

\begin{equation*}
\begin{split}
& (c\tau+d) \sum_{k=0}^{N-1} e^{\frac{2\pi i l k}{N}} \biggl[ \zeta \left( z- \frac{2\pi}{N}((a\tau+b)l + k (c\tau + d)) ;\tau \right) + 2\frac{d \eta_{1}+c \eta_{2}}{N} k \biggr] = \\ \\ 
&(c\tau+d) \sum_{k=0}^{N-1} e^{\frac{2\pi i l k}{N}}  \left[ \zeta \left( z-\frac{2\pi}{N}bl- \frac{2\pi}{N}(l\tau+k) ;\tau \right) +\frac{2\eta_{1}k}{N}  \right]= \\ \\ & (c\tau+d)  \partial_{z} W^{(N,l)}\left( z-\frac{2\pi}{N}bl ;\tau\right),
\end{split}
\end{equation*}

where we have used the fact that $a,d=1$ mod $N$, $c=0$ mod $N$ and the modular properties of the Weierstrass function: 

\begin{equation}\label{weierstrass}
\begin{split}
& \zeta \left(\frac{z}{c\tau + d} ; \frac{a\tau +b}{c\tau + d} \right)= (c\tau+d)\zeta(z;\tau), \\ \\ & \eta_{1}\left( \frac{a\tau +b}{c\tau + d}\right) = (c\tau+d) (d \eta_{1}+c \eta_{2}). 
\end{split}
\end{equation}

We see that $\Gamma_{1}(N)$ preserves only one of the torsion point. The transformation $\frac{2\pi\tau}{N} \rightarrow  \frac{2\pi(a\tau+b)}{N}$ results in a shift of poles and vacua by $\frac{2\pi}{N}bl$. If we set $b=0$ mod $N$, i.e. $\gamma \in \Gamma(N)$, also the order of vacua is preserved modulo periodicity.\\ On the other hand, a matrix $\gamma = \begin{pmatrix} a & b \\ c & d \end{pmatrix} \in \Gamma_{0}(N)$  acts on $\partial_{z} W^{(N,l)}(z;\tau)$ as 

\begin{equation}\label{dw}
\begin{split}
& \partial_{z} W^{(N,l)}\left( \frac{z}{c\tau + d};\frac{a\tau +b}{c\tau + d}\right) = \\ \\  
& (c\tau+d) \sum_{k=0}^{N-1} e^{\frac{2\pi i l k}{N}} \left[ \zeta \left( z- \frac{2\pi}{N}((a\tau+b)l + k (c\tau + d)) ;\tau \right) +2\frac{d \eta_{1}+c \eta_{2}}{N} k \right]= \\ \\ 
& (c\tau+d) \sum_{k=0}^{N-1} e^{\frac{2\pi i al k}{N}} \left[ \zeta \left( z-\frac{2\pi}{N}bl- \frac{2\pi}{N}(al\tau+k) ;\tau \right) +\frac{2\eta_{1}k}{N}  \right]=
\\ \\ & (c\tau+d) \partial_{z} W^{(N,al)}\left( z-\frac{2\pi}{N}bl ;\tau\right),
\end{split}
\end{equation}

where in the third line we used the fact that $a=d^{-1}$ mod $N$. As discussed in section \ref{modular}, $ \Gamma_{0}(N)$ changes the co-level with the map $l \rightarrow al$, acting as the Galois group of the real cyclotomic extension.

\subsubsection{Modular Transformations of the Physical Mass}\label{modphysmass}

It turns out that the superpotential is not invariant under a transformation of $\Gamma(N)$, but it shifts by a constant. On one side, it is not wrong to say that the model is left invariant, since a constant is irrilevant in determining the physics of a system. But, at the same time,  the `physical mass' $W^{(N,l)}(p_{0};\tau)$ is the parameter that we have used to write the $tt^{*}$ equations. This means that $\Gamma(N)$ changes the coordinate of the model on the spectral cover, with a consequent transformation of the ground state metric. An analogous constant shift is induced on the superpotential also by $\Gamma_{1}(N)$ and $\Gamma_{0}(N)$, in addition to the effects discussed in the previous paragraph. Before giving an interpretation of such phenomena and solving the apparent contraddiction, we want to compute these constants.\\ The fact that $W^{(N,l)}(p_{0};\tau)$ transforms non trivially under $\Gamma(N)$ is connected to the geometry of the modular curves. These spaces are not simply connected and a critical value can be defined only on the universal cover, i.e. the upper half plane, where $\Gamma(N)$ plays the role of deck group. This is also strictly related to the fact that the superpotential is defined on the universal cover of the target space, where we have more than $N$ vacua. On the other hand, variations of $W^{(N,l)}(p_{0};\tau)$ with respect to some coordinate on the modular curve must be modular functions of $\Gamma(N)$. Let us give some examples. Among the curves $X(N)$, the cases with $N\leq 5$ are the only ones with genus $0$ and therefore the easiest to describe \cite{rif51}. Genus $0$ means in particular that the modular function field has a unique generator, also called Hauptmodul. This function is unique up to Mobius transformations and modular with respect to $\Gamma(N)$. Moreover, it is the projective coordinate which realizes the isomorphism between $X(N)$ and the Riemann sphere punctured with the cusps. Denoting with $x$ the coordinate on the sphere, we can define the one-form $dW(x)=C(x)dx$, where $C(x)=\partial_{x}W(p_{0};x)$ is the chiral ring coefficient which describes variations of the critical value with respect to the Hauptmodul. Except the case of $N=5$, for these curves we have only the co-level $1$ (and the inverse), so we suspend the notation $(N,l)$. From the theory of projective algebraic curves, $C(x)$ must be a rational function with coefficients in the cyclotomic field. Moreover, the poles must be located in the free IR cusps, where the physical mass is expected to diverge. Thus, $C(x)$ can be fixed up to a multiplicative constant by requiring the correct transformation properties under $\Gamma_{1}(N)$ and $\Gamma_{0}(N)$. In particular, from the transformations of $\partial_{z} W^{(N,l)}(z;\tau)$ under $\Gamma_{1}(N)$, we learn that  $T: \tau \rightarrow \tau +1$ acts on the one form as an automorphism of the sphere with the formula 
\begin{equation*}
T^{*}dW(x)= dW(p_{0};T^{*}x)=dW(\sigma^{-1}(p_{0});x)=e^{-\frac{2\pi i }{N}} dW(x),
\end{equation*}

where $l=1$ is understood.\\ Starting from $X(2)$, we have $3$ cusps which can be viewed as the $3$ vertices of the equatorial triangle of a double triangular pyramid inscribed in $\mathbb{P}^{1}$. A set of representatives for the cusps is given by 

\begin{equation*}
C_{\Gamma(2)}=\left\lbrace 0,1,\infty \right\rbrace 
\end{equation*}

where $\infty$, being a fixed point of $\tau \rightarrow \tau +1$, is a UV cusp and $0,1$ identify two decoupled vacua. Moreover, the quotient group $PSL(2,\mathbb{Z})/\Gamma(2)$ is isomorphic to $\mathbb{S}_{3}$, which is the symmetry group of a triangle. We can use $x(\tau)=1-\lambda(\tau)$ as coordinate on the sphere, where $\lambda(\tau): \mathbb{H}/\Gamma(2) \overset{\sim}{\longrightarrow} \mathbb{P}^{1}/\lbrace 0,1, \infty \rbrace$ is the modular lambda function defined by 

\begin{equation*}
\lambda(\tau)= \frac{\theta_{2}^{4}(0;\tau)}{\theta_{3}^{4}(0;\tau)}.
\end{equation*}

The $3$ cusps fall in the point $x(\infty)=1, x(0)=0,x(1)= \infty$ and $T$ acts on $x$ by the transformation of $PSL(2,\mathbb{C})$ 

\begin{equation*}
T^{*}x =\frac{1}{x}.
\end{equation*}

From what we have said $C(x)$ must have a simple pole in $x=0$ and $x=\infty$. Asking that $T^{*}dW(x)= -dW(x)$, we find 

\begin{equation}
dW(x)= \alpha \frac{dx}{x}
\end{equation}

where $\alpha$ is a complex constant. We see that there is a unique generator of $\Gamma(2)$ which acts non trivially on the critical value. This generates loops around the cusp in $0$ (as well as in $\infty$) and produces the constant shift of the superpotential.\\
In the case of $N=3$ the cusps are located at the $4$ vertices of a regular tetrahedron inscribed in the Riemann sphere. It is not a coincidence that $PSL(2,\mathbb{Z})/\Gamma(3)$ is isomorphic to the symmetry group of this solid figure, i.e. $\mathbb{A}_{4}$. The set of cusps is 

\begin{equation*}
C_{\Gamma(3)}=\left\lbrace 0,1/2,1,\infty\right\rbrace .
\end{equation*}

Also in this case $\infty$ is the only UV cusp, while the other rationals are in the same orbit of $T$ and identify free theories. The integers coprime with $3$ are $1$ and $2 =-1$ mod $3$ which describe the same theory. The generator of the function field is 

\begin{equation*}
J_{3}(\tau)= \frac{1}{i\sqrt{27}}q_{\tau}^{-\frac{1}{3}} \left( \prod_{n=1}^{\infty} \frac{1-q_{\tau}^{n/3}}{1-q_{\tau}^{3n}} \right)^{3} = \frac{1}{i\sqrt{27}} \left( \frac{\eta(\tau/3)}{\eta(3\tau)}\right)^{3} 
\end{equation*}

where $q_{\tau}=e^{2\pi i \tau}$ and 

\begin{equation}
\eta(\tau)= q_{\tau}^{1/24}  \prod _{n=1}^{\infty} (1-q_{\tau}^{n})
\end{equation}

is the Dedekind eta function. The generator of $\Gamma_{1}(3)/ \Gamma(3)$ acts on the Hauptmodul as 

\begin{equation*}
T^{*} J_{3}=\zeta_{3} + \zeta_{3}^{2}J_{3}
\end{equation*}

where we use the notation $ \zeta_{k}= e^{2\pi i/k}$. Denoting $x=J_{3}$, the one-form with values in the chiral ring is 

\begin{equation*}
dW(x)= \alpha\sum_{k=0}^{2} \frac{\zeta_{3}^{-k} dx}{x-x_{k}}
\end{equation*}

where $x_{0}=J_{3}(0)=0,x_{1}=J_{3}(1)=\zeta_{3}$ and $x_{2}=J_{3}(1/2)=1+\zeta_{3}$, with $x_{k+1}=T^{*}x_{k}$. The cusp at $\tau=\infty$ is instead sent to $x=\infty$ on the sphere. It is straightforward to check that $dW(x)$ satisfies $T^{*}dW(x)= \zeta_{3}^{-1}dW(x)$. \\ Another spherical version of a platonic solid appears for $N=4$. In this curve the cusps are the $6$ vertices of a regular octrahedron in $\mathbb{P}^{1}$ with symmetry group  $PSL(2,\mathbb{Z})/\Gamma(4)\sim \mathbb{S}_{4}$. Also in this case the co-levels are just $\pm 1$. The critical points are given by 

\begin{equation*}
C_{\Gamma(4)}=\left\lbrace 0,1/3,1/2,2/3,1,\infty\right\rbrace .
\end{equation*}

In this curve we have two UV cusps, i.e. $\infty$ and $1/2$, which are both fixed points of $T$. The remaining ones represent instead free IR critical points. The Hauptmodul of level $4$ is 

\begin{equation*}
J_{4}(\tau)=\zeta_{8}^{3}\sqrt{8}q_{\tau}^{1/4} \prod_{n=1}^{\infty} \frac{(1-q_{\tau}^{4n})^{2}(1-q_{\tau}^{n/2})}{(1-q_{\tau}^{n/4})^{2}(1-q_{\tau}^{2n})},
\end{equation*}

which transforms under $T$ as 

\begin{equation*}
T^{*}J_{4}= \frac{\zeta_{4}J_{4}}{1-J_{4}}.
\end{equation*}

The fixed points of this map are $0$ and $1-\zeta_{4}$, which correspond respectively to $\tau= \infty, \frac{1}{2}$. The $4$ IR cusps are all in the same orbit of $T$ and fall in the points $J_{4}(0)=\infty, J_{4}(1)=-\zeta_{4}, J_{4}(2/3)=1/(1+\zeta_{4}), J_{4}(1/3)=1$. These must be simple poles for $dW(x)$, with $x=J_{4}$, which reads

\begin{equation}
dW(x)= \alpha \left( \frac{1}{x-1} -\frac{1}{x+\zeta_{4}}+ \frac{\zeta_{4}}{x-1/(1+\zeta_{4})}\right)dx,
\end{equation}

and satisfies $T^{*}dW(x)=\zeta_{4}^{-1}dW(x)$. \\ $X(5)$ is the last case of genus $0$. This modular curve has $12$ cusps which identify the vertices of a regular icosahedron. The quotient $PSL(2,\mathbb{Z})/\Gamma(5)$ acts on the Riemann sphere as $\mathbb{A}_{5}$, the symmetry group of this platonic solid. Among the $12$ cusps

\begin{equation*}
C_{\Gamma(5)}=\left\lbrace 0,2/9,1/4,2/7,1/3,2/5,1/2,5/8,2/3,3/4,1,\infty \right\rbrace 
\end{equation*}

the UV fixed points are $2/5$ and $\infty$, while $\left\lbrace 0,2/9,1/4,3/4,1\right\rbrace $ and $\left\lbrace 2/7,1/3,1/2,5/8,2/3   \right\rbrace $ represent the $5$ decoupled vacua in two inequivalent IR limits. This curve has two inequivalent co-levels, i.e. $l=1,2$ and the Galois group acts on the set of cusps by exchanging the two UV limits and the two groups of IR theories. As in the previous cases, we can use as projective coordinate on $\mathbb{P}^{1}(\mathbb{C})$ the Hauptmodul 

\begin{equation*}
J_{5}(\tau)= \zeta_{5} q_{\tau}^{-1/5} \prod_{n=1}^{\infty} \frac{(1-q_{\tau}^{5n-2})(1-q_{\tau}^{5n-3}) }{(1-q_{\tau}^{5n-4})(1-q_{\tau}^{5n-1}) }
\end{equation*}

which transforms under $T$ as 

\begin{equation*}
T^{*}J_{5}= \zeta_{5}^{-1}J_{5}.
\end{equation*}

The UV cusps $2/5,\infty$ are sent by $J_{5}$ respectively to $x=0,\infty$, which are the fixed points of $T$. 
We know from \ref{dw} that a transformation $\gamma \in \Gamma_{0}(N)$ with $b=0$ mod $N$ acts on the cyclotomic units through the Galois group and permutes the residue of the poles in $dW(x)$. It follows that if two IR cusps are in the same orbit of such $\gamma$, their residue must be related by the correspondent Galois transformation. It is the case for instance of $0$ and $5/8$ if we set $b=5$ and $c=8$. Imposing also the correct transformation property under $T$ for $l=1$, one can repeat the procedure and fix all the coefficients up to an overall constant. Setting $x_{1;k}=\zeta_{5}^{k}J_{5}(0)$ and  $x_{2;k}=\zeta_{5}^{k}J_{5}(5/8)$, with $J_{5}(0)=1+\zeta_{5}+\zeta_{5}^{2}$ and $J_{5}(5/8)=(\zeta_{5}^{2}-\zeta_{5}^{4})/(1+\zeta_{5}-\zeta_{5}^{3}-\zeta_{5}^{4})$, we find

\begin{equation*}
dW(x)= \alpha \sum_{k=0}^{4} \left( \frac{1}{x-x_{1;k}} +  \frac{1}{x-x_{2;k}}\right)\zeta_{5}^{k} dx.
\end{equation*}

where the residue of $J_{5}(0)$ and $J_{5}(5/8)$ are both normalized to $1$.\\ By summarizing, the generators of loops around the IR cusps on the sphere correspond to the subset of generators of $\Gamma(N)$ which act non trivially on the critical value. Once the normalization is fixed, the constant generated by modular transformations can be computed with the residue formula. Moreover, the free IR cusps are all in the same orbit of $\Gamma_{0}(N)$, which acts by multiplication on the coefficients of $dW(x)$. This implies in particular that all the poles have the same order, which must be $1$ from the non trivial monodromy of $W(p_{0};x)$, and their residue are related by Galois transformations.\\ A similar procedure could be carried on in principle also for modular curves of higher genus, but it is more complicated. It is instead much more convenient to find a general expression of the critical value as function on the upper half plane and study its modular properties. Let us take the multi-valued function in \ref{symb}. Since the constants generated by the modular transformations are indipendent from the point, we can set $z=0$. The expression of $W^{(N,l)}(0;\tau)$ reads

\begin{equation}
W^{(N,l)}(0;\tau)= \sum_{k=0}^{N-1} e^{\frac{2\pi i l k}{N}} \log \left[ \Theta \begin{bmatrix} \frac{1}{2}-\frac{l}{N} \\ \frac{1}{2}-\frac{k}{N}\end{bmatrix}\left( 0;\tau \right) e^{- 2 \pi i \left( \frac{l}{N}-\frac{1}{2}\right)\left( \frac{k}{N}-\frac{1}{2}\right) }\right] .
\end{equation}

where the phase $e^{- 2 \pi i \left( \frac{l}{N}-\frac{1}{2}\right)\left( \frac{k}{N}-\frac{1}{2}\right)}$ is a convenient normalization constant. This function remains ill defined as long as we do not specify the determination of the logarithm. This is equivalent to choose, for a fixed $\tau$, a representative of $z=0$ and determine its critical value.  We first set the notations

\begin{equation*}
\begin{split}
 &  q_{\tau}= e^{2\pi i \tau}, \hspace{2cm} q_{z}= e^{2\pi i z}, \\ \\ & \hspace{1.3 cm} z= u_{1} \tau+ u_{2}, 
\end{split}
\end{equation*}

with $ u_{1},u_{2} \in \mathbb{Z}/ N $.  Then, we recall the definition of Siegel functions:

\begin{equation}\label{siegel}
g_{u_{1},u_{2}}(\tau)=- q_{\tau}^{B_{2}(u_{1})/2} e^{2\pi i u_{2}(u_{1}-1)/2}(1-q_{z}) \prod _{n=1}^{\infty} (1-q_{\tau}^{n}q_{z})(1-q_{\tau}^{n}/q_{z}),
\end{equation}

where $B_{2}(x)= x^{2}-x+\frac{1}{6}$ is the second Bernoulli polynomial. Because of their modular properties, these objects are a sort of `building blocks' for the modular functions of level $N$. In particular, all the Hauptmoduls defined previously can be expressed in terms of $ g_{u_{1},u_{2}}(\tau)$ and the Dedekind eta function \cite{rif48,rif49,rif50}.
The theta functions  $  \Theta \begin{bmatrix} \frac{1}{2}-u_{1} \\ \frac{1}{2}-u_{2}\end{bmatrix}\left( 0;\tau \right)$ have the $q$-product representation 

\begin{equation}
\begin{split}
\Theta \begin{bmatrix} \frac{1}{2}-u_{1} \\ \frac{1}{2}-u_{2}\end{bmatrix}\left( 0;\tau \right)= & - q_{\tau}^{B_{2}(u_{1})/2}q_{\tau}^{1/24}  e^{2\pi i (u_{1}-1/2)(u_{2}-1/2)} (1-q_{z}) \\ & \mathrm{x} \prod _{n=1}^{\infty}(1-q_{\tau}^{n}) (1-q_{\tau}^{n}q_{z})(1-q_{\tau}^{n}/q_{z})
\end{split}
\end{equation}

and can be written in terms of $g_{u_{1},u_{2}}(\tau) $  and $\eta (\tau) $ as 

\begin{equation}
 \Theta \begin{bmatrix} \frac{1}{2}-u_{1} \\ \frac{1}{2}-u_{2}\end{bmatrix}(0;\tau)= i g_{u_{1},u_{2}}(\tau) \eta (\tau) e^{ 2 \pi i u_{1}\left( u_{2}-1\right) /2}.
\end{equation}

Since the Siegel and Dedekind functions have neither zeroes nor poles, there is a single-valued branch of $\log \Theta \begin{bmatrix} \frac{1}{2}-u_{1} \\ \frac{1}{2}-u_{2}\end{bmatrix}\left( 0;\tau \right)$ on the upper half plane. Therefore, the critical value can be consistently defined as holomorphic function of $\tau \in \mathbb{H}$. Provided the above relations, we can rewrite $W^{(N,l)}(0;\tau)$ as  

\begin{equation}\label{mass}
W^{(N,l)}(0;\tau)= \sum_{k=0}^{N-1} e^{\frac{2\pi i l k}{N}} \log E_{\frac{l}{N},\frac{k}{N}}(\tau),
\end{equation}

where $E_{\frac{l}{N},\frac{k}{N}}(\tau)$ is the Siegel function of characters $ u_{1}=\frac{l}{N}, u_{2}=\frac{k}{N}$ without the root of unity $e^{2\pi i u_{2}(u_{1}-1)/2} $. Under an integer shift of the characters, these functions satisfy \cite{rif48}

\begin{equation}\label{shift}
E_{u_{1}+1,u_{2}}(\tau)=-e^{-2\pi i u_{2}} E_{u_{1},u_{2}}(\tau), \hspace{2cm} E_{u_{1},u_{2}+1}(\tau)=E_{u_{1},u_{2}}(\tau).
\end{equation}

Moreover, being the Siegel functions up to a multiplicative constant, they have good modular properties. For $\gamma= \begin{pmatrix}   a & b \\ c & d \end{pmatrix} \in SL(2,\mathbb{Z})$, they transform with a phase:

\begin{equation}\label{prop}
\begin{split}
& E_{u_{1},u_{2}}(\tau+b)=  e^{\pi i b B_{2}(u_{1})} E_{u_{1},u_{2}+b u_{1}}(\tau) , \ \ \ \ \mathrm{for } \ c = 0 , \\ \\ 
& E_{u_{1},u_{2}}(\gamma (\tau))= \varepsilon (a,b,c,d) e^{\pi i \delta} E_{u_{1}^{\prime},u_{2}^{\prime}}(\tau) , \ \ \ \ \mathrm{for } \ c\neq 0,
\end{split}
\end{equation}

where

\begin{equation}
\begin{split}
\varepsilon(a,b,c,d)= &
   \begin{cases}
  e^{i\pi (bd(1-c^{2})+c(a+d-3))/6}, \ \ \ \ \mathrm{if} \ c \ \mathrm{is \ odd},
   \\  -ie^{i\pi (ac(1-d^{2})+d(b-c+3))/6}, \ \ \ \ \mathrm{if} \ d \ \mathrm{is \ odd},
   \end{cases} \\ \\ 
    \delta= & u_{1}^{2}ab + 2 u_{1}u_{2}bc + u_{2}^{2}cd - u_{1}b-u_{2}(d-1),
    \end{split}
\end{equation}

and 

\begin{equation}\label{character}    
u_{1}^{\prime}= a u_{1}+c u_{2}, \hspace{1cm}  u_{2}^{\prime}= b u_{1}+d u_{2}.
\end{equation}

In order to compute the constants generated by the modular transformations, we need to evaluate the difference 

\begin{equation}\label{diff}
\chi_{u_{1},u_{2}}(\gamma)=\log E_{u_{1},u_{2}}(\gamma (\tau))-\log E_{u_{1}^{\prime},u_{2}^{\prime}}(\tau),
\end{equation}

for $\gamma \in SL(2,\mathbb{Z})$. Here we assume the characters of the Siegel functions to be normalized such that $0 < u_{1},u_{2},u_{1}^{\prime},u_{2}^{\prime} < 1$. This can always be achieved by the property \ref{shift}. The computation for the case of $\Gamma(N)$ has already been done in \cite{rif49}. In \ref{modulartrnasofrmationmass} we follow closely that derivation, adapting it to the general case. For $c=0$ the transformations belong to the coset group $\Gamma_{1}(N)/\Gamma(N) \simeq \mathbb{Z}_{N}$ and we obtain 

\begin{equation}
\chi_{u_{1},u_{2}}(\gamma)= 2\pi i \frac{1}{2}B_{2}(u_{1}).
\end{equation}

It is clear that in this case we cannot appreciate a modular shift of the critical value. Indeed, these transformations simply translate the vacua:

\begin{equation}\label{trivial}
W^{(N,l)}(0;\tau+b)=e^{-\frac{2\pi i b l^{2}}{N}} W^{(N,l)}(0;\tau).
\end{equation}

On the other hand, for $c \neq 0$, we get the formula  

\begin{equation}\label{formula} 
\begin{split}
 \chi_{u_{1},u_{2}}(\gamma)= & \ 2\pi i \frac{1}{2}  \left( B_{2}(u_{1}) \frac{a}{c} + B_{2}(u_{1}^{\prime}) \frac{d}{c}  -\frac{2}{c} B_{1}(u_{1}^{\prime}) B_{1}( \langle d u_{1}^{\prime}- u_{2}^{\prime} c \rangle ) \right)  \\ \\ & -\frac{2 \pi i}{c}  \sum_{\overset{x \in \mathbb{Z}/c\mathbb{Z},}{ \ x \neq 0} }  [ x, u_{1}^{\prime}, u_{2}^{\prime} ]_{d,c},
 \end{split}
\end{equation}

where $B_{1}(x)=x-1/2$ is the first Bernoulli polynomial, $\langle x \rangle $ represents the fractional part of $x$ and the symbol $ [ x, u_{1}^{\prime}, u_{2}^{\prime} ]_{d,c}$ denotes

 \begin{equation}
 [ x, u_{1}^{\prime}, u_{2}^{\prime} ]_{d,c}=  \frac{e^{2\pi i x \bigl( \frac{ \langle d u_{1}^{\prime}-cu_{2}^{\prime}\rangle-du_{1}^{\prime} }{c}+u_{2}^{\prime} \bigr)}  }{(1-e^{-2\pi i x d/c})(1-e^{2\pi i x/c})}.
 \end{equation}

This result turns out to be indipendent from the branch of the logarithm and in particular from $\tau$. This is consistent with the fact that the modular shift of the critical value is indipendent from the vacuum that we choose. From the general formula we can reduce to the cases of the congruence subgroups. Let us consider $\gamma \in \Gamma_{0}(N)$. Using the fact that $c=0$ mod $N$ and $ad=1$ mod $N$, we have to plug in the above expression:

\begin{equation}
\begin{split}
u_{1}^{\prime}= \langle  a u_{1} \rangle ,&  \hspace{2cm} u_{2}^{\prime}=  \langle  d u_{2}+b u_{1}  \rangle, \\ \\
& \langle d u_{1}^{\prime}-u_{2}^{\prime}c  \rangle= u_{1}.
\end{split}
\end{equation}

If $\gamma \in \Gamma_{1}(N)$, these becomes 

\begin{equation}
\begin{split}
u_{1}^{\prime}= u_{1},&  \hspace{2cm} u_{2}^{\prime}=  \langle  u_{2}+ b u_{1}  \rangle, \\ \\
& \langle d u_{1}^{\prime}-u_{2}^{\prime}c  \rangle= u_{1}.
\end{split}
\end{equation}

The case of $\gamma \in \Gamma(N)$ follows from this by requiring further $b=0$ mod $N$.\\  Setting $u_{1}= \frac{l}{N}$, $u_{2}= \frac{k}{N}$ and summing over $k$ with the residue $e^{\frac{2\pi i l k}{N}}$, we find the modular transformations of the physical mass. In sequence

\begin{equation}\label{transf}  
\begin{split}
& \gamma \in \Gamma_{0}(N): \\ 
& W^{(N,l)}(0;\gamma(\tau))=  e^{-\frac{2\pi i al^{2}b}{N}} W^{(N,al)}(0;\tau) + \Delta W^{(N,l)}_{\Gamma_{0}(N)}(\gamma), \\ \\ 
  & \gamma \in \Gamma_{1}(N): \\ 
& W^{(N,l)}(0;\gamma(\tau))=  e^{-\frac{2\pi i l^{2}b}{N}} W^{(N,l)}(0;\tau) + \Delta W^{(N,l)}_{\Gamma_{1}(N)}(\gamma), \\ \\ 
& \gamma \in \Gamma(N): \\ 
& W^{(N,l)}(0;\gamma(\tau))=   W^{(N,l)}(0;\tau) + \Delta W^{(N,l)}_{\Gamma(N)}(\gamma), 
\end{split}
\end{equation}

with 

\begin{equation}\label{delta}
\begin{split}
& \Delta W^{(N,l)}_{\Gamma_{0}(N)}(\gamma)= -\frac{2 \pi i N}{c} \ e^{-\frac{2\pi i al^{2}b}{N}}\sum_{\overset{x \in \mathbb{Z}/c\mathbb{Z},}{ \ x= -al \ \mathrm{mod} \ N}} \frac{e^{2\pi i x \bigl( \frac{l/N-d 
\langle al/N \rangle }{c} \bigr)} }{(1-e^{-2\pi i x d/c})(1-e^{2\pi i x/c})}, \\ \\ 
& \Delta W^{(N,l)}_{\Gamma_{1}(N)}(\gamma)= -\frac{2 \pi i N}{c} \ e^{-\frac{2\pi i l^{2}b}{N}}\sum_{\overset{x \in \mathbb{Z}/c\mathbb{Z},}{ \ x= -l \ \mathrm{mod} \ N}} \frac{e^{2\pi i x \bigl( \frac{l}{N}\frac{1-d}{c} \bigr)  }}{(1-e^{-2\pi i x d/c})(1-e^{2\pi i x/c})},
\end{split}
\end{equation}

\begin{equation*}
\Delta W^{(N,l)}_{\Gamma(N)}(\gamma)= -\frac{2 \pi i N}{c} \sum_{\overset{x \in \mathbb{Z}/c\mathbb{Z},}{ \ x= -l \ \mathrm{mod} \ N}} \frac{e^{2\pi i x \bigl( \frac{l}{N}\frac{1-d}{c} \bigr)  }}{(1-e^{-2\pi i x d/c})(1-e^{2\pi i x/c})},
\end{equation*}

where the constraints on $x$ follow from the summation over $k$. These formulas are coherent with the transformations of $\partial_{z}W^{(N,l)}(z;\tau)$ that we found in the previous paragraph. \\ We can check in the simple case of $N=2$ that the formulas above give the results obained with the geometrical approach. The physical mass has the expression 

\begin{equation}
W(0;\tau)= \log \frac{\Theta_{4}(0;\tau)}{\Theta_{3}(0;\tau)} = \frac{1}{4}\log(1- \lambda(\tau)),
\end{equation}

where $1-\lambda(\tau)= \left( \Theta_{4}(0;\tau)/ \Theta_{3}(0;\tau) \right) ^{4}$ is the Hauptmodul of level $2$ that we have defined previously. Although $1-\lambda(\tau)$ is invariant under transformation of $\Gamma(2)$, the logarithm is not. We can use the expression for $\Delta W_{\Gamma(2)}$ to derive the modular transformations of $\log(1-\lambda(\tau))$. $\Gamma(2)$ is freely generated by the matrices 

\begin{equation}
T_{1}= \begin{bmatrix}
1 & 2 \\ 0 & 1
\end{bmatrix}, \hspace{2cm}   T_{2}= \begin{bmatrix}
1 & 0 \\ -2 & 1
\end{bmatrix} .
\end{equation}

Using respectively the \ref{trivial} and the \ref{transf}, \ref{delta} with $N=2, l=1$, one finds 

\begin{equation}
\begin{split}
& \log(1-\lambda(\tau+2))= \log(1-\lambda(\tau)), \\ \\ & \log\left( 1-\lambda\left( \frac{\tau}{1-2\tau}\right)\right)  = \log(1-\lambda(\tau))+2\pi i.
\end{split}
\end{equation}

It is clear that $T_{2}$ is the generator of anticlockwise loops around the IR cusp in $\tau=0$. Indeed, the constant is the same we obtain with the residue formula.

\subsubsection{Modular Transformations of the Ground State Metric}

The modular shift of the superpotential seems to contraddict the statement that the model is invariant under transformations of $\Gamma(N)$. But, if we assume the perspective of the universal cover, there is no contraddiction at all. Indeed, the physical mass parametrizes not only models, but also vacua. The modular transformation simply changes the initial choice of the vacuum $p_{0}$ with another one of the same fiber in the universal cover. Therefore, the new coordinate on the spectral curve describes the same model, but a different vacuum. \\ The $tt^{*}$ equations for these class of theories are manifestly covariant under transformations of the congruence subgroups. In particular, the covariance under $\Gamma_{1}(N)$ implies that the equation naturally descends on the space of models. However, as a consequence of the modular shift, the ground state metric is not left invariant. Indeed, matrices of $\Gamma(N)$ and $\Gamma_{1}(N)$ change the basis of lattice generators and consequently the representation of the homology group. Besides this effect, a transformation of $\Gamma_{0}(N)$ changes also the torsion point of the $\mathbb{Z}_{N}$ symmetry, resulting in a permutation of the metric components. Let us consider this more general case. From the transformation of $\partial_{z} W^{(N,l)}(z;\tau)$ under $\Gamma_{0}(N)$ one can read how the generators of the symmetry group are modified. We note that the new function is still periodic of $2\pi \tau$. Therefore, we can consider again $B$ as a generator of $H_{1}(\mathcal{S};\mathbb{Z}) $ in the new model.
The operators $L_{k}$ are left invariant by the transformation as well, since the corresponding homology cycles have the same definition in the model of co-level $al$. Differently, $\sigma$ and $A$ change in relation to the transformation of the torsion point of the vacua. For a $\gamma \in \Gamma_{0}(N)$, we can write

\begin{equation*}
\begin{split}
& \gamma^{*}B=B, \\ 
& \gamma^{*}L_{k}= L_{k} \\
& \frac{2\pi}{N} \longrightarrow \frac{2\pi}{N} (c\tau + d )  \Longrightarrow      \begin{cases} \gamma^{*}\sigma= \sigma^{d} B^{\frac{c}{N}}= \tilde{\sigma} B^{\frac{c}{N}} \\  \gamma^{*}A = A^{d} B^{c}= \tilde{A} B^{c},
\end{cases}
\end{split}
\end{equation*}

where we denote with $\tilde{\sigma}=\sigma^{d}$ the operator associated to the torsion point of co-level $al$, and with $\tilde{A}=A^{d}$ the homology cycle satisfying $\tilde{A}=\tilde{\sigma}^{N}$. The fact that the generator of loops are not involved in the modular transformations is consistence with the truncation of the $tt^{*}$ equation that we discussed in section $3$.\\ We have also to include the shift $z\rightarrow z -\frac{2\pi}{N}bl$, which implies the vacuum transformation

\begin{equation*}
p_{0} \longrightarrow \sigma^{-bl}(p_{0})  \Longrightarrow   \ket{p_{0}} \mapsto \sigma^{-bl}\ket{p_{0}} .
\end{equation*}

Let us study what these transformations mean at the level of vacuum states. We consider again trivial representations of $L_{k}$. The action of $\gamma$ on the point basis is 

\begin{equation*}
\gamma \ket{k;\alpha, \beta}=\  e^{\frac{-i \alpha k}{N}} \sum_{m,n \in \mathbb{Z}} e^{-i( m\alpha + n\beta )} (A^{d} B^{c})^{m} B^{n}  (\sigma^{d} B^{\frac{c}{N}})^{ k} \sigma^{-bl}\ket{p_{0}}
\end{equation*}

\begin{equation*}
\begin{split}
& = \ e^{-i\frac{kd}{N}\left( \frac{\alpha-\beta c }{d}\right) } \sum_{m,n\in \mathbb{Z}} e^{-i \left(  m \left( \frac{\alpha-\beta c}{d}\right)  + n\beta \right)} A^{m} B^{  n} \sigma^{kd-bl}\ket{p_{0}} \\  \\ & = \ e^{-i\frac{bl}{N} \left( \frac{\alpha-\beta c}{d} \right) } \ket{kd-bl; (\alpha-\beta c)/d, \beta}.
\end{split}
\end{equation*}

We note that, in the case of $ b=0$ mod $N$ and $d=1$ mod $N$, the fiber index $k$ is left invariant. This follows from the fact that a transformation of $\Gamma(N)$ preserves the torsion point up to a shift of a lattice vector, which moves the base point $p_{0}$ without changing the fiber. The $\sigma$-eigenstates transform consequently as 

\begin{equation*}
\gamma \ket{j;\alpha, \beta}=\  e^{-i\frac{bl}{N} \left( \frac{\alpha-\beta c}{d} \right) } \sum_{k=0}^{N-1}  e^{- \frac{2 \pi i l k j}{N}}  \ket{kd-bl;(\alpha-\beta c)/d , \beta} 
\end{equation*}

\begin{equation*}
\begin{split}
= & \ e^{-i\frac{bl}{N} \left(  2\pi a l j + \frac{\alpha-\beta c}{d}\right)} \sum_{k=0}^{N-1}  e^{- \frac{2 \pi i al k j}{N}}  \ket{k;(\alpha-\beta c)/d , \beta} \\ \\ 
= & \ e^{-i\frac{bl}{N} \left(  2\pi a l j + \frac{\alpha-\beta c}{d}\right)} \ket{aj;(\alpha-\beta c)/d, \beta}.
\end{split}
\end{equation*}

The action of the symmetry group generators $\tilde{\sigma},\tilde{A},B$ on the transformed states is given by 

\begin{equation*}
\begin{split}
& \tilde{\sigma}\ket{aj;(\alpha-\beta c)/d, \beta}= e^{\frac{2\pi i l j}{N}} e^{\frac{i}{N}(\alpha-\beta c)}\ket{aj;(\alpha-\beta c)/d, \beta}, \\ \\ 
& \tilde{A}\ket{aj;(\alpha-\beta c)/d, \beta}= e^{i(\alpha-\beta c)}\ket{aj;(\alpha-\beta c)/d, \beta}, \\ \\ 
& B\ket{aj;(\alpha-\beta c)/d, \beta}= e^{i\beta}\ket{aj;(\alpha-\beta c)/d, \beta}.
\end{split}
\end{equation*}

From the eigenvalues of the new basis, we learn that the ground state metric transforms in the following way:

\begin{equation}\label{metrictransformation}
\gamma^{*}\varphi_{j}(w;\alpha,\beta)= \varphi_{j}(\gamma^{*}w;\alpha,\beta)= \varphi_{aj}(w;\alpha-\beta c,\beta).
\end{equation}

As anticipated, we see that a transformation of $\Gamma(N)$ changes the representation of the homology, resulting in the character shift $\alpha\rightarrow \alpha-\beta c$. In the case of $\Gamma_{0}(N)$, since the operators $\tilde{\sigma}$ and $\sigma$ are inequivalent and have a different set of eigenstates, we appreciate also the exchange of the metric components along the diagonal. This effect takes place specifically for $N>2$, where we have a non trivial co-level structure. We note further that transformations of the coset $\Gamma_{1}(N)/\Gamma(N)$, i.e. with $c=0$, leave the metric completely invariant.

\subsection{Physics of the Cusps}\label{final}

\subsubsection{Classification of the Cusps}

Now that we have discussed the modular properties of the solution, we want to describe its behaviour around the boundary regions of the domain. These are represented by the cusps of the modular curve $\mathbb{H}/\Gamma(N)$, i.e. the equivalence classes of $\Gamma(N)$ in $\mathbb{Q} \cup \lbrace\infty \rbrace$. First of all, we have to understand which type of model each cusp corresponds to. Let us begin with the cusp at $\tau= i\infty$. It is convenient to come back to the initial lattices of poles and vacua with the shift $z\rightarrow z + \frac{2\pi l \tau}{N}$. Moreover, we rewrite the derivative of the superpotential in terms of $\Theta_{1}(z;\tau)$ as 

\begin{equation}\label{new}
\partial_{z} W^{(N,l)}(z;\tau)= \  \frac{1}{2} \sum_{k=0}^{N-1} e^{\frac{2\pi i lk}{N}} \frac{\Theta_{1}^{\prime}\left( \frac{1}{2}\left( z-\frac{2\pi k}{N}\right) ;\tau\right) }{\Theta_{1}\left(\frac{1}{2}\left( z-\frac{2\pi k}{N}\right) ;\tau\right)}. 
\end{equation}

Using the relation \cite{rif47}

\begin{equation*}
\frac{\Theta_{1}^{\prime}(z,\tau)}{\Theta_{1}(z ;\tau)}= \cot z 
+4 \sum_{n=1}^{\infty} \frac{q_{\tau}^{n}}{1-q_{\tau}^{n}}\sin 2nz,
\end{equation*}

with $q_{\tau}=e^{2\pi i \tau} $, the expression above becomes 
\begin{equation*}
\begin{split}
\partial_{z} W^{(N,l)}(z;\tau)= & \frac{1}{2} \sum_{k=0}^{N-1} e^{\frac{2\pi i lk}{N}} \biggl[ \cot\left(\frac{1}{2}\left( z-\frac{2\pi k}{N}\right) \right) +  \\ & 4 \sum_{n=1}^{\infty} \frac{q_{\tau}^{n}}{1-q_{\tau}^{n}}\sin \left( 2n\left( \frac{1}{2}\left( z-\frac{2\pi k}{N}\right) \right) \right)  \biggr]. 
\end{split}
\end{equation*}

The theta function in these expressions is normalized with quasi-periods $\pi,\pi\tau$. Let us denote with $S$ the infinite sum in $n$. Manipulating the expression, we get 

\begin{equation*}
\begin{split}
S= & \ 2 \sum_{k=0}^{N-1} e^{\frac{2\pi i lk}{N}} \sum_{n=1}^{\infty} \frac{q_{\tau}^{n}}{1-q_{\tau}^{n}}\sin \left( n \left( z-\frac{2\pi k}{N}\right)\right)   \\ \\ = & -i \sum_{n=1}^{\infty} \frac{q_{\tau}^{n}}{1-q_{\tau}^{n}}  \left( e^{inz}\sum_{k=0}^{N-1} e^{\frac{2\pi i k}{N}(l-n)} - e^{-inz}\sum_{k=0}^{N-1} e^{\frac{2\pi i lk}{N}(l+n)} \right) .
 \end{split}
\end{equation*} 

The two sums over $k$ are not $0$ if and only if $n$ satisfies respectively $n=l$ mod $N$ and $n=-l$ mod $N$. Therefore, the derivative of the superpotential becomes 

\begin{equation*}
\begin{split}
\partial_{z} W^{(N,l)}(z;\tau) = & \ \frac{1}{2} \sum_{k=0}^{N-1} e^{\frac{2\pi i lk}{N}} \cot \left( \frac{1}{2}\left( z-\frac{2\pi k}{N}\right) \right)  \\ \\ & -iN \left(  \sum_{\overset{n=1}{n=l \ \mathrm{mod} \ N}}^{\infty} \frac{q_{\tau}^{n}}{1-q_{\tau}^{n}}  e^{inz}-  \sum_{\overset{n=1}{n=-l \ \mathrm{mod} \ N}}^{\infty}\frac{q_{\tau}^{n}}{1-q_{\tau}^{n}}  e^{-inz} \right). 
\end{split}
\end{equation*}

Taking the limit $\tau\rightarrow i\infty$, the series in $q_{\tau}$ are truncated at the leading order. Moreover, since the vacua $-\frac{2\pi l \tau}{N} + \frac{2\pi k}{N} + \Lambda_{\tau}$ escape to infinity for large $\tau$, we have also to take $z \rightarrow  i\infty$. Thus, we obtain 

\begin{equation*}
\partial_{z}W^{(N,l)}(z;\tau) \xrightarrow{ \tau\rightarrow i\infty} -iN \left( q_{\tau}^{l } \ e^{ilz} - q_{\tau}^{(N-l)} \ e^{-i(N-l)z}  \right). 
\end{equation*}

Integrating this expression, we find 

\begin{equation}\label{limit}   
W^{(N,l)}(z;\tau) \xrightarrow{ \tau\rightarrow i\infty} -N\left( q_{\tau}^{l }\  \frac{e^{ilz}}{l} + q_{\tau}^{(N-l)} \ \frac{e^{-i(N-l)z}}{N-l}  \right) ,
\end{equation}

wich we recognize as the superpotential of a $\hat{A}_{N-1}$ model of co-level $l$. In particular, the case of $N=2$ corresponds to the Sinh-Gordon model 

\begin{equation}\label{sinhgordon}
W(z;\tau) \sim  q_{\tau} \cos z .
\end{equation}

Now let us consider the rationals. We associate to a cusp $\frac{a}{c}$ with $\mathrm{gcd}(a,c)=1$ a modular transformation $\gamma_{\frac{a}{c}}=\begin{pmatrix} a & b \\ c & d \end{pmatrix}$ which sends $i\infty$ to $\frac{a}{c}$. In this definition $b,d$ are integers such that $\gamma_{\frac{a}{c}} \in SL(2,\mathbb{Z})$ and clearly the case of $c=0$ corresponds to take again $\tau=i\infty$. One can study the behaviour of the model around $ \tau=\frac{a}{c}$ by acting on $ \partial_{z} W^{(N,l)}(z;\tau)$ with $\gamma_{\frac{a}{c}}$ and then taking the limit $\tau \rightarrow i\infty $. Using the modular properties \ref{weierstrass} of the zeta function, we have 

\begin{equation*}
\begin{split}
& \partial_{z} W^{(N,l)}\left( \frac{z}{c\tau + d};\frac{a\tau + b}{c\tau + d}\right) = \\ \\ &
(c\tau+d) \sum_{k=0}^{N-1} e^{\frac{2\pi i l k}{N}} \left[ \zeta \left( z-  \frac{2\pi}{N} k (c\tau + d) ;\tau \right) +2\frac{d \eta_{1}+c \eta_{2}}{N} k \right],
\end{split}
\end{equation*}

where the torsion point of the poles is now $\frac{2\pi}{N}(c\tau + d)$. Let us introduce the integers $Q= \mathrm{gcd} (c,N)$, with $1 \leq Q \leq \mathrm{min} \left\lbrace c,N \right\rbrace$, $j=\frac{N}{Q}$ and $r=\frac{c}{Q}$. Clearly we have $\mathrm{gcd} (r,j)=1$. By these definitions, we can split the sum over $k$ by writing $k=m+jp$, with $m=0,...,j-1$ and $p=0,...,Q-1$. Let us consider for the moment the cusps with divisor $Q>1$. One obtains

\begin{equation*}
\begin{split}
 & \partial_{z} W^{(N,l)}\left( \frac{z}{c\tau + d};\frac{a\tau + b}{c\tau + d}\right) = \\ \\ 
& (c\tau +d)\sum_{m=0}^{j-1} e^{\frac{2\pi i l m}{N}} \sum_{p=0}^{Q-1} e^{\frac{2\pi i l p}{Q}} \biggl[ \zeta \left( z-  \left( \frac{2\pi r}{j} \tau +\frac{2\pi d}{N} \right)m -\frac{2\pi}{Q} d p;\tau \right)  \\ \\ & + 2 \left( \frac{r\eta_{2} }{j} +  \frac{d\eta_{1} }{N}\right)m + \frac{2 d \eta_{1}}{Q}p \biggr].
\end{split}
\end{equation*}

As $\tau$ becomes very large, for $m\neq 0$ the torsion point approaches $\frac{2\pi r}{j}\tau$.\\  Moreover, by the formulas \cite{rif53}

\begin{equation*}
2\eta_{1}= \frac{G_{2}(\tau)}{2\pi}, \hspace{2 cm} 2\eta_{2}= \frac{\tau G_{2}(\tau)-2\pi i }{2\pi},
\end{equation*}

where $G_{2}(\tau)$ is the Eisenstein series 

\begin{equation*}
G_{2}(\tau)= \sum_{c ,d \in \mathbb{Z}\setminus \left\lbrace 0 \right\rbrace } \frac{1}{(c\tau +d)^{2}},
\end{equation*}

and the asymptotic behaviour 

\begin{equation}
G_{2}(\tau)\xrightarrow{ \tau\rightarrow i\infty} 2\zeta(2),
\end{equation}

where $\zeta(z)$ is the Riemann zeta function, one has

\begin{equation}
\frac{\eta_{2}}{\eta_{1}} \xrightarrow{ \tau\rightarrow i\infty}  \tau.
\end{equation}

Thus, we get the limit 

\begin{equation*}
\begin{split}
 & \partial_{z} W^{(N,l)}\left( \frac{z}{c\tau + d};\frac{a\tau + b}{c\tau + d}\right) \xrightarrow{ \tau\rightarrow i\infty}  
 (c\tau +d) \sum_{p=0}^{Q-1} e^{\frac{2\pi i l p}{Q}} \biggl[ \zeta \left( z -\frac{2\pi}{Q} d p;\tau \right)+ \frac{2d\eta_{1}}{Q} p \biggr]_{\tau \rightarrow i\infty} +    \\ \\ &  (c\tau +d)\sum_{m=1}^{j-1} e^{\frac{2\pi i l m}{N}} \sum_{p=0}^{Q-1} e^{\frac{2\pi i l p}{Q}} \biggl[ \zeta \left( z-   \frac{2\pi r}{j} \tau m ;\tau \right) + \frac{2 r\eta_{2}}{j}m \biggr]_{\tau \rightarrow i\infty} = \\ \\ & (c\tau +d) \sum_{p=0}^{Q-1} e^{\frac{2\pi i l p}{Q}} \biggl[ \zeta \left( z -\frac{2\pi}{Q} d p;\tau \right)+ \frac{2d\eta_{1}}{Q} p \biggr]_{\tau \rightarrow i\infty} = \\ \\ & (c\tau +d) \sum_{p=0}^{Q-1} e^{\frac{2\pi i al p}{Q}} \biggl[ \zeta \left( z -\frac{2\pi}{Q} p;\tau \right)+ \frac{2\eta_{1}}{Q}p \biggr]_{\tau \rightarrow i\infty},
\end{split}
\end{equation*}

where in the last line we have used the fact that $ad=1$ mod $Q$. So, we learn from this expression and the limit  \ref{limit} that the cusp $\frac{a}{c}$ with divisor $Q=\mathrm{gcd} (c,N)$ correspond to an $\hat{A}_{Q-1}$ model of co-level $al$:

\begin{equation}\label{ahat}
W^{(N,l)}(z;\tau) \xrightarrow{ \tau\rightarrow \frac{a}{c}} -Q\left( q_{\tau}^{al }\  \frac{e^{ialz}}{al} + q_{\tau}^{(Q-al)} \ \frac{e^{-i(Q-al)z}}{Q-al}  \right),
\end{equation}

which is a theory with $Q$ vacua up to periodicity $z \sim z+2\pi$.\\
In the case of $Q=1$, the poles $\frac{2\pi k}{N}(c\tau +d) + \Lambda_{\tau}$ are all pushed to infinity when $\tau$ becomes large except for $k=0$. Thus, using the \ref{new} we can write symbolically

\begin{equation*}
W \sim `` \log \Theta_{1}(z/2;\tau\rightarrow i\infty)  ",
\end{equation*}

and using the asymptotics 

\begin{equation*}
\Theta_{1}(z;\tau) \xrightarrow{ \tau\rightarrow i\infty} 2 q_{\tau}^{\frac{1}{8}}\sin z,
\end{equation*}

one finds that these cusps are decribed by the multi-valued superpotential

\begin{equation}\label{free}
W(z)=`` \log \sin \left( z/2 \right) ".
\end{equation}

This model represents the free version of our class of theories, with a $1$ dimensional lattice of poles and one of vacua.\\ We know that the cusps of $X_{1}(N)$ are ramification points of the cover $X(N)\rightarrow X_{1}(N)$. Denoting with $\Gamma_{\frac{a}{c}}$ the stability group of the cusp $a/c$ in $\Gamma_{1}(N)$, one can write the equality \cite{rif45}

\begin{equation}\label{cond}
 \gamma_{\frac{a}{c}}^{-1}\Gamma_{\frac{a}{c}}\gamma_{\frac{a}{c}}= \left\langle \pm \begin{pmatrix} 1 & h \\ 0 & 1 \end{pmatrix}     \right\rangle,
\end{equation} 

which is satisfied with one of the two signs. This relation means that the generator of $\Gamma_{\frac{a}{c}}$ is conjugated to $\pm \begin{pmatrix} 1 & h \\ 0 & 1 \end{pmatrix}$ in $ \gamma_{\frac{a}{c}}^{-1} \Gamma_{1}(N) \gamma_{\frac{a}{c}} $. \\ The number $h$ is called width of the cusp and represents the minimal integer such that $a/c +h \sim a/c$ in $\Gamma(N)$. The absolute value can be equal or less than $N$ and count the number of degenerate vacua of the model labelled by the cusp $\left[ \frac{a}{c} \right]$ of $\Gamma_{1}(N)$. The cusps which satisfy the relation with the plus sign are called regular. From the theorems \ref{width1}, \ref{width2}, the stability condition can be written as:

\begin{equation*}  
\begin{bmatrix} a+ch \\ c \end{bmatrix}= \begin{bmatrix} a \\ c \end{bmatrix} \ \mathrm{mod} \ N.
\end{equation*}

It is clear that the minimal integer $h$ satisfying this condition is $h=j=\frac{N}{Q}$. Thus, for the cusps with divisor $Q$, the $N$ vacua split in $j$ decoupled theories which appear on $X(N)$ as $Q$-degenerate points. These are described by the $\hat{A}_{Q-1}$ models \ref{ahat} for $1<Q\leq N$, or by the free theories \ref{free} if $Q=1$. In particular, the cusps with $Q=N$, or equivalently $c=0$ mod $N$, are the UV limits, since the vacua tend to a unique point on the spectral curve. \\ The exceptions to this picture are represented by the so called irregular cusps, i.e. those which satisfy the \ref{cond} with the mignus sign. In this case the stabilizer of the cusp belongs to $-\Gamma_{1}(N)$ and we have

\begin{equation*}
\begin{bmatrix} a+ch \\ c \end{bmatrix}= -\begin{bmatrix} a \\ c \end{bmatrix} \ \mathrm{mod} \ N.
\end{equation*}

If we exclude the trivial case of $N=2$ where $1\sim-1$ and require $a$ and $c$ to be coprime, we find that the stability condition is satisfied only by cusp $1/2$ for the curve of level $4$. This is known to be the unique irregular cusp for $\Gamma_{1}(N)$. Despite we have $Q=N/Q=2$, the width of the cusp is $h=1$, and the corresponding theory is actually a $\hat{A}_{3}$ model with $4$ vacua. The superpotential is the Sinh-Gordon one in \ref{sinhgordon} as for the cusps with divisor $2$, but we have to impose the identification $z \sim z+ 4\pi$.\\
Now we want to determine the positions of the cusps on the W-plane. Using the expression \ref{mass} for the critical value, we have 

\begin{equation*}
\begin{split}
W^{(N,l)}(0;\tau\rightarrow \frac{a}{c})=& \ W^{(N,l)}(0;\gamma_{\frac{a}{c}}(\tau\rightarrow i\infty))= 
\sum_{k=0}^{N-1} e^{\frac{2\pi i l k}{N}} \log E_{\frac{l}{N},\frac{k}{N}}(\gamma_{\frac{a}{c}}(\tau\rightarrow i\infty)) \\ \\=  &   \sum_{k=0}^{N-1} e^{\frac{2\pi i l k}{N}} \log E_{\big<\frac{al+ck}{N}\big>,\big<\frac{dk+bl}{N}\big>}(\tau\rightarrow i\infty) + \sum_{k=0}^{N-1} e^{\frac{2\pi i l k}{N}} \chi_{\frac{l}{N};\frac{k}{N}}(\gamma_{\frac{a}{c}}),
\end{split}
\end{equation*}

where $ \chi_{\frac{l}{N};\frac{k}{N}}(\gamma_{\frac{a}{c}})$ is given by the formula in \ref{formula}. \\ Let us consider the limit of the first piece. The leading order of $E_{u_{1},u_{2}}(\tau)$ for $\tau \rightarrow i\infty$ is 

\begin{equation*}
\mathrm{ord}_{i \infty} E_{u_{1},u_{2}}(\tau)= \frac{1}{2} B_{2}\left( \langle u_{1} \rangle \right) .
\end{equation*}

Therefore, we find 

\begin{equation*}
\begin{split}
W^{(N,l)}(0;\tau\rightarrow \frac{a}{c})= & \sum_{k=0}^{N-1} e^{\frac{2\pi i l k}{N}} \log E_{\big<\frac{al+ck}{N}\big>,\big<\frac{dk+bl}{N}\big>}(\tau)
\xrightarrow{ \tau\rightarrow i\infty}  \sum_{k=0}^{N-1} e^{\frac{2\pi i l k}{N}} \log q_{\tau}^ {\frac{1}{2} B_{2}\left( \big<\frac{al+ck}{N} \bigr> \right) }  \\ \\ =& \log q_{\tau}^{K^{(N,l)}_{\frac{a}{c}}},
 \end{split}
\end{equation*}

where 

\begin{equation*}
{K^{(N,l)}_{\frac{a}{c}}}= \frac{1}{2}\sum_{k=0}^{N-1} e^{\frac{2\pi i l k}{N}} B_{2}\left( \bigg<\frac{al+ck}{N}\bigg> \right). \hspace{2cm} 
\end{equation*}

Let us develop this expression. Using the Fourier expansion of the second Bernoulli periodic polynomial 

\begin{equation*}
B_{2}(\langle x \rangle)= -\frac{2!}{(2\pi i)^{2}} \sum_{\overset{m=-\infty}{m \neq 0}}^{\infty} \frac{e^{2\pi i m x}}{m^{2}},
\end{equation*}

we get 

\begin{equation*}
\begin{split}
{K^{(N,l)}_{\frac{a}{c}}}= & -\frac{1}{(2\pi i)^{2}} \sum_{k=0}^{N-1} e^{\frac{2\pi i l k}{N}} \sum_{\overset{m=-\infty}{m \neq 0}}^{\infty} \frac{e^{2\pi i m \left( \frac{al+ck}{N}\right) }}{m^{2}} \\ \\  = & -\frac{1}{(2\pi i)^{2}} \sum_{\overset{m=-\infty}{m \neq 0}}^{\infty} \frac{e^{2\pi i m \frac{al}{N}}}{m^{2}} \sum_{k=0}^{N-1} e^{\frac{2\pi i k}{N}(l+mc)}.
\end{split}
\end{equation*}

The sum $\sum_{k=0}^{N-1} e^{\frac{2\pi i k}{N}(l+mc)} $ is not vanishing if and only if $l+mc=0$ mod $N$, which admits solution only when $c$ is coprime with $N$. Thus, we have 

\begin{equation*}
{K^{(N,l)}_{\frac{a}{c}}}= \begin{cases} 
-\frac{N}{(2\pi i)^{2}}^{-\frac{2\pi i a l^{2}r}{N}}  \sum\limits_{\overset{m=-\infty}{m= -lr \ \mathrm{mod} \ N}}^{\infty} \frac{1}{m^{2}} \neq 0, \ \ \ \mathrm{if} \ \mathrm{gcd}(c,N)=1, \\ \\ 
0, \ \ \  \mathrm{otherwise}, \end{cases}
\end{equation*}

where $r= c^{-1}$ mod $N$. So, we learn that if $\mathrm{gcd}(c,N)=1$ the cusp order is not vanishing and therefore the critical value is divergent. Coherently with our analysis, these are the IR fixed points described by free theories. In this limits all the vacua decouple and the solitons connecting them become infinitely massive. On the contrary, the cusps with $1< Q \leq N $ have a finite coordinate on the W-plane:

\begin{equation*}
W^{(N,l)} \left( 0;\frac{a}{c}\right)=\sum_{k=0}^{N-1} e^{\frac{2\pi i l k}{N}} \chi_{\frac{l}{N};\frac{k}{N}}(\gamma_{\frac{a}{c}}).
\end{equation*}

In the case of $c=0$ mod $N$ this becomes 

\begin{equation*}
W^{(N,l)} \left( 0;\frac{a}{c}\right)^{\mathrm{UV}}= \Delta W^{(N,l)}_{\Gamma_{0}(N)}(\gamma_{\frac{a}{c}}).
\end{equation*}

In particular, the cusp $\tau= i \infty$ is located at the origin and provides the boundary condition for the critical limit $\mu\rightarrow 0$. \\ It is clear from the last formula that the UV cusps are all in the same orbit of the Galois group of the modular curve. Counting the equivalence classes of $\Gamma_{1}(N)$, these models are labelled by the co-levels $\pm l$ and their number is equal to $\phi(N)/2$. From this point of view, choosing the co-level is equivalent to pick which UV cusp to put in the origin. Also the free IR cusps are all in the same orbit of the Galois group, since we can map $\tau=0$ to a generic rational $a/c$ such that $\mathrm{gcd}(c,N)=1$ with a matrix of $\Gamma_{0}(N)$. Instead, the other IR cusps with divisor $1< Q < N$ can split in different equivalence classes of $\Gamma_{0}(N)$. In general the Galois group maps a cusp $a/c$ with $\mathrm{gcd}(c,N)=Q$ and $\mathrm{gcd}(a,Q)=1$ to another cusp $\frac{a^{\prime}}{c^{\prime}}$ with $\mathrm{gcd}(c^{\prime},N)=Q$ and $\mathrm{gcd}(a^{\prime},Q)=1$. A computation in \cite{rif45} shows that for a given divisor $Q$ we have $\phi(\mathrm{gcd}(Q,N/Q))$ cusps of $\Gamma_{0}(N)$.

\subsubsection{Boundary Conditions}

In this last section we provide the boundary conditions for the $tt^{*}$ equations and describe the solution around the cusps. Approaching a critical point, the solution has to match the asymptotic behavour required by the physics of the corresponding cusp. The deformations in the space of couplings near these points regard the overall parameter $\mu$ rescaling the superpotential. Near the UV fixed point the ground state metric can be diagonalized in a basis of vacua with definite Ramond charge:

\begin{equation*}
g_{i\bar{i}} \xrightarrow{ \mu \rightarrow 0} (\mu\bar{\mu})^{-q_{i}^{R}-n/2 },
\end{equation*}

where $n$ is the complex dimension of the target manifold. In the present case we have $n=1$. So, the solution to the $tt^{*}$ equations near the critical point is given in terms of the Ramond charges of the vacua, or equivalently the charges of the vector R-symmetry. These are given by the scaling dimensions of the chiral primary operators of the CFT.\\ Let us start with the cusps corresponding to $\hat{A}_{N-1}$ models. These are Landau-Ginzburg theories with superpotential \cite{rif12}

\begin{equation*}
W^{(N,l)}(z;t)= \mu \left( \frac{e^{-lz}}{l} +  \frac{e^{(N-l)z}}{N-l}\right) ,
\end{equation*}

and $\mathrm{gcd}(l,N)=1$. These are integrable models with a $\mathbb{Z}_{N}$ symmetry generated by $\sigma: z\rightarrow z + \frac{2\pi i}{N}$ and $N$ vacua, provided the periodic identification $z \sim z+2\pi i$. The symmetry acts transitively on the critical points, which are given by the condition $e^{Nz}=1$. In the UV limit these theories tend to $\sigma$-models over abelian orbifolds of $\mathbb{CP}^{1}$. These are known to be asymptotically free theories with central charge $\hat{c}=1$ \cite{rif12}. The $tt^{*}$ equations in canonical form are the Toda equations in \ref{toda} with vanishing $\beta$. Indeed, in this limit the unique non trivial homology operator is $A=\sigma^{N}$ and the $U(1)$ charges defining the solution can depend only on $\alpha$. Since $U(1)$ is broken by the superpotential to $\mathbb{Z}_{N}$, it is clear that the generators of the two symmetry groups have a common basis of eigenstates. Let us first consider the case of $\alpha=0$. A basis of $U(1)$ eigenstates in the chiral ring can be generated with the operators $e^{-z}, e^{z}$. Since the superpotential must have R-charge $1$, these have respectively charge $\frac{1}{l}$ and $\frac{1}{N-l}$. Given that $e^{-lz}= e^{(N-l)z}$ in the chiral ring from the vacua condition, we find that the set of eigenstates split in two `towers'

\begin{equation*}
\begin{split}
 & e^{-z} \hspace{1cm} e^{-2z} \hspace{1cm}\cdot \hspace{1cm} \cdot \hspace{1cm} \cdot \hspace{1cm} \cdot \hspace{1cm} \cdot \hspace{1cm} e^{-(l-1)z} \\ 
&  e^{z} \hspace{1.2cm} e^{2z} \hspace{1.25cm} \cdot \hspace{1cm} \cdot \hspace{1cm} \cdot \hspace{1cm} \cdot \hspace{1cm} \cdot \hspace{1cm} e^{(N-l-1)z}  
\end{split}
\end{equation*}

with $U(1)_{V}$ charges 

\begin{equation*}
\begin{split}
 & \hspace{0.45cm} \frac{1}{l} \hspace{1.8cm} \frac{2}{l}\hspace{1.5cm} \cdot \hspace{1cm} \cdot \hspace{1cm} \cdot \hspace{1cm} \cdot \hspace{1cm} \cdot \hspace{1.5cm} \frac{l-1}{l} \\ \\ 
&  \frac{1}{N-l} \hspace{1cm} \frac{2}{N-l}\hspace{1.13cm} \cdot \hspace{1cm} \cdot \hspace{1cm} \cdot \hspace{1cm} \cdot \hspace{1cm} \cdot \hspace{1cm} \frac{N-l-1}{N-l}.
\end{split}
\end{equation*}

Approximately, we can say that the theory splits in two, with a set of operators dominant on the other one according to how we take the limit. We complete the basis by adding the identity $I$ and $e^{-lz}$, which have respectively charge $0$ and $1$. Near the critical point these two operators correspond to a unique marginal degree of freedom which gets a logarithmic correction to the scaling \cite{rif12}. We point out that in this language the Galois group acts directly on the $U(1)$ charges with the map $l\rightarrow al$ and puts in relation the solutions of the different $\hat{A}_{N-1}$ models. To see that the set of charges is invariant under this map we have to use the chiral ring condition $e^{Nz}=1$. The operatorial equality $e^{-kz}=e^{(N-k)z}$ for a generic $k \in \mathbb{Z}$ implies the equivalence $\frac{k}{l} \sim \frac{N-k}{N-l}$ at the level of corresponding charges. In general, the integer $k$ and the co-level $l$ are periodic of $N$ in the chiral ring. So, one can recast all the charges above as $q_{k}=\frac{k}{l}$, $k=0,...,N-1$ and write the action of the Galois group as $q_{k}\rightarrow\frac{k}{al}$. The relation $\frac{k}{l}=\frac{ak}{al}\sim\frac{k^{\prime}}{al}$, with $k^{\prime}= ak$ mod $N$, shows that the set of charges is left invariant by this map.\\ We can include the dependence from the angle $\alpha$ by multiplying the basis above by $e^{\frac{\alpha}{2\pi}z}$. In this way the operators have the correct eigenvalues under $\mathbb{Z}_{N}$ when $\alpha \neq 0$. The $U(1)$ charges as functions of the angle are 

\begin{equation*}
\begin{split}
 & \frac{1}{l}\left( k-\frac{\alpha}{2\pi}\right) , \ \ \ k=1,...,l-1, \\ \\ 
  & \frac{1}{N-l}\left( k+\frac{\alpha}{2\pi}\right), \ \ \ k=1,...,N-l-1, \\ \\ 
 & \frac{\alpha}{2\pi(N-l)}, \hspace{1cm} 1-\frac{\alpha}{2\pi l} .
\end{split}
\end{equation*}

It is clear from \ref{metrictransformation} that for $\beta=0$ a transformation of $\Gamma_{0}(N)$ does not change the dependence on $\alpha$ of the metric components. This can be seen at the level of charges by the fact that the map $l\rightarrow al$ is compensated by the rescaling of the angle $\alpha\rightarrow\alpha/d$.\\ We note further that, since $\beta$ is vanishing, the UV cusps turn out to be fixed points of $\Gamma(N)$. This is consistent with the fact that $A$ is the unique generator of the homology in this regime. \\ The irregular cusp $1/2$ of the modular curve of level $4$ is described by the superpotential

\begin{equation*}
W(z)= \mu \left( e^{2z} + e^{-2z} \right) 
\end{equation*}

with the identification $z \sim z+2\pi i$. This theory has a $Z_{2}$ symmetry generated by $\sigma: z\rightarrow z+\frac{i\pi}{2}$, but $4$ vacua determined by the condition $e^{2z}=e^{-2z}$. This model belongs to $\hat{A}_{3}$ family and is asymptotically a $\sigma$-model on the $\mathbb{CP}^{1}/\mathbb{Z}_{2} $ orbifold. The $tt^{*}$ equations are the Toda ones with $N=4$ and a basis of $U(1)$ eigenstates is given by 

\begin{equation*}
 e^{\frac{\alpha}{2\pi}z} \hspace{1cm} e^{\left( 1+\frac{\alpha}{2\pi} \right) z} \hspace{1cm} e^{-\left( 1-\frac{\alpha}{2\pi}\right) z} \hspace{1cm} e^{-\left( 2-\frac{\alpha}{2\pi} \right) z}
\end{equation*}

with charges respectively

\begin{equation*}
\frac{\alpha}{4\pi} \hspace{1cm} \frac{1}{2}\left(  1+\frac{\alpha}{2\pi}\right) \hspace{1cm} 
\frac{1}{2}\left(  1-\frac{\alpha}{2\pi} \right) \hspace{1cm} \frac{1}{2}\left(  2-\frac{\alpha}{2\pi} \right) . 
\end{equation*}

We conclude by saying that the solution of the $tt^{*}$ equation is singular in the UV cusps:

\begin{equation}
\varphi_{i}(t;\alpha) \xrightarrow{t\rightarrow 0} -2\left( q_{i}(\alpha) -\frac{1}{2}\right) \log t.
\end{equation}

A solution in terms of regular trascendents can be given only on the upper half plane, which is a simply connected space.\\ 
The discussion for the $\hat{A}_{Q-1}$ models for $1< Q < N$ is pretty much the same of the previous paragraph. So, we focus on the free massive theories corresponding to the case of $Q=1$. These IR cusps are Landau-Ginzburg models described by the derivative

\begin{equation}
\partial_{z}W(z;\tau)= \mu \cot\left(  \frac{z}{2}  \right). 
\end{equation}

This function is periodic of $2\pi$ and has simple poles and simple zeroes respectively in $ 2k\pi $ and $\pi + 2k\pi$, $\k \in \mathbb{Z}$. Moreover, it is odd with respect to the parity transformation $\iota: z\rightarrow -z$. Since the target space is not simply connected we need to pull-back the model on the abelian universal cover. A natural basis for the homology is given by the cycles $B,B^{\prime}$ in figure \ref{algebr}. From the residue formula and the parity properties of $\partial_{z}W(z)$ one gets the transformations of the superpotential

\begin{equation}
\begin{split}
& B^{*}W(p)= W(p) - 2\pi i \mu, \\ 
& B^{\prime *}W(p)= W(p) + 2\pi i \mu.
\end{split}
\end{equation}

Proceeding as in \ref{point} we can construct the unique theta-vacua of this theory: 

\begin{equation}
\ket{\phi, \psi}= \sum_{n,m \in \mathbb{Z}} e^{-i(m\phi + n\psi)} B^{m}B^{\prime n} \ket{0},
\end{equation}

where we denote with $\ket{0}$ some vacuum state of the covering model. Setting to $0$ the corresponding critical value, the whole set is simply

\begin{equation}
W_{n,m}= 2\pi i \mu (n-m).
\end{equation}

We want to derive the $tt^{*}$ equation in the parameter $\mu$. The chiral ring operator $C_{\mu}(\phi, \psi)$ acts on the theta-vacuum as differential operator in the angles 

\begin{equation}
C_{\mu}\ket{\phi, \psi}=  \sum_{n,m \in \mathbb{Z}} e^{-i(m\phi + n\psi)}2\pi i(n-m) B^{m}B^{\prime n} \ket{0}=
2\pi \left( \frac{\partial}{\partial \phi}-\frac{\partial}{\partial \psi}\right)  \ket{\phi, \psi}.
\end{equation}

We define the ground state metric 

\begin{equation}
g(t,\phi, \psi)= \braket{\overline{\phi, \psi}}{\phi, \psi} = e^{L(t ,\phi,\psi)},
\end{equation}

where $L(t,\phi,\psi)$ is a real function of the angles and the RG scale $t=\vert \mu \vert$. We can normalize the state so that the topological metric is $1$. Thus, the reality constraint implies

\begin{equation}\label{reality}
\begin{split}
& L(-\phi,-\psi)=-L(\phi,\psi). 
\end{split}
\end{equation}

Moreover, by the commutation relations 

\begin{equation}
\begin{split}
& \iota B=B^{\prime -1}\iota, \\ \\ 
& \iota B^{\prime}=B^{-1} \iota, 
\end{split}
\end{equation}

we have also

\begin{equation}\label{parity}
L(-\psi,-\phi)=L(\phi,\psi).
\end{equation}

The $tt^{*}$ equation for $g(t,\phi, \psi)$ reads

\begin{equation}
\left( \partial_{\mu}\partial_{\bar{\mu}}+ 4\pi^{2}\left( \frac{\partial}{\partial \phi}-\frac{\partial}{\partial \psi}\right) ^{2}\right)  L(t,\phi,\psi)=0.
\end{equation}

We recognize in this expression the equation of a $U(1)$ Bogomolnyi monopole on $\mathbb{R}\times T^{2}$. Abelian $tt^{*}$ monopoles have been studied in \cite{rif13,rif23}. The solution can be expanded in Bessel-MacDonald functions as 

\begin{equation}
L(t,\phi,\psi)= \sum_{ m_{1},m_{2} \in \mathbb{Z} \setminus \left\lbrace 0 \right\rbrace } A(m_{1},m_{2}) K_{0}\left( 4\pi t \vert m_{1}+m_{2} \vert \right)  \exp \left(  i \left(  m_{1}\phi-m_{2}\psi \right)  \right),
\end{equation}

where the coefficients $A(m_{1},m_{2})$ can be determined by imposing appropriate boundary conditions.
One can easily see that the $tt^{*}$ reality constraint \ref{reality} implies

\begin{equation}
A(-m_{1},-m_{2})=- A(m_{1},m_{2}), \hspace{1cm} A(m_{1},m_{2}) \in i \mathbb{R},
\end{equation}

while the parity condition \ref{parity} requires 

\begin{equation}
A(m_{2},m_{1})=A(m_{1},m_{2}).
\end{equation}

Combining these two conditions one gets the further constraint

\begin{equation}
L(t, \psi, \phi)=-L(t,\phi,\psi).
\end{equation}

According to the discussion in section \ref{trunc}, in order to have the abelianity of the solution one should consider trivial representations of the loop generator. If we demand the loop angle to vanish, namely $\phi=\psi$,  we simply find the trivial solution 

\begin{equation}
g(t,\phi, \phi)=1.
\end{equation}

\subsection{Conclusions}

In this chapter we have shown how the $tt^{*}$ geometry of the modular curves is rich of interesting phenomena and outstanding connections between geometry, number theory and physics. These Riemann surfaces parametrize a family of supersymmetric FQHE models in which the usual setting degenerates in a doubly periodic physics on the complex plane. In the subclass of theories of level $N$, the elliptic functions playing the role of superpotentials have $N$ vacua and $N$ poles in the fundamental cell, with the corresponding residues which add up to zero by definition. The cancellation of the total flux between the magnetic field and the quasi-holes guarantees the enhancement of symmetry that makes possible to face analitically these models. In particular, the presence of an abelian symmetry group with a transitive action on the vacua allows to diagonalize the ground state metric, as well as to find the necessary topological data to write the $tt^{*}$ equations. This requires to pull-back the model on the abelian universal cover of the target manifold, where we have seen that the physics is non-abelian. On this space the symmetry group is enlarged with the generators of loops around the poles, which are responsible for the non trivial commutation relations between the generators of the algebra. Hovewer, the abelianity that we have required in the classification can be recovered at the quantum level. In particular, the ansatz of a solution with vanishing loop angles is consistent with all the $tt^{*}$ equations, which can be recasted as Toda equations in the canonical coordinates. \\ Studying the modular properties of these models, we have underlined that the non trivial modular transformations of the superpotential are a natural consequence of the geometry of the modular curves. A critical value as coordinate on the spectral cover can be defined only on the upper half plane, since the F-term variations are rational functions in projective coordinates on the modular curves. This has been studied in the easiest cases of the platonic solids inscribed in the Riemann sphere, but for surfaces of higher genus it is more convenient to parametrize the critical value in terms of the fundamental units of the modular function field. The congruence subgroups have a not trivial effect also on the components of the ground state metric, since they change the representation of the abelian symmetry group.\\ The known results and theorems about the cusps counting and classification have been recovered in a physical language when we have classified the critical limits of this family of theories. One of the main point is that the width of a cusp allows to determine the UV or IR nature of the corresponding RG fixed point.\\  Our investigation has also revealed the algebraic properties of the modular curves. As we pointed out, the most remarkable connection with number theory is that the Galois group of the real cyclotomic extensions acts on the regularity conditions of the $\hat{A}_{N-1}$ Toda equations. This follows from the fact that the $\hat{A}_{N-1}$ models play the role of UV critical limits and belong to the same orbit of the Galois group.

\section{Modular Transformations of $\log E_{u_{1},u_{2}}(\tau)$}\label{modulartrnasofrmationmass}

In section \ref{modphysmass} we have setted the notations

\begin{equation*}
\begin{split}
 &  q_{\tau}= e^{2\pi i \tau}, \hspace{2cm} q_{z}= e^{2\pi i z}, \\ \\ & \hspace{1.3 cm} z= u_{1} \tau+ u_{2}, 
\end{split}
\end{equation*}

with $ u_{1},u_{2} \in \mathbb{Z}/ N $, and defined the modular units 

\begin{equation}\label{expan}  
E_{u_{1},u_{2}}(\tau)= q_{\tau}^{B_{2}(u_{1})/2} (1-q_{z}) \prod _{n=1}^{\infty} (1-q_{\tau}^{n}q_{z})(1-q_{\tau}^{n}/q_{z}),
\end{equation}

which are the Siegel functions up to the root of unity $e^{2\pi i u_{2}(u_{1}-1)/2}$. These objects satisfy \cite{rif48}

\begin{equation}
E_{u_{1}+1,u_{2}}(\tau)=-e^{-2\pi i u_{2}} E_{u_{1},u_{2}}(\tau), \hspace{2cm} E_{u_{1},u_{2}+1}(\tau)=E_{u_{1},u_{2}}(\tau),
\end{equation}

and transform under $\gamma=  \begin{pmatrix}   a & b \\ c & d \end{pmatrix} \in SL(2,\mathbb{Z})$ as 

\begin{equation}\label{prop1}
\begin{split}
& E_{u_{1},u_{2}}(\tau+b)=  e^{\pi i b B_{2}(u_{1})} E_{u_{1},u_{2}+b u_{1}}(\tau) , \ \ \ \ \mathrm{for } \ c = 0 , \\ \\ 
& E_{u_{1},u_{2}}(\gamma (\tau))= \varepsilon (a,b,c,d) e^{\pi i \delta} E_{u_{1}^{\prime},u_{2}^{\prime}}(\tau) , \ \ \ \ \mathrm{for } \ c\neq 0,
\end{split}
\end{equation}

where 

\begin{equation}
\varepsilon(a,b,c,d)= 
   \begin{cases}
  e^{i\pi (bd(1-c^{2})+c(a+d-3))/6}, \ \ \ \ \mathrm{if} \ c \ \mathrm{is \ odd},
   \\  -ie^{i\pi (ac(1-d^{2})+d(b-c+3))/6}, \ \ \ \ \mathrm{if} \ d \ \mathrm{is \ odd},
   \end{cases}
   \end{equation}
   
 \begin{equation*}
  \delta=  u_{1}^{2}ab + 2 u_{1}u_{2}bc + u_{2}^{2}cd - u_{1}b-u_{2}(d-1),
\end{equation*}

and 

\begin{equation}  
u_{1}^{\prime}= a u_{1}+c u_{2}, \hspace{1cm}  u_{2}^{\prime}= b u_{1}+d u_{2}.
\end{equation}

With these definitions, we want to compute the difference 

\begin{equation}
\chi_{u_{1},u_{2}}(\gamma)=\log E_{u_{1},u_{2}}(\gamma (\tau))-\log E_{u_{1}^{\prime},u_{2}^{\prime}}(\tau),
\end{equation}

for $\gamma \in SL(2,\mathbb{Z})$ and generic characters $u_{1},u_{2} \in \mathbb{Z}/N$. From \ref{prop1} we know that there is a power of $E_{u_{1},u_{2}}(\gamma (\tau))/E_{u_{1}^{\prime},u_{2}^{\prime}}(\tau)$ which is equal to one. This number is $12 N$ for $\Gamma(N)$ and $12 N^{2}$ for $\Gamma_{1}(N),\Gamma_{0}(N)$ and the whole $SL(2,\mathbb{Z})$. Therefore, the difference $\chi_{u_{1},u_{2}}(\gamma)$  must be equal to $2\pi i$ times a rational number. Given that the upper half plane is simply connected, this number is independent of $\tau$. Moreover, since $\log E_{u_{1},u_{2}}(\tau) $ is single-valued on the upper half plane, it is also indipendent from the branch of the logarithm. A natural choice, suggested by the $q$-expansion of the Siegel functions, is the principal branch on $\mathbb{C}$ with the negative real axis deleted. From now on we will use this determination. Because $E_{u_{1},u_{2}}(\tau)$ changes by a phase under an integer shift of the characters, we can  assume without loss of generality the canonical normalization  $0 < u_{1},u_{2},u_{1}^{\prime},u_{2}^{\prime} < 1$. \\ Let us first consider the case with $c=0$. These transformations belong to the coset group $\Gamma_{1}(N)/\Gamma(N) \simeq \mathbb{Z}_{N}$ and are generated by $\gamma (\tau)= \tau+1$. Using the expansion of the Siegel function in \ref{expan} we easily obtain 

\begin{equation}
\chi_{u_{1},u_{2}}(\gamma)= 2\pi i \frac{1}{2}B_{2}(u_{1}).
\end{equation}

From now on we assume $c\neq 0$ and write $\gamma(\tau)= \frac{a\tau +b}{c\tau +d}= \frac{a}{c}-\frac{1}{c^{2}\tau + cd}$. \\ Using again the \ref{siegel} we have 

\begin{equation*}
 \log E_{u_{1},u_{2}}(\tau)= 2\pi i B_{2}(u_{1}) \tau + \log(1-q_{z}) + \sum_{n=1}^{\infty} \left( \log(1-q_{\tau}^{n}q_{z})+
 \log(1-q_{\tau}^{n}/q_{z})\right).
\end{equation*}

With $\tau$ in the upper half plane and the characters canonically normalized, the conditions of absolute convergence for the standard series of the principal logarithm are satisfied. Therefore, using series expansions like

\begin{equation*}
\log(1-q_{z})=-\sum_{m=1}^{\infty} \frac{q_{z}^{m}}{m}
\end{equation*}

for the logarithms in the expression, we obtain 

\begin{equation*}
 \log E_{u_{1},u_{2}}(\tau)= 2\pi i \frac{1}{2}B_{2}(u_{1})\tau - \mathcal{Q}(z;\tau),
\end{equation*}

where 

\begin{equation*}
\mathcal{Q}(z;\tau)= \sum_{m=1}^{\infty} \frac{1}{m} \frac{q_{z}^{m}+ (q_{\tau}/q_{z})^{m}}{1-q_{\tau}^{m}}.
\end{equation*}

Then, let us put 

\begin{equation*}
\tau= -\frac{d}{c}+i y, \ \mathrm{with} \ y>0, \hspace{2cm} \gamma(\tau)= \frac{a}{c} + \frac{i}{c^{2}y}.
\end{equation*}

Since it is indipendent of $\tau$, we can calculate $ \chi_{u_{1},u_{2}}(\gamma)$ in the limit $y\rightarrow 0$, i.e. $\tau \rightarrow -\frac{d}{c}$ and $\gamma(\tau) \rightarrow i \infty$, by applying the Abel limit formula. 
Setting 

\begin{equation*}
z_{\gamma}= u_{1} \gamma(\tau)+u_{2}, \hspace{2cm} z^{\prime}= u_{1}^{\prime}\tau + u_{2}^{\prime},
\end{equation*}

and keeping only the immaginary parts, since $ \chi_{u_{1},u_{2}}(\gamma)$ is pure immaginary, we have to evaluate the expression 

\begin{equation}
\begin{split}
 \chi_{u_{1},u_{2}}(\gamma)=&\  2\pi i \frac{1}{2} \left( B_{2}(u_{1}) \frac{a}{c} +   B_{2}(u_{1}^{\prime}) \frac{d}{c}   \right) 
 \\ \\ &- \lim_{\tau\rightarrow -\frac{d}{c}}\left(  \mathrm{Im} \mathcal{Q}(z_{\gamma};\gamma({\tau}))-  \mathrm{Im} \mathcal{Q}(z^{\prime};\tau)\right) .
 \end{split}
\end{equation}

Let us start with $\mathrm{Im} \mathcal{Q}(z_{\gamma};\gamma({\tau}))$. As $\gamma(\tau) \rightarrow i \infty$, $q_{z_{\gamma}}$ and $q_{\gamma(\tau)}/q_{z_{\gamma}}$ approach $0$, therefore

\begin{equation*}
\lim_{\tau\rightarrow -\frac{d}{c}} \  \mathrm{Im} \mathcal{Q}(z_{\gamma};\gamma({\tau}))=0.
\end{equation*}

Now it is the turn of  $\mathcal{Q}(z^{\prime};\tau)$. We can decompose it in two pieces :

\begin{equation*}
\begin{split}
\lim_{\tau\rightarrow -\frac{d}{c}}\ \mathrm{Im}  \mathcal{Q}(z^{\prime};\tau)= & \ \lim_{\tau\rightarrow -\frac{d}{c}} \ \mathrm{Im} \sum_{c \ \nmid \ m} \frac{1}{m} \mathcal{Q}_{m}(z^{\prime};\tau) \  +  \ \lim_{\tau\rightarrow -\frac{d}{c}} \ \mathrm{Im} \sum_{c \mid m} \frac{1}{m}\mathcal{Q}_{m}(z^{\prime};\tau) \\ = & \ L^{\prime} +  L^{\prime \prime} ,
\end{split}
\end{equation*}

where 

\begin{equation*}
\mathcal{Q}_{m}(z^{\prime};\tau)=\frac{q_{z^{\prime}}^{m}+ (q_{\tau}/q_{z^{\prime}})^{m}}{1-q_{\tau}^{m}}.
\end{equation*} 

The symbols $L^{\prime}$ and $L^{\prime \prime}$ denote the sum respectively for $c \nmid m$ and $c \mid m $. We introduce

\begin{equation*}
r= e^{-2\pi y}, \hspace{1.5cm} M=N\vert c \vert, \hspace{1.5cm} \zeta= e^{-2\pi i d/c}, \hspace{1.5cm} \lambda= e^{2\pi i\left(  -\frac{d}{c}u_{1}^{\prime} + u_{2}^{\prime}   \right)}. 
\end{equation*}

It is shown in \cite{rif49,rif50} that the partial sums of these series are uniformly bounded. Therefore, we are allowed to take the limit under the sign of summation. Let us consider first $L^{\prime \prime}$. Using the notation above and taking the immaginary part, we have 

\begin{equation*}
L^{\prime \prime}= \lim_{r\rightarrow 1} \sum_{ c \mid m} \frac{r^{u_{1}^{\prime}m}-r^{(1-u_{1}^{\prime})m}}{1-r^{m}} \frac{1}{2m}(\lambda^{m}-\lambda^{-m}). 
\end{equation*}

Taking the limit under the summation sign, one gets

\begin{equation*}
\begin{split}
L^{ \prime \prime }= & \sum_{ c \mid m}  (1-2u_{1}^{\prime}) \frac{\lambda^{m}-\lambda^{-m}}{2m}= \sum_{m=1}^{\infty}(1-2 u_{1}^{\prime}) \frac{1}{2\vert c \vert m}( \lambda^{\vert c \vert m}-\lambda^{-\vert c \vert m}) \\ \\  = &
\ (1-2u_{1}^{\prime})\frac{1}{2 \vert c \vert}\sum_{m=1}^{\infty}\frac{1}{m}\left(  e^{2\pi i (-d \varepsilon(c)u_{1}^{\prime}+ u_{2}^{\prime} \vert c \vert )m} - e^{-2\pi i (-d \varepsilon(c)u_{1}^{\prime}+ u_{2}^{\prime} \vert c \vert )m} \right),
\end{split}
\end{equation*}

where $ \varepsilon (c)= \vert c \vert / c $. If $t$ is real and not integer, it holds the Fourier expansion 

\begin{equation*}
\sum_{m=1}^{\infty} \frac{1}{m}(e^{2\pi i m t}- e^{-2\pi i m t})= -2\pi i B_{1}\left( \langle t \rangle \right) ,
\end{equation*}

where  $ B_{1}(x)=x-\frac{1}{2}$ is the first Bernoulli polynomial. Thus 

\begin{equation*}
L^{ \prime \prime }= -2\pi i (1-2u_{1}^{\prime})\frac{1}{2 \vert c \vert} B_{1}\left( \langle -d \varepsilon(c)u_{1}^{\prime}+ u_{2}^{\prime} \vert c \vert \rangle\right) = -\frac{2\pi i}{c} B_{1}\left( u_{1}^{\prime}\right)   B_{1}\left(  \langle d u_{1}^{\prime}- u_{2}^{\prime} c \rangle \right) . 
\end{equation*}

Now we turn to the last piece $L^{\prime}$. Taking the limit under the summation sign, we obtain 

\begin{equation*}
L^{\prime}=\lim_{\tau\rightarrow -\frac{d}{c}} \ \mathrm{Im} \sum_{c \ \nmid \  m} \frac{1}{m}\mathcal{Q}_{m}(z^{\prime};\tau) = \sum_{c \ \nmid \ m} \frac{1}{m} \varphi (m),
\end{equation*}

where 

\begin{equation*}
\varphi(m)= \mathcal{Q}_{m}(z^{\prime};\tau)\vert_{\tau= -d/c} = \frac{\lambda^{m}+ (\zeta/\lambda)^{m}}{1-\zeta^{m}}.
\end{equation*}

Since $ \varphi(-m)=-\varphi(m)= \overline{\varphi(m)}$, we note that $\varphi(m)$ is pure immaginary and an odd function of $m$ mod $M=N \vert c \vert$. Now, for each class $x \in \mathbb{Z}/M\mathbb{Z}$ and $2x \not \in M\mathbb{Z}$, we define  

\begin{equation}
f(x)= \sum_{m=1}^{\infty} \frac{a(m,x)}{m}
\end{equation}

where

\begin{equation}
a(m,x)= \begin{cases}
0 \hspace{0.8cm} \mathrm{if} \  m \neq \pm x \ \mathrm{mod} \ M
\\ 1  \hspace{0.8cm} \mathrm{if} \ m = x \ \mathrm{mod} \ M
\\ -1  \hspace{0.5cm} \mathrm{if} \ m =- x \ \mathrm{mod} \ M.
\end{cases}
\end{equation}

Then, $L^{\prime}$ can be rewritten as 

\begin{equation}
L^{\prime}= \frac{1}{2}\sum_{\overset{x \in \mathbb{Z}/M\mathbb{Z},}{ \overset{\ x \neq 0 \ \mathrm{mod} \ c\mathbb{Z},}{ \ 2x \not \in M\mathbb{Z}} }}\varphi(x) f(x).
\end{equation}

In \cite{rif49} is shown that 

\begin{equation}
f(x)= \frac{-i \pi}{M}\left[  \frac{1}{1-e^{2\pi i x/M}} - \frac{1}{1-e^{-2\pi i x/M}} \right].
\end{equation}

Let $\omega= e^{2\pi i /N \vert c \vert}$. Using this expression $L^{\prime}$ becomes 

\begin{equation*}
\begin{split}
L^{\prime}= & \frac{-\pi i}{2M}\sum_{c \  \nmid \ x}\frac{\lambda^{x}+(\zeta/\lambda)^{x}}{1-\zeta^{x}}\left[  \frac{1}{1-\omega^{x}}-\frac{1}{1-\omega^{-x}} \right] \\ \\  
= & \frac{-\pi i}{2M}\sum_{c \ \nmid \ x}   \biggl[ \frac{\lambda^{x}}{(1-\zeta^{x})(1-\omega^{x})} +  \frac{(\zeta / \lambda)^{x}}{(1-\zeta^{x})(1-\omega^{x})}  \\ \\ & -  \frac{\lambda^{x}}{(1-\zeta^{x})(1-\omega^{-x})} - \frac{(\zeta / \lambda)^{x}}{(1-\zeta^{x})(1-\omega^{-x})}  \biggr].                                          
\end{split}
\end{equation*}

Changing $x$ to $-x$ in the last two terms, we find 

\begin{equation*}
L^{\prime}= -\frac{\pi i}{M} \left[    \sum_{c \ \nmid \ x}  \frac{\lambda^{x}}{(1-\zeta^{x})(1-\omega^{x})} + \sum_{c \ \nmid \ x}  \frac{(\zeta / \lambda)^{x}}{(1-\zeta^{x})(1-\omega^{x})} \right] .
\end{equation*}

This expression can be further simplified. We decompose the sum by introducing the variable

\begin{equation*}
\begin{split}
& x= y + k \vert c \vert, \\ \\
0 < y < \vert c \vert , & \hspace{2cm} 0 \leqslant k \leqslant N-1.
\end{split}
\end{equation*}

Let us denote with $S$ the partial sum in the variable $k$ of the first term in $L^{\prime}$. One gets

\begin{equation*}
\begin{split}
S= & \frac{\lambda^{y}}{1-\zeta^{y}}\sum_{k=0}^{N-1} \frac{\lambda^{k \vert c \vert}}{1-\omega^{y+k \vert c \vert}} \\ \\
 = & -\frac{1}{M}\frac{\lambda^{y}}{1-\zeta^{y}} \sum_{r=0}^{N-1}r\omega^{ry} \sum_{k=0}^{N-1} (\lambda \omega^{r})^{k \vert c \vert}.
\end{split}
\end{equation*}

The sum on the right is $0$ unless $ (\lambda \omega^{r})^{\vert c \vert}=1$. Using the definitions of $\lambda$ and $\omega$ in terms of $u_{1}^{\prime}, u_{2}^{\prime}, d,c$, we see that $(\lambda \omega^{r})^{\vert c \vert}=1$ if and only if 

\begin{equation*}
r=  Nd u_{1}^{\prime}-Ncu_{2}^{\prime} \ \ \mathrm{mod} \ N.
\end{equation*}

Letting consequently $r=N  \langle d u_{1}^{\prime}-cu_{2}^{\prime}\rangle + sN$ with $ 0 \leq s \leq \vert c \vert -1$, we have 
\begin{equation*}
\begin{split}
S & =  -\frac{1}{\vert c \vert} \frac{\lambda^{y}}{1-\zeta^{y}} \sum_{\overset{0 \leq r \leq N-1,}{ r= N u_{1}^{\prime}d-Ncu_{2}^{\prime} \ \mathrm{mod} \ N}}  r\omega^{ry} \\ \\ & =   -\frac{1}{\vert c \vert} \frac{\lambda^{y}}{1-\zeta^{y}} \sum_{s=0}^{\vert c \vert -1}(N  \langle d u_{1}^{\prime}-cu_{2}^{\prime}\rangle + sN)e^{2\pi i \frac{y}{Nc} (N  \langle d u_{1}^{\prime}-cu_{2}^{\prime}\rangle + sN)} \\ \\ & =  -\frac{1}{\vert c \vert} \frac{\lambda^{y}}{1-\zeta^{y}}\  e^{2\pi i \frac{y}{c} \langle d u_{1}^{\prime}-cu_{2}^{\prime}\rangle}  \sum_{s=0}^{\vert c \vert -1} s N e^{2\pi i ys/c}
\end{split}
\end{equation*}

\begin{equation*}
\begin{split}
& = N  \frac{\lambda^{y}}{1-\zeta^{y}} e^{2\pi i \frac{y}{c} \langle d u_{1}^{\prime}-cu_{2}^{\prime}\rangle}  \frac{1}{1-e^{2\pi i y/c}} \\ \\ & = N \frac{e^{2\pi i y \bigl( \frac{ \langle d u_{1}^{\prime}-cu_{2}^{\prime}\rangle-du_{1}^{\prime} }{c}+u_{2}^{\prime} \bigr)}  }{(1-e^{-2\pi i y d/c})(1-e^{2\pi i y/c})}. 
\end{split}
\end{equation*} 

In order to write the final result in a more compact way, we introduce the symbol 
 
 \begin{equation}
 [ x, u_{1}^{\prime}, u_{2}^{\prime} ]_{d,c}=  \frac{e^{2\pi i x \bigl( \frac{ \langle d u_{1}^{\prime}-cu_{2}^{\prime}\rangle-du_{1}^{\prime} }{c}+u_{2}^{\prime} \bigr)}  }{(1-e^{-2\pi i x d/c})(1-e^{2\pi i x/c})}.
 \end{equation}

Noting that the second sum in $L^{\prime} $ can be obtained from the first one with the substitution $ u_{1}^{\prime}\rightarrow 1- u_{1}^{\prime}, u_{2}^{\prime} \rightarrow -u_{2}^{\prime}$, we  get 

\begin{equation*}
L^{\prime}= -\frac{\pi i}{c} \left[  \sum_{\overset{x \in \mathbb{Z}/c\mathbb{Z},}{ x \neq 0 } } [ x, u_{1}^{\prime}, u_{2}^{\prime} ]_{d,c}    + \sum_{\overset{ x \in \mathbb{Z}/c\mathbb{Z},}{ x \neq 0} }  [ x, 1-u_{1}^{\prime}, -u_{2}^{\prime} ]_{d,c} \right] .
\end{equation*}

From the property $[ -x,1- u_{1}^{\prime},- u_{2}^{\prime} ]_{d,c}=[ x, u_{1}^{\prime}, u_{2}^{\prime} ]_{d,c} $, one obtains further 

\begin{equation}
L^{\prime}= -\frac{2 \pi i}{c} \sum_{\overset{x \in \mathbb{Z}/c\mathbb{Z},}{  x \neq 0 } } [ x, u_{1}^{\prime}, u_{2}^{\prime} ]_{d,c}  .
\end{equation}

Putting all the pieces together, we finally have 

\begin{equation}
\begin{split}
 \chi_{u_{1},u_{2}}(\gamma)= & \ 2\pi i \frac{1}{2} \left( B_{2}(u_{1}) \frac{a}{c} +   B_{2}(u_{1}^{\prime}) \frac{d}{c}  -\frac{2}{c} B_{1}(u_{1}^{\prime}) B_{1}( \langle d u_{1}^{\prime}- u_{2}^{\prime} c \rangle ) \right)  \\ \\ & -\frac{2 \pi i}{c}  \sum_{\overset{x \in \mathbb{Z}/c\mathbb{Z},}{ \ x \neq 0 }}  [ x, u_{1}^{\prime}, u_{2}^{\prime} ]_{d,c}.
 \end{split}
\end{equation}

\end{document}